\documentclass{osa-article}
\journal{osajournal}
\articletype{Research Article}
\usepackage{soul}

\newcommand{\unitskappa}{${\rm pN \, \upmu m^{-1}}$}
\newcommand{\unitsgamma}{${\rm pN \, ms \, \upmu m^{-1}}$}
\newcommand{\unitsD}{\text{${\rm \upmu m^2 s^{-1}}$}}

\begin{document}

\title{Optical Tweezers: A Comprehensive Tutorial from Calibration to Applications}

\author{
Jan Gieseler,\authormark{1,2} 
Juan Ruben Gomez-Solano,\authormark{3} 
Alessandro Magazz\`u,\authormark{4} 
Isaac P\'erez Castillo,\authormark{3,5} 
Laura P\'erez Garc\'ia,\authormark{4}
Marta Gironella-Torrent,\authormark{6} 
Xavier Viader-Godoy,\authormark{6} 
Felix Ritort,\authormark{6} 
Giuseppe Pesce,\authormark{7}
Alejandro V. Arzola,\authormark{3} 
Karen Volke-Sepulveda,\authormark{3} 
and Giovanni Volpe\authormark{4,*}
}

\address{
\authormark{1}  ICFO - Institut de Ci\`encies Fot\`oniques, The Barcelona Institute of Science and Technology, 08860 Castelldefels (Barcelona), Spain\\
\authormark{2} Harvard University, Department of Physics, 17 Oxford St., Cambridge, MA, USA\\
\authormark{3} Instituto de Física, Universidad Nacional Autónoma de México, Apdo. Postal 20-364, 01000Cd. México, Mexico.\\
\authormark{4} Department of Physics, University of Gothenburg, 41296 Gothenburg, Sweden.\\
\authormark{5} London Mathematical Laboratory, 18 Margravine Gardens, London W6 8RH, United Kingdom\\
\authormark{6} Small Biosystems Lab, Departament de Fsica de la Mat\`eria Condensada, Facultat de F\'isica,
Universitat de Barcelona, 08028 Barcelona, Spain\\
\authormark{7} Department of Physics, Univerisit\`a degli di Studi di Napoli ``Federico II'' Complesso universitario monte S. Angelo, Via Cintia, 80126 Napoli, Italy
}

\email{\authormark{*}giovanni.volpe@physics.gu.se}

\begin{abstract}
Since their invention in 1986 by Arthur Ashkin and colleagues, optical tweezers have become an essential tool in several fields of physics, spectroscopy, biology, nanotechnology, and thermodynamics. 
In this Tutorial, we provide a primer on how to calibrate optical tweezers and how to use them for advanced applications. 
After a brief general introduction on optical tweezers, we focus on describing and comparing the various available calibration techniques. 
Then, we discuss some cutting-edge applications of optical tweezers in a liquid medium,  namely to study single-molecule and single-cell mechanics, microrheology, colloidal interactions, statistical physics, and transport phenomena. 
Finally, we consider optical tweezers in vacuum, where the absence of a viscous medium offers vastly different dynamics and presents new challenges.
We conclude with some perspectives for the field and the future application of optical tweezers. 
This Tutorial provides both a step-by-step guide ideal for non-specialists entering the field and a comprehensive manual of advanced techniques useful for expert practitioners. 
All the examples are complemented by the sample data and software necessary to reproduce them.
\end{abstract}

\tableofcontents

\section{Introduction}\label{sec:1:intro}

In 1970, Arthur Ashkin \cite{ashkin1970acceleration} demonstrated that a focused laser beam can accelerate, decelerate, and even stably trap micrometer-sized neutral particles. 
Later, in 1986, Ashkin and colleagues reported the first realization of an {\em optical tweezers} \cite{ashkin1986observation, ashkin2006optical}: a tightly focused beam of light capable of holding microscopic particles in three dimensions. 
Thanks to this work, Ashkin was awarded a share of the 2018 Nobel Prize in Physics \cite{ashkin2019nobel}. 
Furthermore, one of Ashkin's co-authors, Steven Chu, would go on to use optical tweezing in his work on cooling and trapping atoms; this research earned Chu a share of the 1997 Nobel Prize in Physics \cite{chu1998nobel}.

In the late 1980s, Ashkin and colleagues used optical tweezers to manipulate biological samples, starting from an individual tobacco mosaic virus and {\it Escherichia coli} bacterium \cite{ashkin1990force}. 
In 1993, Lucien Ghislain and Watt Webb invented photonic force microscopy \cite{ghislain1993scanning}, where an optically trapped particle is used as an extremely sensitive cantilever to probe microscopic force fields ranging from femtonewtons ($10^{-15}\,{\rm N}$) to piconewtons ($10^{-12}\,{\rm N}$). 
Since the early 1990s, optical force spectroscopy has been utilized to characterize the mechanical properties of biomolecules and biological motors \cite{block1990bead, bustamante1994entropic, finer1994single}.
Optical tweezers have also been employed  in many fields of physics \cite{grier2003revolution, dholakia2010colloquium, dholakia2011shaping, juan2011plasmon, padgett2011tweezers, padgett2011holographic}, nanotechnology \cite{marago2013optical}, spectroscopy \cite{petrov2007raman}, nanothermodynamics \cite{martinez2017colloidal, gieseler2018levitated}, soft matter \cite{robertson2018optical}, and biology \cite{fazal2011optical}.

\begin{figure}[b!]
	\centering
	\includegraphics[width=12cm]{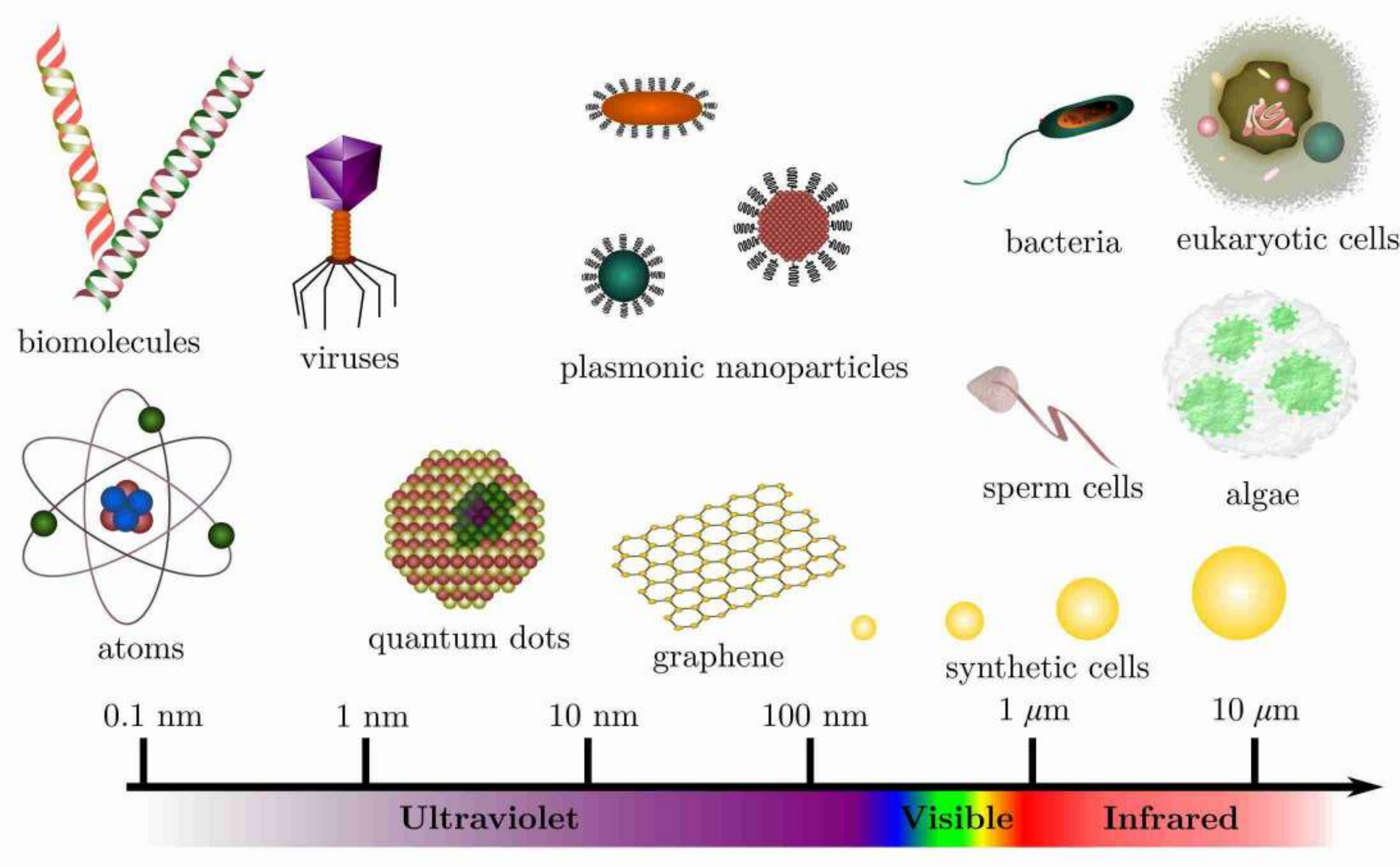}
	\caption{
	{\bf Trappable objects.}
	A broad range of objects have been trapped using optical tweezers, going from single atoms and molecules to microparticles and microorganisms.
	}
	\label{fig:1:trappableobjects}
\end{figure}

An optical tweezers consist of a laser beam focused by a high numerical aperture (NA) microscope objective. 
A microscopic particle (whose refractive index is higher than that of its embedding medium) can then be trapped near the focal spot because of the emergence of scattering and gradient optical forces: the {\em scattering forces} are due to the radiation pressure of the light beam and act along the direction of propagation of the beam; and the {\em gradient forces} pull the particle towards the high-intensity focal spot.
Using optical tweezers, a very broad range of particles has been trapped, as shown in Fig.~\ref{fig:1:trappableobjects}.

There are already several resources to build and operate optical tweezers available in the literature.
In the 2015 book {\em Optical Tweezers: Principles and Applications} \cite{jones2015optical}, Phillip. H. Jones, Onofrio M. Marag\'{o} and Giovanni Volpe have discussed in detail the theory behind optical tweezers as well as how to build and operate optical tweezers setups.
There are also several other tutorials on specific aspects of optical trapping.
In particular, there are computational toolboxes to calculate optical forces using T-matrix methods by Timo Nieminen {\em et al.} \cite{nieminen2007optical} and geometrical optics by Agnese Callegari {\em et al.} \cite{callegari2015computational}.
There is a tutorial on how to simulate the Brownian motion of an optically trapped particle \cite{volpe2013simulation}.
There are several tutorials on how to build optical tweezers setups by Stephen Smith et al. \cite{smith1999inexpensive}, John Bechhoefer and Scott Wilson \cite{bechhoefer2002faster}, Angela Mellish and Andrew Wilson \cite{mellish2002simple}, David Appleyard {\em et al.} \cite{appleyard2007optical}, Woei Lee {\em et al.} \cite{lee2007construction}, Manoj Mathew {\em et al.} \cite{mathew2009multimodal}, and Giuseppe Pesce {\em et al.} \cite{pesce2015step}.

In this Tutorial, we complete and complement these already available resources providing a primer on how to calibrate optical tweezers and how to use them for advanced applications.
In particular, after a brief general introduction on optical tweezers (section~\ref{sec:2:start}), we focus on describing and comparing the various calibration techniques that are available (section~\ref{sec:3:calibration}).
Then, we discuss some of the most exciting cutting-edge applications of optical tweezers in a liquid medium (section~\ref{sec:4:applications}), namely to study single-molecule and single-cell mechanics, microrheology, colloidal interactions, statistical physics, and transport phenomena.
Finally, we consider optical tweezers in vacuum, which features different dynamics than a viscous medium, so that vacuum operation comes with its own experimental challenges and calibration methods (section~\ref{sec:5:vacuum}).
This Tutorial provides both a step-by-step guide ideal for non-specialists entering the field and a comprehensive manual of advanced techniques useful for expert practitioners.
All the examples are complemented by the sample data and the software necessary to reproduce them \cite{SI}.

\section{Getting started}\label{sec:2:start}

In this section we provide some essential background information on optical tweezers. 
First, we will describe a basic optical tweezers setup (section~\ref{sec:2.1:setup}). 
Then, we will briefly discuss the Brownian motion of an optically trapped particle and the most basic optical tweezers calibration techniques (section~\ref{sec:2.2:otc}).
Finally, we will give a brief overview of more advanced optical tweezers and of alternative approaches for the trapping and manipulation of microscopic matter (section~\ref{sec:2.3:alternatives}).

\subsection{A simple setup}\label{sec:2.1:setup}

\begin{figure}[t!]
	\centering
	\includegraphics[width=6cm]{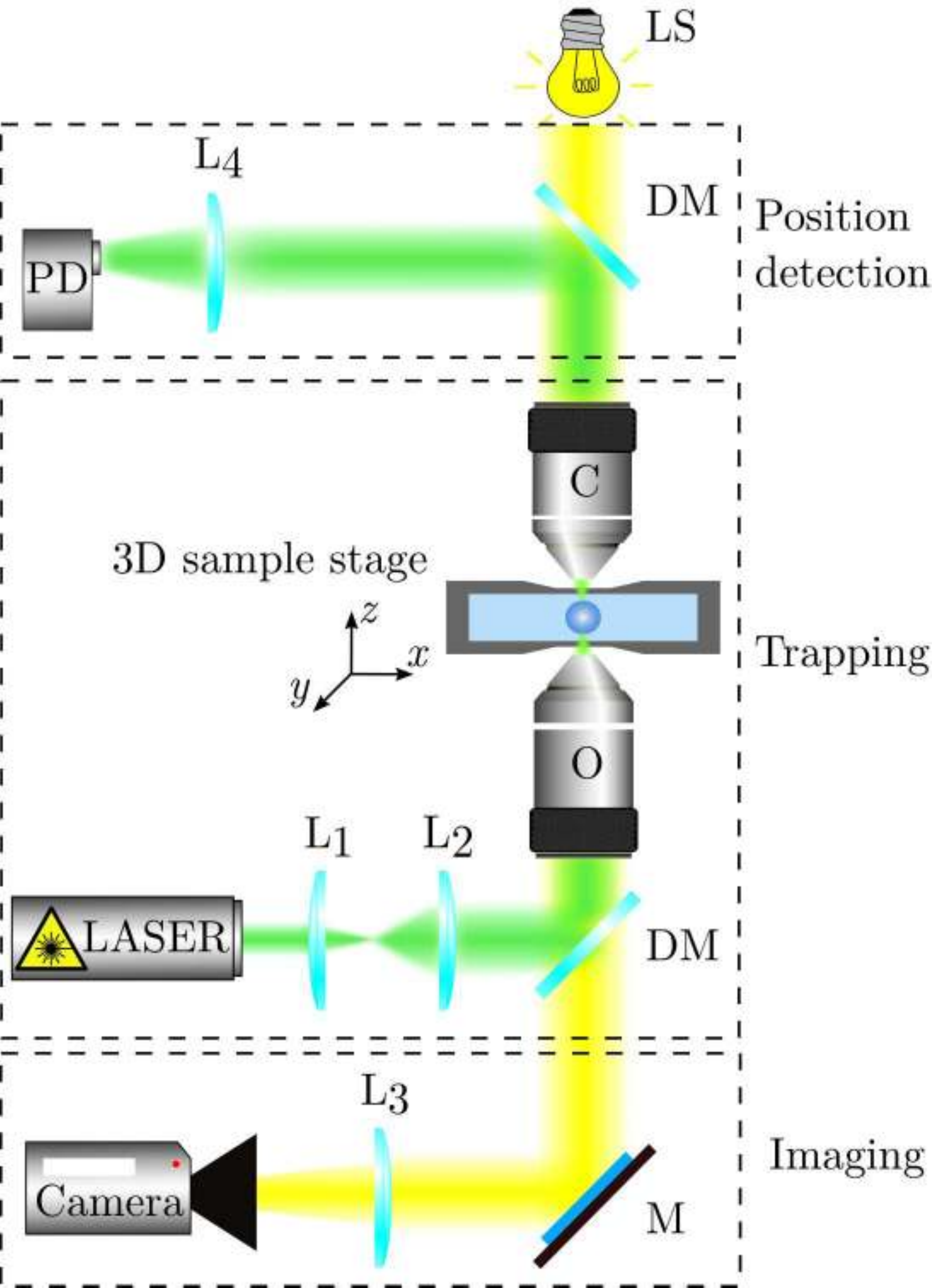}
	\caption{
	{\bf Basic optical tweezers setup.}
	A basic optical tweezers setup is comprised of three parts: the trapping optics, the imaging optics, and the position detection optics.
	DM: Dichroic mirror. M: mirror. O: microscope objective. C: Condenser. L$_1$, L$_2$, L$_3$, L$_4$: lenses. LS: Light source.
	}
\label{fig:2:simplesetup}
\end{figure}

An optical tweezers is a highly-focused laser beam that can trap microscopic particles. 
In practice, an optical tweezers can be built by focusing a laser beam through a microscope, which can be either commercial or homemade.
As illustrated in Fig.~\ref{fig:2:simplesetup}, a basic optical tweezers setup is comprised of three parts: trapping, imaging, and position detection.

\paragraph{Trapping.}

In most applications, the laser is continuous wave (CW), whose wavelength is in the  visible or near-infrared region of the spectrum, and whose power is between $10\,{\rm mW}$ and $1\,{\rm W}$.
The laser beam is expanded by a telescope  to overfill the back-aperture of a high-NA microscope objective, is directed by a dichroic mirror towards the objective, and finally is focused by the objective to a diffraction-limited spot within the sample holder. This focal spot is the optical tweezers, which can trap a microscopic particle. 
Usually, the object to be trapped is suspended in an aqueous solution, which is held between a microscope slide and a cover slip; sometimes the particles to be trapped are suspended in other liquids, or in gases, or even in vacuum (section~\ref{sec:5:vacuum}).

\paragraph{Imaging.} 

Typically, the trapping and imaging objectives are the same. The light for the imaging is provided by a condenser, which is an integral part of any  microscope. It is convenient to image through a dichroic beam splitter, which reflects the trapping light to the microscope, but also transmits some of the imaging light from the condenser to a camera. In both cases, some color filters are typically required in front of the camera to reduce the intensity of the trapping beam to an acceptable level such that the camera is not overexposed.

\paragraph{Position detection.} 

To use the optically trapped particle as a microscopic force transducer, it is necessary to track its position. This can be done by analyzing either the videos of the particle acquired by the camera (digital video microscopy), or the time series acquired through a position detector.
In this latter case, the condenser collects the forward scattered light and projects it onto a position detector (typically either a quadrant photodetector or a position sensing detector). This detects the changes in the forward scattered light pattern due to the changes in the particle's position.

\subsection{Optical tweezers calibration}\label{sec:2.2:otc}

An optically trapped particle can be used as an extremely sensitive microscopic force transducer that can both exert and measure forces ranging from hundreds of piconewtons down to a few femtonewtons.

An optically trapped particle experiences a harmonic force that tends to keep it near a stable equilibrium position. Considering only one dimension, for small displacements from the equilibrium position, the force acting on a colloidal particle positioned at $x$ within a trap centered at $x_{\rm eq}$ is 
\begin{equation}
	F_{\rm ot} 
	= 
	- \kappa_x (x - x_{\rm eq}), 
\end{equation}
where $\kappa_x$ is the trap stiffness along the $x$-direction.
The resulting trapping potential is harmonic: 
\begin{equation}
	U(x) 
	= 
	{1\over2} \kappa_x (x - x_{\rm eq})^2 .
\end{equation}

\begin{figure}[t!]
	\centering
	\includegraphics[width=12cm]{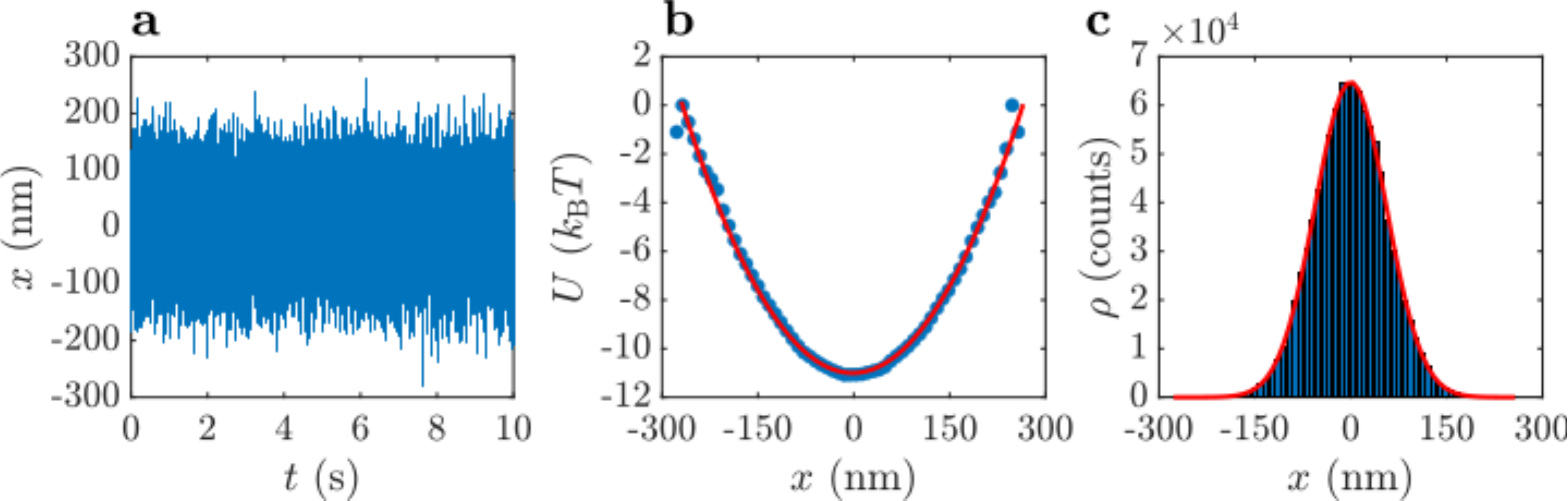}
	\caption{
	{\bf Brownian motion in an optical trap.}
	(a) Trajectory of an optically trapped Brownian particle $x(t)$ and corresponding (b) optical potential and (c) probability density. The equilibrium position is $x_{\rm eq} = 0\,{\rm \upmu m}$.  The experimental parameters are in Tab.~\ref{tab:1:expparam}.
	}
	\label{fig:3:brownian}
\end{figure}

A typical trajectory of an optically trapped particle can be seen in Fig.~\ref{fig:3:brownian}(a).
The particle keeps on moving due to the presence of Brownian fluctuations. Therefore, an optically trapped particle is in a dynamic equilibrium between the thermal noise continuously pushing it out of the trap and the optical forces driving it towards the equilibrium position. For the particle to remain within the optical trap, the optical potential well must be sufficiently deep. The depth of the optical potential is typically characterized in units of the thermal energy $k_{\rm B}T$, where $k_{\rm B}$ is the Boltzmann constant and $T$ is the absolute temperature.
This quantity gives a characteristic energy scale for mesoscopic phenomena. The potential well of an optical tweezers should be at least a few $k_{\rm B}T$ deep to be able to confine a particle, as shown in Fig.~\ref{fig:3:brownian}(b).

The analysis of the thermal motion of the optically trapped particle in the optical potential provides information about the local forces acting on the particle. 
For example, since the particle is held in a harmonic trapping potential, its position distribution is Gaussian, as shown in Fig.~\ref{fig:3:brownian}(c),
\begin{equation}
	\rho(x) 
	= 
	\rho_0 
	\exp 
	\left[ 
		- { \kappa_x (x - x_{\rm eq})^2 \over 2 k_{\rm B} T} 
	\right].
\end{equation}
By sampling this probability distribution, it is possible to measure $\kappa_x$ using the equipartition theorem. The equipartition theorem states that for a system at thermodynamic equilibrium at absolute temperature $T$, the energy associated with each harmonic degree of freedom is equal to ${1\over2}k_{\rm B}T$. Therefore, 
$
\left< U(x) \right> 
=
{1\over2} \kappa_x \left< (x - x_{\rm eq})^2 \right>
=
{1\over2} \kappa_x \sigma_x^2
=
{1\over2}k_{\rm B}T$, where $\sigma_x^2 = \left< (x - x_{\rm eq})^2 \right>$ is the variance of the trapped particle position,
and 
\begin{equation}
	\kappa_x 
	= 
	{k_{\rm B}T \over \sigma_x^2}.
\end{equation}
As we will see in section~\ref{sec:3:calibration}, there are more sophisticated methods to use the thermal fluctuations of an optically trapped particle  to calibrate its stiffness. In particular, it is possible to also exploit the temporal evolution of the particle in the optical trap, for example by calculating its correlation and power spectral density. 

Once an optical tweezers is calibrated, a constant and homogeneous external force $F_{{\rm ext}, x}$  shifts the equilibrium position of the trap. The value of the force can be obtained as:
\begin{equation}
	F_{{\rm ext}, x} 
	= 
	\kappa_x \Delta x_{\rm eq},
\end{equation}
where $\Delta x_{\rm eq}$ is the average particle displacement from the original equilibrium position without the external force.
       
\subsection{Advanced and alternative approaches}\label{sec:2.3:alternatives}

Going beyond the basic setup introduced in the previous section, more advanced setups become necessary to extend the optical manipulation toolkit, introducing, e.g., optical spanners \cite{simpson1997mechanical}, optical rheometers \cite{furst2005applications}, optical stretchers \cite{guck2001optical}, and optical sorters \cite{macdonald2003microfluidic, cizmar2006optical_sorting, ricardez2006modulated}. 
These tools allow one to manipulate multiple particles at once\cite{visscher1996construction, dufresne1998optical, leach20043d, grier2006holographic}, rotate particles \cite{he1995direct, volke2002orbital, arzola2014rotation, brzobohaty2015complex}, measure the mechanical properties of biological samples \cite{guck2005optical}, sort particles \cite{jonavs2008light}, trap particles with refractive index lower than their medium \cite{gahagan1996optical, oneil2000three, garces2002transfer, hernandez2013attractive}, induce optical binding and collective dynamics \cite{burns1989optical, burns1990optical, mohanty2004optical, karasek2008long, dholakia2010colloquium, demergis2012ultrastrong, schmidt2019light}. 
Complex optical potentials, such as periodic and random patterns, provide experimental models to study microscopic and non-linear transport mechanisms \cite{faucheux1995optical, koss2003optical, lee2005observation, arzola2011experimental, volpe2012active, volpe2014brownian, arzola2017omnidirectional, arzola2019spin}. 
Advanced optical manipulation has been employed also to build and study micromachines \cite{galajda2001complex, friese2001optically, maruo2003force} and optofluidic devices \cite{macdonald2003microfluidic, enger2004optical, neale2005all, wang2005microfluidic, leach2006optically, eriksson2007microfluidic, wang2011enhanced}. 
Furthermore, optical manipulation setups have also been successfully employed to investigate the properties of light itself, employing structured light beams with peculiar intensity, phase and polarization distributions \cite{he1995direct, simpson1997mechanical, friese1998optical, oneil2002intrinsic, lopez2006orbital,  nieminen2008forces, baumgartl2008optically}.

In this section, we will provide a brief overview of the main advanced optical manipulation techniques. We will consider only the overdamped regime (i.e., particles immerse in a viscous medium like water), since the case of underdamped systems involves many different experimental aspects that will be treated in (section~\ref{sec:5:vacuum}).
Also, unless otherwise stated, we will assume that the refractive index of the particles under consideration is higher than that of the surrounding medium.

\subsubsection{Beam steering}\label{sec:2.3.1:beamsteering}

Most advanced optical micromanipulation setups require to steer the trapping beam to control the position of the optical tweezers.
The beam steering device can be as simple as a mirror on a gimbal mount (preferred over conventional kinematic mounts to avoid mirror shifting). For example, this simple approach can be used to move the optical trap within the sample instead of moving the sample with a translation stage while keeping the optical trap fixed.
More sophisticated beam-steering devices are also often employed, such as acousto-optic deflectors and spatial light modulators (SLMs).

\begin{figure}[ht!]
	\centering
	\includegraphics[width=12cm]{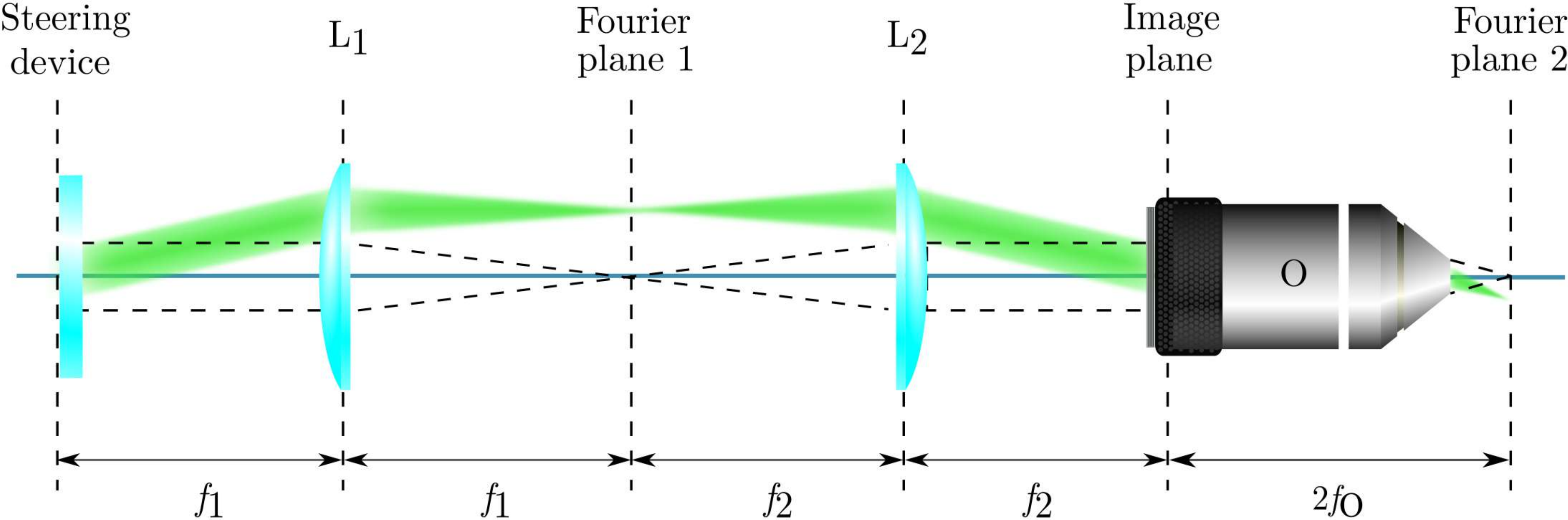}
	\caption{
	{\bf Beam steering mechanism.}
	The trapping beam is steered by the steering device (e.g., a galvo-mirror, an acousto-optic deflector, or a spatial light modulator). To ensure that the trapping beam reaches the back aperture of the microscope objective (O) while being steered, a telescope is added along the beam path (lenses L$_1$ and L$_2$) in a 4f-configuration ($f_1$ and $f_2$ are the focal lengths of L$_1$ and L$_2$ respectively).
	}
\label{fig:4:beamsteering}
\end{figure}

When implementing beam steering, it is essential to ensure that the available light reaches the back aperture of the microscope objective, irrespective of the beam direction. 
To this end, a telescope should be added (4f-system)  to create conjugate planes between the beam steering device and the back aperture of the objective, as shown in Fig.~\ref{fig:4:beamsteering}. 
This telescope can also be used to magnify the beam to slightly overfill the back aperture of the objective (to improve trapping), and a spatial filter can be added at Fourier plane 1 (to remove unwanted higher-order diffraction patterns), which is conjugated with Fourier plane 2.

\subsubsection{Polarization splitting}\label{sec:2.3.2:PS}

\begin{figure}[ht!]
	\centering
	\includegraphics[width=6cm]{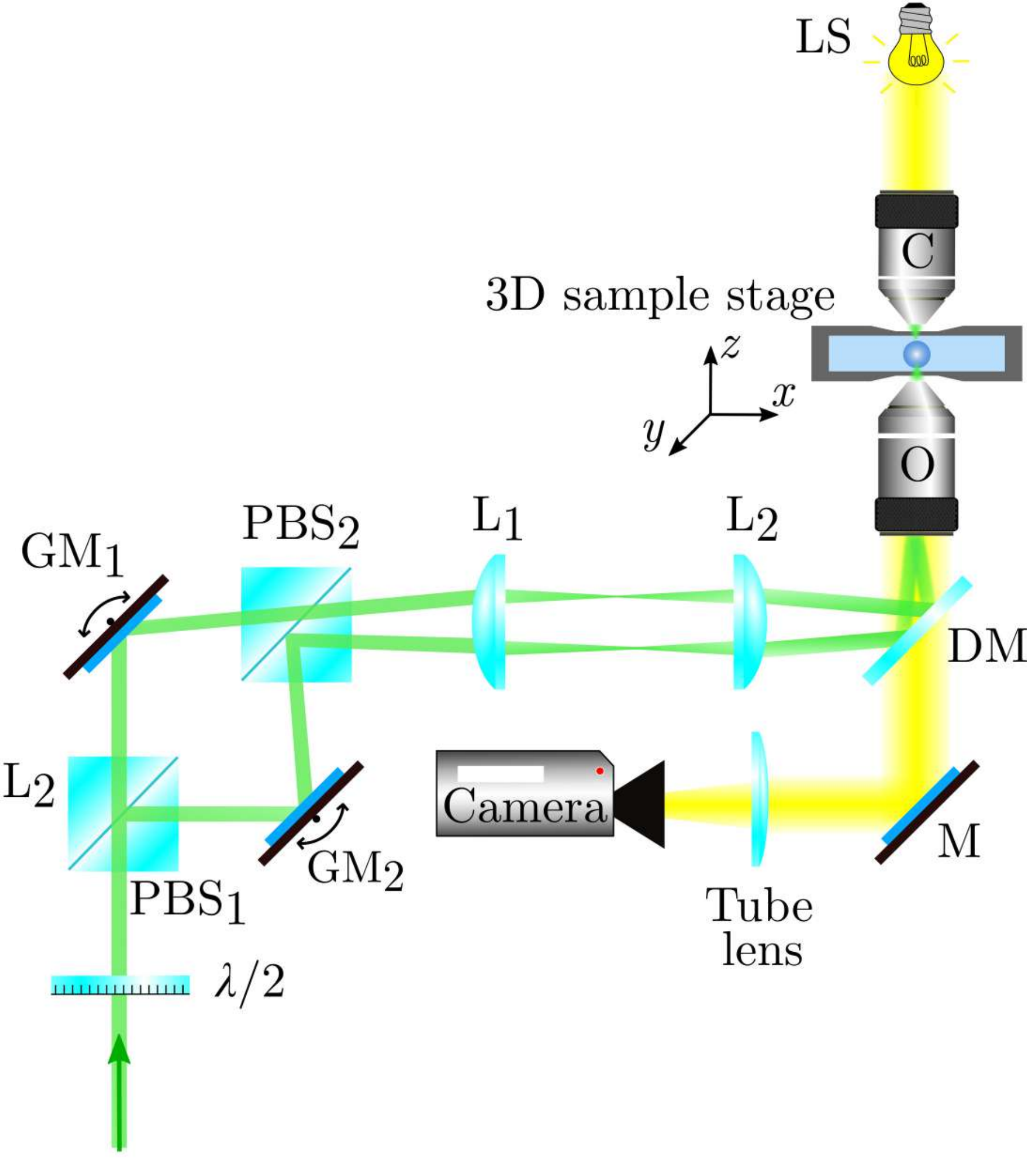}
	\caption{
	{\bf Polarization-splitting optical tweezers.}
	A typical Mach-Zehnder setup with polarizing beam splitters (PBS$_1$ and PBS$_2$) can be used for this purpose. Mirrors
GM$_1$ and GM$_2$ should be mounted on gimbal mounts, and the distance between each of them and the lens L$_1$ should be equal to the focal length of L$_1$ (Fig.~\ref{fig:4:beamsteering}). 
	For alignment purposes, both traps should be initially superimposed, arriving exactly at the same point in the sample plane. 
	The intensity ratio between the traps can also be controlled, by placing a polarizer followed by a $\lambda/2$-retarder at the entrance of the Mach-Zehnder setup. As the retarder is rotated, the proportion of light reflected and transmitted through PBS$_1$ will change. 
	DM: Dichroic mirror. M: mirror. O: microscope objective. C: Condenser.
	}
	\label{fig:5:polarizationsplitting}
\end{figure}

The simplest alternative to have two neighboring traps while avoiding the interference between them is to split the original beam into two orthogonally polarized beams \cite{visscher1996construction,fallman1997design}, as shown in Fig.~\ref{fig:5:polarizationsplitting}. 
If the main control parameter of a given experiment is the distance between the traps, a single steering device can be enough, keeping one trap fixed while the other one is shifted. This works fine for a small distance between the traps. 
If the traps need to be moved over a large area, it might be useful to  use two steering mirrors to minimize the distortion of each trap. 
Dynamical control of each trap can be achieved if the gimbal mounts are operated by actuators along the two axes, giving rise to a continuous motion along arbitrary trajectories in two dimensions in the sample plane. 
Finally, this twin-trap approach can also be used in combination with other advanced techniques, if the aim is to superimpose two complex patterns without interference between them.

\subsubsection{Time-sharing optical traps}

\begin{figure}[ht!]
	\centering
	\includegraphics[width=6cm]{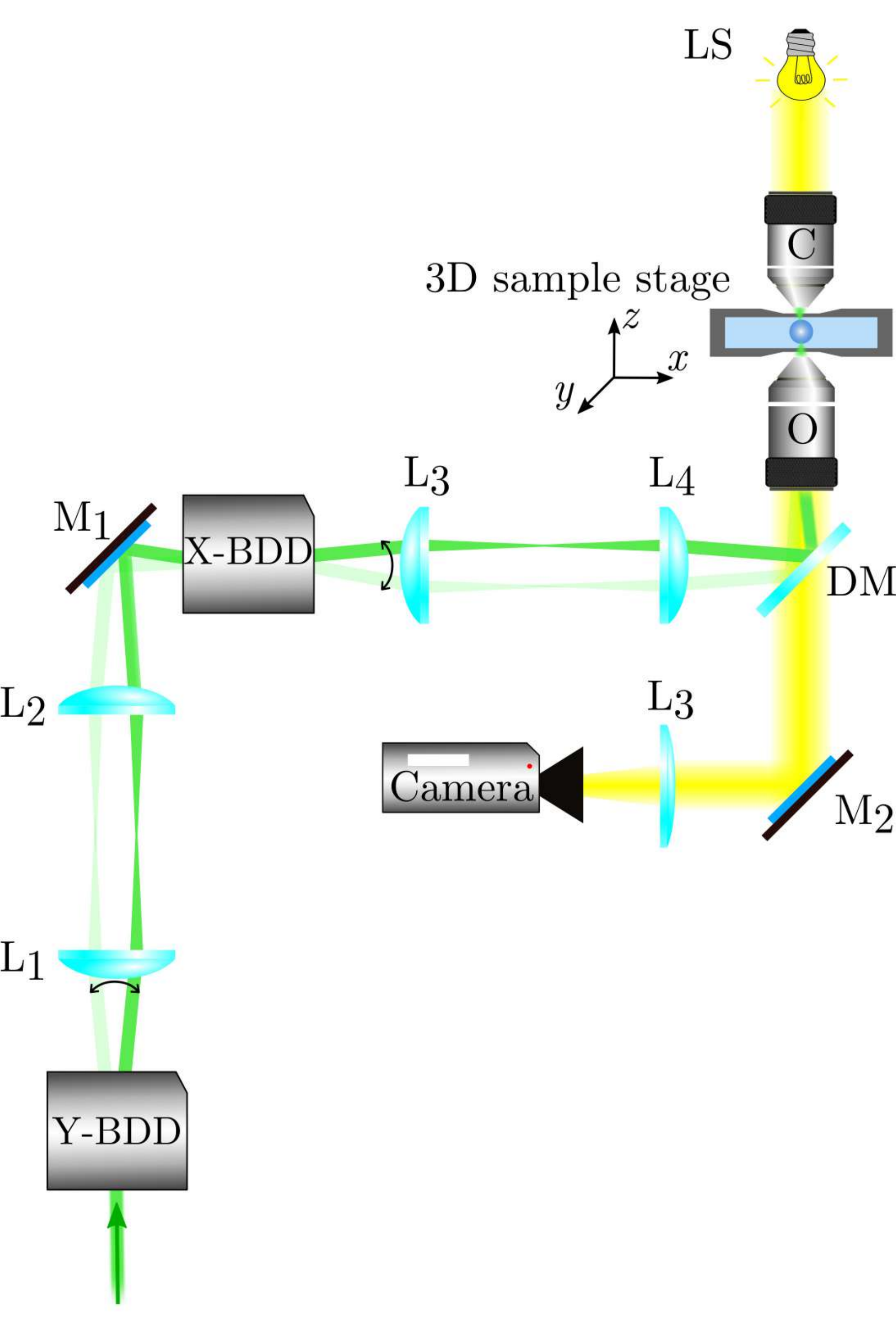}
	\caption{
	{\bf Time-sharing optical tweezers.}
	The trapping beam is time-shared between several optical traps using two beam deflector devices along the $x$- and $y$-directions (X-BDD and Y-BDD, respectively).
	Two telescopes are needed (L$_1$-L$_2$ and L$_3$-L$_4$) to create conjugate planes between the two deflector devices as well as with the back aperture of the microscope objective O. 
	}
	\label{fig:6:timesharing}
\end{figure}

A single beam can be used to generate multiple optical tweezers if it is scanned sufficiently fast, i.e., with a frequency high enough to ensure that an optically trapped particle cannot diffuse away from the optical trap region between consecutive visits of the laser beam.
This technique is known as time-sharing or laser-scanning optical trap. 
It can be performed with different deflecting devices, such as acousto-optic deflectors (AODs), mirror galvanometers (galvo-mirrors), electro-optic deflectors (EODs), and piezo-mirrors  \cite{sasaki1991pattern, visscher1996construction, mio2000design, arai2004synchronized, zaidouny2014periodic}. 
Often, the beam deflecting devices have only one degree of freedom so that, to achieve beam motion over the whole sample plane, it is necessary to use two beam deflecting devices, one for each transverse direction, as illustrated in Fig.~\ref{fig:6:timesharing}.

The angular range and the scanning frequency of the beam deflecting device determines the number of traps that can be created simultaneously and the extent of the trapping area. There is usually a compromise between the scanning speed and the deflection angular range.
The first setup of time-sharing optical tweezers was built with galvo-mirrors \cite{sasaki1991pattern}, which could reach scanning frequencies of the order of $10\,{\rm Hz}$. Nowadays, there are similar commercial systems that can reach up to $100\,{\rm Hz}$, with a angular range of up to $25^\circ$; however, the angular range in the optical trapping setup is limited by the maximum acceptance angle of the microscope objective, which is just a few degrees at most. 
AODs can attain frequencies from tens to hundreds of kilohertz with angular ranges of up to $3^\circ$, which makes them very suitable for trapping applications. For example, these devices have been used in studies of DNA stretching and molecular motors \cite{visscher1996construction}. However, as the AOD working principle is based on the generation of diffraction gratings by means of the propagation of acoustic waves in a glass or quartz cell, resulting in a periodic refractive index modulation, it is worth to keep in mind that the diffraction efficiency can be at most around 70$\%$. 
EODs are even faster than AODs, but with smaller angular ranges of fractions of a degree.
Piezo-mirrors are more limited, reaching frequencies of hundreds of Hertz with angular ranges of only a few hundredths of a degree.

If the position of a given trap is slightly shifted from one visit to the next one, the bead will follow it, making it possible to describe continuous trajectories within the plane. The average intensity in each trap can also be controlled by means of the number of visits of the scanning beam in a full cycle. For example, if a trapping site is visited twice as much as another one, it will experience twice the average intensity.

\subsubsection{Interferometric optical traps}

\begin{figure}[b!]
	\centering
	\includegraphics[width=12cm]{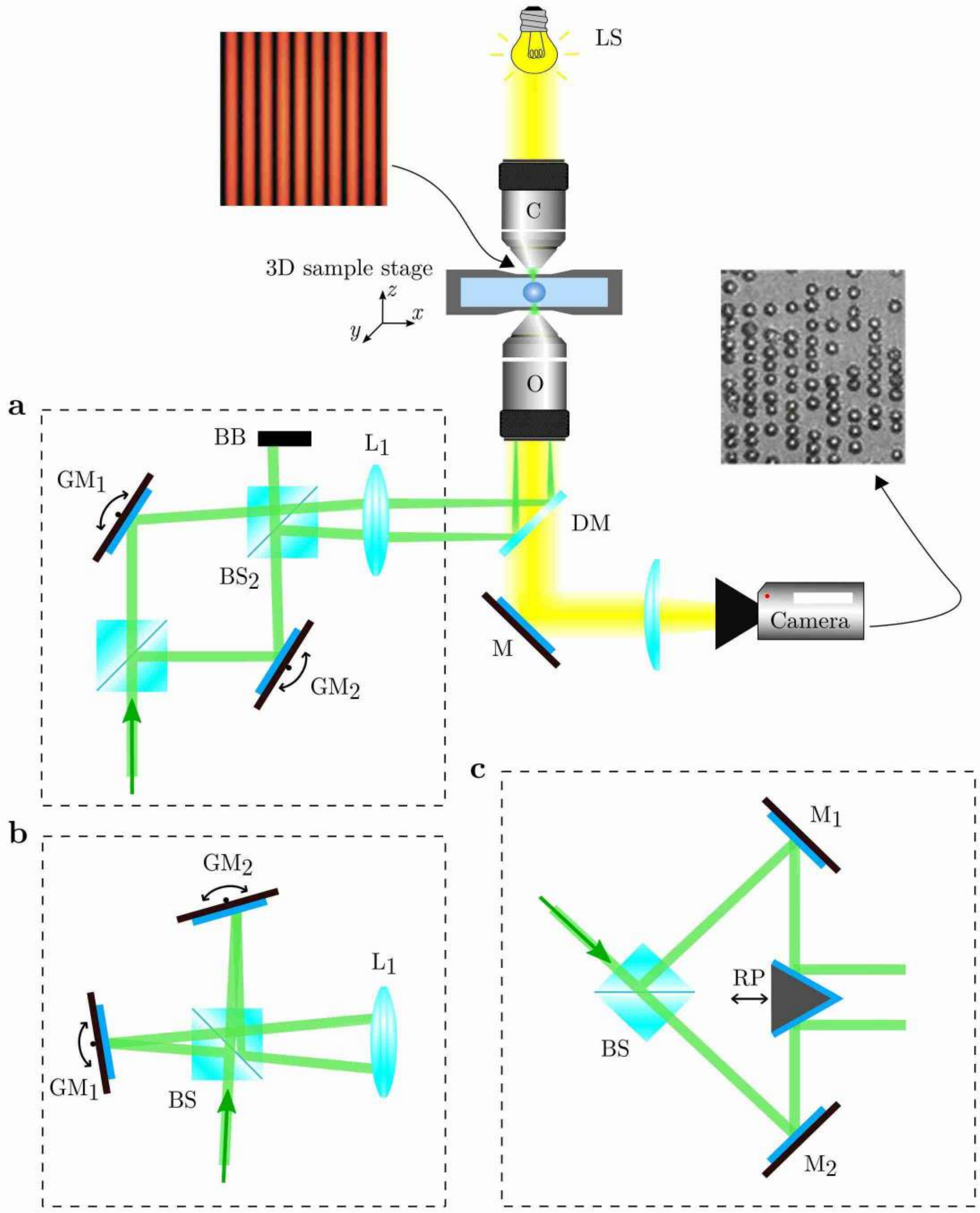}
	\caption{
	{\bf Interferometric optical traps.}
	(a) Interferometric optical trap based on a Mach-Zehnder interferometer.
	(b-c) Different alternatives for the interferometric module. 
	BS: Nonpolarizing beam splitter. RP: Right-angle prism. BB: beam block.
	}
	\label{fig:7:interferometric}
\end{figure}

Interference patterns can trap multiple particles simultaneously and generate ordered structures \cite{burns1989optical, burns1990optical, chiou1997interferometric}.
For example, consider the simplest interference pattern consisting of alternating bright and dark fringes. 
The dynamics of the particles depend on their size with respect to the characteristic size of the fringes.
Particles smaller than the fringes are attracted towards the bright fringes, while remaining free to move along the fringes.
However, much more interesting dynamics emerge when the particle size is comparable or even larger than the fringes \cite{zemanek2004optical, ricardez2006modulated, jonavs2008light}.
This has been called the \textit{size effect} and will be explained in more detail in section~\ref{sec:4.6:transport}.

To generate alternating bright and dark fringes, we need two collimated beams travelling with a relative angle: the larger this angle, the smaller the fringe period (the smallest fringe pattern is half wavelength). 
One possibility is to use a Mach-Zehnder interferometer using non-polarizing beam splitters (Fig.~\ref{fig:7:interferometric}(a)).
A $4f$-system is formed by the lens L$_1$ and the microscope objective that conjugates the mirror on the gimbal mount and the focal plane of the objective (the objective can be replaced by an aspherical lens, since the high-NA requirement may be relaxed in this case). 
While an extended interference pattern generates strong intensity gradients in the transverse plane, the intensity gradient along the propagation direction is usually not enough to trap in 3D so that the particles are pushed by the scattering force against either the bottom or the top surface of the sample cell.\footnote{There are some configurations that generate interference with high-NA objectives, achieving 3D trapping of multiple particles \cite{macdonald2002creation}.}
Fig.~\ref{fig:7:interferometric}(b) shows an alternative configuration for the interferometer module. 
In both configurations (Figs.~\ref{fig:7:interferometric}(a) and \ref{fig:7:interferometric}(b)) a single mirror mounted on a gimbal mount is enough to control the fringe period for small interference angles. 

Alternative configurations are possible, an example of which is illustrated in Fig.~\ref{fig:7:interferometric}(c).
The two beams are redirected by mirrors M$_1$ and M$_2$ towards a right-angle prism, which reflects them towards the objective, which in turn focuses them on the sample. The right-angle prism is mounted on a single-axis translation stage to control the distance between the reflected beams and, therefore, the interference angle. 
A low NA objective is advisable for this configuration to minimize the changes in the intensity distribution along the $z$-axis within the sample depth. 

An extreme case of interference pattern is a speckle light field, which arises from the interference of a multitude of coherent waves with random phases, generated when laser light is, e.g., reflected from a rough surface or transmitted through a turbid medium. Speckle patterns have also been used for optical manipulation \cite{shvedov2010selective, volpe2014speckle, volpe2014brownian}.

\subsubsection{Holographic optical tweezers}\label{sec:2.3.5:hot}

\begin{figure}[bt!]
	\centering
	\includegraphics[width=12cm]{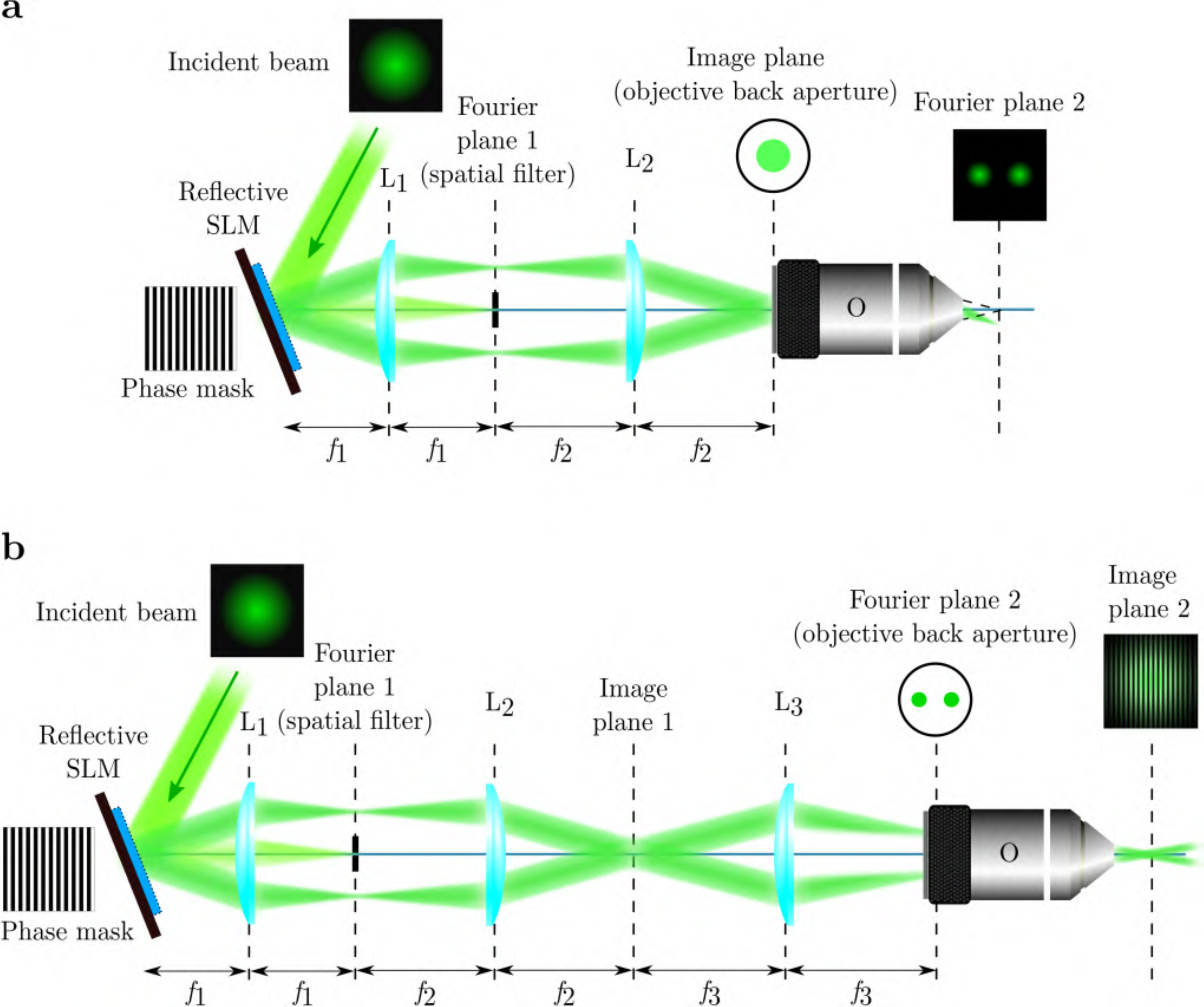}
	\caption{
	{\bf Holographic optical tweezers.}
	A spatial light modulator (SLM) alters the beam profile of an incoming optical beam This beam is then relayed by a series of lenses to an objective that then generates the desired light pattern at the sample. The SLM can be either placed (a) in a plane conjugated to the back focal plane of the objective, or (b) in a plane conjugated with the image plane of the objective.
	}
	\label{fig:8:hot}
\end{figure}

The most versatile optical tweezers setups are arguably based on holographic optical tweezers (HOTs) \cite{dufresne1998optical, tiziani2000multi, curtis2002dynamic, leach20043d, grier2006holographic, jesacher2008full}. 
The basic working principle of HOTs is shown in Fig.~\ref{fig:8:hot}.
The wavefront of an incident beam is reshaped by a hologram projected on a spatial light modulator (SLM) and then relayed by a series of lenses to an objective that then generates the desired light pattern at the sample.
In practice, there are many possible different HOT configurations. Discussing their details is out of the scope of this Tutorial, but a comprehensive review of the topic (including hologram computing algorithms, a numerical toolbox, and a comparison in terms of the main parameters such as efficiency, average intensity, uniformity, and percentage standard error) can be found in Ref.~\cite{jones2015optical}. Here, we will focus our attention on only two approaches.

\begin{figure}[bt!]
	\centering
	\includegraphics[width=12cm]{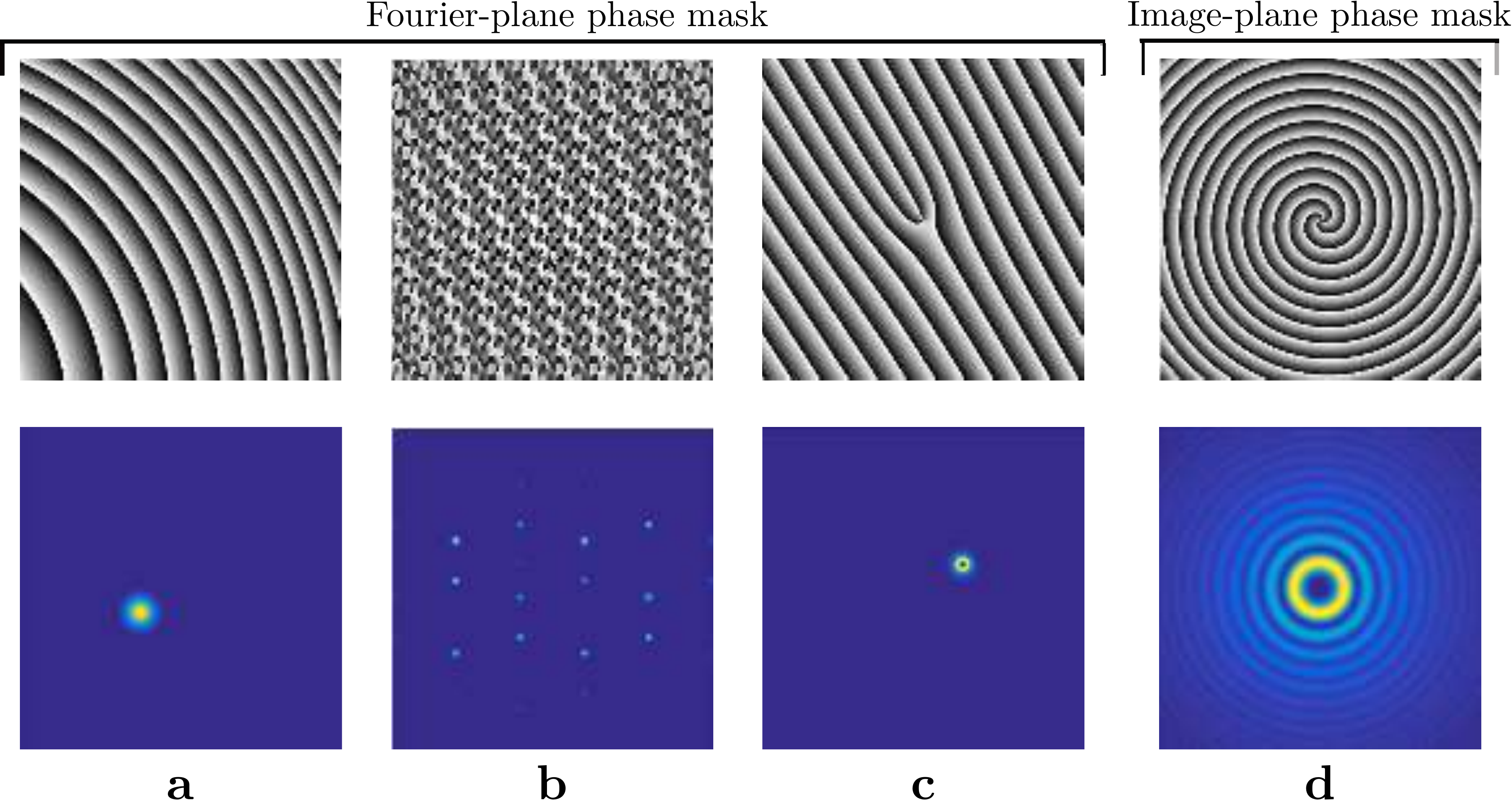}
	\caption{
	{\bf Phase masks and corresponding light intensity.}
	Diffractive optical elements displayed on a phase-only SLM for the generation of: 
	(a) a single optical trap laterally and axially shifted; 
	(b) multiple traps distributed in space (the different brightnesses and sizes of the spots are due to the different positions of the corresponding traps along the vertical $z$-position); 
	(c) a Laguerre-Gaussian beam of topological charge $l=-2$ laterally shifted; 
	(d) a Bessel beam of topological charge $l=3$. 
	In the cases (a)-(c), the light pattern is obtained at the Fourier plane (Fig.~\ref{fig:8:hot}(a)), while in case (d) it is at the image plane (Fig.~\ref{fig:8:hot}(b)). 
	}
	\label{fig:9:masks}
\end{figure}

In the first approach, the desired light pattern at the sample is the Fourier transform of the modulated laser beam right after the SLM.
It is necessary to include a $4f$-system between the SLM and the back aperture of the objective to conjugate the two optical planes and to perform spatial filtering (Fourier Plane 1 in Fig.~\ref{fig:8:hot}(a)).
This approach works particularly well to generate multiple focal spots distributed in predetermined locations in space, which serve as individual traps. 
The simplest case is a single steerable trap (Fig.~\ref{fig:9:masks}(a)), and the corresponding hologram is a superposition of a blazed diffraction grating (which determines the lateral position of the focal spot) and a Fresnel lens (which shifts the focal spot along the propagation axis). 
For the generation of multiple traps, there are different algorithms to calculate the hologram, such as the superposition of gratings and lenses, the random mask encoding, the Gerchberg-Saxton algorithm, and the adaptive-additive algorithm \cite{jones2015optical}. 
An example of a phase mask that produces a set of individual traps can be seen in Fig.~\ref{fig:9:masks}(b). 
Structured beams (e.g., Laguerre-Gaussian beams, Fig.~\ref{fig:9:masks}(c)) can also be straightforwardly created with this approach.
Furthermore, continuous optical potentials with arbitrary shapes can be produced as well by calculating the corresponding hologram with the aforementioned algorithms \cite{jones2015optical}.

In the second approach, a phase profile related to the desired light pattern is directly projected on the SLM and relayed on the objective image plane (this requires an even number of lenses between the two planes, Fig.~\ref{fig:8:hot}(b)).
This configuration is specially suitable when the field of interest is not meant to be tightly focused (low-NA optics), e.g., using phase gratings to split the beam and generate interference fringes at the objective image plane \cite{jakl2014optical}. 
Also, Bessel beams (Fig.~\ref{fig:9:masks}(d)), Mathieu beams, and nondiffracting beams are typically generated at the image plane instead of at the Fourier plane \cite{durnin1987exact, gutierrez2000alternative, arlt2001optical, volke2002orbital, garces2002transfer, bandres2004parabolic, golub2006fresnel, lopez2006orbital, arrizon2009efficient, hernandez2010experimental, ostrovsky2013generation}. 

Typical switching speeds for nematic liquid crystal SLMs can reach $60\,{\rm Hz}$ with a large number of discrete phase levels (typically 256), whereas ferroelectric liquid crystal SLMs can switch at a rate of up to $1\,{\rm kHz}$, but with only two phase levels (0 and $\pi$). 

\section{Calibration}\label{sec:3:calibration}

As we have seen in section~\ref{sec:2.2:otc}, an optical trap acts as a microscopic Hookeian spring with a fixed stiffness so that the restoring force is proportional to the displacement of the optically trapped particle from the center of the optical trap. 
The calibration of an optical tweezers entails the determination of the value of the optical trap stiffness, which in general depends on the properties of both the light beam and the particle. 
The calibration of an optical tweezers is essential to use it as a microscopic force transducer, i.e., to measure and exert forces in the femtonewton and piconewton range.

In principle, it should be possible to determine the value of the stiffness for a given experimental configuration where the properties of the focused laser beam and the optically trapped particle are known. 
In fact, this has been done in certain cases \cite{neto2000theory, rohrbach2005stiffness,  dutra2014absolute, arzola2019spin}. 
However, this is in general a cumbersome task because an exact description of the focused beam and its interaction with particles depends on several parameters that are very difficult to know accurately and may vary drastically even within the same experiment. 
For example, just the theoretical description of the focused beam requires precise information about the focusing lens, the thickness of the coverslip, the refractive index of the media, and the position of the focus inside the sample. 
Moreover, even with all this information at hand, the interaction between a focused beam and a trapped object is very complex, requiring to solve a computationally intensive electromagnetic scattering problem \cite{jones2015optical}. 
Therefore, the development and improvement of in-situ calibration methods have been very important to extend the practical applications of optical tweezers.

The calibration methods can be classified as either passive or active. 
Passive methods (sections \ref{sec:3.2:pot} to \ref{sec:3.10:nonconservative}) use information about the Brownian trajectory of a particle trapped in a static optical tweezers.
Active methods (section~\ref{sec:3.11:active}) analyze the response of the particle to a known time-dependent perturbation to the optical trap exerted by an external force, e.g., generated by a periodic movement of the sample cell or of the position of the optical tweezers. 
The force exerted by the optical tweezers can also be directly determined by measuring how the trapping light is scattered using a digital camera or a photodetector (section~\ref{sec:3.12:direct}).
Most of the mathematical details regarding the non-linear fitting are in section~\ref{sec:3.13:maths}.

To illustrate these calibration methods, we have carried out a series of experiments (Table~\ref{tab:1:expparam}). 
The experimental data and the MATLAB implementation of the calibration techniques explained in this section are provided as supplementary materials \cite{SI}.
An overview and comparison of the calibration methods is in Table~\ref{tab:2:comparison} and the results obtained on the test datasets is in Table~\ref{tab:3:expresults}.

We remark that all methods we are going to describe rely on the knowledge of the particle position in units of length (i.e., meters). 
However, the detection devices used in optical tweezers setups (e.g., CCD cameras, quadrant photodiodes, and position sensitive detectors) provide signals in pixels or volts. 
Therefore, it is necessary to determine the conversion factor from pixels or volts to meters.
For CCD camera, the pixel-to-meter conversion factor can be determined by standard microscopy techniques (e.g., by using a target with know dimensions) \cite{pesce2015step, jones2015optical}.
For quadrant photodiodes or position sensitive detectors, the volt-to-meter conversion factor can be determine by scanning an optically trapped particle across a (much weaker) detection beam, in which case one need to have at least two beams.
It is also possible to perform this calibration self-consistently using only the position signal of an optically trapped particle acquired by a photodetector, when the diffusion coefficient $D$ of the particle is know (e.g., when the particle radius, the medium viscosity and the temperature are known). 
The basic idea is that the knowledge of $D$ determines the scaling factor of the functions of the particle position used in the calibration process. 
The details go beyond this Tutorial and can be found in Ref.~\cite{jones2015optical}.

\begin{table}[t!]
	\begin{center}
	\begin{tabular}{p{6cm}|p{6cm}}
		\textbf{Parameter} 
		& 
		\textbf{Value}
		\\
	\hline
		Particle
		& 
		Polystyrene particle, radius $a = 1.03\,{\rm \upmu m}$
		\\
	\hline
		Laboratory temperature 
		&
		$ T = 20^{\circ}{\rm C} = 293.15\,{\rm K}$
		\\
	\hline
		Sampling frequency
		& 
		$f_{\rm s} = 10^5\,{\rm Hz}$
		\\
	\hline
		Sampling timestep
		& 
		$\Delta t = f_{\rm s}^{-1} = 10^{-5}\,{\rm s}$
		\\
	\hline
		viscosity (water)
		&
		$\eta = 1.00\,{\rm mPa\,s}$
		\\
	\hline
		friction coefficient (Stokes' law~\eqref{eq:stokes})
		& 
		$\gamma^{({\rm th})} = 19.4\,$\unitsgamma
		\\
	\hline
		diffusion coefficient (Einstein-Smoluchowski relation~\eqref{eq:einstein})
		&
		$D^{({\rm th)}}=0.21\,$\unitsD
		\\
	\hline
		number of experimental trajectories
		& 
		$M = 5$
		\\
	\hline
		number data points for each experiment 
		& 
		$N = 10^6$
		\\
	\hline
		objective lens
		& 
		Olympus UPLSAPO $60\times$ 1.2~NA water immersion
		\\
	\hline
		laser wavelength 
		& 
		$1064\,{\rm nm}$
		\\
	\hline
		laser power at the sample for experiment I, II, and III
		&
		$2.3\,{\rm mW}$, $6.0\,{\rm mW}$, and $9.2\,{\rm mW}$
		\\
	\hline
	\end{tabular}
	\caption{
	{\bf Experimental parameters.} 
	A microscopic particle immersed in water is trapped with a tightly-focused laser beam in a standard optical tweezers configuration and tracked using a quadrant photodiode at three different laser powers.
	}
	\label{tab:1:expparam}
	\end{center}
\end{table}

\begin{table}[t!]
	\begin{center}
	\begin{scriptsize}
	\begin{tabular}{p{12cm}}
	\hline
		{\bf Potential analysis} (section~\ref{sec:3.2:pot})
		\\
		$\rho(x)=\rho_0 e^{- \frac{U(x)}{k_{\rm B}T}}$ 
		\\
		{\bf Result.} It infers only  $\kappa_x$. 
		\\
		{\bf Advantages.} 
		Works for any potential (not only harmonic potentials);  it does not require fast acquisition but long acquisition times (in a harmonic optical trap, $T \gg \tau_{{\rm ot},x}$, where $\tau_{{\rm ot},x}$ is the optical trap characteristic time). 
		{\bf Disadvantages}. 
		It does not provide the dynamical properties of the optically trapped particle (e.g., $\gamma$ or $D$).
		\\
	\hline
		{\bf Equipartition method} (section~\ref{sec:3.3:eq})
		\\
		$\kappa_x=\frac{k_{\rm B} T}{\langle (x-x_{\text{eq}})^2\rangle}$ 
		\\
		{\bf Result.} It infers only  $\kappa_x$. 
		\\
		{\bf Advantages.} The simplest and easiest method to implement; it does not require fast acquisition. 
		{\bf Disadvantages}. It is restricted to harmonic potentials; it requires large datasets taken for long times $T\gg\tau_{{\rm ot},x}$; the inferred values are very sensitive to errors in the position detection.
		\\
	\hline	
		{\bf Mean squared displacement (MSD) analysis} (section~\ref{sec:3.4:msd})
		\\
		$\text{MSD}_x(\tau)=\frac{2 k_{\rm B} T}{\kappa_x}\left(1-e^{-\frac{|\tau|}{\tau_{{\rm ot},x}}}\right)$ 
		\\
		{\bf Result.} It infers both $\kappa_x$ and $\gamma$. {\bf Advantages.} it tends to provide accurate values.
		{\bf Disadvantages.} It only applies to harmonic potentials; it is very slow due to the large number of operations in estimating MSD; it requires fitting  a non-linear function; experimental errors are correlated making estimation from least squares more involved.
		\\
	\hline
		{\bf  Autocorrelation function (ACF) analysis} (section~\ref{sec:3.5:acf})
		\\
		$C_x(\tau)=\frac{k_{\rm B} T}{\kappa_x} e^{-\frac{|\tau|}{\tau_{{\rm ot},x}}}$ 
		\\
		{\bf Result.} It infers both $\kappa_x$ and $\gamma$. 
		{\bf Advantages.} Algorithmically faster  than MSD. The fitting can be transformed into a linear one. 
		{\bf Disadvantages.} It only applies to harmonic potentials; it requires fitting a function; experimental errors are correlated making estimation from least squares more involved.	
		\\   
	\hline
	{\bf Power spectral density (PSD) analysis} (section~\ref{sec:3.6:psd})
		\\
		$P_x(f)=\frac{1}{2\pi^2}\frac{D}{f_{c,x}^2+f^2}$ 
		\\
		{\bf Result.} It infers both $\kappa_x$ and $\gamma$. 
		{\bf Advantages.} Algorithmically faster than MSD and ACF; it provides accurate results; it can filter unwanted noise. 
		{\bf Disadvantages.} It only applies to harmonic potentials; one must fit a function, but this can be transformed into a linear problem. 
		{\bf Key reference}: \cite{berg2004power}
		\\
	\hline
		{\bf Force reconstruction via maximum-likelihood-estimator (FORMA) analysis} (section~\ref{sec:3.8:forma})
		\\
		$\mathcal{L}\left(\{x_n\}_{n=1}^N|\{x_{0n}\}_{n=1}^N,\kappa_x/\gamma,D\right)$ 
		\\
		{\bf Result.} It infers both $\kappa_x$ and $\gamma$. 
		{\bf Advantages.} It is  computationally fast to implement; it does not require a large  dataset; the estimates are given by  simple analytical expressions;  it can be extended to any dimension and non-conservative forces. 
		{\bf Disadvantages.} It works for $\Delta t\ll \tau_{{\rm ot},x}$, although this can be generalized; estimates for the friction coefficient are sensitive to the exposure time. 
		{\bf Key reference}: \cite{perez2018high}
		\\
	\hline
		{\bf Bayesian inference analysis} (section~\ref{sec:3.9:bayes})
		\\
		$P(\kappa_x/\gamma,D)=\frac{\mathcal{L}\left(\{x_n\}_{n=1}^N|\{x_{0n}\}_{n=1}^N,\kappa_x/\gamma,D\right)P_0(\kappa_x/\gamma,D)}{Z}$ 
		\\
		{\bf Result.} It infers both $\kappa_x$ and $\gamma$. 
		{\bf Advantages.} It is  computationally fast to implement; it can handle  small datasets and include  information  \emph{a priori}; the estimates are given by  simple analytical expressions; it gives the whole probability of the estimates;  it can be extended to any dimension and non-conservative forces. 
		{\bf Disadvantages.} it works for $\Delta t\ll \tau_{{\rm ot},x}$, although this can be generalized; estimates for the friction coefficient are  sensitive to the exposure time. 
		{\bf Key reference}: \cite{perez2019bayesian}
		\\	
	\hline
	\end{tabular}
	\end{scriptsize}
	\caption{
	{\bf Overview and comparison of calibration methods in one dimension in the passive configuration}. In this table we summarize the practical details of each method, together with advantages, disadvantages and a key reference, whenever possible.}
	\label{tab:2:comparison}
	\end{center}
\end{table}

\newcommand{\kappapotentialExpILF}{13.6\pm0.2}
\newcommand{\kappapotentialExpINLF}{13.6\pm0.2}
\newcommand{\kappapotentialExpIILF}{34.5\pm0.2}
\newcommand{\kappapotentialExpIINLF}{34.6\pm0.2}
\newcommand{\kappapotentialExpIIILF}{56\pm0.4}
\newcommand{\kappapotentialExpIIINLF}{56\pm0.4}

\newcommand{\kappaequiExpI}{13.6\pm0.4}
\newcommand{\kappaequiExpII}{34.9\pm0.6}
\newcommand{\kappaequiExpIII}{56\pm1}

\newcommand{\kappaequiExpIdelta}{13.7\pm0.4}
\newcommand{\kappaequiExpIIdelta}{35.2\pm0.7}
\newcommand{\kappaequiExpIIIdelta}{57\pm1}

\newcommand{\kappaMSDExpINLF}{13.62\pm0.04}
\newcommand{\DMSDExpINLF}{0.216\pm0.001}
\newcommand{\gammaMSDExpINLF}{18.7\pm0.1}
\newcommand{\tauMSDExpINLF}{1.376\pm0.005}

\newcommand{\kappaMSDExpIINLF}{34.95\pm0.09}
\newcommand{\DMSDExpIINLF}{0.2141\pm0.0009}
\newcommand{\gammaMSDExpIINLF}{18.91\pm0.08}
\newcommand{\tauMSDExpIINLF}{1.3761\pm0.001}

\newcommand{\kappaMSDExpIIINLF}{56.4\pm0.2}
\newcommand{\DMSDExpIIINLF}{0.2129\pm0.0008}
\newcommand{\gammaMSDExpIIINLF}{19.02\pm0.08}
\newcommand{\tauMSDExpIIINLF}{0.3371\pm0.0004}

\newcommand{\kappaACFExpILF}{13.5\pm0.4}
\newcommand{\DACFExpILF}{0.22\pm0.01}
\newcommand{\gammaACFExpILF}{18\pm1}
\newcommand{\tauACFExpILF}{1.37\pm0.04}

\newcommand{\kappaACFExpINLF}{13.542\pm0.004}
\newcommand{\DACFExpINLF}{0.219\pm0.005}
\newcommand{\gammaACFExpINLF}{18.5\pm0.4}
\newcommand{\tauACFExpINLF}{1.37\pm0.03}

\newcommand{\kappaACFExpIILF}{35\pm5}
\newcommand{\DACFExpIILF}{0.22\pm0.07}
\newcommand{\gammaACFExpIILF}{19\pm6}
\newcommand{\tauACFExpIILF}{0.54\pm0.09}

\newcommand{\kappaACFExpIINLF}{34.83\pm0.09}
\newcommand{\DACFExpIINLF}{0.216\pm0.004}
\newcommand{\gammaACFExpIINLF}{18.8\pm0.3}
\newcommand{\tauACFExpIINLF}{0.539\pm0.007}

\newcommand{\kappaACFExpIIILF}{56\pm7}
\newcommand{\DACFExpIIILF}{0.22\pm0.06}
\newcommand{\gammaACFExpIIILF}{18\pm5}
\newcommand{\tauACFExpIIILF}{0.33\pm0.05}

\newcommand{\kappaACFExpIIINLF}{55.7\pm0.2}
\newcommand{\DACFExpIIINLF}{0.219\pm0.005}
\newcommand{\gammaACFExpIIINLF}{18.5\pm0.4}
\newcommand{\tauACFExpIIINLF}{0.332\pm0.006}

\newcommand{\kappaPSDExpILF}{13.6\pm0.3}
\newcommand{\gammaPSDExpILF}{19.23\pm0.05}
\newcommand{\DPSDExpILF}{0.2105\pm0.0006}
\newcommand{\fcPSDExpILF}{112\pm2}
\newcommand{\kappaPSDExpIILF}{34.7\pm0.5}
\newcommand{\gammaPSDExpIILF}{19.22\pm0.05}
\newcommand{\DPSDExpIILF}{0.2106\pm0.0006}
\newcommand{\fcPSDExpIILF}{287\pm3}
\newcommand{\kappaPSDExpIIILF}{55.6\pm0.7}
\newcommand{\gammaPSDExpIIILF}{19.26\pm0.05}
\newcommand{\DPSDExpIIILF}{0.2101\pm0.0006}
\newcommand{\fcPSDExpIIILF}{460\pm4}

\newcommand{\kappaPSDExpINLF}{13.5\pm0.5}
\newcommand{\gammaPSDExpINLF}{19.2\pm0.3}
\newcommand{\DPSDExpINLF}{0.211\pm0.004}
\newcommand{\fcPSDExpINLF}{112\pm2}
\newcommand{\kappaPSDExpIINLF}{34.6\pm0.7}
\newcommand{\gammaPSDExpIINLF}{19.2\pm0.3}
\newcommand{\DPSDExpIINLF}{0.211\pm0.003}
\newcommand{\fcPSDExpIINLF}{287\pm2}
\newcommand{\kappaPSDExpIIINLF}{55\pm1}
\newcommand{\gammaPSDExpIIINLF}{19.2\pm0.4}
\newcommand{\DPSDExpIIINLF}{0.211\pm0.004}
\newcommand{\fcPSDExpIIINLF}{460\pm2}

\newcommand{\kappaFORMAExpI}{13.7\pm0.4}
\newcommand{\DFORMAExpI}{0.186\pm0.003}
\newcommand{\gammaFORMAExpI}{21.8\pm0.4}

\newcommand{\kappaFORMAExpII}{35.8\pm0.7}
\newcommand{\DFORMAExpII}{0.178\pm0.003}
\newcommand{\gammaFORMAExpII}{22.7\pm0.3}

\newcommand{\kappaFORMAExpIII}{58\pm1}
\newcommand{\DFORMAExpIII}{0.172\pm0.003}
\newcommand{\gammaFORMAExpIII}{23.6\pm0.5}

\newcommand{\kappaBayesExpI}{13.3\pm0.3}
\newcommand{\DBayesExpI}{0.1875\pm0.0005}
\newcommand{\gammaBayesExpI}{21.59\pm0.05}

\newcommand{\kappaBayesExpII}{34.9\pm0.5}
\newcommand{\DBayesExpII}{0.1817\pm0.0004}
\newcommand{\gammaBayesExpII}{22.27\pm0.05}

\newcommand{\kappaBayesExpIII}{59.6\pm0.7}
\newcommand{\DBayesExpIII}{0.1686\pm0.0004}
\newcommand{\gammaBayesExpIII}{24\pm0.06}

\begin{table}[t!]
	\begin{center}
	\begin{scriptsize}
	\begin{tabular}{p{1.2cm} p{0.35cm}|c|c|c}
		\textbf{Method} 
		& 
		& 
		\textbf{Experiment I} ($2.3\,{\rm mW}$) 
		& 
		\textbf{Experiment II} ($6.0\,{\rm mW}$)
		& 
		\textbf{Experiment III} ($9.2\,{\rm mW}$) 
		\\
	\hline
		& 
		& 
		Linear fit / Non-linear fit
		&  
		Linear fit / Non-linear fit
		&  
		Linear fit / Non-linear fit
		\\
	\hline
		Potential 
		&
		$\kappa^{({\rm ex})}$ 
		& 
		$\kappapotentialExpILF$ / $\kappapotentialExpINLF$ 
		&
		$\kappapotentialExpIILF$ / $\kappapotentialExpIINLF$ 
		&
		$\kappapotentialExpIIILF$ / $\kappapotentialExpIIINLF$
		\\
	\hline
		MSD 
		&
		$\kappa^{({\rm ex})}$ 
		& 
		--- --- --- / $\kappaMSDExpINLF$   
		&
		--- --- --- / $\kappaMSDExpIINLF$   
		& 
		--- --- --- / $\kappaMSDExpIIINLF$    
		\\
		&
		$\gamma^{({\rm ex})}$ 
		& 
		--- --- --- / $\gammaMSDExpINLF$  
		&
		--- --- --- / $\gammaMSDExpIINLF$   
		& 
		--- --- --- / $\gammaMSDExpIIINLF$ 
		\\
		&
		$D^{({\rm ex})}$ 
		& 
		--- --- --- / $\DMSDExpINLF$   
		&
		--- --- --- / $\DMSDExpIINLF$   
		& 
		--- --- --- / $\DMSDExpIIINLF$  
		\\
		&
		$\tau_{{\rm ot},x} $ 
		& 
		--- --- --- / $\tauMSDExpINLF$ 
		&
		--- --- --- / $\tauMSDExpIINLF$ 
		& 
		--- --- --- / $\tauMSDExpIIINLF$		
		\\
	\hline
		ACF 
		&
		$\kappa^{({\rm ex})}$ 
		& 
		$\kappaACFExpILF$ / $\kappaACFExpINLF$ 
		&
		$\kappaACFExpIILF$ / $\kappaACFExpIINLF$ 
		& 
		$\kappaACFExpIIILF$ / $\kappaACFExpIIINLF$
		\\
		&
		$\gamma^{({\rm ex})}$ 
		& 
		$\gammaACFExpILF$ / $\gammaACFExpINLF$ 
		&
		$\gammaACFExpIILF$ / $\gammaACFExpIINLF$ 
		& 
		$\gammaACFExpIIILF$ / $\gammaACFExpIIINLF$
		\\
		&
		$D^{({\rm ex})}$ 
		& 
		$\DACFExpILF$ / $\DACFExpINLF$ 
		&
		$\DACFExpIILF$ / $\DACFExpIINLF$ 
		& 
		$\DACFExpIIILF$ / $\DACFExpIIINLF$
		\\
		&
		$\tau_{{\rm ot},x}$ 
		& 
		$\tauACFExpILF$ / $\tauACFExpINLF$ 
		&
		$\tauACFExpIILF$ / $\tauACFExpIINLF$ 
		& 
		$\tauACFExpIIILF$ / $\tauACFExpIIINLF$		
		\\
	\hline
		PSD 
		&
		$\kappa^{({\rm ex})}$ 
		& 
		$\kappaPSDExpILF$ / $\kappaPSDExpINLF$ 
		&
		$\kappaPSDExpIILF$ / $\kappaPSDExpIINLF$ 
		& 
		$\kappaPSDExpIIILF$ / $\kappaPSDExpIIINLF$
		\\
		&
		$\gamma^{({\rm ex})}$ 
		& 
		$\gammaPSDExpILF$ / $\gammaPSDExpINLF$ 
		&
		$\gammaPSDExpIILF$ / $\gammaPSDExpIINLF$ 
		& 
		$\gammaPSDExpIIILF$ / $\gammaPSDExpIIINLF$
		\\
		&
		$D^{({\rm ex})}$ 
		& 
		$\DPSDExpILF$ / $\DPSDExpINLF$ 
		&
		$\DPSDExpIILF$ / $\DPSDExpIINLF$ 
		& 
		$\DPSDExpIIILF$ / $\DPSDExpIIINLF$
		\\
		&
		$f^{({\rm ex})}_{({\rm c},x)}$ 
		& 
		$\fcPSDExpILF$ / $\fcPSDExpINLF$
		&
		$\fcPSDExpIILF$ / $\fcPSDExpIINLF$ 
		& 
		$\fcPSDExpIIILF$/  $\fcPSDExpIIINLF$
		\\
	\hline
		&
		&
		Direct estimation
		&
		Direct estimation
		&
		Direct estimation
		\\
	\hline
		Equipartition 
		&
 		$\kappa^{({\rm ex})}$ 
		& 
		$\kappaequiExpI$ 
		&
		$\kappaequiExpII$ 
		& 
		$\kappaequiExpIII$
		\\
	\hline
		FORMA 
		&
		$\kappa^{({\rm ex})}$ 
		& 
		$\kappaFORMAExpI$ 
		&
		$\kappaFORMAExpII$ 
		& 
		$\kappaFORMAExpIII$
		\\
		&
		$\gamma^{({\rm ex})}$ 
		& 
		$\gammaFORMAExpI$ 
		&
		$\gammaFORMAExpII$ 
		& 
		$\gammaFORMAExpIII$
		\\
		&
		$D^{({\rm ex})}$ 
		& 
		$\DFORMAExpI$ 
		&
		$\DFORMAExpII$ 
		& 
		$\DFORMAExpIII$
		\\
	\hline
		Bayesian 
		&
		$\kappa^{({\rm ex})}$ 
		& 
		$\kappaBayesExpI$ 
		&
		$\kappaBayesExpII$ 
		& 
		$\kappaBayesExpIII $
		\\
		inference 
		&
		$\gamma^{({\rm ex})}$ 
		& 
		$\gammaBayesExpI$ 
		&
		$\gammaBayesExpII$ 
		& 
		$\gammaBayesExpIII$
		\\
		&
		$D^{({\rm ex})}$ 
		& 
		$\DBayesExpI$ 
		&
		$\DBayesExpII$ 
		& 
		$\DBayesExpIII$
		\\
	\hline
	\end{tabular}
	\end{scriptsize}
	\caption{
	{\bf Experimental results.} 
	Experimental values for the trap stiffness $\kappa$ in [\unitskappa], the friction coefficient $\gamma$ in [\unitsgamma], the diffusion constant $D$ in [\unitsD], the $\tau_{{\rm ot}, x}$ in [${\rm ms}$], and $f^{({\rm ex})}_{({\rm c},x)}$ in [${\rm Hz}$] inferred using various calibration methods on the data from the experiments I, II and III (Table~\ref{tab:1:expparam}, note, in particular, the expected values $\gamma^{({\rm th})}=19.4$ \unitsgamma and $D^{({\rm th})}=0.21$ \unitsD).
	The table is divided in two parts corresponding to the methods requiring fitting a function (linear fit / non-linear fit) and to those not requiring it (direct estimation).}
	\label{tab:3:expresults}
	\end{center}
\end{table}

\subsection{Time series acquisition and notation}

We consider a spherical particle of radius $a$ immersed in a fluid with viscosity $\eta$ at an absolute temperature $T$. According to Stokes' law, the friction coefficient is given by:
\begin{equation}\label{eq:stokes}
	\gamma
	=
	6\pi \eta a,
\end{equation}
while the diffusion constant is given by the  Einstein-Smoluchowski relation
\begin{equation}\label{eq:einstein}
	D
	=
	{k_{\rm B} T \over \gamma},
\end{equation}
where $k_{\rm B}=1.3806 \cdot 10^{-5} \, {\rm pN\,\upmu m\, K^{-1}}$ is Boltzmann's constant.  The trajectory $x(t)$ of an overdamped Brownian particle is described by the overdamped Langevin equation
\begin{equation}\label{eq:8:langevin}
	{d x(t) \over dt}
	=
	{1 \over \gamma} F(x)
	+
	\sqrt{2D} W_x(t),
\end{equation}
where $F(x)$ is a deterministic force exerted on the particle (in our case the optical force) and $W_x(t)$ is a white noise with zero mean and Dirac-delta-correlated variance, i.e., $\langle W_x(t) W_x(t')\rangle=\delta(t-t')$.
Experimentally, the particle's position is recorded at discrete times, yielding a time series $\{x_\ell\}_{\ell=1}^N$, where $x_{\ell}\equiv x(t_\ell)$ at times $t_\ell=\ell \Delta t$ with $\ell = 1,\ldots,N$.
To provide estimates with error bars, we can either repeat the experiment $M$ times or divide the time series of a single experiment of size $S$ into $M$ blocks of $N=S/M$ data points each. 
Either way, we end up with a set of times series $\{x_{\ell}^{(m)}\}_{\ell=1}^N$ for $m=1,\ldots,M$. A summary of the notation is in Table~\ref{tab:4:notation}.

\begin{table}[t!]
	\begin{center}
	\begin{tabular}{p{2cm}|p{10cm}} 
		\textbf{Notation} 
		& 
		\textbf{Explanation}
		\\
	\hline
		$x_{\ell} \equiv x(t_\ell)$ 
		& 
		position of particle at time $t_\ell = \ell\Delta t$ 
		\\
	\hline
		$\{x^{(m)}_\ell\}_{\ell=1}^N$
		& 
		recorded time series of particle's position during the $m$-th experiment with $N$ samples
		\\
	\hline
		$\{\hat{x}^{(m)}_\ell\}_{\ell=1}^L$
		& 
		uncorrelated time series at time intervals $\ell \widehat{\Delta t}$ with $\widehat{\Delta t}=r\Delta t$ and $r$ an integer
		\\
	\hline
		$\kappa_x^{({\rm ex})}$
		& 
		experimental estimate of  the trap stiffness
		\\
	\hline
		$\tau_x^{({\rm ex})}$
		& 
		experimental estimate of characteristic relaxation time
		\\
	\hline
		$D^{({\rm ex})}$, $D^{(\text{th})}$
		& 
		experimental estimate and theoretical value of  diffusion constant
		\\
	\hline
		$\gamma^{({\rm ex})}$, $\gamma^{(\text{th})}$
		& 
		experimental estimate and theoretical value of  friction coefficient
		\\
	\hline
		$T_{\rm s}$ 
		& 
		total acquisition time for each experiment $m$
		\\
	\hline
	\end{tabular}
	\caption{
	{\bf Notation.}
	In this table we summarize the notation related to  the acquisition of the time series and that of theoretical and inferred values of the parameters of the optical trap and that of the particle.
	}
	\label{tab:4:notation}
	\end{center}
\end{table}

\subsection{Potential analysis}
\label{sec:3.2:pot}

\begin{figure}[b!]
	\centering
	\includegraphics[width=12cm]{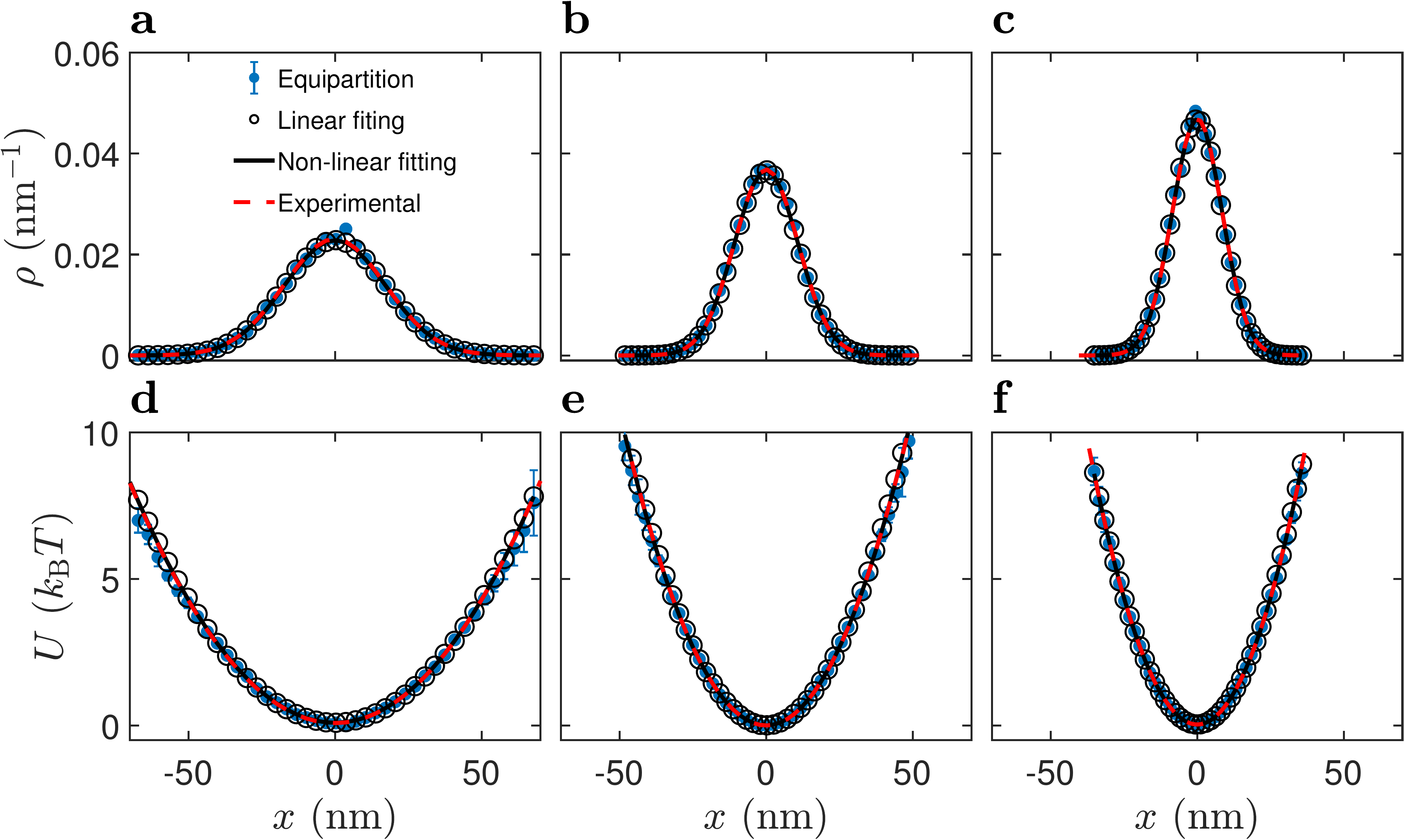}
	\caption{
	{\bf Potential analysis.} 
	Comparison between theoretical expressions and experimental estimates for (a-c) the equilibrium distribution and (d-f) the trap potential for a particle optically trapped at three different powers: (a, d) $2.3\,{\rm mW}$, (b, e) $6.0\,{\rm mW}$, and (c, f) $9.2\,{\rm mW}$ (Table~\ref{tab:1:expparam}). 
	The blue markers are the experimental measurements, the solid black lines are the linear fitting, and the dashed red lines are the non-linear fitting. 
	The black circles are the equilibrium distribution and trap potential obtained with the equipartition method (section~\ref{sec:3.3:eq}).
	The estimates for the trap stiffness $\kappa_x$ are given in Table~\ref{tab:3:expresults}.
	}
	\label{fig:10:potential}
\end{figure}

The potential method is fairly easy to implement and works for any conservative force with a potential that enables equilibrium of the particle in an accessible time. 
For a conservative force $F(x)=-\frac{d U(x)}{dx}$, where $U(x)$ is the equilibrium potential, and the probability density of finding the particle at a given position $x$ at thermal equilibrium follows the Maxwell-Boltzmann distribution:
\begin{equation}\label{eq:9:rho(x)}
	\rho(x)
	=
	\rho_0 e^{- {U(x) \over k_{\rm B}T}},
\end{equation}
where $\rho_0$ is a normalization factor ensuring that $\int \rho(x) dx = 1$. 
The potential $U(x)$ can then be reconstructed as
\begin{equation}
	U(x) 
	= 
	- k_{\rm B} T \ln{\rho(x) \over \rho_0}.
\end{equation}

The expression in equation~\eqref{eq:9:rho(x)} is an equilibrium distribution that theoretically corresponds to the probability of finding a particle at a given position $x$ from an statistical ensemble of independent particles. 
In practice, this distribution is usually computed from data sets coming from the trajectory of a particle, consisting of data points not necessarily uncorrelated. 
For example, in a harmonic potential, two points along the trajectory $x(t)$ and $x(t')$ are uncorrelated if $|t-t'| \gg \tau_{{\rm ot},x}$, where $\tau_{{\rm ot},x}=\gamma/\kappa_x$ is the characteristic relaxation time of the trap.
Thus, to ensure that we have enough independent samples $T_{\rm s} \gg \Delta t \gg \tau_{{\rm ot},x}$ (the second condition ensures that the samples are uncorrelated).
An important limitation of the potential methods is that, to sample rare events (e.g., the anharmonicity of the potential), one requires a number of measurements $N\sim \frac{1}{\rho(x)}$, and depending on the point $x$, $N$ can be extraordinarily large.
On the other hand, being an equilibrium method, the potential method does not rely on the knowledge of the viscosity and the radius of the particle to obtain an estimate for the trap stiffness.

Given a series of trajectories $\{\hat{x}^{(m)}_{\ell}\}_{\ell=1}^N$ with $m=1,\ldots M$, the experimental distribution $\rho^{({\rm ex})}_{m}\left(x;\{\hat{x}^{(m)}_{\ell}\}_{\ell=1}^N\right)$ is constructed as follows:
\begin{enumerate}

\item Determine the common maximum and minimum values of the position variable, denoted  $x_{\rm max}$ and $x_{\rm min}$, respectively. 

\item  Discretize the interval $[x_{\rm min},x_{\rm max}]$ into $P$ bins as:
\begin{equation}
	x_\alpha
	=
	x_{\rm min}+\frac{\alpha}{P}(x_{\rm max}-x_{\rm min})\,,\quad \alpha=0,1,\ldots,P-1
\end{equation} 

\item Use the $m$-th time series  to calculate the frequency of data points falling in each subinterval, obtaining the normalized histogram of $P$ bins for the $M$ experiments, i.e., the estimates  $\{\rho^{({\rm ex})}_{m}\left(x_\alpha\right)\}_{m=1}^{M}$ for $\rho(x)$.

\item Calculate the experimental mean value and variance, viz.
\begin{equation}\label{eq:rhoex}
	\overline{\rho^{({\rm ex})}}(x_\alpha)
	=
	{1 \over M} 
	\sum_{m=1}^{M} \rho^{({\rm ex})}_{m}
	\left(x_\alpha\right),
	\quad
	[\Delta^{(\rho)}_{\alpha}]^2
	=
	{1 \over M-1} 
	\sum_{m=1}^{M}
	\left[
		\rho^{({\rm ex})}_{m}
		\left(x_\alpha\right)
		-
		\overline{\rho^{({\rm ex})}}(x_\alpha)
	\right]^{2}
\end{equation}
for each bin $\alpha=0,1,\ldots, P-1$ in the histogram. 

\item Obtain the experimental potential for each experiment as
\begin{equation}\label{eq:U(x)}
	U_m^{({\rm ex})} \left(x_\alpha\right) 
	=
	-k_{\rm B} T
	\ln{\rho^{({\rm ex})}_{m}\left(x_\alpha\right)},
\end{equation}
from which we can obtain the mean values $\overline{U^{({\rm ex})}}(x_\alpha)$ and their errors $\Delta^{(\rm{U})}_\alpha$ in a similar way as in equation~\eqref{eq:rhoex}. 

\item Using the expression for the harmonic potential (equation~\eqref{eq:rhoextheory}) and the experimental dataset $\{x_\alpha, \overline{\rho^{({\rm ex})}}(x_\alpha), \Delta^{\rm{(\rho})}_\alpha\}$, find the estimates of the parameters  $\kappa_x$ and $x_{\rm eq}$ by means of a weighted non-linear least-square regression. 
Alternatively, the experimental dataset $\{x_\alpha, \overline{U^{({\rm ex})}}(x_\alpha), \Delta^{(\rm{U})}_\alpha\}$ allows us to build a linear model and solve it analytically with linear least-square fitting (section~\ref{sec:3.13:maths}).

\end{enumerate}

As an example, Fig.~\ref{fig:10:potential} illustrates the experimental distributions and potentials for a particle optically trapped at three different powers (corresponding to experiments I, II, and III, Table~\ref{tab:1:expparam}).
By means of either a linear (black solid lines) or non-linear fit (dashed red lines), we are able to estimate the value of $\kappa^{({\rm ex})}_x$ for the experiments I, II and III (using $P=50$ bins and the  $M=5$ experiments to estimate error bars for each bin). 
These estimates of $\kappa^{({\rm ex})}_x$ are given in Table~\ref{tab:3:expresults}. 
 
\subsection{Equipartition method} 
\label{sec:3.3:eq}

The equipartition method is a very simple and easy to implement method, which does not even require fitting a function. 
The equipartition method is a particular case of the potential method where we assume a harmonic potential:
\begin{equation}
	U(x) 
	= 
	\frac{\kappa_x}{2}(x-x_{\rm eq})^2, 
\end{equation}
so that the particle position distribution corresponds to the Gaussian distribution 
\begin{equation}\label{eq:rhoextheory}
	\rho(x)
	=
	\sqrt{\kappa_x \over 2\pi k_{\rm B} T}
	\exp
	\left\{
		-{\kappa_x \over 2 k_{\rm B} T}
		(x-x_{\rm eq})^2
	\right\}.
\end{equation}
Thus, the thermal average of the potential is
\begin{equation}
	\langle U(x)\rangle
	=
	{1\over2}
	\kappa_x
	\langle (x-x_{\rm eq})^2 \rangle
	=
	{\kappa_x\over2}
	\int_{-\infty}^{\infty} 
		\rho(x)(x-x_{\rm eq})^2 
		dx
	=
	{1\over2} k_{\rm B} T,
\end{equation}
which is a restatement of the equipartition theorem, imposing that the energy associated with the harmonic potential degree of freedom is equal to ${1\over2}k_{\rm B}T$.  
This provides a direct method to estimate $\kappa_x$ as
\begin{equation}\label{eq:equi}
	\kappa_x
	=
	\frac{
		k_{\rm B} T
	}{
		\langle (x-x_{\rm eq})^2\rangle
	},
\end{equation}
where we can estimate the equilibrium average $\langle (x-x_{\rm eq})^2\rangle$ by a time average. 
As for the potential method, the trajectory needs to be long enough to probe the equilibrium distribution ($T_{\rm s}\gg \Delta t\gg \tau_{{\rm ot},x}$) and work with the uncorrelated subsample $\{\hat{x}^{(m)}_{\ell}\}_{\ell=1}^L$.\footnote{It is possible also to use a correlated time series as long as the sampling is done at regular time intervals.}

For each trajectory $m=1,\ldots, M$, an estimate of the  ensemble average $\langle (x-x_{\rm eq})^2\rangle$ is given by its sample variance
\begin{equation}
	\sigma^{({\rm ex})2}_{x,m}
	=
	\frac{1}{L}
	\sum_{\ell=1}^L 
		\left(
			\hat{x}^{(m)}_\ell
			-
			x_{\rm eq}^{({\rm ex})}
		\right)^2,
\end{equation}
where $x_{\text{eq},m}^{({\rm ex})} =\frac{1}{L}\sum_{\ell=1}^L \hat{x}^{(m)}_\ell$ corresponds to the sample mean of the $m$-th experiment. Thus, for each experiment we obtain 
\begin{equation}
 \kappa^{({\rm ex})}_{x,m}=\frac{k_{\rm B} T}{\sigma^{({\rm ex})2}_{x,m}}.
\end{equation}
Using the results of the $M$ experiments, we calculate its mean value and variance, viz.
\begin{equation}\label{eq:equipmean}
	\kappa^{({\rm ex})}_{x}
	=
	{1 \over M}
	\sum_{m=1}^{M} 
		\kappa^{({\rm ex})}_{x,m},
	\quad
	\sigma^2_{(\kappa_x)}
	=
	{1 \over M-1}
	\sum_{m=1}^{M} 
		\left(
			\kappa^{({\rm ex})}_{x,m}
			-
			\kappa^{({\rm ex})}_{x}
		\right)^2.
\end{equation}

Using this approach on the experiments I, II and III (Table~\ref{tab:1:expparam}), yields the estimates of $\kappa_x$ reported in Table~\ref{tab:3:expresults}.
These estimates are consistent with those obtained using the potential method albeit with larger errors.
The harmonic potential corresponding to these estimates of the stiffness is plotted with black circles in Fig.~\ref{fig:10:potential} and is also in good agreement with the directly measured potential (blue circles).

Since the estimation of $\kappa_x$ depends on the inverse of the variance of the position, any random error in the position detection will always lead to an underestimation of the stiffness. On the other hand, if the data position is retrieved after a band-pass filter operation, the variance will diminish and hence the stiffness will be overestimated.

\subsection{Mean squared displacement analysis}
\label{sec:3.4:msd}

\begin{figure}[b!]
	\centering
	\includegraphics[width=12cm]{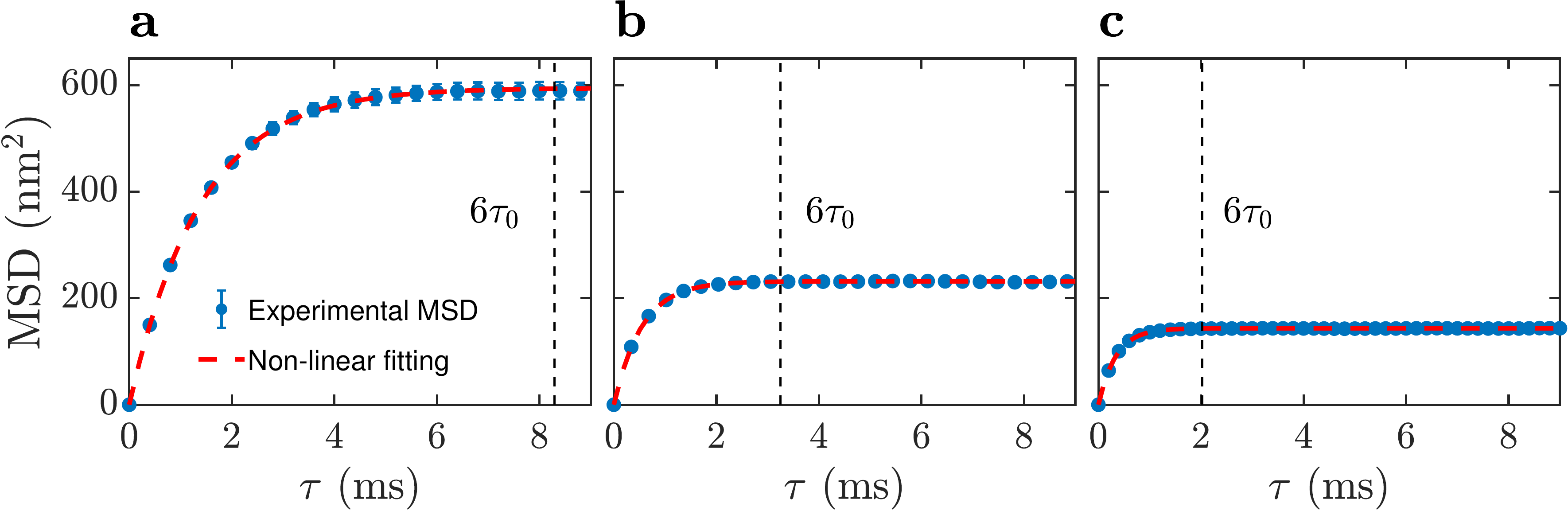}
	\caption{
	{\bf Mean square displacement (MSD) analysis.} 
	Comparison between the theoretical expression~\eqref{eq:MSD} of the MSD and the experimental estimates for a particle optically trapped at (a) $2.3\,{\rm mW}$, (b) $6.0\,{\rm mW}$, and (c) $9.2\,{\rm mW}$ (Table~\ref{tab:1:expparam}).
	The red dashed lines are the non-linear fitting and the blue markers with error bars are the experimental estimates. 
	The data are fitted in the interval $[0.5\tau_{\rm{ot},x},6\tau_{\rm{ot},x}]$ (the upper limit is shown by the dashed vertical lines).
	The estimates are given in Table~\ref{tab:3:expresults}.
	}
	\label{fig:11:MSD}
\end{figure}

Differently from the previous two methods, the mean squared displacement (MSD) analysis takes advantage of the time-correlated properties of the motion of the Brownian particle in the  harmonic potential. 
From the Langevin equation~\eqref{eq:8:langevin} with linear restoring force $F(x)=-\kappa_x x$, the MSD is:
\begin{equation}\label{eq:MSD}
	{\rm MSD}_x(\tau)
	=
	\langle[x(t+\tau) - x(t)]^2\rangle
	=
	{2 k_{\rm B} T \over \kappa_x}
	\left[1-e^{-{|\tau| \over \tau_{{\rm ot},x}}}\right],
\end{equation}
where $\langle(\cdots)\rangle$ denotes time average and $\tau_{{\rm ot},x}=\frac{\gamma}{\kappa_x}$ is the characteristic relaxation time.\footnote{Following Ref.~\cite{risken1996fokker}, we first see that the formal solution of the Langevin equation \eqref{eq:8:langevin} is given by
\begin{equation}
x(t)= x(0)e^{-t/\tau_{{\rm ot},x}}+\sqrt{2D}\int_0^t ds W_x(s) e^{-(t-s)/\tau_{{\rm ot},x}}\,.
\end{equation}
From here one can find, using the properties of white noise of $W_x(s)$, that
\begin{equation}
\overline{x(t_1)x(t_2)}=x^2(0) e^{-(t_1+t_2)/\tau_{{\rm ot},x}}+D\tau_{{\rm ot},x}\left(e^{-|t_1-t_2|/\tau_{{\rm ot},x}}-e^{-(t_1+t_2)/\tau_{{\rm ot},x}}\right)\,.
\end{equation}
This result can be used to derive the expression for the ${\rm MSD}_x(\tau)$ given by equation \eqref{eq:MSD}.} 
${\rm MSD}_x(\tau)$ comprises two different time regimes: 
for $\tau\ll \tau_{{\rm ot},x}$, ${\rm MSD}_x(\tau) \approx \frac{2k_{\rm B} T}{\gamma}|\tau|$, the particle exhibits free diffusion, while  for  $\tau\gg \tau_{{\rm ot},x}$, ${\rm MSD}_x(\tau) \approx \frac{2 k_{\rm B} T}{\kappa_x} $, the particle feels the harmonic potential and we recover the equipartition result given by equation~\eqref{eq:equi}.

If we measure ${\rm MSD}_x(\tau)$ experimentally (blue markers in Figs.~\ref{fig:11:MSD}(a-c)), we can estimate the trap stiffness $\kappa_x$ from a nonlinear fit to equation~\eqref{eq:MSD}. 
This requires that $\Delta t < \tau_{{\rm ot}, x}$, such that the time series $\{x^{(m)}_{n}\}_{n=1}^N$ is correlated for each $m=1,\ldots, M$. 
Denoting $t_j=j\Delta t$ for $j=1,\ldots,N$  and $\tau_\ell=\ell\Delta t$,  the experimental estimate for the MSD is:
\begin{equation}
	{\rm MSD}^{({\rm ex})}_{x,m}(\tau_\ell)
	=
	{1 \over N-\ell}
	\sum_{j=1}^{N-\ell} [x^{(m)}_{j+\ell}-x^{(m)}_j]^2.
\end{equation}
From the estimates $\{{\rm MSD}^{({\rm ex})}_{x,m}(\tau_\ell)\}_{m=1}^{M}$, we obtain their experimental mean and covariance matrix:
\begin{eqnarray*}
	\overline{{\rm MSD}^{({\rm ex})}_{x}(\tau_\ell)}
	&
	=
	&
	{1 \over M} 
	\sum_{m=1}^{M} {\rm MSD}^{({\rm ex})}_{x,m}(\tau_\ell),
	\\
 	\Delta^{[{\rm MSD}^{({\rm ex})}]}_{\ell,\ell'}
	&
	=
	&
	{1 \over M-1}
	\sum_{m=1}^{M} 
		\left(
			{\rm MSD}^{({\rm ex})}_{x,m}(\tau_\ell)
			-
			\overline{{\rm MSD}^{({\rm ex})}_{x}(\tau_\ell)}
		\right)
		\left(
			{\rm MSD}^{({\rm ex})}_{x,m}(\tau_{\ell'})
			-
			\overline{{\rm MSD}^{({\rm ex})}_{x}(\tau_{\ell'})}
		\right).
\end{eqnarray*}
Using the experimental dataset $\{\overline{{\rm MSD}^{({\rm ex})}_{x}(\tau_\ell)}, \Delta^{[{\rm MSD}^{({\rm ex})}]}_{\ell,\ell'}\}$ and the theoretical model of the MSD in equation~\eqref{eq:MSD}, we construct $\chi^2$ according to equation~\eqref{eq:chi2} and minimize it numerically to find the estimates $\kappa_x$ and $\tau_{\rm{ot},x}$ (red shaded lines in Figs.~\ref{fig:11:MSD}(a-c)).  
There are several aspects to consider when proceeding with the non-linear fitting:
First, since the experimental data is time-correlated, we need the covariance matrix when inferring the parameters from the non-linear fitting. 
Second, as the MSD shows two distinct time-regimes, we must be sure to have a good balance of points in both regimes  to avoid overweighting one respect to the other. 
Third, the experimental error of the MSD increases as the lag time grows since the number of data points used to estimate  MSD diminishes. 
Fourth, errors in position detection have a strong impact particularly in the short time regime, because the MSD is proportional to the position squared. 
It is good practice to exclude the short times from the MSD fit, up to a fraction of the expected $\tau_{\rm{ot},x}$, as we did in Fig.~\ref{fig:11:MSD}.

For the experimental data recorded in  experiments I, II and III (Table~\ref{tab:1:expparam}), we perform the non-linear fitting on the interval $[0.5\tau_{\rm{ot},x},6\tau_{\rm{ot},x}]$ and estimate the errors as explained in section~\ref{sec:3.13:maths}.
This provides us with the estimates for $\kappa_x^{({\rm ex})}$ and $\tau^{({\rm ex})}_{{\rm ot},x}$, from which $\gamma = \kappa_x \tau_{{\rm ot},x}$ and $D$ is given by the Einstein-Smoluchowski relation~\eqref{eq:einstein}.
All these estimates are reported in Table~\ref{tab:3:expresults}.

The MSD analysis has a few clear advantages over the equilibrium methods described in the previous two sections.
First, the MSD uses more information so that the trap stiffness is estimated more precisely.
Second, the MSD analysis can infer also dynamic properties, namely the friction coefficient, which is not directly accessible with the equilibrium methods. 
However, care must be taken when fitting a function with correlated errors and to have a good balance of data points between the short-time and long-time regimes.

\subsection{Autocorrelation function (ACF)}
 \label{sec:3.5:acf}

\begin{figure}[b!]
	\centering
	\includegraphics[width=12cm]{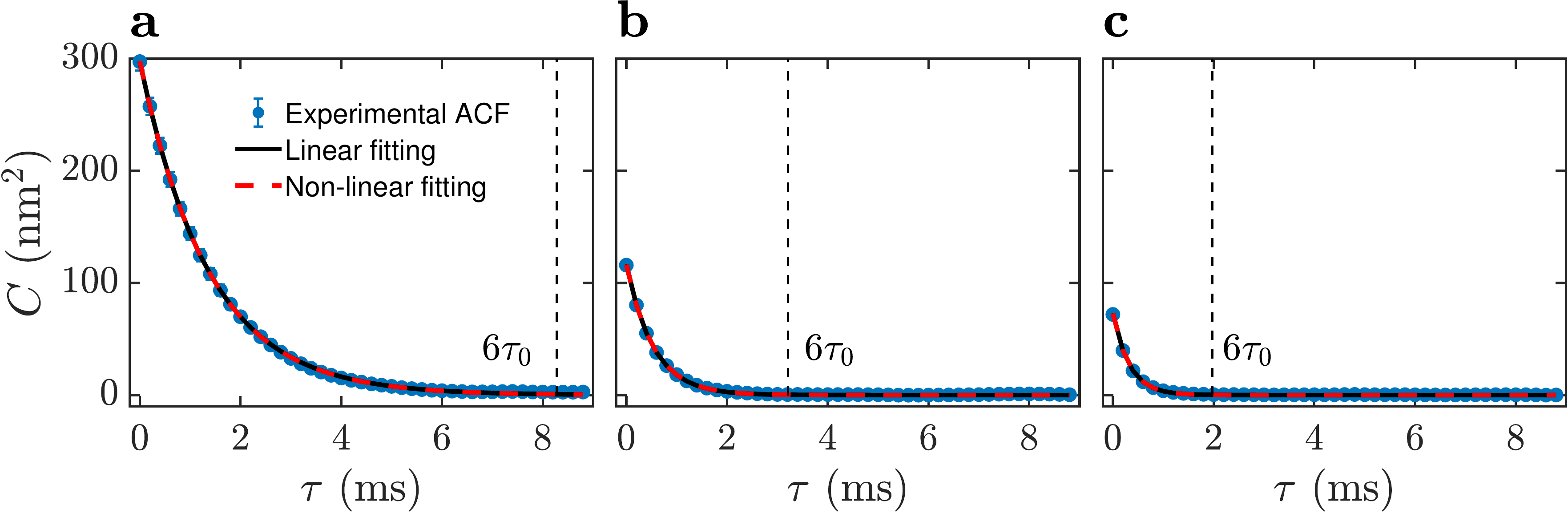}
\caption{ 
	{\bf Autocorrelation function (ACF) analysis.}
	Comparison between the theoretical expression~\eqref{eq:28:ACF} of the ACF and the experimental  estimates for a particle optically trapped at (a) $2.3\,{\rm mW}$, (b) $6.0\,{\rm mW}$, and (c) $9.2\,{\rm mW}$ (Table~\ref{tab:1:expparam}). 
	The black solid lines and red dashed lines are the linear and non-linear fitting, respectively, while the blue markers with error bars are the experimental estimates. 
	The data are fitted on the interval $[0,2.5\tau_{\rm{ot},x}]$ (the upper limit is shown by the dashed vertical lines). 
	The estimates are given in Table~\ref{tab:3:expresults}.}
	\label{fig:12:ACF}
\end{figure}

Starting from the Langevin equation~\eqref{eq:8:langevin} with linear restoring force $F(x)=-\kappa_x x$, the position autocorrelation function (ACF) is \cite{risken1996fokker}
\begin{equation}\label{eq:28:ACF}
	C_x(\tau)
	=
	\langle x(t+\tau)x(t)\rangle
	=
	\frac{
		k_{\rm B} T
	}{
		\kappa_x
	} 
	e^{ -{|\tau| \over \tau_{{\rm ot},x}}}.
\end{equation} 
From a correlated time series sampled at regular time steps $\Delta t$, we can estimate the ACF by using
\begin{equation}
	C^{({\rm ex})}_{x,m}(\tau_\ell )
	=
	{1 \over N-\ell}
	\sum_{j=1}^{N-\ell} 
		x^{(m)}_{j+\ell}
		x^{(m)}_j,
\end{equation}
which gives a set of estimates $\{C^{({\rm ex})}_{x,m}(\tau_\ell)\}_{m=1}^{M}$ for each experiment $m=1,\ldots,M$ (blue markers with error bars in Figs.~\ref{fig:12:ACF}(a-c)). 
From these estimates, we obtain their experimental mean and covariance matrix:
\begin{eqnarray}
	\overline{C^{({\rm ex})}_{x}(\tau_\ell)}
	&
	=
	&
	{1 \over M}
	\sum_{m=1}^{M}
		C^{({\rm ex})}_{x,m}(\tau_\ell),
	\\
	\Delta^{[C^{({\rm ex})}_{x}]}_{\ell,\ell'}
	&
	=
	&
	{1 \over M-1}
	\sum_{m=1}^{M}
		\left(
			C^{({\rm ex})}_{x,m}(\tau_\ell)
			-
			\overline{C^{({\rm ex})}_{x}(\tau_\ell)}
		\right)
		\left(
			C^{({\rm ex})}_{x,m}(\tau_{\ell'})
			-
			\overline{C^{({\rm ex})}_{x}(\tau_{\ell'})}
			\right).
\end{eqnarray}
 
Using the the autocorrelation equation \eqref{eq:28:ACF}, we can build $\chi^2(\kappa_x,\tau_{{\rm ot},x})$ according to equation \eqref{eq:chi2} and minimize with respect to the parameters $\kappa_x$ and $\tau_{{\rm ot},x}$, solving the resulting set of equations by means of non-linear procedures using iterative solutions (red dashed lines in Figs.~\ref{fig:12:ACF}(a-c)).  
Another possibility is to use the logarithm of the autocorrelation instead, i.e., $\ln C_x(\tau)=\ln (k_{\rm B}T/\kappa_x)-\tau/\tau_{{\rm ot},x}$ and perform a linear regression, in which case the experimental dataset is $\{ \overline{\ln{C^{({\rm ex})}_x}}(\tau_\ell)\}$ together with their covariance  matrix for the $M$ experiments (black solid lines in Figs.~\ref{fig:12:ACF}(a-c)).
As in the MSD analysis, from the estimates for $\kappa_x$ and $\tau_{{\rm ot},x}$ we can determine estimates for $\gamma = \kappa_x \tau_{{\rm ot},x}$ and $D = k_{\rm B} T / \gamma$.

As in the MSD analysis, the ACF analysis requires $\Delta t\ll \tau_{{\rm ot},x}$ and $T_{\rm s}\gg\tau_{{\rm ot},x}$. 
Furthermore, since the ACF function decays exponentially with time, only the values corresponding to short times, around the relaxation time, are relevant to get the information of the confining potential. 

The estimates for $\kappa_x$, $\tau_{{\rm ot},x}$, $\gamma$ and $D$ performing the fittings on the interval $[0,2.5\tau_{\rm{ot},x}]$ are provided in Table~\ref{tab:1:expparam}.
The ACF analysis results are comparable to those obtained with the MSD analysis, both using non-linear and analytical linear regression. 
However, the errors for $\kappa_x^{({\rm ex})}$ are bigger than those obtained from the MSD analysis, which can be  expected from the expression of the ACF (equation~\eqref{eq:28:ACF}), whose initial value is solely determined by the trap stiffness and not the decaying behavior so that we only have one data point to estimate $\kappa^{({\rm ex})}_x$.
An important difference is also the computation time: while the ACF is computed with very efficient algorithms based on the Fast Fourier Transform, the MSD requires a rather inefficient numerical computation.

\subsection{Power spectral density (PSD)}
\label{sec:3.6:psd}

\begin{figure}[b!]
	\centering
	\includegraphics[width=12cm]{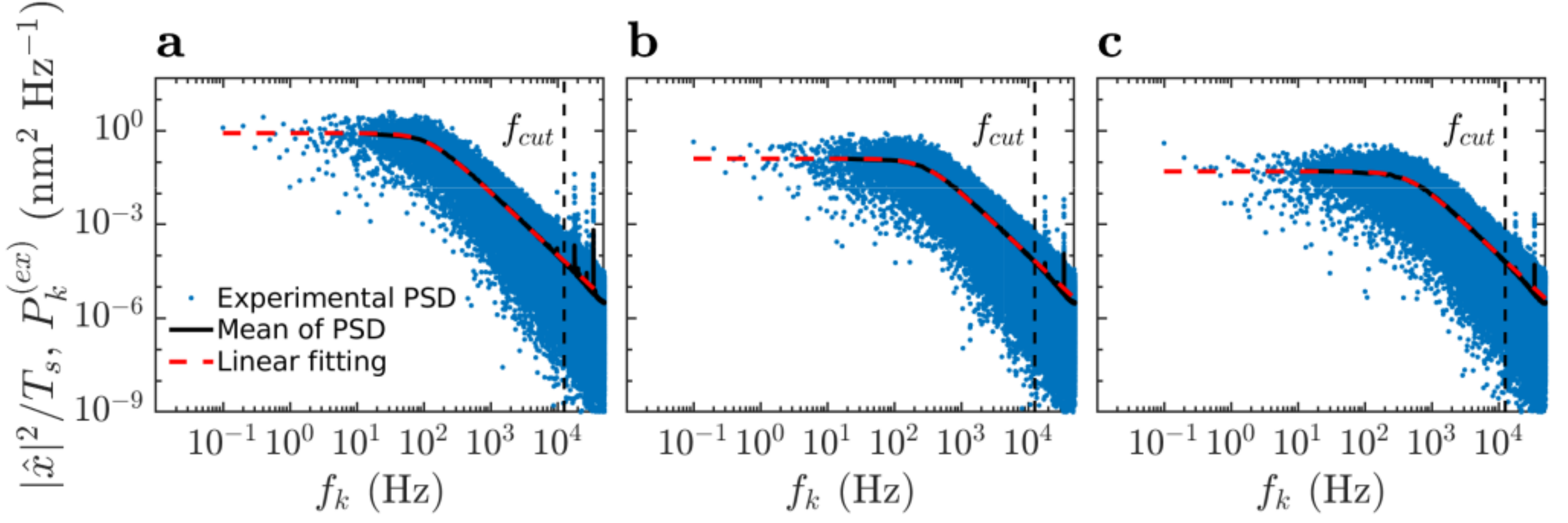}
	\caption{
	{\bf Power spectral density analysis.} 
	Comparison between the theoretical expression~\eqref{eq:32:lorentzian} and experimental estimates for a particle optically trapped at (a) $2.3\,{\rm mW}$, (b) $6.0\,{\rm mW}$, and (c) $9.2\,{\rm mW}$ (Table~\ref{tab:1:expparam}). 
	The blue dots are the experimental estimates $\frac{|\hat{x}_k|^2}{T_{\rm s}}$ given by equation~\eqref{eq:fft}, the solid red lines correspond to its expected value $P_\alpha^{({\rm ex})}=\langle|\hat{x}_k|^2/T_{\rm s}\rangle_{n_{\rm w}}$ over the $n_{w}$ frequencies belonging to the subdomain $\alpha$, and the black dashed lines are the result of  the linear fitting.  
	The vertical dotted lines indicate the maximum frequency used to perform the fitting; in this example, $f_{{\rm max}}\equiv\textrm{max}{\{f_\alpha\}}=f_{{\rm Nyq}}/4=f_{\rm s}/8 $.
	The estimates are given in Table~\ref{tab:3:expresults}.
	}
	\label{fig:13:PSD}
\end{figure}

Calibration of an optical tweezer by means of the power spectral density (PSD) is considered to be one of the most reliable methods due to its ability  to remove common sources of noise with relative ease since one works in frequency space.
In contrast to other methods, the PSD analysis is very robust to unwanted  periodic noise in the extracted trajectory of the particle, due to, for example, illumination, electronic noise, fan noise of different electronic components inside the laboratory, or even some resonant vibration that is difficult to isolate with optical tables.
The method explained in this section is based on Ref.~\cite{berg2004power}.

Starting from the Langevin equation~\eqref{eq:8:langevin} with $F(x)=-\kappa_x x$ and performing a Fourier transform, one obtains the PSD, which has a Lorentzian form and is the expected value of the average energy in the frequency domain,
\begin{equation}\label{eq:32:lorentzian}
	P(f)
	=
	\langle|\widetilde{x}(f)|^2\rangle
	=
	{1 \over 2\pi^2}
	{D \over f_{{\rm c},x}^2+f^2},
\end{equation}
where $\widetilde{x}(f)$ is
\begin{equation*}
	\widetilde{x}(f)
	=
	{1 \over \sqrt{T}}
	\int_0^{T} 
		x(t)
		e^{-i2\pi f t}
		dt, 
\end{equation*}
and $f_{{\rm c},x}=\kappa_x/(2\pi\gamma)=1/(2\pi \tau_{{\rm ot},x})$ is the corner frequency. 
To derive equation~\eqref{eq:32:lorentzian}, we have used the fact that the spectrum of the zero-mean white noise, $|\widetilde{W}_x(f)|^2$, has an exponential distribution with an expected value $\langle |\widetilde{W}_x(f)|^2\rangle=1$. 

Experimentally, we have a trajectory sampled with frequency $f_{\rm s}=1/\Delta t$ up to time $T_{\rm s}$, i.e., a time series $x_j=x(t_j)$ with $t_j=j\Delta t$ and  $j=1,\ldots,N$, which is used to perform the discrete Fourier transform (DFT):
\begin{equation}\label{eq:fft}
	\hat{x}_k
	=
	\Delta t 
	\sum^{N}_{j=1}
		e^{i2\pi f_k t_j}
		x_j
	=
	\Delta t 
	\sum^{N}_{j=1} 
		e^{i2\pi jk/N}
		x_j.
\end{equation}
It is expected that $\langle |\hat{x}_k|^2\rangle/T_{\rm s}\approx\langle |\tilde{x}(f_k)|^2\rangle$ when $f_{\rm s}\gg f_{{\rm c},x}$, so it is important to sample the trajectory at frequencies some orders of magnitude higher than the corner frequency. 
Furthermore, the discrete PSD will suffer from aliasing. 
A simple formula for the expected value of the aliased PSD can be derived from the discretized Einstein-Ornstein-Uhlenbeck process \cite{berg2004power}, yielding
\begin{equation}
	P^{({\rm aliased})}_k
	=
	\frac{
		\langle |\hat{x}_k|\rangle
	}{
		T_s
	}
	=
	\frac{
		(\Delta x)^2\Delta t
	}{
		1+c^2-2c\cos(2\pi f_k\Delta t/N)
	},
\end{equation}
where $\Delta x=((1-c^2)D/2\pi f_{{\rm c},x})^{1/2}$ and $c=e^{-2\pi f_{{\rm c},x}/f_{\rm s}}$. 
As long as we sample at frequencies some orders of magnitude higher than the corner frequency, we can mitigate the aliasing by simply cropping the spectrum at a frequency below the Nyquist frequency $f_{{\rm Nyq}}=f_{\rm s}/2$. 

In principle, the expected value in the PSD can be estimated by averaging the experimental values $|\hat{x}_k|^2/T_{\rm s}$ over a large set of $M$ experiments. 
In practice, one would need to acquire hundreds of experiments to have a good estimation. 
Therefore, it is common to apply a compression process to obtain a good estimation of the expected value with only one or a few experiments. 
Windowing is the most common compression process: it divides the frequency domain in $N_{\rm w}$ narrow and equidistant subdomains with $n_{\rm w}$ values, whose average represents the expected value at the corresponding mean frequency of the subdomain.
This gives rise to the reduced experimental dataset for the PSD $P_\alpha^{({\rm ex})}=\langle|\hat{x}_k|^2/T_{\rm s}\rangle_{n_{\rm w}}$ with average frequency $f_{\alpha}$,  with $\alpha=0,1,2,\ldots,N_{\rm w}-1$. With this procedure, the error in the expected value is $\sigma^{({\rm ex})}_\alpha= P^{({\rm ex})}_\alpha/\sqrt{n_{\rm w}}$. 
If multiple realizations of the experiment are available, it is also possible to first compress the experiments and then average the results, giving an error $\sigma^{({\rm ex})}_\alpha= P^{({\rm ex})}_\alpha/\sqrt{M n_{\rm w}}$. 

Once the experimental PSD is estimated, the parameters $f_{{\rm c},x}$ and $D$ are estimated by least-square fitting. 
Using directly the model ~\eqref{eq:32:lorentzian} and the data set $\{f_\alpha, P_\alpha^{({\rm ex})}\}_{\alpha=0}^{N_{{\rm w}}-1}$ with their associated weights, $W_\alpha=1/(\sigma^{({\rm ex})}_\alpha)^2$, we can build $\chi^2(f_{{\rm c},x}, D)$ according to equation~\eqref{eq:chi2} and then apply the non-linear fitting procedure obtaining a numerical solution for the estimates. 
Alternatively, an analytical linear fitting can be performed using the experimental dataset $\{f_\alpha^2, 1/ P^{({\rm ex})}_\alpha\}_{\alpha=0}^{N_{\rm w}-1}$ with weights given by $W_\alpha=1/(P^{({\rm ex})\,-1}_\alpha/\sqrt{n_{\rm w}})^2$ and the linear model $1/P(f)=2\pi^2f^2/D+2\pi^2 f^2_{c,x}/D$. 
Finally, an analytical solution can also be found \cite{berg2004power} which minimizes
\begin{equation}\label{eq:psdanalytic}
	\chi^2(f_{{\rm c},x}^2,D)
	=
	{1 \over N_{\rm w}}
	\sum^{N_{\rm w}-1}_{\alpha=0}
		\left(
			\frac{
				P^{({\rm ex})}_\alpha
				-
				P(f_\alpha;f_{{\rm c},x}^2,D)
			}{
				P(f_\alpha;f_{{\rm c},x}^2,D)
				/
				\sqrt{n_{\rm w}}
			}
		\right)^2.
\end{equation}
The analytical solutions for the estimates using $\chi^2$ defined by equation~\eqref{eq:psdanalytic} are
\begin{align}
	f^{({\rm ex})}_{c,x}
	&
	=
	\sqrt{
		\frac{
			S_{0,1}S_{2,2}-S_{1,1}S_{1,2}
		}{
			S_{1,1}S_{0,2}-S_{0,1}S_{1,2}
		}
	},
	\label{eq:fittingpsd1}
	\\
	 D^{({\rm ex})}
	 &
	 =
	 2\pi^2
	 \left(
	 	\frac{
			S_{0,2}S_{2,2}-S_{1,1}^2
		}{
			S_{1,1}S_{0,2}-S_{0,1}S_{1,2}
		}
	\right),
	\label{eq:fittingpsd2}
\end{align}
where $S_{p,q}=\sum^{N_{\rm w}}_k f^{2p}_k  P^{({\rm ex})\,q}_k$.
The errors for these estimates are
\begin{align}
	\frac{
		\sigma(f^{({\rm ex})}_{c,x})
	}{
		f^{({\rm ex})}_{c,x}
	}
	&
	=
	\frac{
		1
	}{
		\sqrt{(u-v)f^{({\rm ex})}_{c,x}T_{\rm s}}
	},
	\\
	\frac{
		\sigma(D^{({\rm ex})})
	}{
		D^{({\rm ex})}
	}
	&
	=
	\sqrt{
		\frac{
			u
		}{
			(x_{\rm max}-x_{\rm min})
			\pi 
			f^{({\rm ex})}_{c,x}
			T_{\rm s}
		}
	},
\end{align}
where 
$x_{\rm min}=f_{\rm min}/f^{({\rm ex})}_{c,x}$,
$x_{\rm max}=f_{\rm max}/f^{({\rm ex})}_{c,x}$,
$u = \frac{2x_{\rm max}}{1+x^2_{\rm max}} - \frac{2x_{\rm min}}{1+x^2_{\rm min}} + 2\,{\rm arctan}\left(\frac{x_{\rm max}-x_{\rm min}}{1+x_{\rm max}x_{\rm min}}\right)$, 
$v = \frac{4}{x_{\rm max}-x_{\rm min}}\,{\rm arctan}^2\left(\frac{x_{\rm max}-x_{\rm min}}{1+x_{\rm max}x_{\rm min}}\right)$,
and $f_{\rm min}$ and $f_{\rm max}$ correspond to the minimum and maximum on the list of frequencies $\{f_\alpha\}$.

The blue dots in Figs.~\ref{fig:13:PSD}(a-c) show the values $|\hat{x}_k|^2/T_{\rm s}$ computed by a fast Fourier transform for the trajectories acquired in experiments I, II, and III (Table~\ref{tab:1:expparam}):
\begin{equation}\label{eq:fft}
	\frac{|\hat{x}_k|^2}{T_{\rm s}}
	=
	{\Delta t^2 \over T_{\rm s}}
	|\mathrm{FFT}\{\{x_j\}^N_{j=1}\}_k|^2 
	\ \ \ 
	\mbox{with} 
	\ \ \ 
	f_k
	=
	{k \over T_{\rm s}} 
	\ \ \ \
	\mbox{for} 
	\ 
	k=0,1,2,\ldots,N/2.
\end{equation}
Their expected values are obtained by compression using $N_{\rm w}=500$ (solid red lines in Figs.~\ref{fig:13:PSD}(a-c)) and used to fit the Lorentzian function given by equation~\eqref{eq:32:lorentzian} using equations~\eqref{eq:fittingpsd1} and \eqref{eq:fittingpsd2} (dashed black lines in Figs.~\ref{fig:13:PSD}(a-c)). 
To avoid aliasing (which gives rise to the plateau at high frequencies), the fitting was performed only with data corresponding to frequencies smaller than $f_{\rm max}=f_{{\rm Nyq}}/4=f_{\rm s}/8$. 
From this fitting, we obtain the estimates for $f_{{\rm c}, x}$ and $D$ and, from these, the estimates for $\kappa_x$, $\tau_{{\rm ot},x}$, and $\gamma$.
Repeating this procedure for $M=5$ experiments, we obtain the mean values of the estimates and their errors.
All these estimates are reported in Table~\ref{tab:3:expresults}.

\subsection{Drift method}
\label{sec:3.7:drift}

The drift method \cite{ryter1980properties, wu2009direct, volpe2010influence, brettschneider2011force} relies on the fact that the drag force must be equal to the optical force. 
From the Langevin equation \eqref{eq:8:langevin}, this implies that
\begin{equation}\label{eq:Fstat}
	F(x_0)
	=
	6\pi \eta a
	\frac{
		\langle\Delta x \rangle_{x_0}
	}{
		\Delta t
	},
\end{equation}
where $\langle\Delta x \rangle_{x_0}$ is the local drift, which is the average displacement of a particle when located at initial position $x_0$ and under the external force $F_{x_0}$ during a time interval $\Delta t$.

From an experimental trajectory the local drift can be estimated as the conditional average
\begin{equation}
	\langle\Delta x \rangle_{x_0}
	=
	{1 \over S}
	\sum_{i=1}^S
		\left[
			x_{i}(t+\Delta t)
			-
			x_i(t)
		\mid|
			x_i(t) \approx x_0
		\right],
\end{equation}
where the $S$ corresponds to the number of times the initial position of the particle is close to $x_0$.

When applying this method there are several caveats.
First, the size of the volume element near $x_0$ must be small enough to probe correctly the spatial variations of the force field.
Second, the position of the particle must be measured with a precision much smaller than that of the volume element.
Finally, equation~\eqref{eq:Fstat} is valid when the diffusion coefficient is constant, while the presence of a space-dependent diffusion coefficient (e.g., due to hydrodynamics) leads to the emergence of \emph{spurious drift} \cite{volpe2010influence, brettschneider2011force}. 

In the presence of a spurious drift, a diffusion gradient term need to be added to equation~\eqref{eq:Fstat} leading to
\begin{equation}\label{eq:Fstatc}
	F(x_0)
	=
	\gamma (x_0) 
	\frac{
		\langle\Delta x\rangle_{x_0}
	}{
		\Delta t
	}  
	-
	\alpha
	\gamma(x_0)
	{dD(x_0) \over dx_0},
\end{equation}
where $\alpha$ may take values in the interval $\left [ 0,1 \right ]$, depending on  the specific stochastic process being modelled. 
In particular, $\alpha=1$ corresponds to  a colloidal particle in a thermal bath \cite{lau2007state, hottovy2015smoluchowski, volpe2016effective}. 

\subsection{Force reconstruction via maximum-likelihood-estimator analysis (FORMA)}
\label{sec:3.8:forma}

The force reconstruction via maximum-likelihood-estimator analysis (FORMA) was introduced recently to retrieve the force field acting on a Brownian particle from the analysis of its displacements \cite{perez2018high}. This method is able to estimate  accurately the conservative and non-conservative components of the force field with important advantages over the previously discussed techniques, being parameter-free, requiring ten-fold less data and executing orders-of-magnitude faster \cite{perez2018high}.

Let us consider a particle in a harmonic potential and suppose that after a time interval $\Delta t\ll \tau_{{\rm ot},x}$ the Brownian particle is at a position $x$, having started at position $x_0$. 
Repeating this measurement $N$ times, the likelihood of observing this set of measurements is
\begin{equation}\label{eq:43}
	\mathcal{L}(
		\{x_n\}_{n=1}^N
		|
		\{x_{0n}\}_{n=1}^N,
		\kappa_x/\gamma,
		D
	)
	=
	{1 \over (4\pi D\Delta t)^{N/2}}
	\exp
		\left[
			-{1 \over 4D}
			\sum_{n=1}^N
				\Delta t
				\left(
					{x_n-x_{0n} \over \Delta t}
					+
					x_{0n}
					{\kappa_x \over \gamma}
				\right)^2
		\right],
\end{equation}
which depends on the parameters $\kappa_x/\gamma$ and $D$.
By maximizing the logarithm of the likelihood given by equation~\eqref{eq:43} with respect to the parameters $\kappa_x/\gamma$ and $D$, one obtains the so-called maximum likelihood estimators (MLEs) for a given time series $\{x^{(m)}_{\ell}\}_{\ell=1}^N$:
\begin{equation}\label{eq:42:FORMA}
	{\kappa_{x,m}^{({\rm ex})} \over \gamma}
	=
	-\frac{
		\sum_{n=1}^{N-1}
			x^{(m)}_{n}
			\frac{
				x^{(m)}_{n+1} -x^{(m)}_{n}
			}{
				\Delta t
			} 
	}{
		\sum_{n=1}^{N-1} [x^{(m)}_{n}]^2
	},
	\quad\quad
	D^{({\rm ex})}_{m}
	=
	{\Delta t \over 2 N}
	\sum_{n=1}^{N-1}
		\left(
			\frac{
				x^{(m)}_{n+1}-{x^{(m)}_{n}}
			}{
				\Delta t
			}
			+
			\frac{
				\kappa_{x,m}^{({\rm ex})}
			}{
				\gamma
			} 
			x^{(m)}_{n}
		\right)^2.
\end{equation}
From these equations and the Einstein-Smoluchowski relation~\eqref{eq:einstein}, it is straightforward to estimate $\kappa_{x,m}^{({\rm ex})}$, $\gamma$, and $D^{({\rm ex})}_{m}$. 
The standard errors for these estimates can be obtained from the mean and standard deviations from repeated experiments:
using multiple time series $m=1,\ldots,M$ to obtain the estimates  $\{\kappa_{x,m}^{({\rm ex})}\}_{m=1}^{M}$ and  $\{D^{({\rm ex})}_{m}\}_{m=1}^{M}$ yields the experimental mean and variances:
\begin{equation}\label{eq:43:FORMA2}
	\begin{split}
		\overline{\kappa_{x}^{({\rm ex})}}
		&
		=
		{1 \over M}
		\sum_{m=1}^{M} \kappa_{x,m}^{({\rm ex})},
		\quad\quad 
		\Delta^{(\kappa_{x})}
		=
		{1 \over M-1}
		\sum_{m=1}^{N_{\rm exp}} 
			\left(
				\kappa_{x,m}^{({\rm ex})}
				-
				\overline{\kappa_{x}^{({\rm ex})}}
			\right)^2 ,
	\\
		\overline{D^{({\rm ex})}}
		&
		=
		{1 \over M}
		\sum_{m=1}^{N_{\rm exp}} D_{m}^{({\rm ex})},
		\quad\quad 
		\Delta^{(D)}
		=
		{1 \over M-1}
		\sum_{m=1}^{M}
			\left(
				D_{m}^{({\rm ex})}
				-
				\overline{D^{({\rm ex})}}
			\right)^2 .
	\end{split}
\end{equation}
FORMA provides also an alternative way to get error estimates for the MLEs given by equation~\eqref{eq:42:FORMA} for a single experiment, without having to repeat the experiment.
Indeed, if we perform a Taylor expansion of the log-likelihood around the  MLEs given by equation~\eqref{eq:42:FORMA} up to second order, we obtain a Gaussian approximation for $\mathcal{L}(\{x_n\}_{n=1}^N|\{x_{0n}\}_{n=1}^N)$. This yields the following standard deviations for  $\kappa_x^{({\rm ex})}$ and $D^{({\rm ex})}$
\begin{equation}\label{eq:44:emles}
	\sigma_{D^{({\rm ex})}}
	=
	D^{({\rm ex})}
	\sqrt{2 \over N},
	\quad\quad 
	\sigma_{\kappa_x^{({\rm ex})}}
	=
	\sqrt{
		\frac{
			2D^{({\rm ex})}
		}{
			\Delta t\sum_{n=1}^N x_{0n}^2}
		}.
\end{equation}
When applying FORMA, $\Delta t\ll \tau_{{\rm ot},x}$ and the exposure time should be much smaller than $\Delta t$ (i.e., as close as possible to an instantaneous recording of the particle position).

We report the estimates for the experimental data in Table~\ref{tab:3:expresults}, obtained using formulas~\eqref{eq:42:FORMA} and \eqref{eq:43:FORMA2}.
Comparing these results with the ones obtained by the other methods, we can see some discrepancies in the diffusion coefficient and hence in the friction coefficient.
This is due to the fact that the condition that $\Delta t\ll \tau_{{\rm ot},x}$ and that the exposure time be much smaller than $\Delta t$ does not hold true as the laser power is increased, causing the relaxation time to decrease.

Since FORMA only depends on the local variables of the trajectory, that is the positions and their corresponding displacements, it allows to retrieve the forces of arbitrary optical landscapes, including non-conservative forces, and it does not need a trajectory with regular sampling, commonly required in the other methods. 
This is particularly useful when the problem is to retrieve the forces of extended potentials, where the particle neither reaches the equilibrium nor is able to explore the whole space by itself, such as for a dual optical tweezer or a speckle pattern \cite{perez2018high}. 
The main drawback in using FORMA is the need for high spatial resolution and fast sampling. The latter can be addressed by generalizing the solution for any arbitrary time interval $\Delta t$, as recently introduced in Ref.~\cite{singh2018fast}. 

\subsection{Bayesian inference}
\label{sec:3.9:bayes}

While FORMA uses a non-Bayesian linear regression to infer the properties of the trap and the Brownian particle \cite{perez2018high}, other methods have used Bayesian inference to infer optical forces \cite{turkcan2012bayesian, richly2013calibrating, el2016primer, bera2017fast, singh2018fast}. 
Recently, we have derived a Bayesian approach using conjugate priors, which renders very simple formulas for the posterior distribution, with known moments at all orders and marginal distributions \cite{perez2019bayesian}. Here, we provide an overview of this method.

Let us suppose that we want to infer $D$ and $\kappa_{x}/\gamma$. 
Let $P_0(\kappa_x/\gamma,D)$ be their \emph{a priori} probability distribution. 
Then, given a time series, the posterior probability is given by Bayes' rule:
\begin{equation}\label{eq:47}
	P(\kappa_x/\gamma, D)
	=
	\frac{
		\mathcal{L}(
			\{x_n\}_{n=1}^N
			|
			\{x_{0n}\}_{n=1}^N,
			\kappa_x/\gamma,
			D)
			P_0(\kappa_x/\gamma, D)
	}{
		Z
	}
\end{equation}
with $Z$ being a normalization factor. 
Choosing the {\it a-priori} probability $P_0(\kappa_x/\gamma,D)$ to be a so-called conjugate prior for the model's likelihood,\footnote{A conjugate prior is a particular type of a priori distribution, chosen so that the posterior distribution belongs to the same family of distributions of such prior. One may intuitively think of conjugate priors as eigenfunctions of equation~\eqref{eq:47}.}
we obtain the following marginal posterior distribution for the trap stiffness and the diffusion constant
\begin{equation}\label{eq:48}
	\begin{split}
		P_D(x)
		&
		=
		{\rm InvGamm}\left(x|\alpha_N,\beta_N\right),
		\\
		P_{\kappa_x/\gamma}(x)
		&
		=
		\frac{
			\Gamma\left(\frac{2\alpha_N+1}{2}\right)
		}{
			\Gamma(\alpha_N)
		}
		\frac{
			1
		}{
			\sqrt{2\pi\alpha_N}
		}
		\sqrt{
			\frac{
				\alpha_N
			}{
				\tilde\gamma_N\beta_N
			}
		}
		\left(
			1
			+
			{1 \over 2\alpha_N}
			{\alpha_N \over \beta_N\tilde\gamma_N}
			(x-K_N)^2
		\right)^{-(2\alpha_N+1)/2},
	\end{split}
\end{equation}
where ${\rm InvGamm}(x)$ is the inverse Gamma distribution, and the parameters $\alpha_N$, $\beta_N$, $\gamma_N$, and $K_N$ are determined by the measurement record:
\begin{equation}
	\begin{split}
	\beta_N
	&
	=
	\beta_0
	+
	{1 \over 2}
	\sum_{n=1}^{N}
		\frac{
			[x_{n+1}-x_{n}(1-K_N\Delta t)]^2
		}{
			\Delta t
		}
		+
		{1 \over 2\gamma_0}
		(K_N-K_0)^2, 
	\\
	\alpha_N
	&
	=
	\alpha_0
	+
	{N \over 2}, 
	\\
	K_N 
	&
	=
	-{1 \over \Delta t}
	\left(
		-1
		+
		\frac{
			\Delta t
			\sum_{n=1}^{N} 
				x_{n+1}
				x_{n}
				+
				{1-K_0\Delta t \over \gamma_0}
		}{
			{1 \over \gamma_0}
			+
			\Delta t
			\sum_{n=1}^N x_{n}^2
		}
	\right), 
	\\
	{1 \over \gamma_N}
	&
	=
	{1 \over \gamma_0}
	+
	\Delta t
	\sum_{n=1}^{N} x_{n}^2,
	\end{split}
\end{equation}
The experimental estimates $D^{({\rm ex})}$ and $\kappa^{({\rm ex})}_x$ are the mean values and variances  of the posterior distributions $P_{D}$ and $P_{\kappa_x/\gamma}$:
\begin{equation}
	\begin{split}
		D^{({\rm ex})}
		&
		=
		\int dx ~ xP_D(x)
		=
		\frac{
			\beta_N
		}{
			2(\alpha_N-1)
		},
		\quad\quad 
		\sigma^2_{D^{({\rm ex})}}
		=
		\frac{
			\beta_N^2
		}{
			4(\alpha_N-1)^2 
			(\alpha_N-2)
		},
		\\
		\frac{
			\kappa^{({\rm ex})}_x
		}{
			\gamma
		}
		&
		=
		\int dx ~xP_{\kappa_x}(x)
		= 
		K_N,
		\quad\quad 
		\sigma^2_{\kappa_x/\gamma}
		=
		\frac{
			\gamma_N \beta_N
		}{
			\alpha_N-1
		}.
	\end{split}
\end{equation}
Formulas for higher cumulants can also be calculated analytically from equations~\eqref{eq:48}.

The hyper-parameters $\alpha_0$, $\beta_0$, $\gamma_0$, and $K_0$, determine the \emph{a priori} information about the model's parameters. 
For example, $\alpha_0$, $\beta_0$, $\gamma_0$, and $K_0$  can be chosen such that for $N=0$ (i.e., without experimental data), we obtain their theoretical values and systematic uncertainties (i.e.,  $D^{({\rm th})}\pm \sigma_{D^{({\rm th})}}$ and $\kappa^{({\rm th})}_x/\gamma\pm \sigma_{\kappa^{({\rm th})}_x/\gamma}$):
\begin{equation}
	\begin{split}
		\alpha_0
		&
		=
		2
		+
		\frac{
			(D^{({\rm th})})^2
		}{
			\sigma^2_{D^{({\rm th})}}
		},
		\quad\quad 
		\beta_0
		=
		\frac{
			2D^{({\rm th})}[(D^{({\rm th})})^2+\sigma^2_{D^{({\rm th})}}]
		}{
			\sigma^2_{D^{({\rm th})}}
		},
		\\
		K_0
		&
		=
		\frac{
			\kappa^{({\rm th})}_x
		}{
			\gamma
		},
		\quad\quad
		\gamma_0
		=
		\frac{
			\sigma^2_{\kappa^{({\rm th})}_x/\gamma}
		}{
			2D^{({\rm th})}
		}.
	\end{split}
\end{equation}
From Table~\ref{tab:1:expparam}, $D^{({\rm th)}}=0.21$ \unitsD and, adding an error of 10 \%, allows us to fix the values $\alpha_0$ and $\beta_0$. 
For the theoretical value of the inverse relaxation time, we are more in the dark, as we do not know the trap stiffness before the calibration procedure; thus, we can use an estimate obtained by another method to determine the values of $K_0$ and $\gamma_0$.
Using Bayesian inference on one of the time series $\{x^{(m)}_{n}\}$, we obtain the results reported in Table~\ref{tab:3:expresults}.
For the estimates regarding the friction and diffusion coefficients, we notice that the present method has the same issue concerning the exposure time as FORMA.

The Bayesian inference method extracts all statistical information from the measured data, so that it provides an unbiased estimator of the experimental parameters. 
Moreover, it can provide reasonable estimates of the parameters also when little data is available, thanks to the use of prior information. 
The Bayesian inference method can also be generalized to more than one dimension for any time interval $\Delta t$.

\subsection{Measurement of non-conservative force fields}
\label{sec:3.10:nonconservative}

While the previous methods rely on the forces being conservative, non-conservative forces often play also a role in experiments. 
The drift method can be straightforwardly applied to measure any force field \cite{wu2009direct}.
Also, the ACF method has been adapted to measure non-conservative forces using cross-correlations \cite{volpe2006torque, volpe2007brownian, pesce2009quantitative}. 
Here, we briefly show how FORMA can be adapted to measure non-conservative forces \cite{perez2018high}, while the details for the Bayesian method can be found in \cite{perez2019bayesian}.

Let us consider the two-dimensional overdamped Langevin equation
\begin{equation}\label{eq:lehd}
	\dot{{\bf r}}
	=
	{1 \over \gamma}
	{\bf F}({\bf r})
	+
	\sqrt{2D}{\bf w},
\end{equation}
where ${\bf F}({\bf r})$ is the force field and ${\bf w}$ is a vector of independent white noise. 
The Taylor expansion of $F({\bf r})$ around ${\bf r}={\bf 0}$ yields
\begin{equation}
	{\bf F}({\bf r})
	=
	{\bf F}_0
	+
	{\bf J}_0 \cdot {\bf r}
	+
	\mathcal{O}({\bf r}^2),
\end{equation}
where ${\bf F}_0={\bf F}({\bf 0})$ and ${\bf J}_0={\bf J}({\bf 0})$ are the force and the Jacobian at the point ${\bf r}={\bf 0}$, respectively. 
We further assume ${\bf r}={\bf 0}$ to be an equilibrium point so that ${\bf F}_0={\bf 0}$.

Discretizing the Langevin equation~\eqref{eq:lehd}, we obtain
\begin{equation}\label{eq:dlehd}
	\mathbf{f}_n
	=
	\gamma
	\frac{
		\Delta {\bf r}_n
	}{
		\Delta t
	}
	=
	{\bf J}_0 \cdot {\bf r}_n
	+
	\sigma{\bf w}_n,
\end{equation}
where $\sigma=\sqrt{\frac{2D\gamma^2}{\Delta t}}$ and ${\bf w}_n$ is a vector of independent Gaussian numbers with zero mean and unit variance. 
Equation~\eqref{eq:dlehd} can be understood as a multivariate linear regression model: for a given set of pairs of data points $\{({\bf r}_n,\mathbf{f}_n)\}_{n=1}^N$, 
we construct the likelihood for \eqref{eq:dlehd}, similarly as the expression~\eqref{eq:43} for the one-dimensional case, and maximize it with respect to ${\bf J}_0$ to obtain the corresponding MLE ${\bf J}_0^\star$, which is given by the Moore-Penrose pseudo-inverse:
\begin{equation}
	{\bf J}_0^\star
	=
	\left[
		\mathbf{R}^{\rm T}
		\mathbf{R}
	\right]^{-1}
	\mathbf{R}^{\rm T}
	{\bf F},
\end{equation}
where $\mathbf{R}$ and ${\bf F}$ are $N \times 2$ matrices given by
\begin{equation}
	\mathbf{R}
	=
	\begin{pmatrix}
		{\bf r}^T_1
		\\
		\vdots
		\\
		{\bf r}^{T}_N
	\end{pmatrix},
	\quad\quad 
	{\bf F}
	=
	\begin{pmatrix}
		\mathbf{f}^T_1
		\\
		\vdots
		\\
		\mathbf{f}^T_N
		\end{pmatrix}.
 \end{equation}
From the MLE ${\bf J}_0^\star$, we can estimate the force as ${\bf F}^\star={\bf J}_0^\star\cdot{\bf r}$. 
This result can be naturally split into conservative and non-conservative parts, corresponding to the symmetric and antisymmetric decompositions of ${\bf J}_0^\star={\bf J}_{\rm c}^\star+{\bf J}_{\rm r}^\star$, respectively. 
The symmetric part can be rewritten as
\begin{equation}
	{\bf J}_{\rm c}^\star
	=
	{1 \over 2}
	\left(
		{\bf J}^\star_0
		+
		{\bf J}^{\star {\rm T}}_0
	\right)
	=
	\mathbb{R}(\theta^\star)
	\begin{pmatrix}
		 -\kappa_1^\star
		 &
		 0
		 \\
		0
		&
		-\kappa_2^\star
	\end{pmatrix}
	\mathbb{R}^{-1}(\theta^\star),
\end{equation}
with $\mathbb{R}$ a rotation matrix and $\kappa_1^\star$ and $\kappa_2^\star$ the trap stiffness along the principal axes; $\theta^\star$ represent the orientation of these axes with respect to the Cartesian system of reference.
The rotational part is in turn given by
\begin{equation}
	{\bf J}_{\rm r}^\star
	=
	{1 \over 2}
	\left(
		{\bf J}^\star_0
		-
		{\bf J}^{\star T}_0
	\right)
	=
	\begin{pmatrix}
		0
		&
		-\gamma\Omega^\star
		\\
		\gamma\Omega^\star
		&
		0
	\end{pmatrix},
\end{equation}
with $\Omega^\star$ being the angular frequency at which the particle would orbit around the center of the trap in the absence of Brownian noise.

\subsection{Active calibration techniques}
\label{sec:3.11:active}

All calibration methods discussed so far are \emph{passive}, i.e., they are based on the determination of the trap stiffness from the stochastic motion of the particle in thermal equilibrium with the surrounding fluid within a harmonic potential. 
Alternatively, the optical tweezers calibration can be performed by \emph{actively} applying an external force on the particle and measuring its response.  

A simple active method consists of exposing an optically trapped particle to an external uniform flow in the $x$-direction with a known velocity $v_{\rm fluid}$ so that the particle experiences a mean drag force $F_{\rm drag}=-\gamma v_{\rm fluid}$.
Consequently, the particle undergoes a mean displacement $\Delta x = \langle x - x_{\rm eq}\rangle$ from the trap center $x_{\rm eq}$ so that the restoring optical force balances the viscous drag.
The stiffness can then be measured as 
\begin{equation}\label{elastic}
	\kappa 
	=
	\frac{
		\gamma v_{\rm fluid}
	}{
		\Delta x
	}.
\end{equation}
In practice, a uniform flow can be applied by moving the sample at constant speed $-v_{\rm fluid}$, e.g., by means of a piezoelectric translation stage, while keeping the particle far away from the cell walls to avoid hydrodynamics interactions which change the value of friction coefficient $\gamma$ with respect to the value given by Stokes' law \cite{simmons1996quantitative}.

Another active method requires moving sinusoidally the position of the sample stage $x_{\rm stage}$, i.e., 
\begin{equation}\label{eq:stagemotion}
	x_{\rm stage}(t)
	=
	A\sin(2\pi f_{\rm stage} t),
\end{equation} 
while keeping the optical tweezers at fixed $x_{\rm eq} = 0$ \cite{tolic2006calibration}.
This method not only provides the value of the trap stiffness, but also the drag coefficient and the spatial conversion factor of the measured particle position $x$ (typically in volts or pixels) into actual distances.

The Langevin equation for the motion of the particle in the case of the oscillating stage is 
\begin{equation}\label{eq:Langevinflow}
	{dx(t) \over dt} 
	-
	v_{\rm stage}(t)
	= 
	-2\pi f_{{\rm c},x} x(t) 
	+ \sqrt{2D}W_x(t),
\end{equation}
where $v_{\rm stage}(t) = 2\pi f_{\rm stage} A \cos (2\pi f_{\rm stage} t)$ is the velocity of the stage and $W_x$ is a white Gaussian noise.
The corresponding PSD of the particle position is
\begin{equation}\label{eq:PSDflow}
	P(f)
	=
	\underbrace{
		{D \over \pi^2}
		{1 \over f^2+f_{{\rm c},x}^2}
	}_{\mbox{thermal contribution}}
	+
	\underbrace{
		\frac{
			A^2
		}{
			2
			\left(
				1
				+
				{f_{{\rm c},x}^2 \over f_{\rm stage}^2}
			\right)
		}
	\delta
	(f-f_{\rm stage})
	}_{\mbox{mechanical response}},
\end{equation}
where the thermal contribution is equal to that given by equation~\eqref{eq:32:lorentzian}, while the mechanical response is that of the particle to the excitation frequency $f = f_{\rm stage}$. 
Because the measurement time in an experiment is finite, $t_{\rm msr} < \infty$ , the Dirac-delta peak in equation~\eqref{eq:PSDflow} becomes a spike with finite height $P_{\rm stage} = P(f_{\rm stage})$ and width $\Delta f$. 
If the total measurement time is chosen to be an integer multiple of the driving frequency of the stage, then the spike reduces to a single datum $P_J$ at frequency $f_J = f_{{\rm stage}}$ for some integer $J$, whereas  $\Delta f = 2\delta f$, where $\delta f$ corresponds to the frequency resolution of the PSD. 
In this case, the discrete version of equation~\eqref{eq:PSDflow} for the $j$-th data point is
\begin{equation}\label{eq:PSDflowdiscrete}
	P_j 
	=
	{D \over \pi^2}
	{1 \over f_j^2+f_{{\rm c},x}^2} 
	+ 
	\frac{
		A^2 t_{\rm msr}
	}{
		2
		\left(
			1
			+
			{f_{{\rm c},x}^2 \over f_j^2}
		\right)
	}
	\delta_{j,J},
\end{equation}
where $\delta_{j.J}$ represents the Kronecker delta. 
Therefore, disregarding the single point $(f_J, P_J)$, a fit $P_{\rm fit}(f)$ to the one-sided Lorentzian function in equation~\eqref{eq:PSDflowdiscrete} allows to compute the diffusion coefficient $D^{(1)}$ and the corner frequency $f_{{\rm c},x}$ (section~\ref{sec:3.6:psd}), from which the trap stiffness can be determined as
\begin{equation}\label{eq:kappafit}
	\kappa^{(1)} 
	=  
	\frac{
		2\pi f_{{\rm c},x} k_{\rm B} T
	}{
		D^{(1)}
	}.
\end{equation}
If $x$ has not been yet converted into units of length, the values and units of $D^{(1)}$ and $\kappa^{(1)}$ are not correct. 
A further step using the active component of the spectrum in equation~\eqref{eq:PSDflowdiscrete} allows to get the calibration factor $\alpha$ of $x$ into units of length: $x \rightarrow \alpha x$, where $\alpha$ can have units of m$/$V or m$/$pixel depending on whether the particle position was recorded with a quadrant photodiode, a position sensitive detector or a camera. 
The power contained in the triangular spike centered at $j = J$, $W^{\rm exp} = \frac{1}{2} [P_J - P_{\rm fit}(f_J)] 2 \delta f$, where $ P_{\rm fit}(f_J) =  \frac{D^{(1)}/\pi^2}{f_J^2+f^{2}_{c,x}} $, and the theoretical value given in equation~\eqref{eq:PSDflow}, $W^{\text{th}} = \frac{A^2}{2\left(1+{f_{{\rm c},x}^2}/{f_{\rm stage}^2}\right)}$, must be equal.  Then, $\alpha^2 W^{\rm exp} = W^{\text{th}}$, from which 
\begin{equation}\label{eq:calibfactor}
	\alpha 
	= 
	\frac{
		A
	}{
		\sqrt{
			2
			\left(
				1
				+
				\frac{
					f_{{\rm c},x}^2
				}{
					f_{J}^2
				}
			\right) 
			[P_J - P_{\rm fit}(f_J)] 
			\delta f 
		}
	}.
\end{equation}
Hence, the corrected values with the right units of the trap stiffness and the friction coefficient are
\begin{equation}\label{eq:kappa}
	\kappa 
	=  
	\frac{
		2\pi f_{{\rm c},x} k_{\rm B} T
	}{
		\alpha^2 D^{(1)}
	}
\end{equation}
and
\begin{equation}\label{eq:gamma}
	\gamma  
	= 
	\frac{
		k_{\rm B} T
	}{
		\alpha^2 D^{(1)}
	}.
\end{equation}

An alternative to the aforementioned method is to move the position of the center of the trap, while keeping the sample cell at rest, i.e. $x_{\rm stage} = 0$ and $x_{\rm eq}(t) = A\sin(2\pi f_{\rm eq} t)$. 
This can be achieved for example with an acousto-optic deflector or a galvomirror.
In this case, unlike the case of an external flow, an effective oscillatory force $k x_{\rm eq}(t) = k A\sin(2\pi f_{\rm eq} t)$ is exerted on the trapped particle and the corresponding Langevin equation is 
\begin{equation}\label{eq:Langevintrap}
	\frac{dx(t)}{dt} 
	= 
	- 2\pi f_{{\rm c},x} 
	\left[
		x(t) 
		- 
		x_{\rm eq}(t) 
	\right] 
	+ 
	\sqrt{2D}W_x(t).
\end{equation}
The corresponding power spectral of the particle position is given in this case by
\begin{equation}\label{eq:PSDtrap}
	P(f)=\frac{D}{\pi^2}\frac{1}{f^2+f_{{\rm c},x}^2}+\frac{A^2}{2\left(1+\frac{f_{\rm eq}^2}{f_{{\rm c},x}^2}\right)}\delta\left(f-f_{\rm eq}\right).
\end{equation}
while its discretized form is 
\begin{equation}\label{eq:PSDtrapdiscrete}
	P_j =  \frac{D}{\pi^2}\frac{1}{f_j^2+f_{{\rm c},x}^2} + \frac{A^2t_{\rm msr}}{2\left( 1+\frac{f_j^2}{f_{{\rm c},x}^2}  \right)}\delta_{j,J},
\end{equation}
Thus, once the PSD of the particle position is computed, the calibration of $\alpha$, $\kappa$ and $\gamma$ proceeds similar to the case of the oscillatory flow created by moving the stage (equations~\eqref{eq:kappafit}-\eqref{eq:gamma}). 

Active calibration methods can also be applied to less trivial situations, for instance for particles trapped in viscoleastic fluids. 
In such a case, the instantaneous drag force on the particle at time $t$ is not simply given by $\gamma \frac{dx(t)}{dt}$ as in equations~\eqref{eq:Langevinflow} and \eqref{eq:Langevintrap}, but it is related to the frequency-dependent rheological properties of the medium, which are \emph{a priori} unknown. This will be discussed in detail in section~\ref{sec:4.3:rheology}.

\subsection{Direct optical force measurement}
\label{sec:3.12:direct}

All the calibration methods discussed until now provide an indirect measurement of the optical forces, relying on the properties of the motion of the particle in the optical potential, and therefore depend on the assumption of a spherical particle and a harmonic optical potential.
It is also possible to measure directly the optical force\cite{smith2003optical, farre2010force, thalhammer2015direct}.

Direct measurement of optical forces is based on recording the change of momentum between the ingoing and outgoing light, since the exerted optical force is equal to the change in the momentum flux of the trapping light: let $I(\theta,\varphi)$ be the intensity scattered into the direction $(\theta,\varphi)$ given in spherical coordinates, then the momentum flux density $\dot{\bf p}$ of the outgoing light is
\begin{equation}
	\dot{\bf p}
	=
	{1 \over c}
	I(\theta, \varphi)
	\begin{pmatrix}
		\sin\theta\cos\varphi
		\\
		\sin\theta\sin\varphi
		\\
		\cos\theta
	\end{pmatrix}.
\end{equation}
The optical force acting on the particle corresponds to the difference between the total momentum flux of ingoing and outgoing light, where the incoming  momentum flux is given by precisely the same formula but with opposite sign. 
To correctly apply this method, one needs to detect all of the scattered light. 
Further details of how this can be achieved experimentally can be found in Refs.~\cite{smith2003optical, farre2010force, thalhammer2015direct}.
Strickly speaking this is a rather unfeasible task. However, for most practical applications in which the refractive index of the observed particle is similar to that of the solvent, light scatters only slightly, making its experimental collection possible.

\subsection{Mathematical details on error analysis and fitting}
\label{sec:3.13:maths}

Most of the calibration techniques described above use a physical observable $\Psi$ that depends on a single trajectory. 
Let us denote its experimental estimate for trajectory $m$ as $\Psi^{({\rm ex})}_m(\alpha)\equiv\Psi^{({\rm ex})}_m(\alpha,\{x_{\ell}^{(m)}\}_{\ell=1}^N)$, where $\alpha$ may denote either a time dependence or a spatial one. 
For example, in the potential method $\alpha$  represents the position associated to the histogram bins, while for the MSD $\alpha$  corresponds to the discrete time. 
We can then use the $M$ experiments to calculate the experimental mean and the experimental covariance of $\Psi$. These are:
\begin{eqnarray}
	\overline{\Psi^{({\rm ex})}}(\alpha)
	&
	=
	&
	{1 \over M}
	\sum_{m=1}^{M}\Psi^{({\rm ex})}_{m}\left(\alpha\right),
	\label{eq:meanvar1}
	\\
	\Delta^{(\Psi)}_{\alpha,\beta}
	&
	=
	&
	{1 \over M-1}
	\sum_{m=1}^{M}
		\left[
			\Psi^{({\rm ex})}_{m}\left(\alpha\right)
			-
			\overline{\Psi^{({\rm ex})}}(\alpha)
		\right]
		\left[
			\Psi^{({\rm ex})}_{m}\left(\beta\right)
			-
			\overline{\Psi^{({\rm ex})}}(\beta)
		\right]
	+
	\epsilon^2_{\eta_1,\eta_2,\ldots}(\alpha).
	\label{eq:meanvar2}
\end{eqnarray}
Importantly, the experimental covariance matrix $\Delta^{(\Psi)}_{\alpha,\beta}$ in equation \eqref{eq:meanvar2} has two terms. 
The first term corresponds to the experiment-to-experiment fluctuations that arise naturally due to the intrinsic randomness of the physical problem as well as due to the correlation between different points $\alpha$ and $\beta$. 
The second term ($\epsilon^2_{\eta_1,\eta_2,\ldots}$) captures the variances due to some known error in some variables ($\eta_1,\eta_2, \ldots$) that are needed to compute $\Psi^{({\rm ex})}_q(\alpha)$ as, for instance, uncertainties $\delta\alpha$ in the value of $\alpha$, which can be estimated by standard error propagation techniques.\footnote{Using propagation of errors, the contribution to the total variance by the errors in the uncorrelated variables $\eta_1, \eta_2, \dots$ is 
\begin{equation}\label{eq:errorprop}
	\epsilon^2_{\eta_1,\eta_2,\ldots}(\alpha)
	\approx
	{1 \over M}
	\sum_{m=1}^{M}
		\left[
			\left|
				\frac{
					\partial \Psi^{({\rm ex})}_m(\alpha)
					}{
					\partial \eta_1
					}  
			\right|^2
			\delta^2_{\eta_1}
			+
			\left|
				\frac{
					\partial  \Psi^{({\rm ex})}_m(\alpha)
					}{
					\partial \eta_2
					}
			\right|^2
			\delta^2_{\eta_2} 
			+
			\ldots
		\right],
\end{equation}
where $\delta_{\eta_1}, \delta_{\eta_2}, \ldots$ are the errors associated to the corresponding variables \cite{young2015everything}.} 

Once $\overline{\Psi^{({\rm ex})}}(\alpha)$ and $\Delta^{(\Psi)}_{\alpha,\beta}$ have been calculated, they are fitted to the corresponding physical observable $\Psi(\alpha;\boldsymbol{\theta})$  to infer the unknown parameters $\boldsymbol{\theta}=(\theta_1,\ldots, \theta_q)$ of the model.  
There are several approaches to perform this task. 
The most popular are based on least-square regressions (LSR), which defines the best estimates of the parameters as those that minimize the mean squared error.\footnote{Less common, but very powerful methods, are the robust least-squares, including the least absolute regression (LAR) and the Bisquare weights \cite{huber1981robust}, both very useful to minimize the effect of outliers.}
A fantastic book dealing with these matters is that by Peter Young \cite{young2015everything}.

The general least-square method used in this Tutorial relies on the weighted $\chi^2$ function 
\begin{equation}\label{eq:chi2}
	\chi^2(\boldsymbol{\theta})
	=
	\sum_{\alpha,\beta=1}^{P} 
		\left(
			\overline{\Psi^{({\rm ex})}}(\alpha)
			-
			\Psi(\alpha;\boldsymbol{\theta})
		\right)
		R_{\alpha,\beta}
		\left(
			\overline{\Psi^{({\rm ex})}}(\beta)
			-
			\Psi(\beta;\boldsymbol{\theta})
		\right),
\end{equation}
where $R_{\alpha\beta}$ is a symmetric semi-positive definite matrix that captures the correlation in the experimental dataset. 
The goal is to find the set of parameters $\boldsymbol{\theta}^{(\star)}$ that minimizes $\chi^2(\boldsymbol{\theta})$. 
When $R$ is taken to be $[\Delta^{(\Psi)}]^{-1}$, this is the so-called generalized least-square method or the correlated $\chi$-squared method.  
The standard weighted least-square method is recovered when the off-diagonal terms of the experimental covariance are assumed to be zero so that $[\Delta^{(\Psi)}]^{-1}_{\alpha\beta}=\text{diag}\left(1/\Delta^{(\Psi)}_{11},\ldots,1/\Delta^{(\Psi)}_{PP}\right)$. 
If we do not have access to experimental errors, we can perform the standard least-square method by assuming the matrix $\Delta^{(\Psi)}$ to be equal to the identity matrix. 
In some cases, the inverse of the sample covariance matrix is an ill-posed problem and it is highly recommended to use the standard method of weighted least squares but including the experimental covariance matrix in the error estimation  of the parameters \cite{fogelmark2018fitting}. This is the approach we have used to estimate errors for the MSD and ACF methods.

If the expression $\chi^2(\boldsymbol{\theta})$ is a simple function of the parameters, the minimization can be carried out analytically.  
For example, this is the case when $\Psi(\alpha;\boldsymbol{\theta})$ is linear with respect to the model's parameters, that is
\begin{equation}\label{eq:linearmodel}
	\Psi(\alpha;\boldsymbol{\theta})
	=
	\sum_{j=1}^q \theta_j\Phi_j(\alpha),
\end{equation}  
where $\{\Phi_j(\alpha) \}$ is a set of functions, giving rise to the well established linear least-square fitting. In some cases, a simple redefinition of variables can transform a non-linear fitting into a linear one, as, for instance,  in the potential method.  

When it is not possible to obtain explicit forms for $\boldsymbol{\theta}^{(\star)}$, one must rely on numerical minimizers, being the standard ones the Levenberg-Marquardt-Girard-Wynne-Morrison and the Trust-Region, implemented already in most of the standard computational packages for mathematical analysis, such as Python,  MATLAB, Mathematica, and R.
However, the computational methods given in these packages do not normally allow to use the experimental covariance matrix to estimate the covariance matrix of the estimated parameters. 
In this case, it is recommended to use the packages provided in \cite{fogelmark2018fitting}, which we also have implemented in this Tutorial.

\section{Applications}
\label{sec:4:applications}

We will now describe how optical tweezers have been used in several advanced applications. 
For each field of application, we provide a brief introduction and a detailed description of some paradigmatic experiments. 
In this section, we focus on experiments with particles optically trapped in a liquid environment, while we discuss experiments with particles optically trapped in air or vacuum in section~\ref{sec:5:vacuum}.

\subsection{Single-molecule mechanics}
\label{sec:4.1:molecule}

The study of single-molecule mechanics using force spectroscopy techniques has rapidly expanded over the past three decades \cite{ritort2006single , deniz2007single, neuman2007single, moffitt2008recent, neuman2008single, zhang2013high, capitanio2013interrogating, miller2017single}. 
Many labs worldwide use single-molecule methods to manipulate and track individual molecules. 
Broadly speaking, single-molecule methods can be classified into two large families depending on whether they manipulate molecules by exerting forces or they visualize molecules in action using appropriate markers. 
The latter include single-molecule fluorescence used to monitor the time evolution of fluorescently labeled molecular complexes {\it in vitro} and {\it in vivo}. 
Manipulation methods are capable of exerting forces from femtonewton to nanonewton. 
For example, they have been used to study the elastic properties of biopolymers \cite{camunas2016elastic}, the thermodynamics and kinetics of intramolecular (e.g., molecular folding)\cite{woodside2014reconstructing, schonfelder2016power} and intermolecular (e.g., ligand binding) interactions\cite{manosas2017single}, molecular motors\cite{spudich2011optical}, and fundamental problems in nonequilibrium physics\cite{ritort2008nonequilibrium, ciliberto2017experiments}. 
The field of force spectroscopy includes techniques such as laser optical tweezers \cite{jones2015optical}, magnetic tweezers \cite{sarkar2016guide}, atomic force microscopy\cite{zlatanova2000single}, acoustic force spectroscopy\cite{sitters2015acoustic}, electrostatic trapping \cite{ruggeri2017single}, microfluidics \cite{streets2014microfluidics}, microneedles\cite{yanagida2011single}, biomembrane force probe \cite{evans1995sensitive}, and centrifugal force\cite{wong2011massively}. 
These techniques tend to compromise between measurement accuracy and high-throughput capabilities. 
While magnetic tweezers and acoustic force spectroscopy achieve high throughput, optical tweezers and atomic force microscopy remain the most accurate techniques in the low force (piconewton) and high force (nanonewton) regimes, respectively. Piconewton forces are the most relevant at the molecular level, making optical tweezers an excellent technique to accurately characterize molecular reactions. 

In this section, we describe in detail how optical tweezers can be used to investigate molecular folding using short DNA hairpins (a double-helix stem of a few tens of base pairs terminated by a loop) as model systems. 
We will describe how to synthesize a short DNA hairpin, how to unzip it by stretching it while holding it between a micropipette and an optical tweezers, and how to analyze the obtained data. 
This experiment provides an ideal example of the use of single-molecule mechanics to determine elastic behaviour of biopolymers, to measure the intramolecular weak interactions that stabilize native molecular structures, and to extract the free energy of biomolecule formation from out-of-equilibrium pulling experiments. 
The experimental setup we use is based on the miniTweezers setup developed by Smith, Cui and Bustamante \cite{smith2003optical}.

\subsubsection{Microfluidics chamber}
\label{sec:4.1.1:chamber}

\begin{figure}[h!]
	\centering
	\includegraphics[width=10cm]{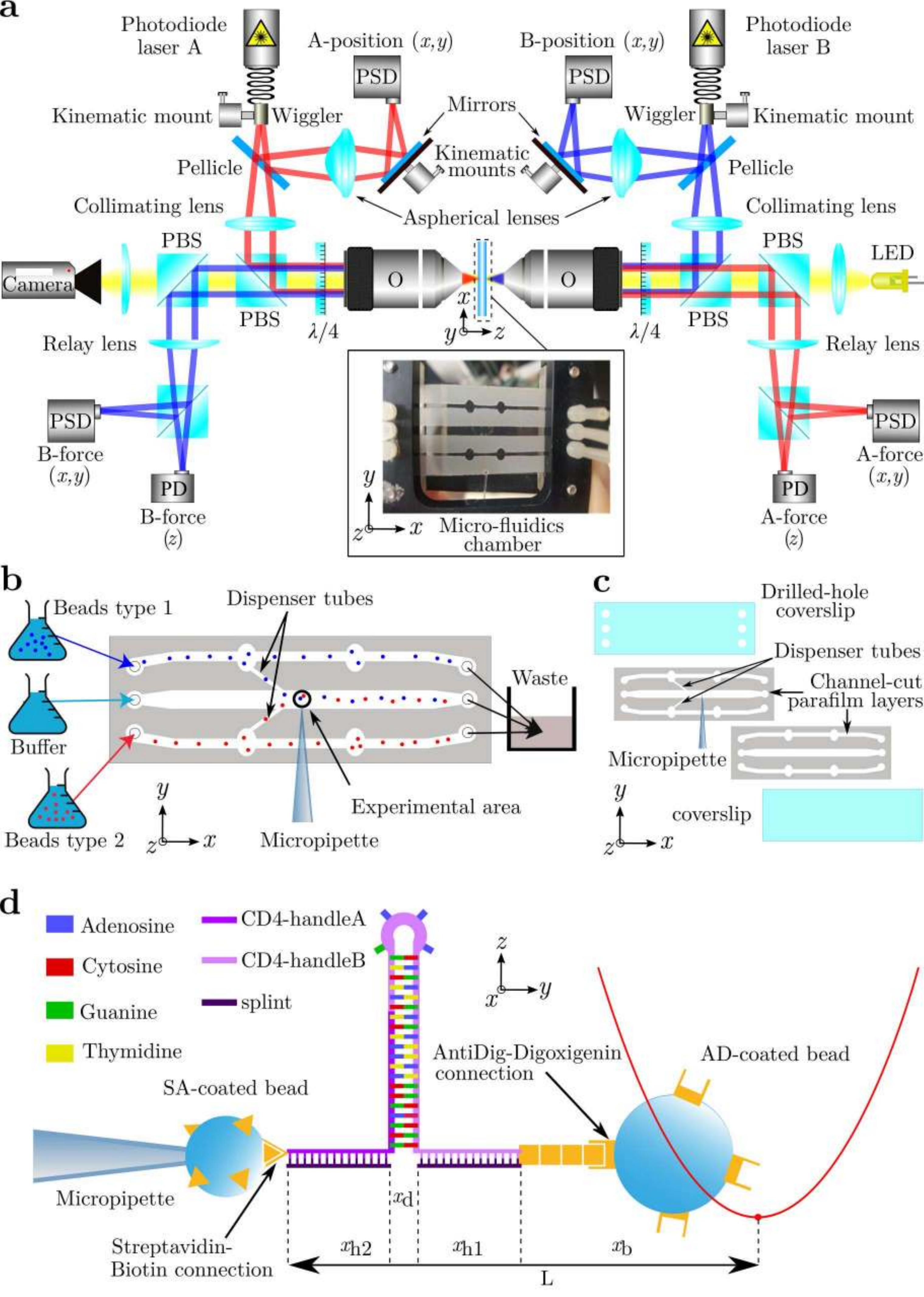}
	\caption{
	{\bf Single-molecule pulling experiments:  miniTweezers setup.}
	Schematic representations (a) of the miniTweezers setup,
	(b) of the microfluidics chamber (the flow goes from left to right, and the laser beams propagate perpendicularly to the surface of the chamber),
	(c) of the assembly procedure for the construction of the chamber, 
	and (d) of the molecular configuration for the DNA hairpin pulling experiments (where the DNA hairpin is held between a bead held by a micropipette and a bead captured by the optical trap, depicted as an harmonic potential). Note that  $x_{\rm h2}$, $x_{\rm d}$, $x_{\rm h1}$, and $x_{\rm b}$ are distances along the $y$-axis.
	}
	\label{fig:14:miniTweezers}
\end{figure}

The schematic of the experimental setup is presented in Fig.~\ref{fig:14:miniTweezers}(a).
The experiments are performed in a microfluidics chamber placed vertically within the setup.
As schematically shown in Figs.~\ref{fig:14:miniTweezers}(b) and \ref{fig:14:miniTweezers}(c), this microfluidics chamber has three channels: we will refer to the three channels as the upper, central and lower channel. 
The propagation of both laser beams is perpendicular to the chamber surface (z-axis).
The experimental area is restricted to the central channel, where the object of study (biomolecule or cell) is held by two beads, one held by the optical trap and the other held by a glass micropipette, as shown in Fig.~\ref{fig:14:miniTweezers}(d).
The upper and lower channels are used to supply the two types of coated beads used in the experiments.

The microfluidics chamber is realized by sandwiching two layers of parafilm (Parafilm M, Bemis) between two coverslips (No. 2, dimensions $24\,{\rm mm} \times 60\,{\rm mm} \times 200\,{\rm \upmu m}$), as shown in Fig.~\ref{fig:14:miniTweezers}(c). 
In detail, the steps to prepare the chamber are:
\begin{enumerate}
	\item The entrance and exit holes are drilled in one of the glass coverslips using a laser cutter. The coverslips are cleaned with a solution of 70\% ethanol.\footnote{Although 100\% ethanol can also be used, a 70\% solution is preferred because it prevents the sporulation of some microorganisms.}
	\item The three channels are drawn in the two parafilm layers using a laser cutter.
	\item One parafilm layer is attached on the drilled glass coverslip. 
	\item A glass micropipette with a diameter of $\sim 1\,{\rm \upmu m}$ is produced by heating and pulling a glass tube (King precision glass, Inc., inside diameter=$0.04\,{\rm mm}$, outside diameter=$0.08\,{\rm mm}$, length=$6.00\,{\rm mm}$, glass type KG-33), as described in section \ref{sec:4.1.3:pipette}.
\item This glass micropipette is placed on top of the parafilm layer perpendicular to the channels with the tip positioned in the central channel, as shown in Figs.~\ref{fig:14:miniTweezers}(b) and \ref{fig:14:miniTweezers}(c).
	\item The upper and lower channels are connected to the central one via glass dispenser tubes (King precision glass, Inc., inside diameter=$0.04\,{\rm mm}$, outside diameter=$0.10\,{\rm mm}$, length=$6.00\,{\rm mm}$, glass type KG-33) cut using a diamond-tip cutter to obtain a clean cut.
	\item The second parafilm layer is placed on top and the chamber is closed using the second glass coverslip (previously cleaned with 70\% ethanol), as shown in Fig.~\ref{fig:14:miniTweezers}(c).
	\item The chamber is sealed by heating it on a hot plate at $120^\circ{\rm C}$ while exerting a pressure of about $1\,{\rm Kg}$ in all the chambers' surface (either by placing a weight on top of the microfluidics chamber or, more simply, by exerting pressure by hand). To prevent the glass from breaking, it is recommended to sandwich the chamber between two thicker glass slides to homogenize the pressure applied to their surfaces.
	\item The chamber is placed in the metallic mount, as shown in the inset in Fig.~\ref{fig:14:miniTweezers}(a). The correct alignment between the holes drilled in the glass coverslip and the holes in the mount is critical to ensure the flow in the chamber. It is also very important to tighten the screws to avoid buffer losses between the plastic tubes and the glass chamber, since this would cause flows inside the chamber. Do this carefully, because tightening the screws too much could break the glass coverslip.
	\item The side of the micropipette coming out of the chamber is cut to get rid of the excess tube, leaving only about $\sim 3\,{\rm cm}$, and is introduced into a polyethylene tube (polyethylene tubing, Warner Instruments, PE-10/100; outside diameter = $0.61\,{\rm mm}$, inside diameter = $0.28\,{\rm mm}$). This tube is then fixed to the mount with tape. The pipette is easily breakable, so the tube needs to be placed as straight as possible. The connection is then sealed by using a special glue (Norland, NOA-61; UV Curing Optical Adhesives). The glue is placed between the the glass tube of the pipette and the polyethylene tube to fill the void outer space between tube and pipette via capillarity. After observing that the glue has entered into the tube, the UV-curation is performed by leaving the whole chamber for $\sim 25\,{\rm min}$ under a UV lamp. Since the glue could reach the end of the glass tube of the micropipette and block it, the chamber has to be placed under the UV-radiation as fast as possible. After the curation, a syringe ($1\,{\rm ml}$ Luer Lock, HSW SOFT-JECT U100 Insulin Henke Sass Wolf) is inserted using a needle (BD Microlance $30{\rm G} \times 1/2"$ -- $0.30\,{\rm mm} \times 13\,{\rm mm}$) at the end of the tube to create suction at the tip of the micropipette.
	\item The holes of the mount are connected with silicone-rubber tubes (Tygon 3350, Saint-Gobain, .031 ID X .093 OD X 50 FT TYGON 335) via nylon socket screws (Nylon set $8-32 \times 3/8"$, Product-Components, previously drilling a hole with a number 45 drill bit, of $0.82"$), as shown in Fig.~\ref{fig:14:miniTweezers}(c). A segment of $\sim20\,{\rm cm}$ polyethylene tube (Polyethylene tubing, Warner Instruments, PE-50/100: outside diameter = $0.97\,{\rm mm}$, inside diameter = $0.58\,{\rm mm}$) is inserted into the silicone-rubber tubes. The three exit tubes are connected to a trash (any small plastic container with a capacity volume $\sim 100\,{\rm ml}$), while the three entry channels are connected to a syringe using a polyethylene tube (Polyethylene tubing, Warner Instruments, PE-50/100: outside diameter = $0.97\,{\rm mm}$, inside diameter = $0.58\,{\rm mm}$) which is inserted into the silicone-rubber tubes on one end, and a needle (HSW FINE-JECT $23{\rm G} \times 1"$ -- $0.6\,{\rm mm} \times 25\,{\rm mm}$).
\end{enumerate}

\subsubsection{The miniTweezers setup}
\label{sec:4.1.2:experimental}

The miniTweezers \cite{smith2003optical} (Fig.~\ref{fig:14:miniTweezers}(a)) consists of two focused counter-propagating laser beams ($P=200\,{\rm mW}$, $\lambda=845\,{\rm nm}$) that create a single optical trap \cite{ribezzi2013counter}. 
It employs high-NA objectives (${\rm NA} = 1.2$), but underfills them  to be able to collect almost all scattered light to measure the change of light momentum using position sensitive detectors (PSDs), as described in section~\ref{sec:3.12:direct}. 
Before the objectives, a pellicle diverts $\sim$8\% of each laser beam to a secondary PSD to determine the optical trap position. 
The remaining $\sim 92\%$ is collimated by using a lens and it is introduced into the optical axis by using a polarizing beam-splitter that selects the horizontally polarized light. Then the linearly polarized light of each laser is circularly polarized by a quarter waveplate\footnote{The use of quarter-wave plates ensures that the light coming from the two laser beams do not interact with each other and guarantees that the light reflected from the particle is not returned to the laser but is reflected to the opposite PSD \cite{huguet2010statistical}.}. The laser beams are focalized by their corresponding objectives and form the optical trap, which can hold dielectric objects with a refraction index higher than the surrounding aqueous medium (e.g., polystyrene beads) and exert forces to them. The exiting light is collected by the opposite objective and is converted to vertically polarized light by another quarter-wave plate. The vertically polarized light can be extracted from the optical path using two polarizing beam-splitters and relay lenses that redirect the light to the PSDs that measure the intensity of the beam (i.e., its exerted force). The miniTweezers has a resolution of $0.1\,{\rm pN}$ and $1\,{\rm nm}$ at a $1\,{\rm kHz}$ acquisition rate. 

The chamber where the experiments are performed (section~\ref{sec:4.1.1:chamber}) is placed between the two objectives and held with a metallic mount with three stages to permit movement along the $x$-, $y$- and $z$-directions.
There are two possible ways to manipulate the position of the optical trap with respect to the chamber depending on the precision that is needed.
For large displacements (up to hundreds of micrometers), the whole chamber is displaced along the $x$-, $y$- and $z$-directions using stepmotors.
For fine displacements (less than a few micrometers, but with a resolution of $\sim 1\,{\rm nm}$ and rates of displacement up to $\sim 50\,{\rm nm\,s^{-1}}$), the optical trap is displaced along the $x$- and $y$-directions by a 2D piezoelectric motor attached to the tip of the optical fiber (wiggler) used to inject the laser into the optical setup. 

Before any experiment, the setup needs to be aligned going through the following steps:
\begin{enumerate}
	\item Remove any trapped object.
	\item Move the chamber with the stepmotors so that the micropipette is a few micrometers away from the optical trap, in the so-called ``working zone''.
	\item Remove any voltage applied to the piezoelectric motor.
	\item Turn off the LED and remove the light filter, to be able to see the optical trap on the camera.
	\item Decrease the laser power until it is possible to distinguish both lasers. 
	\item Using the screws of the kinematic mount, move the B laser (the one that has the real image shown on the screen) until it is superposed on the A laser.
	\item Put back on the light filter, turn on the LED, and increase the laser power to its working value.
	\item Set the current values recorded by the PSD as the zero force baseline along the $x$- and $y$-directions (center of the light spot) and along the $z$-direction (size of light spot).
	\item Trap a bead with the optical trap by moving the chamber with the stepmotors close to the appropriate dispenser tube.
	\item Bring the trapped bead to the working zone.
	\item Fine-tune the alignment of the B laser using the screws of the kinematic mount controlling the wiggler of its fiber so that the xy-force signals from both force-PSDs are as close as zero as possible, i.e., both lasers exert an xy-force as close to zero as possible.
	\item Fine-tune the position of one of the objectives along the $z$-directions so that the z-force from both force-PSDs is as close to zero as possible.
	\item For both lasers, move the kinematic mount of the mirror that deflects the light that gets towards the position-PSD (between the aspherical lenses and the PSD) until its signal is zeroed (i.e., the light hits the center of the position-PSD). 
\end{enumerate}

\subsubsection{Pipette making}
\label{sec:4.1.3:pipette}

\begin{figure}[b!]
	\centering
	\includegraphics[width=8cm]{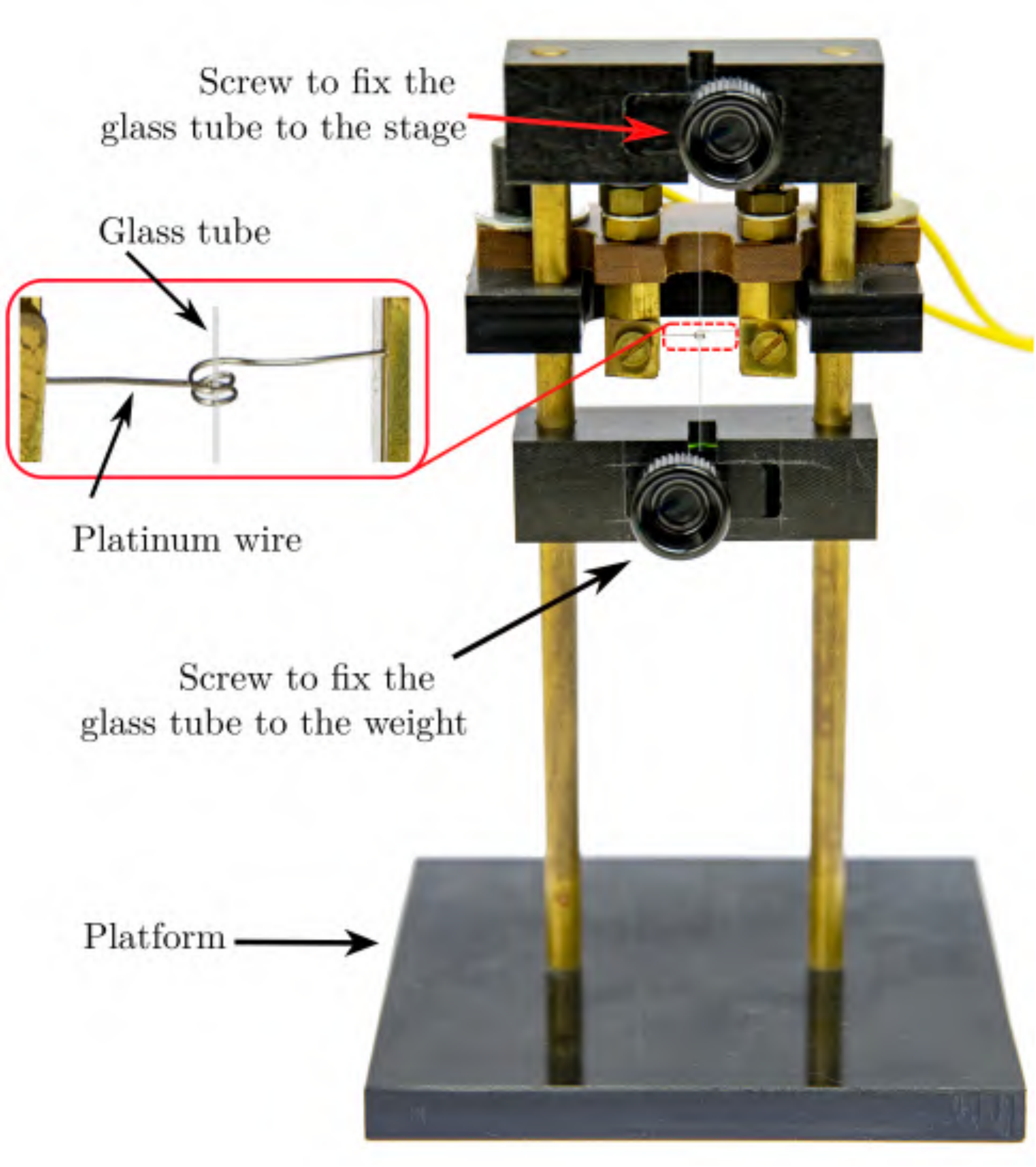}
	\caption{
	{\bf Pipette puller.} 
	First, the glass tube is carefully centered within a coiled platinum wire, as shown in the inset. 
	Then, one end of the tube is attached to the puller, while the other is attached to a weight that pulls the tube down. 
	Finally, an electric current intensity ramp is applied through the platinum wire, heating the adjacent glass tube, which during the melting is pulled down, creating a micropipette.
	}
	\label{fig:15:puller}
\end{figure}

A micropipette is used to hold a microparticle by air suction, as shown in Fig.~\ref{fig:14:miniTweezers}(b). The tip of this micropipette needs to have an inside diameter $\sim 1\,{\rm \upmu m}$, large enough to exert sufficient suction and small enough not to let the microparticles flow inside it. 
While one might use a commercial pipette puller to produce the pipettes, here, we will explain how a glass tube can be pulled to produce such micropipette using the homemade device shown in Fig.~\ref{fig:15:puller} \cite{smith2003optical, pipettepuller}.

This homemade pipette puller consists of a plastic platform with two parallel metallic bars, which hold a stage and along which a plastic weight can slide up and down. 
The two ends of the glass tube are fixed with screws to the stage and the weight, respectively. Thus, the glass tube is held in tension due to the pull exerted by gravity on the weight. 
The central part of the glass tube passes through a platinum wire; the two ends of the wire are connected to an electric supply, which provides a current ramp from $0$ to $\sim6\,{\rm A}$ in $\sim 8\,{\rm s}$. The exact maximum intensity and time of the ramp have to be tuned to get the appropriate diameter and shape of the pipette (in general, longer times and higher intensities correspond to smaller tips). 
Specifically, the steps required to produce a micropipette are:
\begin{enumerate}
	\item Unplug the wire from the electric supply and place the pipette puller horizontally.
	\item Insert the glass tube through the stage, the platinum coil, and the plastic weight.
	\item Carefully center the glass tube within the platinum wire.
	\item Screw the glass tube at the stage and plastic weight.
	\item Place the platform vertically (as shown in Fig.~\ref{fig:15:puller}), carefully to avoid any sudden hit that may break the glass tube.
	\item Apply the current ramp. The weight will drop into the platform. The geometry of the platinum wire is of a critical importance to obtain the proper shape and size of the micropipette. For that reason, when the weight drops into the platform, we have to ensure that the remaining glass tube of the stage is not touching the platinum wire, otherwise it would get stuck to it.
	\item Unscrew the plastic weight and (with extreme caution and preferably with ethanol-cleaned tweezers) take the micropipette and place it into the chamber being built.\footnote{The micropipette easily breaks or gets blocked by dust particles at the minimum contact of the tip of the glass tube with anything, so it needs to be handled with extreme care.}
\end{enumerate}

\subsubsection{Preparation of the beads}
\label{sec:4.1.4:bead}

Biomolecules such as DNA, RNA and proteins are chains whose units are either nucleic bases or aminoacids \cite{nelson2004biological}. 
These molecules can be conveniently manipulated through beads attached at their ends. 
Antigen-antibody connections can be used to attach a specific bead at each end, taking advantage of the fact that these connections are extremely specific \cite{sela2013structural}. 
In particular, avidin/streptavidin--biotin\footnote{The interaction between streptavidin and biotin is one of the strongest non-covalent interactions in Nature \cite{chivers2011biotin}.} and digoxigenin--anti-digoxigenin connections are often employed in single-molecule force-spectroscopy experiments, because they can hold forces up to $\sim 100\,{\rm pN}$ at typical optical-tweezers loading rates of $\sim 1\,{\rm pN\,s^{-1}}$ \cite{pincet2005solution}.

Each bead is coated with a different molecule, which specifically binds to the cognate tails.
Streptavidin-coated (SA) beads can be directly purchased (SPHERO streptavidin -- polystyrene particles, $0.5\%\,{\rm w/y}$, $2.17\,{\rm \upmu m}$, $5\,{\rm ml}$). 
Anti-digoxigenin-coated (AD) beads are purchased as G-protein-coated polystyrene beads (Kisker Biotechnologies -- G-coated polystyrene particles, $0.5\%\,{\rm w/y}$, $3.18\,{\rm \upmu m}$, $5\,{\rm ml}$), which must then be activated with anti-digoxigenin. 
The difference in size between these beads permits one to easily distinguish them by microscopy.

\paragraph{Buffers.}

The following buffers are required for the preparation of the beads: 
\begin{description}
	\item[PBS (pH~7.0)]
NaCl $0.14\,{\rm M}$, KCl $2.7\,{\rm mM}$, K$_2$HPO$_4$ $\cdot$ 3H$_2$O $61\,{\rm mM}$, KH$_2$PO$_4$ $39\,{\rm mM}$, NaN$_3$ (sodium azide) $0.02\%$. To prepare $50\,{\rm ml}$: fill $\sim 40\,{\rm ml}$ of a $50\,{\rm ml}$ Falcon tube with Milli-Q water; add $0.406\,{\rm g}$ of NaCl, $0.01\,{\rm g}$ of KCl, $0.696\,{\rm g}$ of K$_2$HPO$_4$ $\cdot$ 3H$_2$O, $0.265\,{\rm g}$ of KH$_2$PO$_4$, and $1\,{\rm g}$ of NaN$_3$; dissolve using a magnetic mixer; add Milli-Q water until reaching $50\,{\rm ml}$; check the pH and add NaOH until the solution reaches a pH~7.0.
	\item[PBS (pH~7.4)]
NaCl $0.14\,{\rm M}$, KCl $2.7\,{\rm mM}$, K$_2$HPO$_4$ $\cdot$ 3H$_2$O $80.2\,{\rm mM}$, KH$_2$PO$_4$ $20\,{\rm mM}$, NaN$_3$ (Sodium azide)  $0.02\%$. For preparing $50\,{\rm ml}$: follow the same procedure as for the previous buffer, adjusting to pH~7.4.
	\item[Antibody crosslinker buffer (pH~7.4)]
Na$_2$HPO$_4$ $100\,{\rm mM}$, NaCl $100\,{\rm mM}$.  To prepare $10\,{\rm ml}$: fill $\sim 7\,{\rm ml}$ of a $10\,{\rm ml}$ Falcon tube with Milli-Q water; add $0.142\,{\rm g}$ of Na$_2$HPO$_4$ and $0.058\,{\rm g}$ of NaCl; shake well until the salts have dissolved; add Milli-Q water until the $10\,{\rm ml}$ of total volume is reached; check the pH and add NaOH until the solution reaches pH~7.4.
\end{description}
All salts can be acquired from any chemical distributor (e.g., Sigma Aldrich). All products have to be of a biomolecular grade of purity. The water for preparing all buffers has to be Milli-Q water.\footnote{Milli-Q water is obtained by filtering the source water (usually distilled water) first through mixed bed ion exchange and organics (activated charcoal) cartridges, and then through a filter which removes any intact organisms. Usually a UV lamp completes the purification process.}
For the RNA experiments, use RNAse-free water for all preparations (and make sure the pHmeter is also cleaned with RNAse-free water).

\paragraph{SA beads.}

For the already coated SA beads, the protocol consists in exchanging the buffer where they are preserved with PBS (pH~7.4).
The procedure ($\sim 30\,{\rm min}$) for the preparation of $1\,{\rm ml}$ of SA beads (which should be sufficient to perform experiments for several months) is:
\begin{enumerate} 
	\item Homogeneously resuspend the SA beads. To do so, place the container with the purchased particles into a vortex mixer for several seconds. An additional step of several seconds of sonication further improves the resuspension.
	\item Take $1\,{\rm ml}$ of the resuspended SA beads and place it in a new Eppendorf tube.
	\item Centrifuge the Eppendorf tube at $10000\,{\rm rpm}$ for $5\,{\rm min}$. The dense SA beads will precipitate, while the buffer will float above them.
	\item Extract the overnatant buffer and add $1\,{\rm ml}$ PBS (pH~7.4). Mix the SA beads and buffer. Sonicate a few seconds and centrifuge at $10000\,{\rm rpm}$ for $5\,{\rm min}$, again. Extract the overnatant, resuspend in $1\,{\rm ml}$ PBS (pH~7.4) and sonicate during several seconds. This step is performed to exchange the buffer.
	\item Aliquote in 20 tubes ($50\,{\rm \upmu l}$ each). This allows for an optimal sterile preservation of the beads.
	\item Store at $4^\circ{\rm C}$ for up to $\sim6$ months.
\end{enumerate}

\paragraph{AD beads.}

The protocol for the preparation of the AD beads consists of three steps: 
(1) exchange the preservation buffer; 
(2) attach the anti-digoxigenin to the G-protein;
(3) exchange the crosslinking buffer.
The procedure ($\sim 90\,{\rm min}$) for the preparation of $1\,{\rm ml}$ of AD beads (which should be sufficient to perform experiments for several months) is:
\begin{enumerate}
	\item (The first time the anti-digoxigenin batch is dissolved.) Add $200\,{\rm \upmu l}$ of PBS (pH~7.4) to dilute the dry anti-digoxigenin (sheep polycolonal anti-dig antibody, Roche 1333 089).
	\item Prepare the dimethyl pimelimidate (DMP) crosslinker buffer by mixing $50\,{\rm mg}$ DMP and $200\,{\rm \upmu l}$ antibody crosslinker buffer (pH~7.4). The DMP crosslinker buffer needs to be freshly prepared every time the AD beads are synthesized.
	\item Homogeneously resuspend the G-coated beads. To do so, place the container with the purchased particles into a vortex mixer for several seconds. An additional step of several seconds of sonication improves the resuspension.
	\item Centrifuge at $5000\,{\rm rpm}$ for $2\,{\rm min}$. The dense beads will precipitate, while the buffer float above them.
	\item Extract the overnatant buffer and add $1\,{\rm ml}$ antibody crosslinker buffer (pH~7.4). Mix the beads and the buffer, sonicate for a few seconds, and centrifuge at $5000\,{\rm rpm}$ for $2\,{\rm min}$.
	\item Repeat the previous step (washing) and resuspend within $1\,{\rm ml}$ antibody crosslinker buffer (pH~7.4).
	\item Add $60\,{\rm \upmu l}$ of anti-DIG antibody and $30\,{\rm \upmu l}$ of freshly dissolved DMP crosslinker buffer.\footnote{The volume of added anti-DIG antibody will depend on the coating of the beads: it may need to be higher if the vendor supplies the beads with a higher density of coating. Nevertheless, the proportions of anti-DIG and DMP crosslinker buffer have to be preserved (2:1).}
	\item Tumble at room temperature for $60\,{\rm min}$. 
	\item Centrifuge at $5000\,{\rm rpm}$ for $2\,{\rm min}$. 
	\item Wash twice with $1\,{\rm ml}$ PBS (pH~7.0), resuspend with the same buffer, and sonicate during several seconds. 
	\item Aliquote in 20 tubes ($50\,{\rm \upmu l}$ each). 
	\item Store at $4^\circ{\rm C}$ for up to $\sim6$ months.
\end{enumerate}

\paragraph{Molecular buffer.}

A typical molecular buffer for DNA pulling experiments consists of $10\,{\rm mM}$ Tris pH~7.5\footnote{TRIS is used in the formulation of buffer solutions in the pH range between 7.5 and 8.5.} (Trizma, Sigma Aldrich), $10\,{\rm mM}$ EDTA\footnote{EDTA is widely used for scavenging metallic ions, including divalent ones, which most enzymes need to be active. For this reason, it is widely used as a food preservative or stabilizer. In our case, it inactivates DNAses and RNAses, preventing nucleic acid degradation.} (EDTA, Sigma Aldrich), $1\,{\rm M}$ NaCl, and $0.01\%$ NaN$_3$ (sodium azide, to avoid bacterial growth). To facilitate the molecular buffer preparation, it is convenient to prepare stocks of:
\begin{description}
	\item[1 M Tris pH~7.5] 
Fill $\sim 40\,{\rm ml}$ of a $50\,{\rm ml}$ Falcon tube with Milli-Q water. Add $6.05\,{\rm g}$ of Tris-HCl. Dissolve using a magnetic mixer. Add Milli-Q water until reaching $50\,{\rm ml}$. Check the pH and add a solution of $25\%$ HCl until the solution reaches pH~7.5. For longer storage, a final auto-cleavage can be performed.
	\item[EDTA 0.5 M pH~8.0]
Fill $\sim 40\,{\rm ml}$ of a $50\,{\rm ml}$ Falcon tube with Milli-Q water. Add $7.306\,{\rm g}$ of EDTA. Add NaOH to the solution (the EDTA does not dissolve in water if pH$<$7.5). Dissolve using a magnetic mixer. Add Milli-Q water until reaching $50\,{\rm ml}$. Check the pH and add NaOH until the solution reaches pH~8.0.
	\item[5 M NaCl] 
Fill $\sim 40\,{\rm ml}$ of a $50\,{\rm ml}$ Falcon tube with Milli-Q water. Add $14.49\,{\rm g}$ of NaCl. Dissolve using a magnetic mixer and heat the tube to facilitate it. Add Milli-Q water until reaching $50\,{\rm ml}$.
	\item[1\% NaN$_3$]
Fill $\sim 40\,{\rm ml}$ of a $50\,{\rm ml}$ Falcon tube with Milli-Q water. Add $0.5\,{\rm g}$ of NaN$_3$. Dissolve using a magnetic mixer. Add Milli-Q water until reaching $50\,{\rm ml}$.
\end{description}
After the stocks have been prepared, for preparing $50\,{\rm ml}$ of the molecular buffer: Pour $\sim 30\,{\rm ml}$ of Milli-Q water in a $50\,{\rm ml}$ Falcon tube. Add $10\,{\rm ml}$ of the $5\,{\rm M}$ NaCl stock, $1\,{\rm ml}$ of the $0.5\,{\rm M}$ EDTA pH~8.0 stock, $0.5\,{\rm ml}$ of the $1\,{\rm M}$ Tris pH~7.5 stock, and $0.5\,{\rm ml}$ of the 1\% NaN$_3$ stock. Add Milli-Q water until reaching $50\,{\rm ml}$ of volume. Mix the molecular buffer and filter it (Sterile Syringe Filter, ${\rm w/0.2\,\upmu m}$ Cellulose, Acetate Membrane, VWR International) introducing the filtered solution in a new $50\,{\rm ml}$ Falcon tube.

\paragraph{Incubation of the beads.}

To finalize the preparation of the beads, incubate in an Eppendorf tube ($\sim1.5\,{\rm ml}$) $1\,{\rm \upmu l}$ of a solution of the biomolecule of interest\footnote{Typically, the molecule is concentrated and needs to be diluted 1:10 to 1:100.} mixed with $14\,{\rm \upmu l}$ of buffer (where the experiments are going to be performed) and $5\,{\rm \upmu l}$ of the previously prepared AD beads. After $\sim 25\,{\rm min}$ have passed, add $1\,{\rm ml}$ of the molecular buffer. 

For the SA beads, no incubation is required, dilute $1\,{\rm \upmu l}$ of SA beads in $1\,{\rm ml}$ of the molecular buffer in an Eppendorf tube.

\subsubsection{Synthesis of a short DNA hairpin}

As shown in Fig.~\ref{fig:14:miniTweezers}(d), a typical single-molecule experiment is performed using a short DNA hairpin of a few tens of basepairs (bp). A DNA hairpin is a single-stranded DNA molecule that forms a double helix stem ended by a loop.  
The hairpin is flanked by two double stranded helices that act as handles. Handles are molecular spacers used to manipulate the DNA hairpin. These spacers also avoid non-specific interactions between the DNA hairpin and the beads when performing the experiments. 
The handles are tagged with biotin and digoxigenin at one end to specifically attach them to the coated beads. The 29-bp handles are chosen because they provide higher rigidity than longer handles and higher signal-to-noise ratio measurements \cite{forns2012improving}.

To synthesize short DNA hairpins with 29-bp dsDNA handles, the desired DNA sequence (that we will denote as CD4) is purchased in a series of oligonucleotides that self-assemble into the hairpin and handles\cite{forns2012improving}. Briefly, a 29-bp sequence is used to make the handles, with the hairpin sequence flanked by the handles at both sides (Table~\ref{tab:5:oligos}). For very short hairpins ($\leq 13\,{\rm bp}$), the hairpin sequence can be ordered as a single oligonucleotide; however, for longer hairpins, it is useful to split the molecular construct into two oligonucleotides that are annealed and ligated together. Finally, the dsDNA handles are created by annealing a third oligonucleotide that is complementary to the handle region (``splint''). The sequences of the oligonucleotides used for the CD4 DNA hairpin used in this section are shown in Table~\ref{tab:5:oligos}. 
One oligonucleotide (Table~\ref{tab:5:oligos}, CD4-handleA) is purchased with a biotin at its beginning (5'-biotinylated), whereas the oligonucleotide containing the opposite handle (Table~\ref{tab:5:oligos}, CD4-handleB) is purchased 5'-phosphorylated and then tailed at its 3'-end with multiple digoxigenins. All three oligonucleotides can be purchased from companies such as Eurofins, Fisher Scientific or Sigma Aldrich.

\begin{table}[h!]
	\footnotesize
	\begin{center}
	\begin{tabular}{p{2cm}|p{10cm}} 
		\textbf{Name} 
		& 
		\textbf{Sequence} 
		\\
	\hline
		CD4-handleA 
		& 
		5$'$-biotin-AGT TAG TGG TGG AAA CAC AGT GCC AGC GC\textcolor{blue}{G CGA GCC ATA A}T-3$'$  
		\\
	\hline 
		CD4-HandleB 
		& 
		5$'$-Phos- CTC ATC TG\textbf{G AAA} CAG ATG AGA TTA TGG CTC G\textcolor{blue}{CG ACT TCA CTA ATA CGA CTC ACT ATA GGG A} -3$'$
		\\
	\hline 
		splint 
		& 
		5$'$-\textcolor{blue}{GCG CTG GCA CTG TGT TTC CAC CAC TAA CT}-3$'$
		\\
	\hline 
	\end{tabular}
	\caption{
	{\bf Oligonucleotides.}
	Oligonucleotides used to synthesize the hairpin by a tailing, annealing and ligation reaction. The splint is used to the create the 29-bp double-stranded DNA handles. The hairpin loop is highlighted in bold and the sequences corresponding to the handles are highlighted in blue.
	}
	\label{tab:5:oligos}
	\end{center}
\end{table}

To complete the DNA hairpin synthesis, three main reactions have to be performed:
\begin{enumerate}
	\item Tailing reaction: As we have already noted, one oligonucleotide is already purchased with the biotin at its 5' end. However, to be able to stretch the hairpin, digoxigenins need to be added to the other end of the hairpin by performing the tailing reaction described in Table~\ref{tab:6:tailing}.

\begin{table}[h!]
	\footnotesize
	\begin{center}
	\begin{tabular}{lc}
	\hline 
		Milli-Q water 
		& 
		$8\,{\rm \upmu l}$  
		\\
		5X CoCl$_2$ solution* 
		&
		$4\,{\rm \upmu l}$ 
		\\
		5X reaction buffer* 
		& 
		$4\,{\rm \upmu l}$  
		\\
		10 mM dATP (Sigma Aldrich) 
		& 
		$1\,{\rm \upmu l}$  
		\\
		1 mM DIG-dUTP (Sigma Aldrich) 
		& 
		$1\,{\rm \upmu l}$  
		\\
		100 $\upmu$M CD4-HandleB 
		& 
		$1\,{\rm \upmu l}$  
		\\
		Terminal transferase ($400\,{\rm U\,\upmu l^{-1}}$, Sigma Aldrich) 
		& 
		$1\,{\rm \upmu l}$  
		\\
	\hline 
		Total Volume 
		& 
		$20\,{\rm \upmu l}$ 
		\\
	\hline 
	\end{tabular}
	\caption{
	{\bf Oligonucleotide tailing reaction.} 
	Mix all the items in an Eppendorf tube and keep it at $37^\circ{\rm C}$ for $15\,{\rm min}$. After the reaction is finished, add $1$ to $2\,{\rm \upmu l}$ of $0.5\,{\rm M}$ EDTA pH~8.0 to quench the reaction.
	(*) Included in the Terminal transferase kit from Sigma Aldrich.
	}
	\label{tab:6:tailing}
	\end{center}
\end{table}

After the tailing steps, the oligos are purified using the QIAquick Nucleotide Removal Kit (Qiagen) and eluted in $50\,{\rm \upmu l}$ Tris-Cl $10\,{\rm mM}$, giving rise to a final concentration, assuming a 100\% efficiency, of $\sim 2\,{\rm \upmu M}$. 

	\item Annealing reaction: The final construct is then assembled in an equimolar reaction. Since the sequences are complementary, at equal concentration for all the oligonucleotides, the equilibrium structure is the formed hairpin with dsDNA handles (Fig.~\ref{fig:14:miniTweezers}(c)). The annealing is performed in two steps : (a) CD4-handleA + splint and CD4-handleB + splint; (b) mix of the two previous reactions. The annealing protocol is described in Table~\ref{tab:7:annealing}.

\begin{table}[h!]
	\footnotesize
	\begin{center}
	\begin{tabular}{lcc}
		& 
		Reaction I 
		& 
		Reaction II 
		\\
	\hline	
		Milli-Q water 
		& 
		$12\,{\rm \upmu l}$ 
		&  
		$8\,{\rm \upmu l}$ 
		\\
		DIG-tailed CD4-HandleB ($\sim 2\,{\rm \upmu M}$)  
		& 
		- 
		&  
		$5\,{\rm \upmu l}$ 
		\\
		CD4-HandleA ($5\,{\rm \upmu M}$)  
		& 
		$1\,{\rm \upmu l}$ 
		&  
		-
		\\
		splint ($100\,{\rm \upmu M}$) 
		& 
		$1\,{\rm \upmu l}$ 
		& 
		$1\,{\rm \upmu l}$ 
		\\
		$1\,{\rm M}$ Tris pH~7.5 
		& 
		$0.5\,{\rm \upmu l}$ 
		&  
		$0.5\,{\rm \upmu l}$  
		\\
		$5\,{\rm M}$ NaCl 
		& $0.5\,{\rm \upmu l}$ 
		&  
		$0.5\,{\rm \upmu l}$  
		\\
	\hline 
		Total Volume 
		& 
		$15\,{\rm \upmu l}$ 
		&  
		$15\,{\rm \upmu l}$  
		\\
	\hline 
	\end{tabular}
	\caption{
	{\bf Annealing reaction.} 
	Mix the items for each reaction in a separate Eppendorf tube. Keep both reactions at $95^\circ{\rm C}$ for $1\,{\rm min}$, $80^\circ{\rm C}$ for $10\,{\rm min}$, decrease by $0.5^\circ{\rm C}$ every $10\,{\rm min}$ down to $40^\circ{\rm C}$. Hold the temperature, mix both reactions and decrease by $0.5^\circ{\rm C}$ every $20\,{\rm min}$ down to $10^\circ{\rm C}$. The decrease in temperature can also be achieved by letting the tubes inside a thermal bath cooling down with the heater off; however, the usage of a thermocycler is recommended, since it allows for a better temperature control.
	}
	\label{tab:7:annealing}
	\end{center}
\end{table}

	\item Ligation: The molecular assembly is almost completed. All the Watson-Crick bonds of the structure have been formed. However, the backbone of the two oligos of the stem of the hairpin are still not covalently bonded. The final step is to connect the phosphorylated 5' end of CD4-handleB oligo with the C3' end of CD4-handleB. To do so, the assembly is then ligated in an overnight reaction using T4 DNA ligase (New England Biolabs) at $16^\circ{\rm C}$. The ligase is heat inactivated at $65^\circ{\rm C}$ for $10\,{\rm min}$, as shown in Table~\ref{tab:8:ligation}.

\begin{table}[h!]
	\footnotesize
	\begin{center}
	\begin{tabular}{lc}
	\hline 
		Milli-Q water 
		& 
		$52\,{\rm \upmu l}$  
		\\
		Annealing product 
		& 
		$30\,{\rm \upmu l}$ 
		\\
		$10\times$ T4 DNA ligase buffer 
		& 
		$10\,{\rm \upmu l}$  
		\\
		$10\,{\rm mM}$ ATP 
		& 
		$5\,{\rm \upmu l}$  
		\\
		T4 DNA ligase ($400\,{\rm U\, \upmu l^{-1}}$) 
		& 
		$3\,{\rm \upmu l}$  
		\\
	\hline 
		Total Volume
		&
		$100\,{\rm \upmu l}$ 
		\\
	\hline 
	\end{tabular}
	\caption{
	{\bf Ligation reaction.} 
	Mix all the items in an Eppendorf tube and keep them at $16^\circ{\rm C}$ overnight. Afterwards, keep the Eppendorf tube at $65^\circ{\rm C}$ for $10\,{\rm min}$  to inactivate the ligase.}
	\label{tab:8:ligation}
	\end{center}
\end{table}

\end{enumerate}

\subsubsection{Pulling experiments}

Paradigmatic manipulation experiments one can perform with a DNA hairpin are pulling experiments. The DNA hairpin is tethered between the two beads and the force it experiences is cyclically varied between a minimum and a maximum force value. 
Starting at a low force (typically $0$ to $5\,{\rm pN}$) with the hairpin folded, the trap is moved away from the pipette at a constant pulling speed and the exerted force is steadily increased. 
At a given point, the hairpin will reach a force high enough to unravel the double-stranded DNA helix structure reaching the unfolded state. This mechanically induced denaturation process is called unzipping.
The stretching of the unfolded hairpin keeps on until a maximum force is reached. 
Afterwards, the same reverse protocol is applied: the trap is moved backwards at the same pulling speed and the applied force decreased until reaching the minimum force value. In this reverse process (called rezipping) the DNA will refold to the hairpin native state when the applied force is low enough. Once the applied force reaches the initial minimum value a new pulling cycle starts again. The CD4 DNA hairpin is a good molecular marker that can be used for force calibration. It reversibly unfolds and folds at $14.7\pm0.3\,{\rm pN}$ at standard conditions ($T=298\,{\rm K}$, $1\,{\rm M}$ NaCl).

\begin{figure}[h!]
	\begin{center}
	\includegraphics[width=12cm]{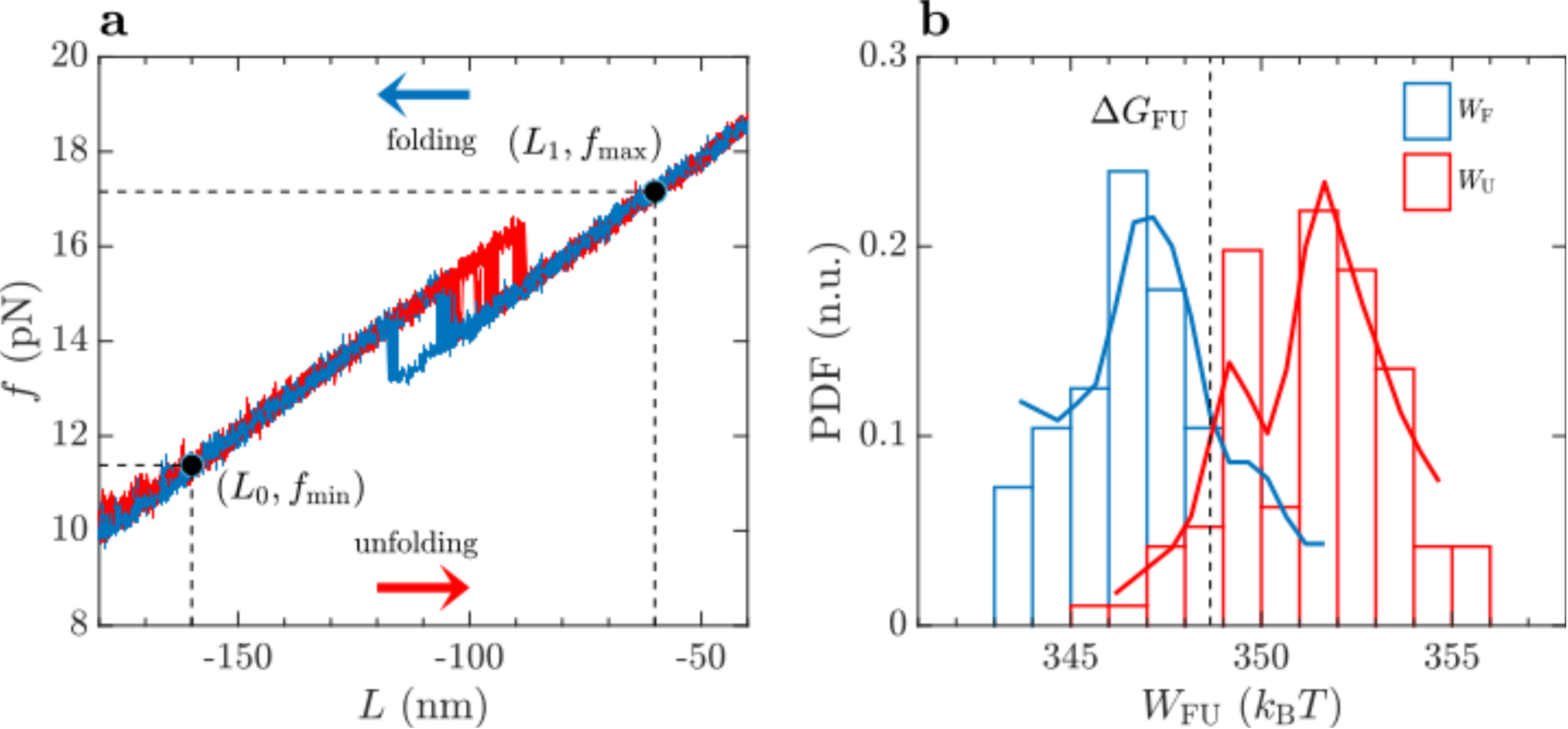}
	\caption{
	{\bf Free-energy of hairpin formation.}
	(a) Hairpin pulling trajectories shown as force-distance curves. The unfolding cycles are plotted in red and the folding cycles in blue. The black dots indicate the initial ($L_0,f_{\rm min}$) and final ($L_1,f_{\rm max}$) points considering for computing the work $W_{\rm FU}$.
	Since the $L$ values are obtained from the light position at the position-PSD, the $L_0$ and $L_1$ stand as relative trap positions with respect to an arbitrary zero (which corresponds to the center of the position-PSD).
	(b) Probability density functions (PDFs) of the work, computed using equation~\eqref{eq:work}, for the unfolding ($W_{\rm U}$, red histograms) and folding ($W_{\rm F}$, blue histograms) cycles. The value at which the probabilities cross (dashed vertical line) is the free energy difference between the folded and unfolded states $\Delta G_{\rm FU}$. This value is obtained as the crossing point between the lines interpolating the histograms (blue and red solid lines).
	}
	\label{fig:16:single-molecule}
	\end{center}	
\end{figure}

The detailed procedure to perform the pulling experiment is:
\begin{enumerate}
	\item The three channels are filled with the molecular buffer described in section~\ref{sec:4.1.4:bead} through the syringes connected to the channels. The buffers have to be kept in the fridge and filtered using a $10\,{\rm ml}$ syringe and a filter (Sterile Syringe Filter, w/$0.2\,{\rm \upmu m}$ Cellulose, Acetate Membrane, VWR International) to prevent microorganism growth. 	\item The chamber (already held in the metallic mount as described in section~\ref{sec:4.1.1:chamber}) is placed between the objectives.
	\item The buffer containing SA beads is flowed in the upper channel and AD beads in the lower channel (the channels can be interchanged).
	\item The microfluidics chamber is moved using the stepper motors to bring the optical trap close to the dispenser tube of the upper supplier channel to capture a SA bead. 
	\item After a SA bead is captured by the optical trap, the microfluidics chamber is moved back to position the trapped SA bead on the tip of the micropipette. The syringe is then pulled and the SA bead is held by air suction at the tip of the pipette.
	\item The microfluidics chamber is then moved close to the dispenser tube of the lower channel supplying AD beads.
	\item Once an AD bead is captured (paying special attention not to trap more than a single bead), the AD bead is brought close to the SA bead, i.e., the chamber is moved so that the optical trap is in the working zone.
	\item The tether is formed by poking the SA bead held by the micropipette with the optically trapped AD bead, until a non-zero force is recorded upon retracting the AD from the SA bead.
	\item Check that only one molecule has been attached to the bead. To do so, the easiest way is to move the optical trap away from the bead in the pipette and verify that a sudden decrease of the force by $\sim 1.5\,{\rm pN}$ occurs at $\sim 15{\rm pN}$ (red lines in Fig.~\ref{fig:16:single-molecule}(a)), which indicates the unfolding of the DNA molecule. Then, by moving inwards the optical trap, a similar increase of the force should occur at  $\sim 14{\rm pN}$ (blue lines in Fig.~\ref{fig:16:single-molecule}(a)), which indicates that the hairpin is folding back. This equilibrium unzipping-rezipping force has been found to be $14.7\pm 0.3$pN at standard conditions (see above).
	\item Now, the pulling experiment can start by cyclically varying the distance $L$ between the pipette-held particle and the center of the trap (as shown in Fig.~\ref{fig:14:miniTweezers}(d)). In this way, repeated unfolding and refolding pulling cycles ($\sim 100$) are performed over the same molecule while recording the force and the trap position. Some representative data are shown in Fig.~\ref{fig:16:single-molecule}(a), with a typical pulling speed of $100\,{\rm nm\,s^{-1}}$. To extract the free energy of formation of the hairpin, as we will see in the next section, it is key that the hairpin is folded at $L=L_0$  and unfolded at $L=L_1$.
\end{enumerate}

\subsubsection{Extracting the free-energy of hairpin formation}

One of the main applications of single molecule experiments is the possibility to extract free energy differences from irreversible work measurements using fluctuation theorems \cite{seifert2012stochastic}. From the force-distance curves (Fig.~\ref{fig:16:single-molecule}(a)) the work $W$ exerted on the molecule between the starting trap position, $L_0$, and the final one, $L_1$ can be directly measured as the area below the curve:
\begin{equation}\label{eq:work}
	W
	=
	\int_{L_0}^{L_1} f\,d L.
\end{equation}
For quasi-static processes (i.e., sufficiently slow pulls) the work equals the free energy difference $\Delta G_{\rm FU}$ between the initial $L_0$ (folded) and final $L_1$ (unfolded) trap positions. However, this is not true for irreversible pulls which exhibit hysteresis between the stretching (forward) and releasing (reverse) processes. The difference in the area between consecutive unfolding and folding cycles is the work dissipated by the system, which is positive in average according to the second law of thermodynamics. Noteworthy, since we are not in the thermodynamic limit (as we are dealing with small systems), there may be trajectories that violate the second law of thermodynamics.

Fluctuation theorems state that the work distributions measured along the forward and reverse processes fulfil a symmetry relation \cite{seifert2012stochastic}. One of the main consequences of such a relation is that forward and reverse work distributions, despite being different due to the hysteresis, cross each other at the value of $\Delta G_{\rm FU}$, independently of how irreversible the pulling process is. To extract the work distributions we use equation~\eqref{eq:work} to compute the work for all the forward (stretching) and reverse (releasing) cycles. The probability distributions of the work for the forward and reverse processes are shown in Fig.~\ref{fig:16:single-molecule}(b) by the red and blue histograms, respectively. The crossing point of the forward and reverse work probability distributions corresponds to the difference in free energy $\Delta G_{\rm FU}$ between the folded ($L_0$) and unfolded ($L_1$) states. The work histogram in Fig.~\ref{fig:16:single-molecule}(b) is obtained by taking $L_0 = -140\,{\rm nm}$ ($f_{\rm min}=11.35\,{\rm pN}$) and $L_1=-60\,{\rm nm}$ ($f_{\rm max}=17.16\,{\rm pN}$). Note that $L_0$ and $L_1$ stand as relative trap positions with respect to an arbitrary zero. The crossing point of the histograms gives $\Delta G_{\rm FU}=348.67\,k_{\rm B}T$ (with $1k_{\rm B}T=4.114\,{\rm pN\,nm}$ at $T=25^\circ{\rm C}$).

The free energy difference between the folded and unfolded states can be written as:
\begin{equation}\label{eq:free_energy_contributions}
	\Delta G_{\rm FU} 
	=
	\Delta G_{\rm FU}^0 
	+ \Delta W_{\rm FU}^{\rm ssDNA} 
	- \Delta W_{\rm FU}^{\rm dipole} 
	+ \Delta W_{\rm FU}^{\rm handles} 
	+ \Delta W_{\rm FU}^{\rm bead},
\end{equation}
where the term $ \Delta G_{\rm FU}^0$ is the free energy of formation of the molecule, which is the quantity we are interested in obtaining, and the other contributions are the reversible work differences related to the different parts of the experimental setup (ssDNA, handles, bead, hairpin dipole, as shown in Fig.~\ref{fig:14:miniTweezers}(d)). These contributions can be computed as follows:
\begin{enumerate}
	\item  $\Delta W_{\rm FU}^{\rm ssDNA}$ is the reversible work needed to stretch the unfolded molecule, which is modeled with an inextensible Worm-like chain \cite{marko1995stretching}:
\begin{equation}\label{eq:WLC}
	f_{\rm ssDNA}(x)
	=
	{k_{\rm B}T \over p}
	\left(
		{1 \over 4
			\left(
				1-\frac{x}{n l_0} 
			\right)^2
		}
		- {1 \over 4}
		+ {x \over n l_0}
	\right),
\end{equation}
where $f_{\rm ssDNA}$ is the force, $x$ is the molecular extension, $n$ is the number of monomers of the molecule ($n=44$ bases for our case), $l_0 = 0.59\,{\rm nm}$ is the contour length per monomer, and $p = 1.35\,{\rm nm}$ is the persistence length (parameters from Ref.~\cite{alemany2014determination}). 
The stretching contribution of this released ssDNA is then computed as
\begin{equation}\label{wlc_stretching}
\Delta W_{\rm FU}^{\rm ssDNA}=\int_{0}^{x(f_{\rm max})}f_{\rm ssDNA}(x)dx.
\end{equation}
The value we obtain is $\Delta W_{\rm FU}^{\rm ssDNA}=19.0\,k_{\rm B}T$.
	\item $\Delta W_{\rm FU}^{\rm dipole}$ is the reversible work needed to stretch the folded molecule, which is modelled as a single dipole of length $d=2\,{\rm nm}$\cite{alemany2014determination}, corresponding to the double-helix diameter\cite{sinden1994dna}:
\begin{equation}\label{eq:FJC}
	x_{\rm dipole}(f)
	=
	d
	\left[
		\tanh{f\,d \over k_{\rm B}T}
		- {k_{\rm B}T \over f\,d}
	\right]
\end{equation}
and
\begin{equation}\label{work}
	\Delta W_{\rm FU}^{\rm dipole}
	=
	\Delta f\Delta x
	- \int_{f_{\rm min}}^{f_{\rm max}} x_{\rm dipole}(f')df',
\end{equation}
where $\Delta f=f_{\rm max}-f_{\rm min}$ and $\Delta x=x(f_{\rm max})-x(f_{\rm min})$. The value we obtain is $\Delta W_{\rm FU}^{\rm dipole}=4.6\,k_{\rm B}T$.
	\item Finally, the last two contributions to equation~\eqref{eq:free_energy_contributions} can be computed by considering the slope of the force-distance curves just before the rip, where the distance corresponds to the optical trap position ($L$), as shown in Fig.~\ref{fig:16:single-molecule}(a). The slope $\kappa_{\rm eff}$ can be easily obtained by linearly fitting the data points right before the unfolding event. This yields the effective stiffness of two serially connected springs: the trapped bead of stiffness $\kappa_{\rm B}$, and the handles stiffness, $\kappa_{\rm handles}$: $\kappa_{\rm eff}^{-1}=\kappa_{\rm B}^{-1}+\kappa_{\rm handles}^{-1}$. 
Assuming that the hairpin orientation has a much higher stiffness than the handles and the optical trap, the combined work can be written as \cite{alemany2014determination}:
\begin{equation}\label{eq:work_effective}
	\Delta W_{\rm FU}^{\rm handles} + \Delta W_{\rm FU}^{\rm bead}
	\approx
	\frac{f_{\rm max}^2-f_{\rm min}^2}{2\,\kappa_{\rm eff}},
\end{equation}
where we have considered that the stiffness of the folded molecule is very large and consequently does not contribute to $\kappa_{\rm eff}$. The value we obtain is $\Delta W_{\rm FU}^{\rm handles}+\Delta W_{\rm FU}^{\rm bead}=282\,k_{\rm B}T$.
\end{enumerate}
From equations~\eqref{wlc_stretching}, \eqref{work} and \eqref{eq:work_effective}, and from the crossing point of the forward and reverse work probability distributions ($\Delta G_{\rm FU}$), we obtain all the terms needed to calculate the free energy of formation of the hairpin, $\Delta G_{\rm FU}^0$, from equation~\eqref{eq:free_energy_contributions}. The free energy of formation at zero force we obtain for this molecule is $\Delta G_{\rm FU}^0=51.9\,k_{\rm B}T$, which is in agreement with that of the sequence of the CD4 hairpin, $\Delta G_{FU}^0=50.6\,k_{\rm B}T$, using the values provided in Ref.~\cite{santalucia1998unified}. 
Typical errors of $\Delta G_{\rm FU}^0$ are on the order of several $k_{\rm B}T$, corresponding, in this case, to a 5\% to 10\% relative error.

\subsubsection{Other applications to single-molecule mechanics and outlook} 

``Take a single DNA molecule and pull from its extremities while recording the force-extension curve until it gets fully straightened.''  
This thought experiment, which was just a dream a few decades ago, has now become standard in many research labs worldwide.  
Since the seminal works on single-molecule manipulation almost 30 years ago \cite{block1990bead, bustamante1994entropic, finer1994single}, the number of applications of single-molecule mechanics has kept on growing. 
Applications might be classified in two large sets: those aimed to address specific biological questions (e.g., the mechanics of maintenance and regulatory enzymes in the genome, the energetics and kinetics of DNA hybridization, RNA and protein folding, molecular footprinting of proteins and ligands binding DNA), and those aimed to enlighten fundamental physico-chemical questions (e.g., polymer physics, chemical reactions, nonequilibrium physics) using biomolecular matter as model systems. 
The reader can have a look at published reviews specifically addressing these fields to discover that the future of the single-molecule field has a long way ahead to run \cite{ritort2006single , deniz2007single, neuman2007single, moffitt2008recent, neuman2008single, zhang2013high, capitanio2013interrogating, miller2017single}. 
Most optical-tweezers and single-molecule technology is still instrumental craftwork not widely available in research institutes and labs worldwide, yet it is becoming steadily more popular among biologists aiming to enter the field.  
Future developments in the field will be the parallelization of measurements in optical tweezers making it high-throughput either using holographic techniques \cite{grier2006holographic} (recently extended also for acoustic tweezing \cite{marzo2019holographic}), combining single-molecule fluorescence with force-measurement techniques  \cite{candelli2011combining,cordova2014combining, whitley2017high}, controlled force resistive pulse sensing\cite{keyser2006direct, bulushev2014measurement}, and plasmonic trapping \cite{juan2011plasmon}. 
Finally, a major step would be to apply force spectroscopy techniques to {\it ex-vivo} and {\it in-vivo} conditions.  

\subsection{Single-cell mechanics}
\label{sec:4.2:cell}

Biological microscopic objects such as organelles and living cells have been optically manipulated since the late 1980s  \cite{ashkin1987opticala, ashkin1987opticalb}. 
In Ref.~\cite{block1989compliance}, the swimming motion of optically trapped bacteria was recorded while the trapping forces were measured providing the first example of single-cell optical manipulation.  
Since then, biomechanics at the cellular scale has kept on growing in terms of techniques and objects of study\cite{roca2017quantifying, basoli2018biomechanical}. 
Techniques such as atomic force microscopy \cite{muller2011force, dufrene2013force}, optical tweezers \cite{hormeno2006exploring, verdeny2011optical,arbore2019probing}, particle tracking microscopy \cite{wirtz2009particle, shen2017single}, traction microscopy \cite{schwarz2015traction, colin2018future, ferrari2019recent }, magnetic twisting cytometry \cite{puig2001measurement}, acoustic force spectroscopy\cite{sorkin2018probing}, and micropipette aspiration \cite{lee2014application} have become commonly employed to investigate the response of cellular objects to mechanical stresses. 
Essential cellular processes (e.g., cell division) and collective effects (e.g., cell signaling and migration in tissues) are currently being investigated with such techniques. 

In this section, we will demonstrate one of the simplest cell manipulation experiments with optical tweezers:
the mechanical stretching of a single red blood cell (RBC) to measure its rigidity to an externally applied deformation or strain.

\subsubsection{RBC buffer preparation}

The RBC buffer should mimic the physiological conditions found in human blood. 
It can be prepared as a solution of $130\,{\rm mM}$ NaCl, $20\,{\rm mM}$ K/Na phosphate buffer at 7.4~pH, $10\,{\rm mM}$ glucose, and $1\,{\rm mg\,ml^{-1}}$ BSA (Bovine Serum Albumin) \cite{betz2009atp}. 
It is convenient to prepare higher concentration stocks ($1\,{\rm M}$) of every solution and store them in the fridge (higher concentration of salts together with low temperature prevent microbial contamination), keeping them in $50\,{\rm ml}$ Falcon tubes for repeated use. 
To prepare a $50\,{\rm ml}$ Falcon tube containing the RBC buffer, follow these steps:
\begin{enumerate}
	\item Add $6.5\,{\rm ml}$ of $1\,{\rm M}$ NaCl, $1\,{\rm ml}$ of $1\,{\rm M}$ K/Na phosphate buffer at 7.4~pH,\footnote{The $1\,{\rm M}$ K phosphate buffer at 7.4~pH is obtained by mixing $40.1\,{\rm ml}$ of $1\,{\rm M}$ K$_2$HPO$_4$ and $9.9\,{\rm ml}$ of $1\,{\rm M}$ KH$_2$PO$_4$. The $1\,{\rm M}$ Na Phosphate buffer at 7.4~pH is obtained by mixing $38.7\,{\rm ml}$ of $1\,{\rm M}$ Na$_2$HPO$_4$ and $11.3\,{\rm ml}$ of $1\,{\rm M}$ NaH$_2$PO$_4$. All salts are available from Sigma-Aldrich.} and $0.5\,{\rm ml}$ of $1\,{\rm M}$ glucose into an empty $50\,{\rm ml}$ Falcon tube.
	\item Add $50\,{\rm g}$ of BSA.
	\item Add $42\,{\rm ml}$ of milli-Q water. 
	\item Mix the RBC buffer and filter it (Sterile Syringe Filter, w/$0.2\,{\rm \upmu m}$ Cellulose, Acetate Membrane, VWR International).
	\item Place the filtered solution into a new $50\,{\rm ml}$ Falcon tube. 
\end{enumerate}
All remaining unused RBC buffer must be re-filtered on a daily basis to prevent molecular aggregation and bacterial growth. After three days, it is recommended to dispose of the remaining RBC buffer because the re-filtering will not be enough to ensure proper conservation.

\subsubsection{Blood extraction}

To obtain human blood, the easiest way is to prick the finger of a healthy donor. Use of diabetic testing lancets will ensure sterilization of the pricking device. The extraction is made following these steps:
\begin{enumerate}
	\item  Disinfect the finger of the healthy donor with ethanol and dry it with handkerchief paper to remove any remaining liquid.
	\item Prick the donor's finger with the lancet and apply some pressure to extract a drop of blood. A single drop of blood will be enough to perform experiments during an entire day.
	\item With a $200\,{\rm \upmu l}$ pipette take $10\,{\rm \upmu l}$ of blood from the finger.
	\item Dilute as fast as possible the blood drop in $1\,{\rm ml}$ of RBC buffer inside a $1.5\,{\rm ml}$ Eppendorf tube to prevent degradation of the RBCs.
\end{enumerate}

\subsubsection{Bead preparation}

The simplest option to attach the beads to the RBCs is to take advantage of non-specific interactions between polystyrene beads and RBCs.\footnote{There are also several options to treat the beads and RBC to specifically attach them, e.g., using the lectin protein Concanavalin A \cite{sorkin2018probing} or biotinylating the RBC \cite{turlier2016equilibrium}} To do so:
\begin{enumerate}
	\item Add $1\,{\rm ml}$ RBC buffer in a $1.5\,{\rm ml}$ Eppendorf tube.
	\item Homogeneously resuspend the polystyrene beads by placing the purchased beads (Microbead NIST Traceable Particle Size Standard, $3.00\,{\rm \upmu m}$, Polysciences, Inc.) into a vortex mixer for several seconds. An additional step of several seconds of sonication improves the resuspension.
	\item Add $3\,{\rm \upmu l}$ of the bead stock solution to the Eppendorf tube.
\end{enumerate}
It is recommended to have two Eppendorf tubes with bead solution available to ensure continued supply of the chamber with beads during the experiment.

\subsubsection{Setup preparation}

The experiment can be performed using the same microfluidics chamber and miniTweezers setup presented in section~\ref{sec:4.1:molecule}. In particular, the procedure to build the chamber and to place it into the mount is the same as described in section~\ref{sec:4.1.1:chamber}, the only difference being a longer plastic tube ($40\,{\rm cm}$) to connect one of the syringes to the inlet of the central channel. Once the chamber is fixed into the mount with the syringes connected through the plastic tubes, the next steps are:
\begin{enumerate}
	\item Flow $3\,{\rm ml}$ of RBC buffer in each of the three channels to produce an homogeneous environment in the chamber.
	\item Identify which dispenser tube (upper/lower) has the most clear exit (i.e., the one displaying the most sharp perpendicular cut in the video image) to the central channel. This will simplify trapping the beads.
	\item Fill the syringe connected to the dispenser tube selected in the previous point with the bead solution. Suck about half of the Eppendorf tube content with the syringe.
	\item Fill the other syringe (upper/lower channel) with RBC solution. Suck about half of the Eppendorf tube content with the syringe.
	\item Introduce the mount with the chamber between the two objectives of the optical tweezers.
	\item Connect the syringe with the larger plastic tube (the one of the central channel) to a pump to apply a controlled flow to the central channel when it is needed. We will use this flow to stretch the RBC-bead dumbbell captured in the optical trap, thereby preventing thermal- and photo-damage of the cell.
\end{enumerate}

\subsubsection{Construction of the cell-bead configuration}

\begin{figure}[b!]
	\begin{center}
	\includegraphics[width=12cm]{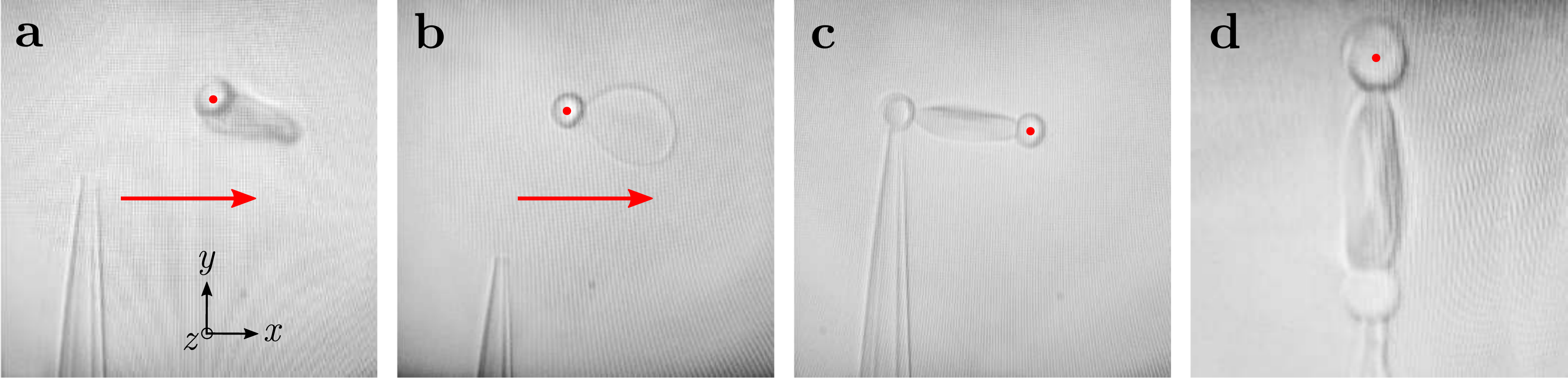}
	\caption{
	{\bf Construction of the cellular configuration.}
	Images of different steps to obtain the RBC  configuration required for the experiment:
	(a) the RBC is attached to a bead and partially trapped by the lasers; 
	(b) the bead is trapped, while the RBC is not trapped;
	(c) the second bead is attached to the other side of the RBC;
	(d) final experimental configuration at zero reading for the force along the $y$-direction.
	The diameter of the beads is $3\,{\rm \upmu m}$. 
	The red dot represents the position of the optical trap.
	The red arrow in (a) and (b) represents the presence and direction of the fluid flow.
	}
	\label{fig:17:RBC_config}
	\end{center}
\end{figure}

To perform cellular pulling experiments, the RBC need to be held between two beads attached to the cell membrane on opposites sides, as shown in Fig.~\ref{fig:17:RBC_config}(d).
One bead is held by the micropipette by air suction, while the other bead is held by the optical trap.
The use of beads as handles to manipulate the RBC minimizes the direct exposure of the RBC to the laser light. 
The steps of the procedure to obtain this cellular configuration are:
\begin{enumerate}
	\item Align the lasers without a bead (steps~1--8 described section~\ref{sec:4.1.2:experimental}).
	\item Identify the position of the pipette and of the two dispenser tubes that connect the upper and the lower channels with the central one.
	\item Move the chamber using the stepmotors and place the optical trap close to the exit of the dispenser tube that contains the bead solution.
	\item Flow some bead solution pushing the syringe either by hand or using a pump, and trap a bead with the optical trap.
	\item Move the trapped bead next to the pipette tip in the working zone.
	\item Align the lasers with the trapped bead (steps~9--13 described section~\ref{sec:4.1.2:experimental}).
	\item With the bead already trapped, try different flow values with the pump connected to the syringe of the central channel until the force reaches $\sim10\,{\rm pN}$ due to the RBC buffer flow. This will be the approximate flow that we will need to keep the optically trapped bead separated from the RBC (Figs.~\ref{fig:17:RBC_config}a and \ref{fig:17:RBC_config}b) after we obtain an attachment.
	\item Turn off the pump.
	\item Move the trapped bead to the area of the blood dispenser tube exit. It is important not to put the bead exactly at the exit of the tube to prevent losing the bead due to the RBC solution flow.
	\item Flow some RBC solution applying a small pressure to the syringe. RBCs will flow into the central channel.
	\item Move the trapped bead close to a RBC and try to make a connection by poking the surface of the RBC with the bead. The connection is formed when the RBC follows the trapped bead when moved.
	\item Move the chamber using the stepmotors to place the bead and the attached RBC next to the micropipette. 
	\item Turn on the pump. The configuration should look like Fig.~\ref{fig:17:RBC_config}(a). The RBC is attached to the bead and partially trapped by the lasers.
	\item If the flow is not enough to separate the RBC from the trap and, therefore, to obtain the configuration of Fig.~\ref{fig:17:RBC_config}(b), it is necessary to move the chamber using the stepmotors along the horizontal direction to exert an extra force and separate the RBC from the optically trapped bead.
	\item Move the bead close to the tip of the pipette.
	\item Carefully apply air suction to immobilize the bead. This step is critical because a large enough air suction force could absorb the RBC inside the pipette.
	\item Move the optical trap away from the bead. If the bead is properly fixed to the tip of the pipette, the bead will stay there. If this is not the case, when the optical trap is moved the bead will follow it.
	\item Do not turn off the pump flow in the central channel. To minimize the possibility of losing the bead attached to the RBC fixed on the tip of the pipette, do the following steps as quickly as possible.
	\item Halve the flow to facilitate the trapping of the second bead. Check that the RBC stays separated from the bead, if this is not the case, increase the flow again to separate the cell from the bead.
	\item Move the optical trap to the exit of the bead dispenser tube.
	\item Flow the bead solution and try to trap a bead. It will be harder due to the RBC buffer flow on the central channel. To catch a bead with the pump on, it is easier to stay at the exit of the dispenser tube without moving the trap. Wait for the bead to fall into the optical trap instead of actively moving the trap around.
	\item Once you have trapped a bead, move the chamber to the working area.
	\item Try to make a connection between the trapped bead and the RBC as in Fig.~\ref{fig:17:RBC_config}(c). Notice that the two beads are on opposite sides of the RBC.
	\item Check that the new connection between the bead and the RBC is done by moving the optical trap to the same direction as the pump flow. If the RBC is deformed while moving the trap, the attachment is done. 
	\item Turn off the pump to remove the flow of the central channel.
	\item Nip very carefully the plastic tube that is connected to the entrance of the central channel with a clamp to prevent flows due to the difference of pressure at both ends of the plastic tubes.
	\item Move the stepmotors to obtain the configuration shown in Fig.~\ref{fig:17:RBC_config}(d), aiming at having a zero reading for the force along the $y$-direction. 
\end{enumerate}

\begin{figure}[b!]
	\begin{center}
	\includegraphics[width=8cm]{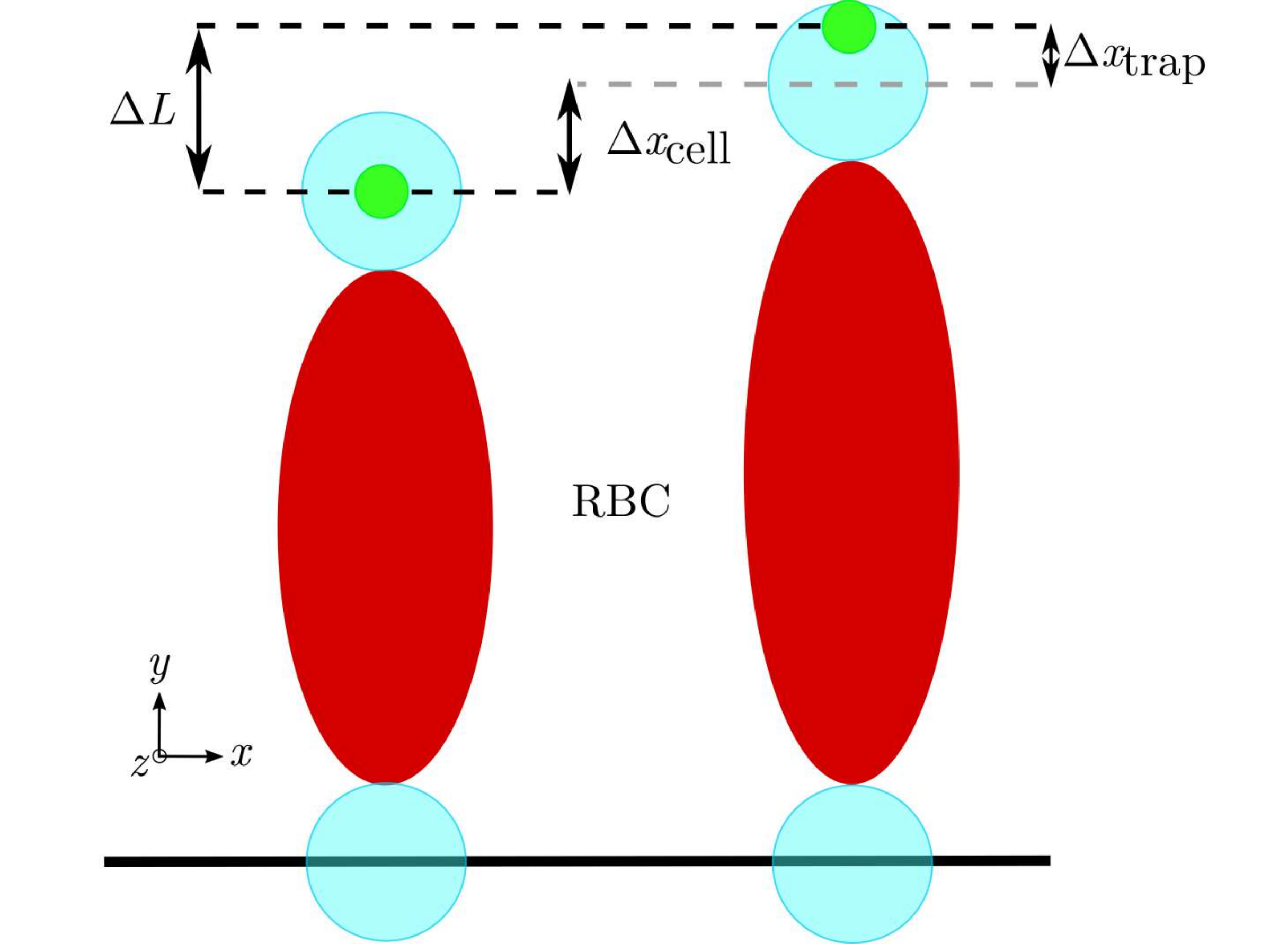}
	\caption{
	{\bf Sketch of RBC extension under applied force.}
	The red ellipse represents the RBC, the light blue circles represent the beads, and the small green circle represents the optical trap.
	The cell extension is $\Delta x_{\rm cell}$, while $\Delta x_{\rm trap}$ is the bead displacement respect to the trap center, and $\Delta L$ is the displacement of the optical trap. Note that $\Delta x_{\rm cell}$ and $\Delta x_{\rm trap}$ are distances along the $y$-axis.
	} 
	\label{fig:18:RBC_sketch}
	\end{center}	
\end{figure}

\begin{figure}[t!]
	\begin{center}
	\includegraphics[width=12cm]{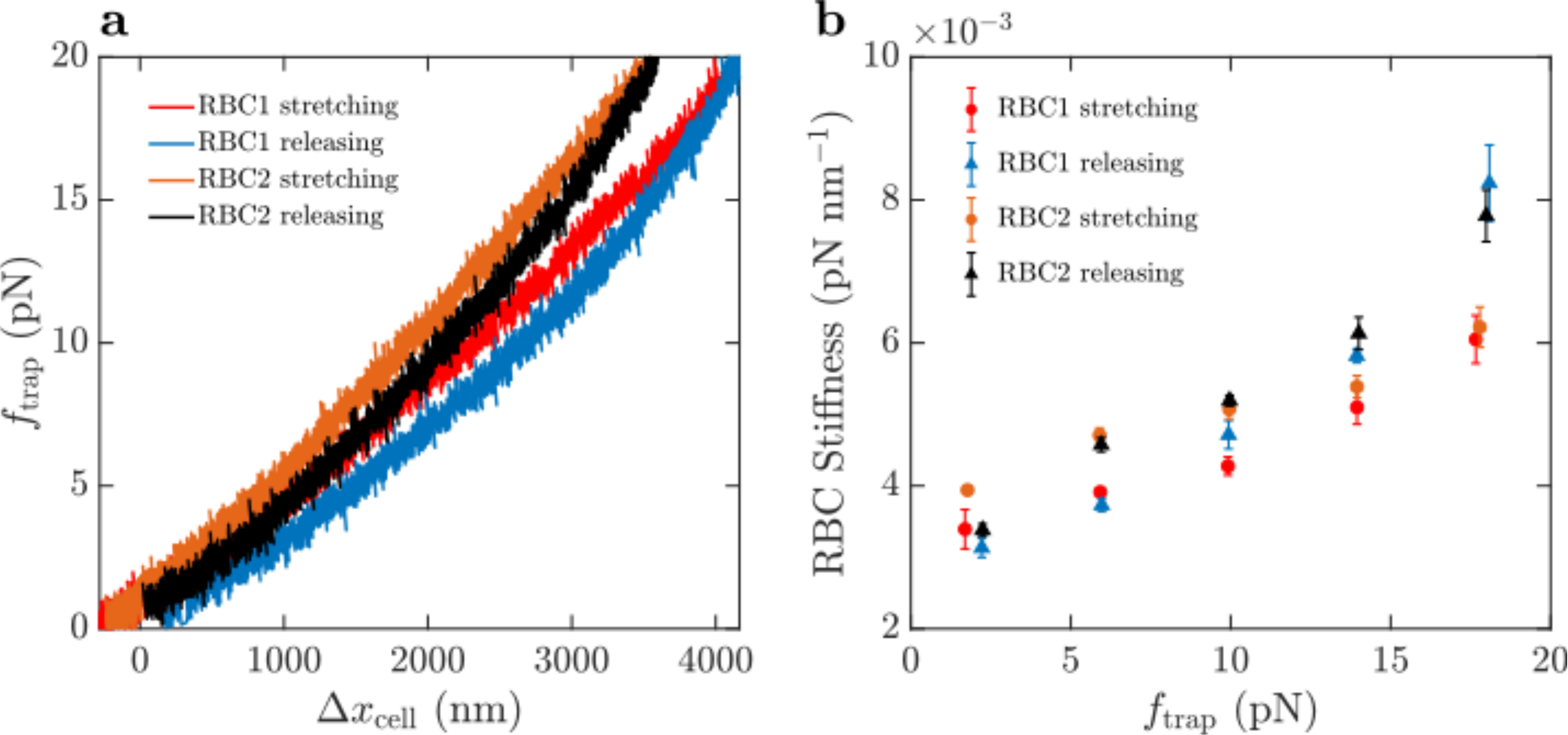}
	\caption{
	{\bf Force--cellular-extension curves and RBC stiffness.} 
	(a) Force--cellular-extension curves for two different RBCs. 
	(b) Stiffness as a function of force obtained from the slope of the force--cellular-extension curves. 
	} 
	\label{fig:19:RBC_data}
	\end{center}	
\end{figure}

\subsubsection{Experiments}

By cyclically stretching and releasing the RBC, it is possible to obtain the force-extension curves from which the stiffness and deformability of the RBC can be extracted. 
Starting with a RBC in the configuration described in the previous section (Fig.~\ref{fig:17:RBC_config}(d)), it is possible to stretch the cell by moving the optically trapped bead outwards (i.e., away from the micropipette) and release it by moving it inwards (i.e., towards the micropipette).\footnote{A range for the optical trap displacement of at least $\sim 8\,{\rm \upmu m}$ is necessary to perform the RBC stretching. Thus, it is recommended to move the optical trap towards the lower limit of its y-displacement, before starting the pulling protocol.} 
Follow these steps to define the stretching-realising protocol:
\begin{enumerate}
	\item Select the \emph{minimum force} from which to start stretching the RBC (e.g., $0\,{\rm pN}$).\footnote{Depending on the laser power, it could be very hard to reach the $0\,{\rm pN}$ force as the RBC will be attracted to the optical trap.}
	\item Select the \emph{maximum force} at which to stop the RBC stretching (e.g., $20\,{\rm pN}$).\footnote{Above $20\,{\rm pN}$, as we have not applied any specific coating to the beads, a very thin membrane tube (theter) forms and alters the force-extension curve.}
	\item Select the \emph{pulling velocity} (e.g., $140\,{\rm nm\,s^{-1}}$).
	\item Select the \emph{refolding time} (i.e., the time to wait at the minimum force before starting a new cycle; e.g., $0\,{\rm s}$).
\end{enumerate}
Starting from the minimum force, the RBC is pulled at the constant velocity until the maximum force. After reaching this maximum force, the RBC is pushed at the same constant velocity to the minimum force. 
During this process, the displacements and forces along the RBC stretching direction ($y$-direction) are recorded for both the A- and B-trap in the miniTweezers setup (Fig.~\ref{fig:14:miniTweezers}(a)); then, the two displacements are averaged to measure the trap displacement, while the two forces are summed to obtain the total force acting on the RBC.
The whole process is called \emph{pulling cycle}. 
Recording 5 pulling cycles for each RBC is typically enough for the subsequent analysis. 

\subsubsection{Data analysis}

The data analysis aims at computing the RBC stiffness knowing the stiffness of the optical trap $\kappa_{\rm trap}$.  
The experiment measures the trap displacement $\Delta x_{\rm trap}$, not the cellular extension $\Delta x_{\rm cell}$. 
Thus, $\Delta x_{\rm cell}$ can be computed from $\Delta x_{\rm trap}$, $\kappa_{\rm trap}$, and the measured optical force $f_{\rm trap}$.
For each intermediate trap position from $L_0$ to $L_{\rm f}$, we are able to transform $L$ into $x_{\rm cell}$ substracting the trap contribution (Fig.~\ref{fig:18:RBC_sketch}), as we show in the following expression:\footnote{This is the same analysis done to obtain the molecular extension in long nucleic acids ($\approx10\,{\rm kbp}$ double-stranded nucleic acids or $\approx1\,{\rm kbp}$ single-stranded nucleic acid \cite{bosco2013elastic}).}
\begin{equation}
	\Delta x_{\rm cell} 
	= 
	\Delta L - \Delta x_{\rm trap}
	=
	\Delta L - {f_{\rm trap} \over \kappa_{\rm trap}},
\end{equation}
where $\Delta L$ is the variation of the trap position between that at minimum force and that at maximum force.
The resulting force-cellular-extension curve is shown in Fig.~\ref{fig:19:RBC_data}(a) for the second pulling cycle for two different RBCs.\footnote{It is best to analyze the second pulling curve because the first one could exhibit a special transient behavior.}

The RBC stiffness can be measured from the slope of the force--cellular-extension curves. 
As can be observed in Fig.~\ref{fig:19:RBC_data}(a), the force--cellular-extension curve is not a straight line. 
This indicates that the stiffness of the RBC depends on the force applied to the cell. 
For this reason, the slope is computed for five different force windows whose stiffness values are represented in Fig.~\ref{fig:19:RBC_data}(b). 
The RBC stiffness increases as we increase the applied force, the stiffness curves of the stretching and releasing cross each other between $6$ and $10\,{\rm pN}$, and the releasing stiffness is higher than the stretching stiffness at high forces while it is lower at low forces.
The expected values for RBC stiffness in this specific geometry are in the range between $3\,{\rm pN\,\upmu m^{-1}}$ and $10\,{\rm pN\,\upmu m^{-1}}$. 
The dispersion between different RBCs, in terms of force--cellular-extension curves, can be very large. 
This can be due to the fact that the RBCs are not separated by density so that the RBCs can have very different ages and, thus, very different physiological properties \cite{piomelli1993mechanism}. 

\subsubsection{Other applications to single-cell mechanics and outlook}

Until now, the manipulation of single cells with optical tweezers has not been extensively investigated as single-molecule manipulation. 
The main reason is the difficulty inherent to the geometry of the pulling assays when working with cells. 
Indeed, while the specific labeling of molecules makes it possible to design the most diverse pulling geometries, this is more difficult to achieve when manipulating single cells which ultimately require the physical contact between the optically trapped bead and the lipid cell membrane.  
Additionally, single cell manipulation with optical tweezers is feasible on cells in suspension (such as RBC, lymphocytes, senescent, stem cells), whereas adherent cells are harder to manipulate, often requiring confocal microscopy setups with optical traps very close to the glass substrate. 
For the latter class of cells, other techniques (e.g., atomic force microscopy, traction microscopy, magnetic twisting cytometry) turn out to be more informative. 
Finally, let us mention that while most single cell manipulation studies are carried out in {\it in-vitro} conditions (e.g., on the Petri dish), it will be necessary to expand the use of these manipulation techniques to more relevant environments such as {\it ex-vivo} tissues. 
In this case, the mechanical response of a single cell in a tissue-like environment is expected to differ a lot with respect to the isolated cell.  
Overall, this makes the future of single cell mechanics very promising beyond biology, with implications in physiology and medicine.

\subsection{Microrheology}
\label{sec:4.3:rheology}

Rheology investigates the mechanical response of soft materials such as simple liquids, polymer solutions, micellar fluids, colloidal suspensions, liquid crystals, gels, and foams \cite{waigh2016advances, robertson2018optical}.
Specifically, it is concerned with flow and deformation of such materials under applied stress or strain. Depending on the specific details of their microstructure, different kinds of rheological behavior can be observed. 
For instance, for most simple liquids, the rheological response is \emph{Newtonian}: upon imposing a given strain rate, the resulting stress is proportional to it, thereby dissipating energy instantaneously by viscous resistance. 
Instead, complex fluids and gels, whose components are usually long macromolecules suspended in a viscous solvent, exhibit \emph{viscoelasticity}, i.e., both viscous (liquid-like) and elastic (solid-like) responses are possible \cite{bird1987dynamics}; therefore, they are able to store and dissipate energy depending on the time-scale of observation. 
The goal of rheology is to provide quantitative parameters, e.g., viscosities, relaxation moduli and creep compliances, that relate the stress and the strain (or the strain rate) of the material under deformation. To this end, the most straightforward techniques consist in imposing stresses and strains in a controlled manner by means of a rheometer to the material sample of interest and then measuring the corresponding mechanical response. 
Despite their relative simplicity, these methods require typically milliliter-sized sample volumes, thus being unsuitable for soft matter systems that are expensive or difficult to find in abundance, such as biological fluids and newly synthesized materials.

Microrheology overcomes this major drawback by measuring mechanical properties of a soft material sample using colloidal probes directly embedded in it. The rheological information of the material is fully inferred from the motion of the probe, which requires only the detection and tracking of its  position over time. Therefore, very small sample amounts, typically ranging from pico- to microliters, are needed. This in turn allows to carry out  \emph{in-situ} rheological measurements which are inaccessible to macroscopic methods, e.g., biofluids within living tissues and cells, soft interfaces and membranes. Furthermore, microrheology offers several advantages over conventional macroscopic rheometry. For instance, unlike bulk rheology, which provides quantities averaged over the entire macroscopic sample, microrheology permits to explore local mechanical properties at the length-scale of the micron-sized probe. This is particularly useful for investigating heterogeneous materials and fluids close to interfaces, where submillimetric spatial variations of the rheological parameters occur. In addition, undesirable artifacts that can arise due to inertial effects when using a macroscopic rheometer are automatically ruled out in microrheology due to the extremely low Reynolds numbers (${\rm Re} \lesssim 10^{-3}$) of the flows induced by the motion of a micron-sized probe. This expands significantly the available frequency range over which the linear viscoelastic spectrum of the system can be determined without inertial corrections, typically from the range $[10^{-1}\,{\rm Hz}, 10^2\,{\rm Hz}]$ up to the range $[10^{-1}\,{\rm Hz}, 10^6\,{\rm Hz}]$. Besides, conventional rheometers are usually designed for either highly elastic or highly viscous materials, but they are not sensitive enough to measure low viscosities and elastic moduli; then, another advantage of microrheology is its capability to detect very weak viscoelastic parameters. Moreover, the typical forces and energies involved in microrheology, either from the thermal motion of the material molecules or those externally exerted by optical tweezers, are of the order of $\sim$pN and $\sim$$k_{\rm B} T$, respectively. This guarantees that the microstructure of the sample  is not irreversibly destroyed by the motion of the probe. 

\subsubsection{Basic setup and concepts}

\begin{figure}[ht!]
	\begin{center}
	\includegraphics[width=12cm]{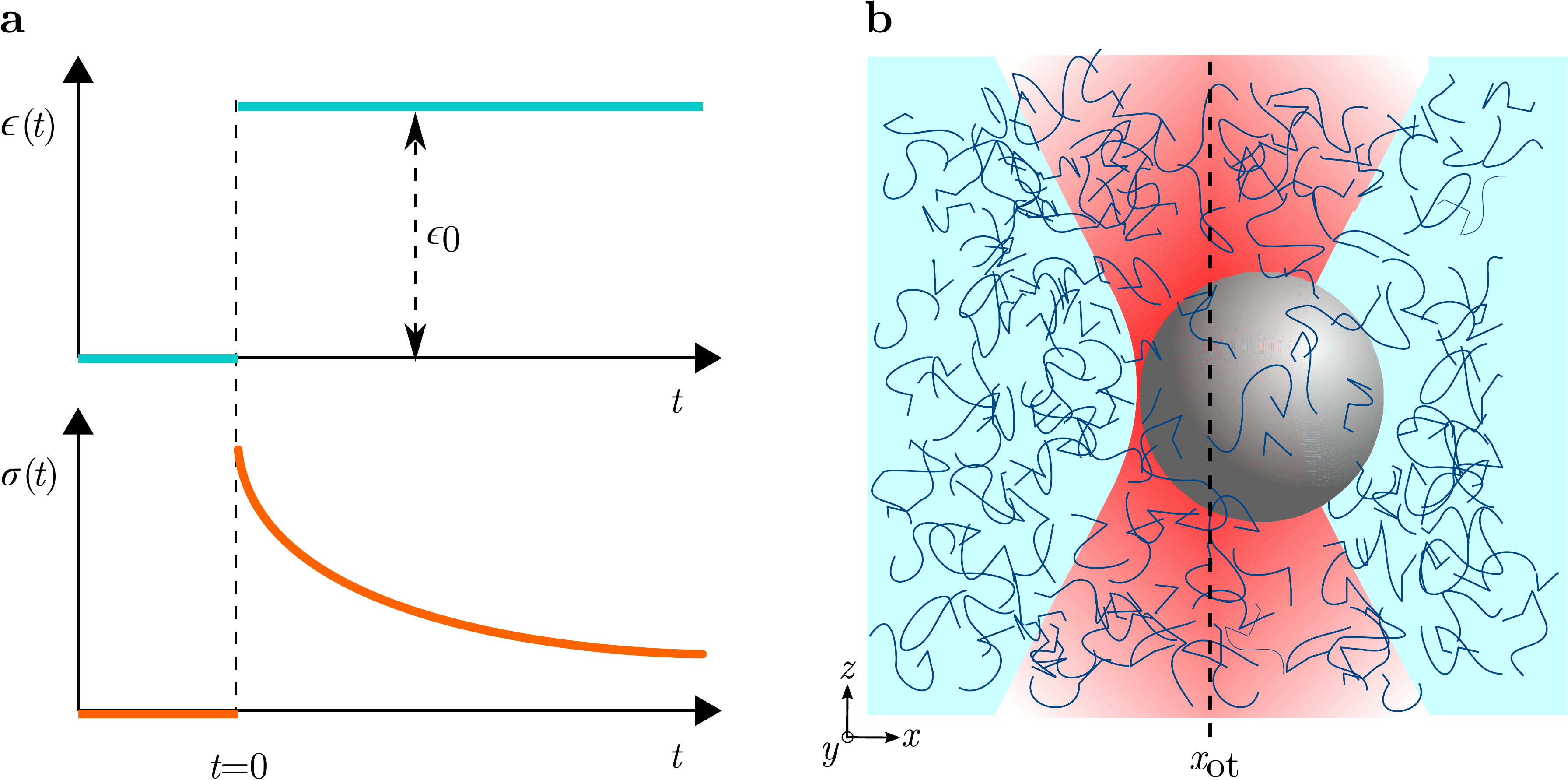}
	\caption{
	{\bf Microrheology with an optically trapped particle.}
	(a) Sketch of the general response of the shear stress $\sigma(t)$ of a viscoelastic material to a step-like shear strain $\epsilon(t)$. For $t \ge 0$, the ratio $\sigma(t)/\epsilon_0$ defines the stress relaxation modulus $G(t)$.
	(b) Sketch of the typical experimental system used in microrheology: a spherical colloidal bead (radius $a$), embedded in a soft material, is trapped by optical tweezers.
	If the position of the optical trap is fixed ($x_{\rm ot}(t) \equiv 0$, \emph{passive microrheology}), the particle is only subject to the thermal collisions of the molecules of the surrounding fluid microstructure. 
	If $x_{\rm ot}(t)$ is varied in time, an additional time-dependent force is exerted on the probe particle (\emph{active microrheology}). 
	}
	\label{fig:20:microrheology}
	\end{center}
\end{figure}

Optical tweezers provide the ideal tool to manipulate probe particles in microrheology \cite{fusco2007high, pesce2009microrheology, squires2010fluid, robertson2018optical, tassieri2012microrheology}, because they permit one to tailor the forces that can be exerted to locally strain the sample.

To extract meaningful information by means of standard microrheological methods, some experimental conditions must be fulfilled. First of all, the material must be homogeneous and isotropic at the scale of the probe, which is usually valid when the relevant length-scale of the material, e.g., the mesh size of gels and semidilute polymer solutions, is much smaller than the probe size. 
In this case, the environment behaves as a continuum and its mechanical response can be fully characterized by a single scalar function, e.g., the stress relaxation modulus $G(t)$.  
For example, upon applying a step-like shear strain at $t=0$ to a material, $\epsilon(t) = \epsilon_0 \Theta(t)$, where $\Theta$ is the Heaviside function, the stress relaxation modulus is given by
\begin{equation}\label{eq:85111}
G(t) = {\sigma(t) \over \epsilon_0} \quad {\rm for} \quad t \ge 0 ,
\end{equation}
where $\sigma(t)$ is the resulting shear stress (Fig.~\ref{fig:20:microrheology}(a)).
For a \emph{viscoelastic liquid}, $G(t)$ decays to zero over a finite time, while for a \emph{viscoelastic solid}, it decays to a constant non-zero value. In both cases, the drag force experienced by an embedded particle at time $t$ can be written as 
\begin{equation}\label{eq:86111}
{\bf F}_{\rm drag}(t) = -\int_{-\infty}^t \mathrm{d}t' \Gamma(t - t') \dot{\bf r}(t'),
\end{equation}
where $\dot{\bf r}(t')$ is the instantaneous particle velocity at time $t' < t$ and the minus sign indicates that the drag force is opposed to the instantaneous velocity, whereas $\Gamma$ is a memory function related to $G(t)$, which weights the role of the previous history of the particle motion on its current drag force due to the temporal correlations induced by the surrounding medium. For example, for a spherical particle of radius $a$, the memory function is $\Gamma(t) = 6\pi a G(t)$,  which can be regarded as a generalization of the Stokes' law (equation~\eqref{eq:stokes}). For a Newtonian fluid with shear viscosity $\eta$, the relaxation modulus is $G(t) = 2\eta\delta(t)$,\footnote{Here, we use the following convention for the Dirac delta $\int_0^{\infty} delta(t') dt' = 1/2$. Thus, the factor 2 is needed in the relaxation modulus G(t) to recover the correct friction coefficient.} which corresponds to an instantaneous response of the drag force ${\bf F}_{\rm drag} (t) = -6\pi \eta a \dot{\bf r}(t)$ to the particle velocity $\dot{\bf r}(t)$. Moreover, for some techniques, it is required that the sample is in thermal equilibrium and that the forces externally applied to the probe are small enough to keep the material in its linear response regime. Under such conditions, the time evolution of a single particle coordinate, $x$, is described by the generalized Langevin equation \cite{kubo1966fluctuation} 
\begin{equation}\label{eq:GLE}
	0 
	= 
	-\int_{- \infty}^t \mathrm{d}t'  \Gamma(t - t') \dot{x}(t')
	- \kappa [ 
		x(t) 
		- 
		x_{\rm ot}(t)
	] 
	+ \zeta(t),
\end{equation}
where $ x_{\rm ot}$ represents the $x$-component of the central position of the optical trap, while $\zeta$ is a Gaussian noise of zero-mean, $\langle \zeta(t) \rangle = 0$, and autocorrelation $\langle \zeta(t) \zeta(t') \rangle = k_{\rm B} T \Gamma(|t - t'|)$.

Depending on how optical tweezers are employed to manipulate a colloidal probe embedded in a soft material (Fig.~\ref{fig:20:microrheology}(b)), two kinds of microrheological techniques exist: \emph{passive} and \emph{active} \cite{squires2010fluid,waigh2016advances}. 
In the following sections, we describe how to prepare a prototypical viscoelastic fluid (section~\ref{susec:4.3.2:viscoelastic}) as well as how to determine rheological parameters of simple liquids using passive and active microrheology (sections~\ref{sec:4.3.3:passive_microrheology}
 and \ref{sec:4.3.4:active_microrheology}).

\subsubsection{Preparation of a viscoelastic fluid and setup}\label{susec:4.3.2:viscoelastic}

An equimolar wormlike micellar solution made of the surfactant cetylpyridinium chloride (CPyCl) and the salt sodium salicylate (NaSal), both dissolved in water, represents a prototypical viscoelastic fluid.
To prepare it, overnight mixing of the two components in deionized water is required at approximately $50^\circ{\rm C}$, after which a homogenous and isotropic viscoelastic fluid results.\footnote{It is recommended to wear gloves and safety googles when handling CPyCl, as direct contact with power can cause skin and eye irritation.}
To get reproducible rheological parameters of the fluid, the conductivity of water should be kept at most at $10\,{\rm \upmu S\,m^{-1}}$ (milli-Q water is a good option for this purpose).
Depending on the surfactant/salt concentration, different kinds of microstructures in the aqueous solvent are formed \cite{handzy2004oscillatory}. For instance, above the first critical micelle concentration but below $4.5\,{\rm mM}$, the surfactant molecules aggregate into spherical and cylindrical micelles of radius $2$ to $3\,{\rm nm}$, which result in a weakly viscoelastic behavior of the dilute solution. 
Increasing further the concentration up to $10\,{\rm mM}$, the surfactant molecules self-organize in long flexible cylindrical micelles, which form, overlap, and deform dynamically in the solvent. Such wormlike micelles have a radius from $2$ to $3\,{\rm nm}$ and a contour length ranging from $100\,{\rm nm}$ to $1\,{\rm \upmu m}$, while their persistence length and mesh size are of order $10\,{\rm nm}$ for concentrations between $5$ and $10\,{\rm mM}$.
To perform optical-tweezers-based microrheology, a  very small amount of micron-sized beads must be suspended in a microliter volume of solution. Special care is needed to avoid air bubbles inside the fluid, which can be easily formed due to the presence of the surfactant. Sonication during a few minutes of the fluid with the dispersed colloidal particles is recommended before being loaded into the sample cell. 
Such a sample cell consist of a standard chamber (cross-sectional area about $1\,{\rm cm^2}$) made of a glass slide and a coverslip, separated by a distance of approximately $100\,{\rm \upmu m}$ by means of, e.g., spacer tape or parafilm strips. The chamber must be laterally sealed with UV-curable optical adhesive after inserting a small volume ($\sim10\,{\rm \upmu l}$) of the suspension, in order to avoid evaporation of the fluid and drift.
During the measurements, the fluid sample must be kept thermally coupled to a thermostat at constant temperature $T$ within an accuracy range of $\pm 1\,{\rm K}$, as the rheological properties of the fluid strongly depend on temperature.
It is crucial to maintain $T$ above  the so-called Krafft temperature \cite{everett1972manual}, which is $18^\circ{\rm C}$ for the specific micellar solution described here. Below the Krafft temperature, the surfactant crystallizes, thus suppressing the viscoelasticity of the fluid.

To illustrate the microrheological techniques described in the following, two Newtonian fluids with distinct viscosities at room temperature are used. The first one is deoinized water, whose conductivity must be less than $10\,{\rm \upmu S\,m^{-1}}$ in order to facilitate the sample cell preparation. The second Newtonian fluid is propylene glycol n-propyl ether (PNP), also known as 1-propoxy-2-propanol, which can be directly purchased with a purity $\approx 99\%$ so that no further purification is needed. On the other hand, two viscoelastic fluids are employed: an equimolar CPyCl/NaSal micellar solution at $4\,{\rm mM}$ and one at $5\,{\rm mM}$. The different fluids and their main properties are summarized in Table~\ref{tab:9:microrheology}.

\begin{table}[h!]
	\footnotesize
	\begin{center}
	\begin{tabular}{l|l|l}
		\textbf{Fluid}
		&
		\textbf{Properties}
		&
		\textbf{Technique}
		\\
	\hline 
		Water
		&
		Newtonian (low viscosity)
		&
		Passive microrheology
		\\
	\hline 
		PnP
		&
		Newtonian (high viscosity)
		&
		Passive microrheology
		\\
	\hline 
		CPyCl/NaSal ($4\,{\rm mM}$) 
		&
		Viscoelastic (small elastic modulus and relaxation time)
		&
		Passive microrheology
		\\
	\hline 
		CPyCl/NaSal ($5\,{\rm mM}$) 
		&
		Viscoelastic (larger elastic modulus and relaxation time)
		&
		Passive and active microrheology
		\\
	\hline 
	\end{tabular}
	\caption{
	{\bf Microrheology fluids.} 
	Fluids used to illustrate microrheology with an optically trapped particle.
	}
	\label{tab:9:microrheology}
	\end{center}
\end{table}

All the experiments described in this section are performed in a standard optical tweezers setup implemented with an oil-immersion microscope objective ($100\times$, ${\rm NA}=1.4$) which tightly focuses a Gaussian laser beam ($\lambda = 1070\,{\rm nm}$) onto the sample. A single spherical particle is trapped and kept at least $10\,{\rm \upmu}$ away from any wall of the sample cell in order to avoid hydrodynamic interactions, which can significantly affect the values of the fluid's rheological parameters with respect to those obtained from bulk measurements. Imaging and position detection can be realized by conventional video microscopy. In particular, for the determination of the linear viscoelastic spectrum by passive microrheology, it is required that the acquisition frequency of the camera must be above about 1000 frames per second. Therefore, the use of a position sensitive detector is an alternative for this purpose. Furthermore, beam steering, which is crucial for active microrheology, is achieved here by means of galvomirrors (section~\ref{sec:2.3.1:beamsteering}).

\subsubsection{Passive microrheology}
\label{sec:4.3.3:passive_microrheology}

Passive microrhelogy is the simplest approach to study the local mechanical response of soft matter. As its name suggests, the embedded colloidal bead is used as a passive element, whose thermal motion provides all the necessary information to extract, via  fluctuation-dissipation relations, the frequency-dependent rheological parameters of the material under investigation. To prevent sedimentation as well as significantly large diffusive displacements of the probe from its initial position, which would otherwise restrict the measurement time of the particle position $x(t)$, a static optical trap can be used to confine the particle motion around a mean fixed position, i.e., $x_{\rm ot}(t) = 0$, as depicted in Fig.~\ref{fig:20:microrheology}(b). Here, we present two direct applications of passive microrheology with optical tweezers for viscous and viscoelastic liquids. 

\paragraph{Determination of zero-shear viscosities.}

\begin{figure}[b!]
	\begin{center}
	\includegraphics[width=12cm]{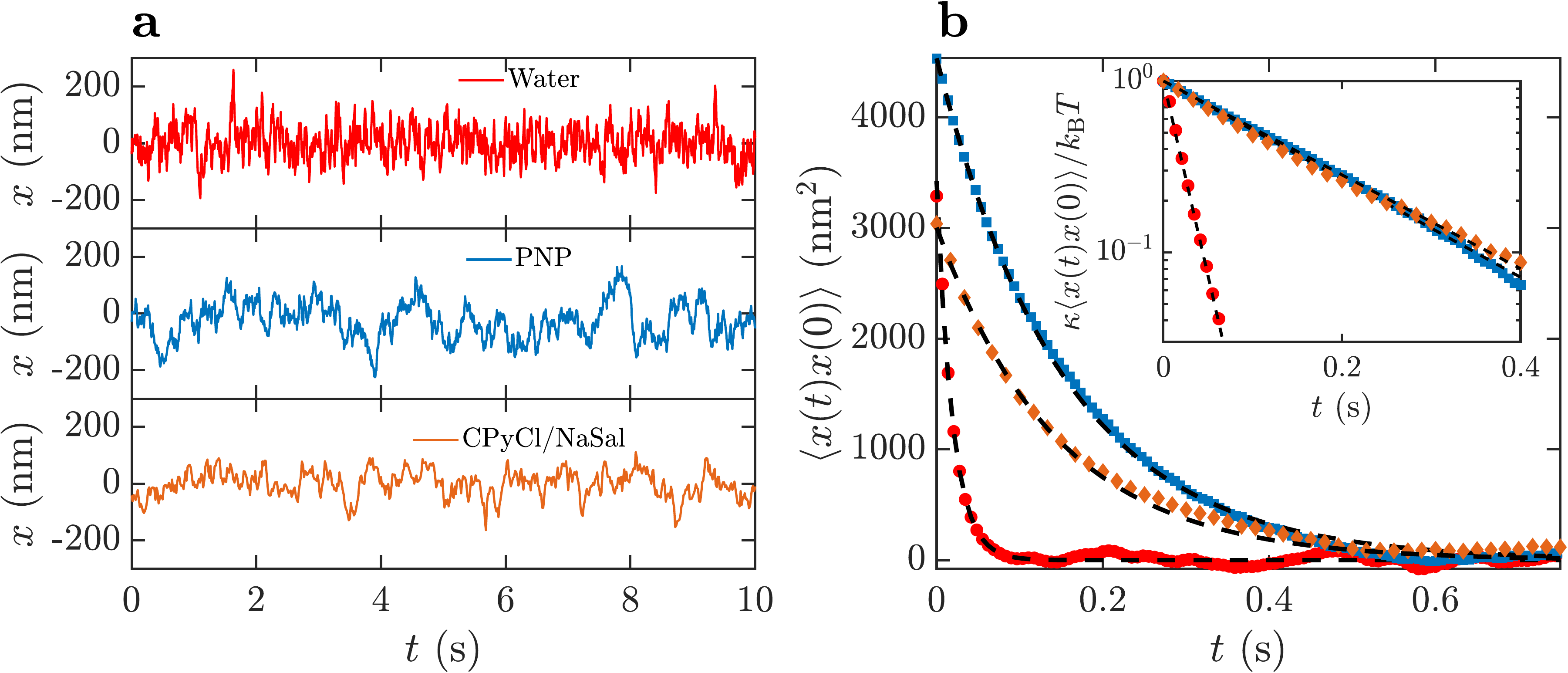}		
	\caption{
	{\bf Determination of zero-shear viscosities.}
	(a) Trajectories of particles of distinct diameters trapped in different fluids. From top to bottom: water ($2a = 2.73\,{\rm \upmu m}$), PNP ($2a = 3.25\,{\rm \upmu m}$), and aqueous micellar solution of CPyCl/NaSal at $4\,{\rm mM}$ ($2a = 3.25\,{\rm \upmu m}$). 
	(b) Autocorrelation functions of the particle position for the cases in water ($\circ$), PNP ($\square$), and micellar solution ($\diamond$). The dashed lines correspond to exponential fits. 
	Inset: semilog representrations of the position autocorrelation function for the three cases, normalized by their corresponding variances $\langle x(t)^2\rangle = \frac{k_{\rm B} T}{\kappa}$. 
	}
	\label{fig:21:PM}
	\end{center}
\end{figure}

For a Newtonian liquid, the autocorrelation function of the particle position provides a straightforward way to determine its viscosity $\eta$. From equation~\eqref{eq:GLE}, using the instantaneous memory function $\Gamma(t) = 2\gamma \delta(t)$ for a purely viscous liquid, the expression for its autocorrelation function can be readily derived (section~\ref{sec:3.5:acf}):
\begin{equation}\label{eq:autocorrx}
	\langle x(t) x(0) \rangle 
	= 
	{k_{\rm B} T \over \kappa} 
	e^{ - t / \tau_{\rm ot}}  
	\quad 
	\mbox{for} 
	\quad 
	t \ge 0,
\end{equation}
where  $\langle \ldots \rangle$ represents an ensemble average over different realizations of the thermal noise, while $\tau_{\rm ot} = \gamma/\kappa$ is a decay rate related to the viscous friction coefficient $\gamma$ of the probe and the trap stiffness $\kappa$. In practice, the ensemble average can be replaced by a time average along the stochastic trajectory $x(t)$, provided that the total measurement time of $x(t)$ is much larger than $\tau_{\rm ot}$. 
For a spherical particle of radius $a$, the friction coefficient is $\gamma = 6 \pi a \eta$, where $\eta$ is the fluid viscosity. Then, once the trap stiffness $\kappa$ is known (e.g., by the equipartition method: $\kappa = \frac{k_{\rm B} T}{\langle x(t)^2 \rangle}$, section~\ref{sec:3.3:eq}), the experimental autocorrelation function of $x(t)$ can be fitted to the right-hand side of equation~\eqref{eq:autocorrx}. Therefore, $\eta$ can be determined from the fitting parameter $\tau_{\rm ot}$ by means of $\eta = {\kappa \tau_{\rm ot} \over 6 \pi a}$. This is illustrated in Fig.~\ref{fig:21:PM}, where the viscosities at $T = 25^{\circ}$C of two Newtonian fluids, water and propylene glycol n-propyl ether (PNP), are computed by means of this passive method. In Fig.~\ref{fig:21:PM}(a), we plot the typical trajectories $x(t)$ of particles embedded in these liquids, trapped by optical tweezers at approximately $40\,{\rm \upmu m}$ away from the cell walls.  The corresponding autocorrelation functions of $x(t)$ are shown in Fig.~\ref{fig:21:PM}(b), where we show that the experimental data (symbols) can be very well fitted to equation~\eqref{eq:autocorrx} (dashed lines).
From the equipartition relation, we first determine the values of the trap stiffness for each case: $\kappa = 1.24\,{\rm pN\,\upmu m^{-1}}$ for the particle in water (diameter $2a = 2.73\,{\rm \upmu m}$) and $\kappa = 0.90\,{\rm pN\,\upmu m^{-1}}$ for the particle in PNP (diameter $2a = 3.25\,{\rm \upmu m}$). Next, from the fitting parameters $\tau_{\rm ot}$, we find the corresponding values of the viscosities of both fluids: $0.9\,{\rm mPa\,s}$ for water and $4.5\,{\rm mPa\,s}$ for PNP, which agree very well with their bulk values.

This technique can be extended to determine steady-state flow properties of viscoelastic fluids, which exhibit  both liquid- and solid-like responses at times shorter than a typical time $\tau$. Such a time-scale reflects the stress relaxation of the elastic material microstructure suspended in the solvent. For a viscoelastic micellar solution, $\tau$ originates from the continuous formation and breaking of the micelles as well as the reptation modes of the wormlike structures, and can range from a few milliseconds to several seconds in the semidilute regime \cite{cates1990statics}. Unlike viscoelastic \emph{solids} (e.g., gels), viscoelastic fluids are able to flow with constant shear rate at sufficiently long times, $t \gg \tau$, after applying constant shear stress at $t = 0$. Therefore, they are characterized by a zero-shear viscosity $\eta_0$, which in absence of a confining potential, leads to a long-time diffusion of an embedded spherical particle, where the diffusion coefficient is given by $\frac{k_{\rm B} T}{6 \pi a \eta_0}$. In the presence of an optical trap, from equation~\eqref{eq:GLE} it can be shown that the position autocorrelation function has the form
\begin{equation}\label{eq:autocorrxVE}
	\langle x(t) x(0) \rangle 
	= 
	{k_{\rm B} T \over \kappa} 
	A 
	e^{ - t / \tau_{\rm ot}} 
	\quad 
	\mbox{for} 
	\quad 
	\tau \ll t < \tau_{\rm ot},
\end{equation}
where $A<1$ is a prefactor that depends mainly on $\kappa$ and $\tau$. Equation~\eqref{eq:autocorrxVE} is valid provided that $\tau_{\rm ot} \gg \tau$ \cite{tassieri2015microrheology}, which must be verified \emph{a posteriori}. To fulfill this condition of large time-scale separation, sufficiently small values of the trap stiffness are needed. In Figs.~\ref{fig:21:PM}(a) and \ref{fig:21:PM}(b), we illustrate this method for a $2a = 3.25\,{\rm \upmu m}$ silica particle suspended in a dilute micellar solution (concentration $4\,{\rm mM}$, $T = 295\,{\rm K}$), trapped by optical tweezers with stiffness $\kappa = 1.34\,{\rm pN\,\upmu m^{-1}}$.
At this concentration, the viscoelastic fluid is composed of non-overlapping micellar structures, thus resulting in typical relaxation times, $\tau$, of order of milliseconds. We check that equation~\eqref{eq:autocorrxVE} is a very good approximation to the experimental position autocorrelation function, where only the first experimental point slightly deviates from the fit. The fitting parameter $\tau_{\rm ot} = 0.16\,{\rm s}$ yields the value $\eta_0 = 6.9\,{\rm mPa\,s}$, which is also in agreement with reported bulk measurements \cite{handzy2004oscillatory}. 

\paragraph{Determination of storage and loss moduli.}

\begin{figure}[b!]
	\begin{center}
	\includegraphics[width=12cm]{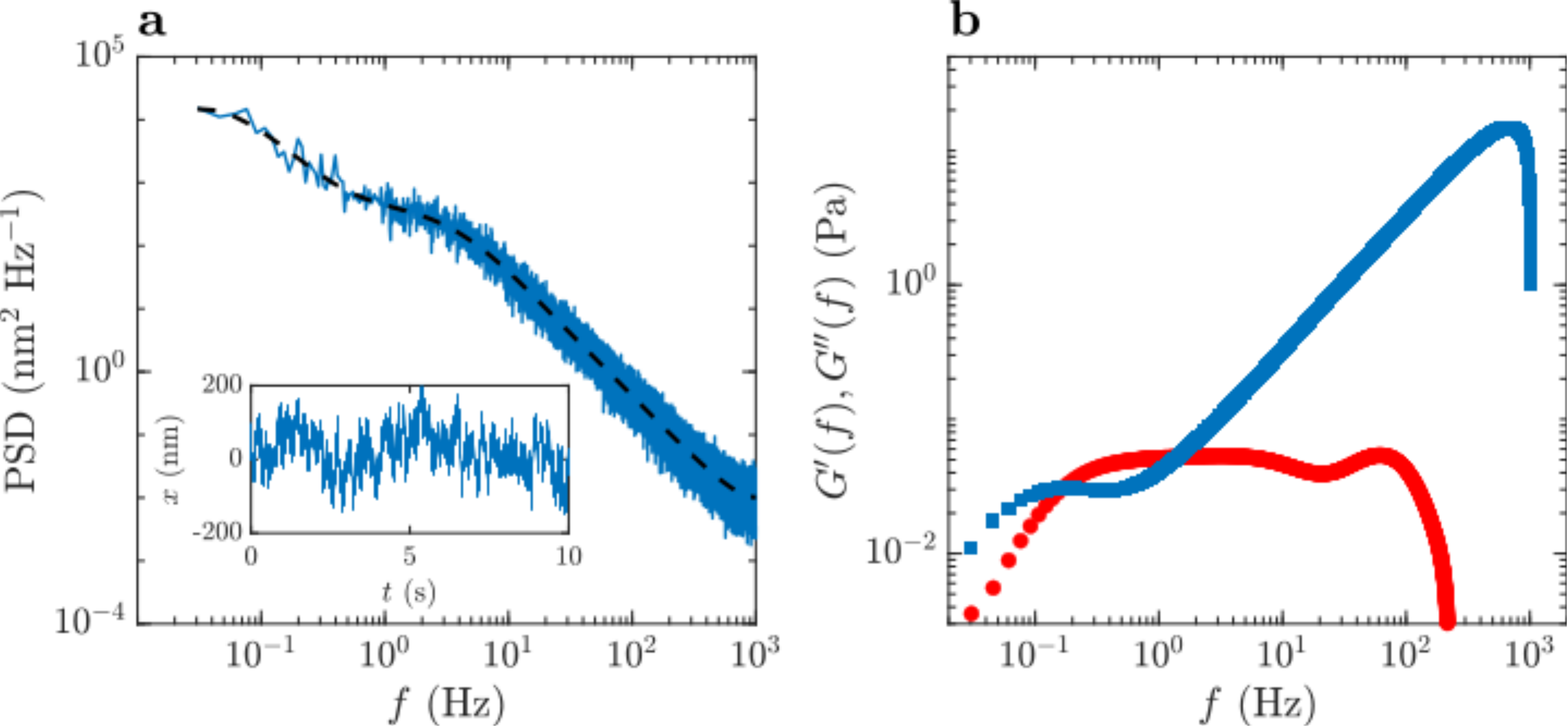}
	\caption{
   	{\bf Determination of storage and loss moduli.}
	(a) Power spectral density of equilibrium fluctuations of $x(t)$ for a particle trapped by optical tweezers in a wormlike micellar solution of CPyCl/NaSal at 5 mM. 
	Inset: time evolution of $x(t)$ over $10\,{\rm s}$. 
	(b) Storage modulus (red $\circ$) and loss modulus (blue $\square$) of the wormlike micellar solution determined by passive microrheology. 
	}
	\label{fig:22:PM2}
	\end{center}
\end{figure}

The complex shear modulus, $G^*(f) = G'(f) +  i  G''(f)$, is a quantity used in rheology to characterize linear viscoelasticity of soft matter in the frequency domain
\cite{larson1999structure}. 
$G'(f)$ is the storage modulus and $G''(f)$ is the loss modulus, which account for the elastic energy in phase with the applied strain and the out-of-phase viscous dissipation of the material, respectively. 
For instance, for a purely viscous liquid (viscosity $\eta$), $G'(f) = 0$ and $G''(f) = 2\pi f \eta$, whereas for an elastic solid (elastic modulus $G_0$), $G'(f) = G_0$, and $G''(f) = 0$. In general, for viscoelastic materials, both storage and loss moduli are non-zero. 
$G^*(f)$ is closely related to the Fourier transform, $\tilde{G}(f)$, of the stress relaxation modulus, $G(t)$, by means of 
\begin{equation}\label{eq:90111}
G^*(f)  = 2\pi  i  f \tilde{G}(f). 
\end{equation}
Therefore, for an applied oscillatory shear strain $\epsilon(f) = \epsilon_0 e^{2\pi i  f t}$ with amplitude $\epsilon_0$ and frequency $f$, the resulting shear stress is also oscillatory and can be expressed as $ G^*(f) \epsilon(f)$. 

A very helpful expression for $G^*(f)$ in terms of the equilibrium fluctuations of the position $x(t)$ of a trapped particle can be derived from the fluctuation-dissipation theorem.
The latter relates the linear response function $\chi(t)$ of $x(t)$ to a perturbative force $F(t)$, defined as
\begin{equation}\label{eq:linearresp}
	\langle x(t) \rangle_F 
	= 
	\langle x(0) \rangle_F 
	+ 
	\int_0^t dt' \chi(t - t')  F(t'),
\end{equation}
with the \emph{equilibrium} ($F(t)=0$) autocorrelation function of $x(t)$, i.e., $\langle x(t) x(0) \rangle$. In equation~\eqref{eq:linearresp}, the \emph{nonequilibrium}  ensemble average $\langle \ldots \rangle_F$ is defined in presence of the time-dependent perturbation. From equations~\eqref{eq:GLE} and \eqref{eq:linearresp},  the Fourier transform of $\chi(t)$, $\tilde{\chi}(f) = \int_{-\infty}^{\infty} dt e^{-2\pi i f t}\chi(t) = \chi'(f) +  i  \chi''(f)$, can be directly related in frequency domain to the complex shear  modulus $G^*(f)$ by means of 
\begin{equation}\label{eq:StokesEinstein}
 	 G^*(f) 
	 =
	 {1 \over 6\pi a} 
	 \left[ 
	 	{1 \over \tilde{\chi}(f)} 
		- 
		\kappa 
	\right].
\end{equation}
Note that equation~\eqref{eq:StokesEinstein} shows that the elastic contribution of the optical tweezers stiffness to the mechanical response of the system must be subtracted for a correct determination of the complex shear modulus of the surrounding medium. Moreover, equation~\eqref{eq:StokesEinstein} involves directly the inverse of $\tilde{\chi}(f)$, which can be determined by either passive or active microrheology, as will be shown. Therefore, this expression is applicable for both types of microrheological techniques. In particular, in the case of passive microrheology, by making use of the fluctuation-dissipation theorem in frequency domain and the Kramers-Kronig relations, one can find both the real, $\chi'(f)$, and the imaginary, $\chi''(f)$, parts of the response function $\tilde{\chi}(f)$:
\begin{eqnarray}
	\chi'(f) 
	& = & 
	{2\pi \over k_{\rm B} T} 
	P 
	\int_0^{\infty} d\nu 
		\frac{
			\nu^2 S(\nu)
		}{
			\nu^2 - f^2
		}, 
	\label{eq:rechi} 
	\\
	 \chi''(f) 
	 & = & 
	 \frac{
	 	\pi f
	}{
		2k_{\rm B} T
	} 
	S(f), 
	\label{eq:imchi} 
\end{eqnarray}
where $S(f) = \langle |\tilde{x}(f)|^2 \rangle$ is the one-sided PSD of $x(t)$ in thermal equilibrium, i.e., the Fourier transform of $\langle x(t) x(0)\rangle$, whereas $P$ denotes the Cauchy principal value of the integral. Thus, once $\chi'(f)$ and $\chi''(f)$ are determined from the experimentally recorded trajectory of the trapped particle, the storage modulus $G'(f)$ and loss modulus $G''(f)$ can be computed over the available frequency range $[0, f_{\rm max}]$ from equation~\eqref{eq:StokesEinstein}. For a given sampling frequency $f_{\rm s}$ of $x(t)$,  $f_{\rm max}$ is fixed by the Nyquist frequency: $f_{\rm max} = {1\over2}f_{\rm s}$. This technique needs a rather high sampling frequency of $x(t)$ to avoid a large underestimation of the integral in equation~\eqref{eq:rechi}.

In Fig.~\ref{fig:22:PM2}, we apply the previous method to characterize the linear viscoelasticity of the wormlike micellar solution in the semidilute regime at $5\,{\rm mM}$ and $T = 20^\circ {\rm C}$. 
First, we compute the equilibrium PSD, $S(f)$, of a long trajectory $x(t)$ measured over $25\,{\rm min}$ at a sampling frequency $f_{\rm s} = 2000\,{\rm Hz}$. Since any discrete Fourier transform involved in the calculation of $S(f)$ leads inexorably to a noisy curve, the spectral profile must be smoothed before numerically computing the integral of equation~\eqref{eq:rechi}. 
In Fig.~\ref{fig:22:PM2}(a), we show the original noisy profile of $S(f)$ (blue line) obtained by means of a discrete Fourier transform computed with $2^{17}$ points. A polynomial fit (dashed black line) is performed to smooth the PSD profile. Then, a direct numerical calculation using equations~\eqref{eq:StokesEinstein}-\eqref{eq:imchi} leads to the values of the storage and loss moduli shown in Fig.~\ref{fig:22:PM2}(b) over the available frequency interval $0\,{\rm Hz} \le  f \le 1000\,{\rm Hz}$. 
At high frequencies approaching ${1\over2}f_{\rm s}$, an abrupt decrease of the numerical values of both $G'(f)$ and $G''(f)$ is observed. This is an artifact due to the finite frequency range, which imposes a cut-off in the required Kramers-Kronig integral transformation \cite{schnurr1997determining}. Therefore, the physically meaningful values of the components of
the complex modulus $G^*(f)$ are only those at frequencies one decade below the Nyquist frequency \cite{nishi2018symmetrical}.

\subsubsection{Active microrheology}
\label{sec:4.3.4:active_microrheology}

\begin{figure}[b!]
	\begin{center}
	\includegraphics[width=12cm]{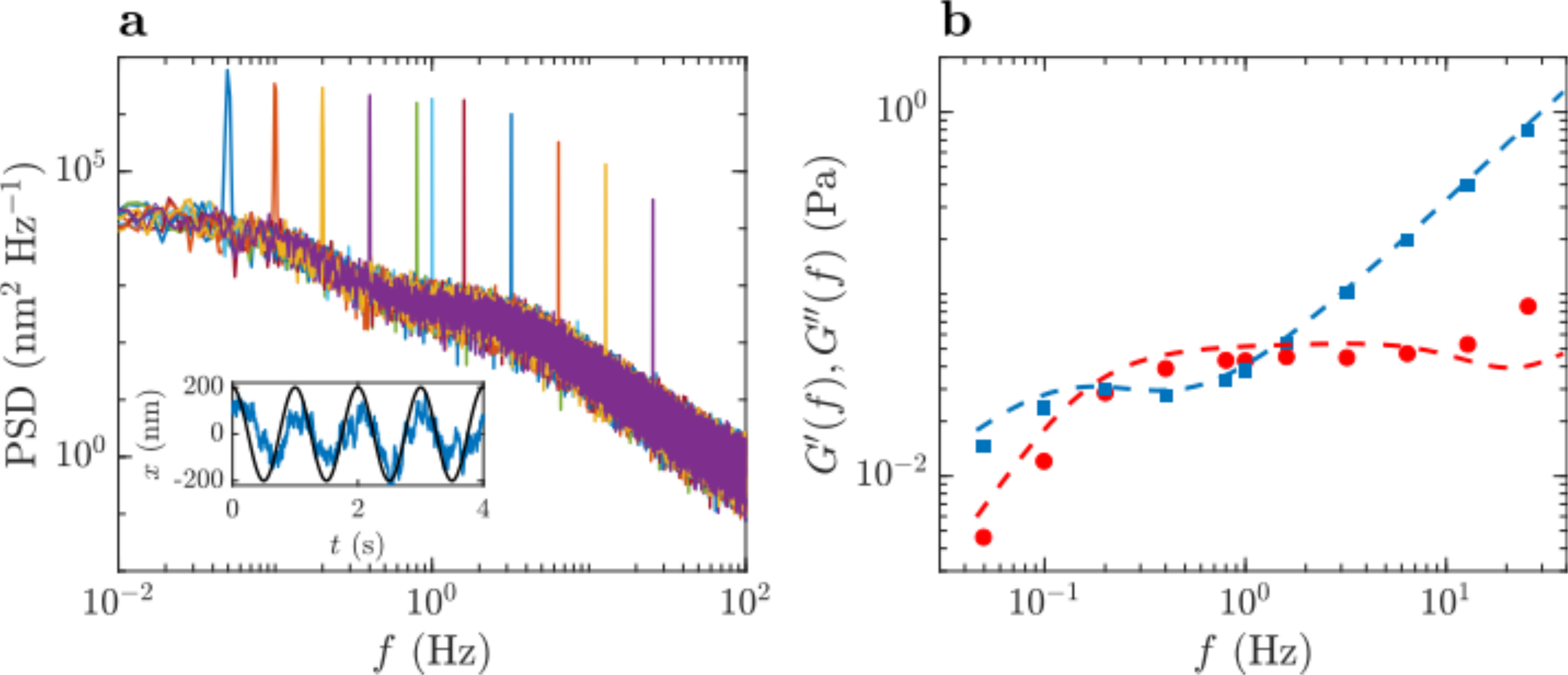}
	\caption{
	{\bf Active microrheology.}
	(a) Power spectral density of the particle position, driven at different frequencies by a sinusoidal motion of the optical trap. The peaks are located at the imposed driving frequencies $f_{\rm d}$.  
	Inset: Time evolution of the perturbative force divided by the trap stiffness at $f_{\rm d} = 1\,{\rm Hz}$ (thick solid line) and resulting particle position (thin solid line). 
	(b) Numerical values of the storage modulus (red $\bullet$) and loss (blue $\blacksquare$) obtained by means of equations~\eqref{eq:StokesEinstein} and \eqref{eq:respactphi}. The dotted and dashed lines depict the corresponding curves shown in Fig.~\ref{fig:22:PM2}(d) obtained by passive microrheology.
	}
	\label{fig:23:AM}
	\end{center}
\end{figure}

In active microrheology, the trapped colloidal probe is externally forced through the material to locally strain it, and its mechanical response is directly measured \cite{wilson2011small}. 
From this response, $G^*(f)$ can be determined by means of equation~\eqref{eq:StokesEinstein}. 
Different kinds of perturbations are possible. For instance, the probe particle can be trapped by static optical tweezers while the whole sample cell is moved by a piezo stage (section~\ref{sec:3.11:active}).
Here, we will focus on a perturbation created by the oscillatory motion of the position $x_{\rm ot}(t)$ of the optical trap, which exerts a time-dependent force $F(t) = \kappa x_{\rm ot}(t)$ on the particle according to equation~\eqref{eq:GLE} (Fig.~\ref{fig:20:microrheology}(b)). 

As active microrheology provides a direct measurement of the linear response function $\chi(t)$ defined by equation~\eqref{eq:linearresp}, it does not necessarily require that the material under investigation is in
thermal equilibrium. Consequently, this technique is suitable for the investigation of rheological properties of general viscoelastic fluids and solids, including  out-of-equilibrium soft matter, such as actin networks \cite{mizuno2008active, gallet2009power}, physical gels \cite{gomez2013nucleation}, and glassy colloidal suspensions \cite{jop2009experimental, senbil2019observation}.
It is particularly useful for the determination of the storage $G'(f)$ and loss $G''(f)$ modulus at a specific driving frequency $f_{\rm d}$, for which a single-frequency oscillatory perturbation force 
\begin{equation}\label{eq:sinforce}
	F(t) 
	= 
	\kappa X_0 
	\cos(2\pi f_{\rm d} t + \phi)
\end{equation}
can be directly applied to the colloidal probe, where $X_0$ is the maximum displacement of the optical trap and $\phi$ is the initial phase. Therefore, a previous passive calibration of $\kappa$ is needed to know the applied force (section~\ref{sec:3:calibration}). In order for both the optical trapping and the material microstructure to remain in the linear response regime, $X_0$ must be sufficiently small. This is generally fulfilled if the typical energy injected by $F(t)$ is at most of the order of $\sim k_{\rm B} T$, i.e., $X_0 \lesssim \sqrt{k_{\rm B} T / \kappa}$.

In Figure~\ref{fig:23:AM}, we show some results to illustrate the implementation of active microrheology to characterize the rheological properties of a wormlike micellar solution at $5\,{\rm mM}$ and $T = 20^{\circ}{\rm C}$ (Table~\ref{tab:9:microrheology}).
In the inset of Figure~\ref{fig:23:AM}(a), the black line shows the time evolution of $F(t)/\kappa$, which results from the sinusoidal motion of the optical trap (equation~\eqref{eq:sinforce} with $X_0 = 200\,{\rm nm}$, $\kappa = 1.0\,{\rm pN\,\upmu m^{-1}}$, $f_{\rm d} = 1\,{\rm Hz}$). 
The blue line plots the trajectory $x(t)$ of an actively driven particle ($2a = 2\,{\rm \upmu m}$), which clearly exhibits an oscillatory behavior with a time delay relative to $F(t)$ due to the resistance of the surrounding viscoelastic fluid. 
Here, the crucial step is to compute the Fourier transform of the linear response function from the measurement of $x(t)$ and $x_{\rm ot}(t)$. 
From equation~\eqref{eq:linearresp}, this is given by 
\begin{equation}\label{eq:respact}
	\tilde{\chi}(f) 
	= 
	\frac{
		\langle \tilde{x}(f) \rangle_F
	}{
		\kappa \tilde{x}_{\rm ot}(f)
	},
\end{equation}
where the numerator is the value at frequency $f$ of the non-equilibrium ensemble average of the Fourier transform of $x(t)$, in presence of $F(t)$ with the same initial phase $\phi$, whereas the denominator involves the Fourier transform of the trap position. Due to the sinusoidal form of ${x}_{\rm ot}(t)$, the denominator is zero for all frequencies $f$ different from $f_{\rm d}$. Therefore, in practice it is customary to compute the inverse of equation \eqref{eq:respact} and to approximate it at $f_{\rm d}$ by
\begin{equation}\label{eq:respactphi}
	{1 \over \tilde{\chi}(f_{\rm d})} 
	= 
	\frac{
		\kappa 
		\langle 
			\tilde{x}_{\rm ot}(f_{\rm d}) 
			\tilde{x}^*(f_{\rm d}) 
		\rangle_{\phi}
	}{
		\langle 
			|\tilde{x}(f_{\rm d})|^2 
		\rangle_{\phi}
	},
\end{equation}
where the numerator is the Fourier transform of the cross-correlation function  between the trap position and the particle position,  while the denominator is the value of the PSD of $x(t)$, both quantities at frequency $f_{\rm d}$, first computed for a given $F(t)$ and then averaged over different realizations of $\phi$. In Fig.~\ref{fig:23:AM}(a), we plot the PSD of the particle position $x(t)$ in presence of the force $F(t)$ at different driving frequencies $f_{\rm d}$. The pronounced peaks at $f = f_{\rm d}$ result from the sinusoidal form of $F(t)$, while for $f \neq f_{\rm d}$, the contribution to $S(f)$ is only due to thermal noise in the fluid. In Fig.~\ref{fig:23:AM}(b), we plot as symbols the values of the storage and loss moduli computed by means of equations~\eqref{eq:StokesEinstein} at the distinct driving frequencies. For comparison, we also represent as dotted and dashed lines the corresponding values determined by passive microrheology, demonstrating the rather good agreement between both methods. 

Although more elaborate to implement in experiments, active microrheology is less prone to numerical artifacts than passive microrheology. This is because it is based on a measurement of the mechanical response of the particle position at a single frequency, which does not require any indirect numerical integration over a discrete finite frequency interval as those involved in the Kramers-Kronig relations.

\subsubsection{Other microrheological applications and outlook}

In the previous sections, we have described some basic applications of microrheology with optical tweezers. More advanced methods, which are beyond the scope of this Tutorial, focus on investigating flow and deformation properties of complex materials which do not trivially fulfill the general conditions described above. Some important examples are the following:
\begin{description}

\item[Interfacial microrheology.] Near a liquid-solid, a liquid-liquid or a liquid-vapor interface, the viscosity of liquids becomes anisotropic and exhibits a dependence on the closest distance to the interface due to the specific hydrodynamic boundary conditions. Passive microrheology can be applied to determine such a spatial dependence, provided that the three spatial coordinates of the probe particle, embedded in the liquid phase of interest, can be tracked \cite{dufresne2001brownian, wang2009hydrodynamic, shlomovitz2013measurement}. 

 \item[Two-point microrheology.] The motion of pairs of colloidal particles not so far from each other is strongly correlated when suspended in a fluid. Since the flow and strain fields around one of them entrain the second particle, the  motion of the latter encodes information of the rheological properties of the fluid at the location of the former. Therefore, a dual-beam optical tweezers can be used to fix their mean separation and perform two-point microrheology. Here, one particle plays the role of a control element, either thermally or externally driven, while the second one is used as a passive probe. This technique is particularly useful to investigate the mechanical response of heterogeneous and anisotropic soft materials, such as gels and polymer solutions \cite{crocker2000two, paul2019active}. 

\item[Transient microrheology.] Using a combination of both active and passive microrheology, the transient behavior of viscoelastic materials either upon flow startup or cessation can be investigated. For example, during an active period, an optically trapped particle can be driven though the viscoelastic fluid, followed by a sudden release from the tweezers by shutting off the laser, during which its position is recorded. During this non-equilibrium passive period, the particle motion is subject to the recovery of the initially strained fluid microstructure. Consequently, this technique allows to measure in a straightforward manner relaxation times and relaxation moduli of  viscoelastic materials \cite{chapman2014nonlinear,gomez2015transient}. 

\end{description}

Microrheology has evolved during the last two decades into a powerful experimental tool to investigate complex soft materials. Most of its current applications rely on a well-established theoretical framework based on equilibrium statistical mechanics, low-Reynolds-number hydrodynamics and linear response theory. Nevertheless, more recent approaches aim to investigate non-linear rheological properties, e.g., shear thinning of polymer solutions, by means of colloidal probes driven at very high velocities or under large-amplitude oscillations \cite{robertson2018optical}. Although the implementation of such microrheological techniques does not represent a big experimental challenge, the interpretation of the data in terms of meaningful parameters and its connection with bulk quantities is not trivial. A complete understanding of such a wealth of  information will certainly rely on current theoretical advancements in non-equilibrium soft matter systems \cite{zia2018active}, which are able to find a direct link between the motion of the probe and the microstructural deformation of the surrounding medium. 

\subsection{Colloidal interactions}\label{sec:4.4:colloidal}

A colloidal system comprises a continuum medium (the \emph{solvent}, such as water) and a disperse phase (the \emph{solute}, such as small colloidal particles from $10\, \rm{nm}$ to $10\, \upmu \rm{m}$).
Due to their dimensions, colloidal particles experience thermal fluctuations and undergo Brownian motion \cite{dhont1996introduction}. Since their properties can be described using statistical physics ensembles, colloidal suspensions represent an ideal model system for statistical physics and they can be considered as ``big atoms'' \cite{poon2004colloids} mimicking atomic systems at different time and length scales, which make their dynamics accessible by optical experiments. 
Therefore, colloidal suspensions have been used to investigate phase transitions between gas, liquid, solid, and crystalline phases \cite{anderson2002insights, pusey1986phase}, crystal nucleation \cite{auer2001prediction,elliot2001direct, gasser2001real}, glassy states \cite{pham2002multiple}, and vapor-liquid interfaces \cite{aarts2004direct}.
Colloidal suspensions are also employed in industry to stabilize emulsions (oil in water and water in oil) and foams \cite{gonzenbach2006ultrastable, zanini2017universal}. 
Colloidal interactions play a key role in all these applications, from food industry \cite{dickinson1995advances} to nanotechnology \cite{zerrouki2008chiral}.
The main colloidal interactions are:
\begin{description}

\item[Van der Waals forces.] 
Van der Waals forces arise between neutral particles such as atoms, molecules or colloids. They have an electromagnetic origin due to the interaction between the dipoles of the singles particles. They are repulsive at short distance range, due to repulsion between the atoms' electron clouds, and attractive at larger distance. Their intensity quickly decreases with the inter-particle distance following a power law \cite{van1873over, parsegian2005van}.

\item[Double-layer forces.]
Double-layer forces are very important for the stability of many biological systems and for the formation of colloidal crystals \cite{mcbride2002diffuse, popa2010importance}. They result from the spontaneous charging of the particle surfaces when immersed in a liquid, due to ionic adsorption or dissociation. They can be both attractive or repulsive.

\item[DLVO theory forces.]
Double-layer forces and van der Waals forces are combined in the DLVO theory (named after Derjaguin, Landau, Verwey und Overbeek) \cite{israelachvili2015intermolecular} to calculate the interaction between spherical colloidal particles \cite{russel1991colloidal, ducker1991direct}. 

\item[Steric forces.] 
Steric forces are repulsive nonbonding interactions influencing the spatial conformation and the reactivity of ions and molecules. They can be used to promote or prevent flocculation (\emph{steric stabilization}) of colloidal solutions. They play an important role in chemistry, biochemistry, pharmacology and food industry, where they are often employed to design the physical properties of food and pharmaceutical products by steric stabilization \cite{dickinson2010flocculation, adams2007modern,cao1990creaming, dickinson1995advances}, as well as to treat waste water and to purify drinkable water by flocculation of metals and airborne particles \cite{moghaddam2010coagulation,bratby2016coagulation}.

\item[Depletion forces.] 
Depletion forces are attractive forces arising in a suspension of large and small particles \cite{asakura1954interaction}. Usually, the large particles are colloids (typically, at least a fraction of a micrometer), while the small particles can be smaller colloids, micelles, solvent macromolecules, or dissolved ions \cite{barrat2003basic, marenduzzo2006depletion, de1981polymer,oosawa1954surface, asakura1954interaction}. 

\item[Critical Casimir forces.] 
Critical Casimir forces emerge between objects (e.g., colloidal particles) immersed in a critical mixture close to its critical point \cite{fisher1978phenomena, gambassi2009casimir, gambassi2011critical}. Althougth these forces were first predicted theoretically in 1978 by M.~E.~Fisher and P.~G.~de Gennes \cite{fisher1978phenomena} in analogy to quantum-electrodynamical (QED) Casimir forces \cite{casimir1948attraction}, they have been measured directly only recently  \cite{hertlein2008direct, paladugu2016nonadditivity} proving their relevance for soft matter and nanotechnological applications thanks to their piconewton strength and nanometric action range  \cite{bonn2009direct, piazza2011depletion, magazzu2019controlling}.

\end{description}

Here, we will focus our attention on how to measure critical Casimir forces. In particular, we will describe a typical experiment to quantify the dynamic effects of critical Casimir forces between two colloidal particles.

\subsubsection{Critical Casimir forces}

Critical Casimir forces take place between objects immersed in a critical mixture when they  confine the mixture composition fluctuations arising near the critical temperature.
These forces are typically in the piconewton and nanometer ranges, they are tunable as a function of temperature, and they feature an exquisite dependence on the surface properties of the involved objects \cite{gambassi2011critical, gambassi2009casimir, soyka2008critical, nellen2009tunability}.
Attractive critical Casimir forces arise whenever the density fluctuations are confined between objects with the same surface properties (e.g., between two hydrophilic or hydrophobic particles). 
Repulsive critical Casimir forces take place between objects with opposite surface properties (e.g., between a hydrophilic and a hydrophobic particle).

\subsubsection{Experiment outline}

In the following sections, we will present in detail how to perform an experiment where the effects of critical Casimir forces on the free dynamic of a pair of colloidal particles are measured using blinking optical tweezers \cite{grier1997optical, pesce2010blinking, pesce2014long, magazzu2019controlling}. 

We employ hydrophilic silica microspheres (diameter $d=2.06\pm0.05\, {\rm \upmu m}$, Microparticles GmbH) dispersed in a water-2.6--lutidine mixture at the critical lutidine mass fraction $c^{\rm c}_{\rm L}=0.286$, corresponding to a lower critical point at the temperature $T_{\rm c} \simeq 34^\circ {\rm C}$ \cite{grattoni1993lower, gambassi2009critical}.
Due to their extremely sensitive dependence on temperature, we have to control and stabilize the sample temperature to within $\pm2\,{\rm mK}$ via a feedback controller \cite{paladugu2016nonadditivity, magazzu2019controlling}. 

Using two holographic optical traps, we trap two microparticles in the bulk of the critical mixture at a surface-to-surface distance of $\approx 300\,{\rm nm}$. This distance is larger than the range of the electrostatic repulsion between the particle surfaces, but still comparable with largest action range of critical Casimir forces. 
Whenever the two optical potentials are switched off (by chopping the laser beam), the colloids diffuse in the bulk of the critical mixture. 

When the temperature $T$ of the solution is sufficiently lower than $T_{\rm c}$ (i.e., for $\Delta T=T_{\rm  c}-T \gtrsim 500\,{\rm mK}$) no critical Casimir forces are observed and the particle diffuses freely in the medium.
Approaching the critical temperature $T_{\rm c}$ (i.e., for $\Delta T \to 0$), density fluctuations start to take place in the critical mixture leading to attractive critical Casimir forces between the two colloids, affecting their diffusion and reducing the average inter-particle distance.
To obtain sufficient statistics of the dynamics of the colloids, we need to repeat the entire blinking process for several times (i.e, about 400 times) for each value of $\Delta T$.

\subsubsection{Experimental setup and feedback temperature controller}

The experimental setup requires holographic optical tweezers, digital video microscopy, and  feedback temperature control \cite{paladugu2016nonadditivity, magazzu2019controlling}. 
The holographic optical tweezers (section~\ref{sec:2.3.5:hot}) is realized using a phase-only spatial light modulator, a laser beam with a  wavelength of $532\,{\rm nm}$, and a oil-immersion objective ($100\times$, ${\rm NA} = 1.30$).
The resulting traps can be periodically switched on and off by a chopper.
 
The most crucial part of the setup is the temperature controller to stabilize the sample temperature to within $\pm2\,{\rm mK}$.
This is realized using a two-stage stabilization protocol.
The first temperature stabilization is achieved by keeping constant the temperature of the sample holder to within $\pm50\,{\rm mK}$ by a circulating water chiller. 
The second and finer temperature stabilization to within $\pm2\,{\rm mK}$ is achieved by a temperature feedback controller applied through the objective. The controller reads in real time the temperature of the objective and stabilizes it by a Peltier element thermically connected to the objective, which is the closest element to the investigated volume. 
An important aspect to take into account is the thermal isolation of the sample area as well as of the entire setup to prevent thermal fluctuations due to air flow and to room temperature fluctuations.

\subsubsection{Potential analysis}

\begin{figure}[t!] 
	\begin{center}
	\includegraphics[width=12cm]{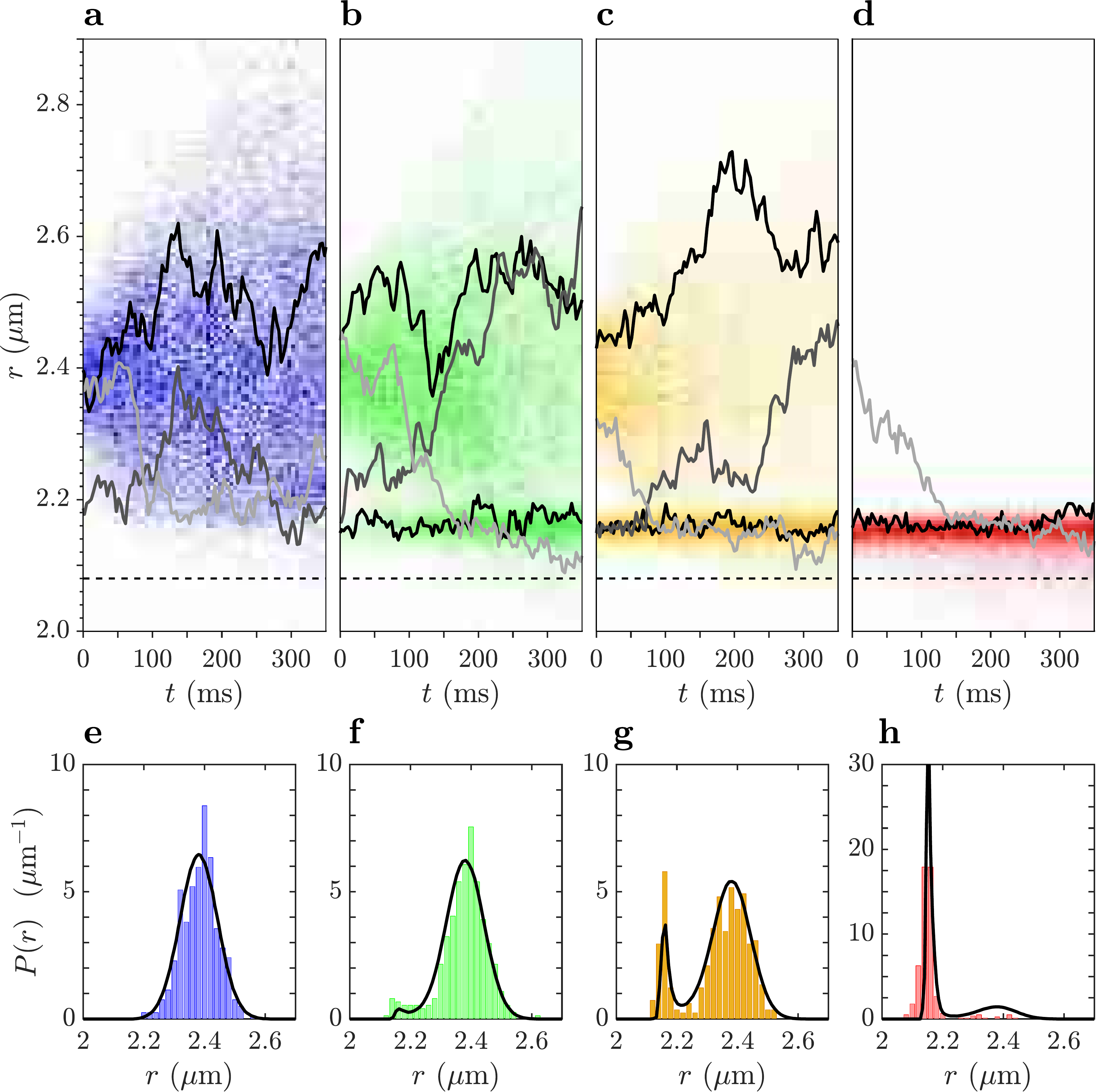}
	\caption{
	{\bf Trajectories and potentials with critical Casimir forces.}
	(a-d) Time evolution of the probability density distribution (coloured background) of the inter-particles distance $r(t)$ obtained from 400 different trajectories after the optical traps are switched off at $t=0\,{\rm ms}$ for decreasing values of $\Delta T$: (a) $\Delta T=456\pm2\,{\rm mK}$, (b) $200\pm2\,{\rm mK}$, (c) $163\pm2\,{\rm mK}$, and (d) $108\pm2\,{\rm mK}$. The solid lines indicate representative individual trajectories.
	The dashed horizontal line indicates the distance $r$ corresponding to the diameter $d$ of the colloids. Sometimes $r(t)$ is smaller than $d$ because a displacement of the colloids along the vertical $z$-axis causes their projections onto the $xy$-plane to overlap. This occurs more frequently in the presence of critical Casimir forces and particularly if the particles are temporarily stuck together.
	(e-h) Equilibrium distribution $P_{\rm eq}(r)$ of the inter-particle distance $r(0)$ (i.e., when the optical tweezers are switched off) for two optically trapped colloids at temperatures (f) $\Delta T=456\pm2\,{\rm mK}$, (g) $200\pm2\,{\rm mK}$, (h) $163\pm2\,{\rm mK}$, and (i) $108\pm2\,{\rm mK}$. Each histogram is obtained from 400 different experimental values. 
	The solid black lines are the theoretical  distribution of $r(0)$, obtained via Monte Carlo integration ($10^6$ samples) of two optically trapped particles subjected to the theoretical total potential $V(\mathbf{R}_1,\mathbf{R}_2)$ (equation~\eqref{eq:98}).
	}
	\label{fig:24:CCF1}
	\end{center}
\end{figure}

The motion of the two colloids is recorded by a camera with acquisition frequency $ \gtrsim 300\,{\rm fps}$ and the acquired videos can be analyzed by digital video microscopy  \cite{crocker1996methods, jones2015optical} to obtain the trajectories ${\bf r}_{1}(t)$ and ${\bf r}_{2}(t)$ of the centers of the two particles projected onto the $xy$-plane, where ${\bf r}_{l}(t) = (x_{l}(t), y_{l}(t))$ with  $l=1, 2$ labelling the particles.
Once these trajectories are obtained, it is possible to calculate their relative distance $r(t)=|{\bf r}_{\rm 2}(t)-{\bf r}_{\rm 1}(t)|$ for different values of $T$ (Fig.~\ref{fig:24:CCF1}).
The time evolution of the probability density distribution, for all the inter-particle distances $r(t)$ at specific values of $\Delta T$, are reported in Figs.~\ref{fig:24:CCF1}(a-d).
For $\Delta T=456\pm2\,{\rm mK}$, far away from the critical point (Fig.~\ref{fig:24:CCF1}(a)), the particles diffuse freely and there is no critical Casimir forces affecting their behaviour.
Increasing the temperature, $\Delta T = 200\pm2\,{\rm mK}$ (Fig.~\ref{fig:24:CCF1}(b)) and $\Delta T =163\pm2\,{\rm mK}$ (Fig.~\ref{fig:24:CCF1}(c)), density fluctuations start to take place in the mixture and critical Casimir forces arise affecting the dynamics of the colloids. We can observe that occasionally they cause adhesion between the colloids as can be inferred from the emergence of a peak in the inter-particle distance probability density at $r\approx 2.16\,{\rm \upmu m}$. 
This peak is due to the gradual emergence of attractive critical Casimir forces between the particles, approaching $T_{\rm c}$, and indicate that critical Casimir forces are strong enough to produce particle adhesion also in the presence of the optical potentials. The peak position indicates the region where the repulsive electrostatic forces and the attractive critical Casimir forces are balanced \cite{magazzu2019controlling}.
When the temperature $T$ is very close to $T_{\rm c}$ ($\Delta T=108\pm2\,{\rm mK}$, Fig.~\ref{fig:24:CCF1}(d)), strong attractive critical Casimir forces inhibit the free diffusion of the particles, which often adhere to each other so that the values of $r$ lie within a small region resulting from the equilibrium between the repulsive electrostatic forces and the attractive critical Casimir forces \cite{magazzu2019controlling}.

For each value of $\Delta T$, all the initial inter-particle distances ${\bf r}_{1,2}(0)$ represent the equilibrium distribution of initial positions $\mathbf{R}_l = (x_l,y_l,z_l)$ with $l=1,2$ of the two colloids subjected to a total potential including the optical potentials $V_{{\rm ot},1}(\mathbf{R}_1)+V_{{\rm ot},2}(\mathbf{R}_2)$, the repulsive electrostatic potential $V_{\rm es}(\rho)$, and eventually the potential of the critical Casimir forces $V_{\rm C}(\rho)$:
\begin{equation}\label{eq:98}
	V(\mathbf{R}_1,\mathbf{R}_2)
	=
	V_{{\rm ot},1}(\mathbf{R}_1)
	+ V_{{\rm ot},2}(\mathbf{R}_2)
	+ V_{\rm es}(\rho)
	+ V_{\rm C}(\rho),
\end{equation}
where $\rho = |\mathbf{R}_2-\mathbf{R}_1| - d$ is the actual surface-to-surface distance between the two colloids. Note that the projected distance $r$ introduced above is generically smaller than the actual center-to-center distance $|\mathbf{R}_2-\mathbf{R}_1|$, due to possible displacements of the colloids along the vertical $z$-direction. 

Assuming that the optical potentials $V_{{\rm ot},1}$ and $V_{{\rm ot},2}$ reported in equation~\eqref{eq:98} are harmonic, we have:
\begin{equation}\label{eq:Vopt}
	V_{{\rm ot},l}(\mathbf{R}_l)
	=
	{1\over2}\,k_l(\mathbf{R}_l 
	- \mathbf{R}_{0,l})^2,
\end{equation}
where the centers $\mathbf{R}_{0,l}$ and the stiffnesses $k_l$ of traps $l=1,2$ can be determined experimentally by the separate calibration of the two optical traps. These initial values are fixed in all the experiment and in the following analysis.

For the electrostatic repulsion potential $V_{\rm es}$, we can consider the simple expression \cite{paladugu2016nonadditivity,gambassi2009critical}
\begin{equation}\label{eq:Ves}
	V_{\rm es}(\rho) 
	= 
	k_{\rm B}\, T_{\rm c} \, {\rm e}^{- (\rho-\rho_{\rm es})/\ell_{\rm D}},
\end{equation}
where $\rho$ is the surface-to-surface distance between the colloids, $\rho_{\rm es}$ is an effective parameter, which depends on the surface charges, while  $\ell_{\rm D}$ is the Debye screening length \cite{paladugu2016nonadditivity,gambassi2009critical}.

For the potential $V_{\rm C}$ of the critical Casimir forces, we can consider the theoretical prediction based on the Derjaguin approximation \cite{gambassi2009critical}:
\begin{equation}\label{eq:VC}
	V_{\rm C}(\rho) 
	= 
	k_{\rm B} T_{\rm c} {d \over 4\rho} \Theta(\rho/\xi), 
\end{equation}
where $\Theta$ is a universal scaling function \cite{vas2007MC, vas2009MC, hasenbusch2012thermodynamic} and $\xi$ is the bulk correlation length of the critical fluctuation of the density fluctuations of the critical mixture.

Whenever the optical potential is switched on, the particles trajectories evolve under the action of the total potential $V(\mathbf{R}_1,\mathbf{R}_2)$, and after a sufficiently long time, they reach the equilibrium distribution:
\begin{equation} \label{eq:peq}
	P_{\rm eq}(\mathbf{R}_1,\mathbf{R}_2) 
	\propto 
	\exp[-V(\mathbf{R}_1,\mathbf{R}_2)/(k_{\rm B}T)].
\end{equation}
The histograms in Figs.~\ref{fig:24:CCF1}(e-h) represent the experimental values of the equilibrium distribution $P_{\rm eq}$ and the solid black lines are the corresponding theoretical results obtained from the Monte Carlo integration of equation~\eqref{eq:peq}.
When the mixure temperature is far away from $T_{\rm c}$ ($\Delta T = 456\pm2\, {\rm mK}$, Fig.~\ref{fig:24:CCF1}(e)), the probability distribution $P_{\rm eq}(r)$ is very well approximated by a Gaussian distribution centered at the value $r\simeq2.40\,\,{\rm \upmu m}$,  corresponding to experimental distance $r_0$ between the centres of the two optical traps \cite{magazzu2019controlling}.
Reducing $\Delta T$ (Figs.~\ref{fig:24:CCF1}(e-h)), a peak arises at $r\simeq2.16\,{\rm \upmu m}$ on the left flank of the Gaussian distribution, becoming more dominant at the expense of the Gaussian distribution. 

Starting from probability distribution $P_{\rm eq}(r)$, we fit its experimental values by Monte Carlo integration of equation~\eqref{eq:peq} based on the theoretical potential $V(\mathbf{R}_1,\mathbf{R}_2)$.
In particular, we use the experimental values of $V_{{\rm ot},l}(\mathbf{R}_l)$ (equation~\eqref{eq:Vopt}) and $\Theta(\rho/\xi)$ (equation~\eqref{eq:VC}) as input functions, and $\rho_{\rm es}$, $\ell_{\rm D}$, and $\xi$ as fitting parameters.
In doing this, we assume that the fitting parameters $\rho_{\rm es}$ and $\ell_{\rm D}$ have the same values ($\ell_{\rm D}\simeq 13\,{\rm nm}$ and $\rho_{\rm es}\simeq 95\,{\rm nm}$) for all $T$, while the correlation length $\xi$ is specific to each $T$.
Once the values of $\xi$ have been obtained, we calculate the real temperature of the solution by fitting the experimental  temperature values measured at the objective with the theoretically expected temperature dependence of $\xi$, according to the method reported in Refs.~\cite{paladugu2016nonadditivity, magazzu2019controlling}.

\subsubsection{Drift analysis}

\begin{figure}[h!]
	\begin{center}
	\includegraphics[width=12cm]{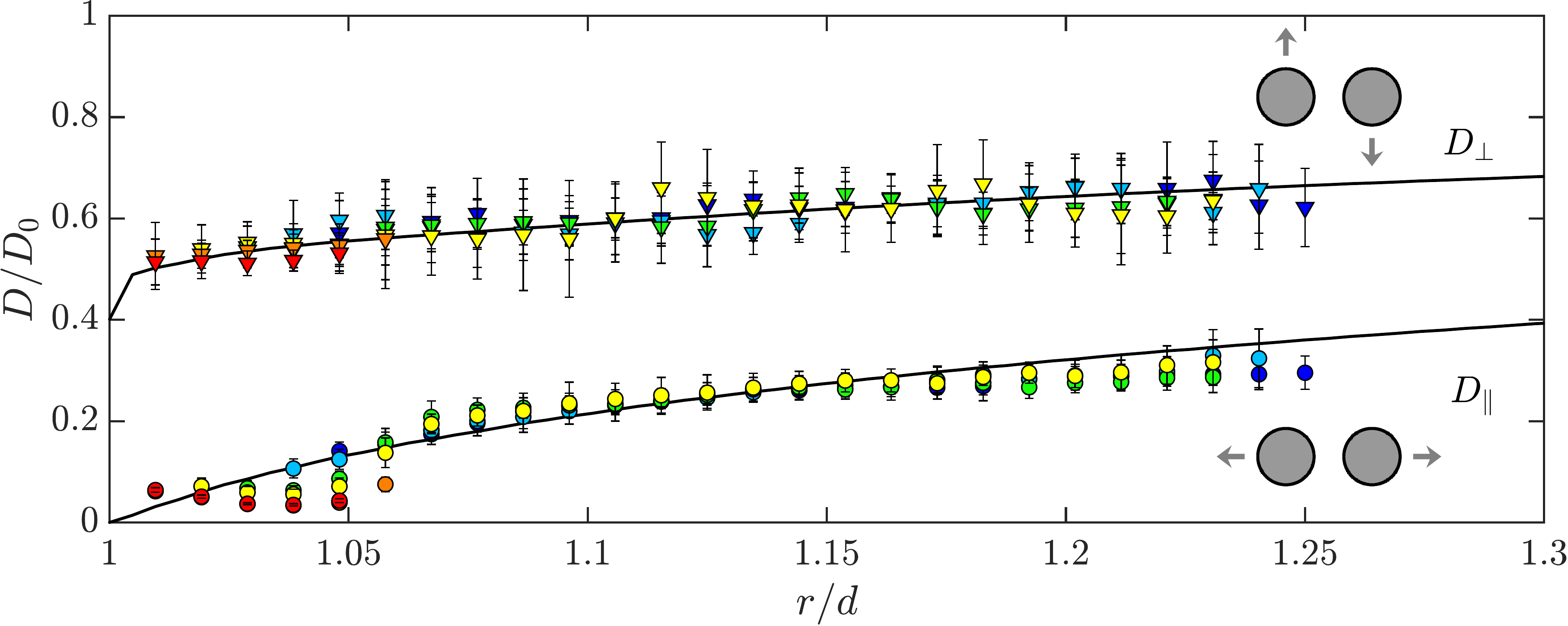}
	\caption{
	{\bf Parallel and perpendicular diffusion coefficients.}
	Experimental values of the normalized parallel $D_\| / D_0$ (circles) and perpendicular $D_\perp / D_0$ (triangles) diffusion coefficients as functions of the normalized inter-particle distance $r/d$.
The parallel and perpendicular directions refer to the line connecting the centers of the two colloids, and $D_0$ is the bulk diffusion constant (equation~\eqref{eq:D0}) fitted to the experimental data.
	The colors refer to data taken at $\Delta T=456\pm2\,{\rm mK}$ (blue),  $273\pm2\,{\rm mK}$ (light blue), $200\pm2\,{\rm mK}$ (green), $163\pm2\,{\rm mK}$ (yellow), $127\pm2\,{\rm mK}$ (orange), and $108\pm2\,{\rm mK}$ (red). Errors bars represent the standard deviation of the experimental values. 
	The solid lines represent the theoretical predictions accounting for the effect of the hydrodynamic interaction between the colloids \cite{batchelor1976brownian}. 
	The nature of the deviations observed in $D_\|$ at short and long distances is discussed in the main text.
	}
	\label{fig:25:CCF2}
	\end{center}
\end{figure}

\begin{figure}[h!]
	\begin{center}
	\includegraphics[width=12cm]{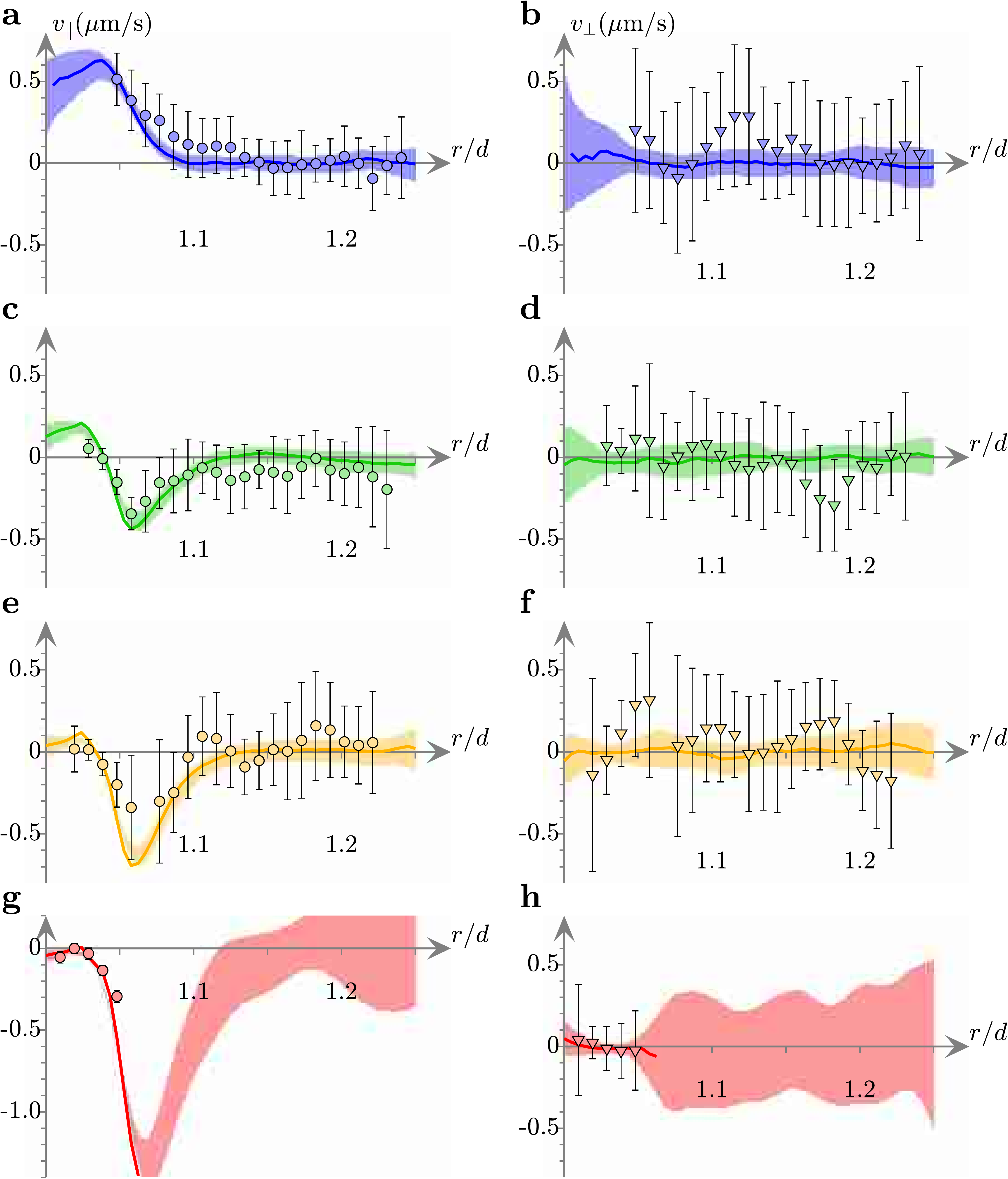}
	\caption{
	{\bf Parallel and perpendicular drifts.}
	Drift velocities parallel  $v_\parallel$ and perpendicular $v_\perp$ to the direction connecting the centers of the two colloids for (a, b) $\Delta T=456\pm2\,{\rm mK}$, (c, d) $200\pm2\,{\rm mK}$, (e, f) $163\pm2\,{\rm mK}$ and (g, h) $108\pm2\,{\rm mK}$.
	The symbols with errorbars represent the experimental data, the colored lines represent the corresponding simulation results, and the shaded areas represent the error of the numerical estimates due to the uncertainties in the fit parameters.
	}
	\label{fig:25:CCF3}
	\end{center}
\end{figure}

The relative position and the distance between the two particles can be used to determine the values of their diffusion coefficient $D(r)$ and their drift velocity $v(r)$ as function of their inter-particle distance $r$.
Due to the hydrodynamic interaction between the two colloids, the particle diffusion coefficient depends on the distance between the particles and is different along the directions parallel and perpendicular to the line connecting the centers of the two particles (see equation~(5.5) in Ref.~\cite{batchelor1976brownian}).
For this analysis, we consider the particle trajectories as a sequence of values ${\bf r}_i$, numbered by $i$ and acquired at times $i t_{\rm s}$, where $t_{\rm s}$ is the time between sampling.
We can decompose the $i$-th displacement $\Delta{\bf r}^{(n)}_{i} = {\bf r}_{i+n} - {\bf r}_{i}$ into its parallel and perpendicular components:
\begin{equation}
	\Delta r^{(n)}_{i \parallel} 
	= 
	\Delta{\bf r}^{(n)}_{i} \cdot \hat{\bf r}_{i}
\end{equation}
and 
\begin{equation}
	\Delta r^{(n)}_{i \perp} 
	= 
	\Delta{\bf r}^{(n)}_{i} \cdot (\hat{\mathbf{z}} \times \hat{\bf r}_{i}),
\end{equation}
where $\hat{\bf r}_{i}={\bf r}_{i}/r_{i}$  and $\hat{\mathbf{z}}$ is the unit vector along the $z$-direction, which is perpendicular to the $xy$-plane of observation where the position vectors ${\bf r}_{i}$ lie.
Then, the parallel and perpendicular diffusion coefficients are \cite{magazzu2019controlling}:  
\begin{equation}\label{eq:def-Dpp}
	D_{\parallel}(r) 
	= 
	{1\over2} 
	\left\langle 
		\frac{|\Delta {r}^{(n=3)}_{i \parallel}|^2}{3 t_{\rm s}} 
		\left| \frac{}{} \right.
		r_{i} 
		\in 
		\left\lbrack r-\delta r, r+\delta r  \right\rbrack 
	\right\rangle,  
\end{equation}
and
\begin{equation}\label{eq:def-Dpe}
	D_{\perp} (r) 
	= 
	{1\over2} 
	\left\langle  
		\frac{|\Delta {r}^{(n=3)}_{i \perp}|^2}{3 t_{\rm s}} 
		\left| \frac{}{} \right. 
		r_{i} 
		\in 
		\left\lbrack r-\delta r, r+\delta r  \right\rbrack 
	\right\rangle,
\end{equation}
where $\delta r$ represent the amplitude of a small interval around $r$.
The measured normalized values of $D_\perp$ and $D_\|$ (symbols) are reported in Fig.~\ref{fig:25:CCF2} as functions of the ratio $r/d$, together with the theoretical prediction obtained in Ref.~\cite{batchelor1976brownian} (solid line) for no-slip boundary conditions. 
In particular, $D_\perp$ and $D_\|$ are normalized by the bulk diffusion constant
\begin{equation}\label{eq:D0}
	D_0 
	= 
	\frac{k_{\rm B} T}{3 \pi \eta d}  
	\simeq 
	0.22 \, {\rm \upmu m^2 s^{-1}},
\end{equation} 
where $\eta \simeq 2 \cdot 10^{-3}\,{\rm Ns\,m^{-2}}$ is the viscosity of the mixture close to $T_{\rm c}$ \cite{clunie1999interdiffusion}. 
The values of $D_{\perp}$ and $D_{\parallel}$ in Fig.~\ref{fig:25:CCF2} shows satisfactory agreement, with a systematic discrepancy emerging only in $D_\|$ for $r/d \lesssim 1.05$ due to the limited experimental acquisition rate, which does not allow us to resolve times shorter than $3\,{\rm ms}$.\footnote{The same discrepancy can be observed on simulated data considering trajectories sampled with the same time step  $t_{\rm s}$ as that used in the experiment. Furthermore, reducing significantly the time step $t_{\rm s}$ in the simulations, the discrepancy with the theoretical line is much less pronounced and eventually disappears as $t_{\rm s} \to 0$.}

The parallel and perpendicular drift velocities can be measured using:
\begin{equation}\label{eq:def-vpp}
	v_{\parallel}(r) 
	= 
	\left\langle  
		\frac{\Delta {r}^{(n=10)}_{i \parallel}}{10 t_{\rm s}} 
		\left| \frac{}{} \right.  
		r_{i} 
		\in 
		\left\lbrack r-\delta r, r+\delta r  \right\rbrack 
	\right\rangle,  
\end{equation}
and
\begin{equation}\label{eq:def-vpe}
	v_{\perp}(r) 
	= 
	\left\langle  
		\frac{\Delta {r}^{(n=10)}_{i \perp}}{10 t_{\rm s}} 
		\left| \frac{}{} \right.  
		r_{i} 
		\in 
		\left\lbrack r-\delta r, r+\delta r  \right\rbrack 
	\right\rangle.
\end{equation}
For large $\Delta T$ (Fig.~\ref{fig:25:CCF2}(a)), $v_\parallel$ is positive at small values of $r/d$ due to the repulsive electrostatic potential $V_{\rm es}$, which is dominant in this range and pushes the particles away from each other, while it rapidly vanishes for increasing $r/d$, because the electrostatic repulsion decays exponentially with the inter-particles distance and no other force is affecting the motion of the colloids.
Decreasing $\Delta T$ (Figs.~\ref{fig:25:CCF2}(c) and \ref{fig:25:CCF2}(e)), $v_\parallel$ becomes negative within a certain range of values of $r/d$, due to the arising of attractive critical Casimir forces; 
however, at smaller values of $r/d$, $V_{\rm es}$ is dominant and $v_\parallel$ is still positive. 
If $\Delta T$ is reduced further ($T$ is very close to $T_{\rm c}$, Fig.~\ref{fig:25:CCF2}(g)), critical Casimir forces became so strong that $v_\parallel$ presents only negative values because the particles move towards each other until their velocity vanishes at contact; at distances larger than the range of the critical Casimir forces action range, instead, $v_\parallel$ vanishes and the particles undergo Brownian diffusion.\footnote{In proximity of $T_{\rm c}$, the range of distances larger than the critical Casimir forces action range can actually be explored only via numerical simulations with sufficiently high statistics. In the experiment, instead, the particles almost always stick together.}

In Figs.~\ref{fig:25:CCF2}(b,d,f,h), we report the experimental and numerical values for the orthogonal component $v_\perp$ of the drift velocity (equation~\eqref{eq:def-vpe}). Here, notice that $v_\perp$ vanishes in all investigated cases and shows no temperature dependence because all the forces at play in this experiment act along the direction which connects the centers of the particles \cite{magazzu2019controlling}.

\subsubsection{Other colloidal interactions and outlook}

Critical Casimir forces can play an important role in nanoscience and nanotechnology thanks to their piconewton strength, nanometric action range, fine tunability as a function of temperature, and particle surface dependence. 
They can be employed to manipulate micro and nano-objects (e.g., by controllable periodic deformations of chains), to assemble micro and nano-devices (e.g., via the self-assembly of colloidal molecules \cite{marino2016assembling, nguyen2017tuning}), and to drive nanomachines (e.g., by powering rotators \cite{schmidt2017microscopic}) at the nano and micro-meter scale. 

Beyond critical Casimir forces, other colloidal interactions have also been studied using optical tweezers. Recently, Van der Waals forces have been directly measured between two isolated optically trapped rubidium atoms \cite{beguin2013direct} opening the path to the atomic modeling of  micro- and nano-structured devices controlled by these forces \cite{woods2016materials}. 
Double layer forces can also play an important role in nanotechnology due to their ability to tune the transport and deposition rate of micro and nano-sized particles. 
Steric forces play an important role in tuning the rates and the activation energies of most chemical reactions, crucial in chemistry, biochemistry, and pharmacology.
Depletion forces are extensively used to stabilize colloidal solutions by flocculation, to produce the programmable self-assembly of colloidal nanostructures \cite{cademartiri2014programmable}, and to drive cellular organization \cite{marenduzzo2006depletion}. 

\subsection{Statistical physics}

The advent of minituarization methods and microscopic manipulation techniques, as those discussed in this Tutorial, has opened the door to understand fundamental concepts and ideas at the foundations of statistical physics, allowing us to explore how macroscopic laws emerge from their microscopic dynamics. 
In this section, we only discuss experiments performed in a liquid. For related work in vacuum, see section~\ref{sec:5:vacuum} and Ref.~\cite{gieseler2018levitated}.
We will start discussing the theory of Kramers transitions and stochastic resonance (section~\ref{sec:4.5.1:kramers}).
We will continue exploring non-equilibrium fluctuation-dissipation theorems (section~\ref{sec:4.5.2:ness}) and introducing stochastic thermodynamics (section~\ref{sec:4.5.3:st}).
Finally, we will explain how the Carnot's cycle and efficiency of macroscopic engines can be scaled down to the microscopic realm, and conclude with a discussion of the concept of Maxwell's demons and Szilard engines (section~\ref{sec:4.5.4:engines}).

\subsubsection{Kramers' transitions and stochastic resonance}\label{sec:4.5.1:kramers}

Using optical trapping techniques, Simon {\em et al.} \cite{simon1992escape} and later McCann {\em et al.} \cite{mccann1999thermally} studied the escape of a Brownian particle from a potential well, visualizing for the first time thermally activated processes and giving in this way a neat experimental proof of Kramers' theory \cite{kramers1940brownian}.

Kramers' theory in the overdamped limit predicts the mean rate for the activation of processes over an energy barrier $\Delta U$ to be
\begin{equation}\label{eq:Kramers}
	\frac{1}{\overline{\tau}}
	\approx
	\frac{1}{\tau_0}
	e^{-\Delta U/k_{\rm B}T},
\end{equation}
where $\overline{\tau}$ is  the mean residence time, also called \emph{Kramers' time} and $\tau_0$ is the overall relaxation time of the well, which can be expressed in terms of the angular frequencies near the stable ($U_{\rm A}(x)=\kappa_{\rm A} x^2/2=m\omega_{\rm A} x^2/2$) and unstable  ($U_{\rm S}(x)=\kappa_{\rm S} x^2/2=m\omega_{\rm S} x^2/2$) critical points as follows,  
\begin{equation}\label{eq:dkdkdod}
	\tau_0
	=
	\frac{
		\pi\gamma
	}{
		m|\omega_{\rm S}|\omega_{\rm A}
	}
	=
	\frac{
		\pi\gamma
	}{
		\sqrt{|\kappa_{\rm S}|\kappa_{\rm A}}
	},
\end{equation}
with $\gamma$ the friction coefficient and $m$ the mass of the particle. For the two-dimensional case, the expression for  $1/\tau_0$ takes the following form \cite{landauer1961frequency}: 
\begin{equation}\label{eq:2drate}
	\frac{1}{\tau_{0}}
	=
	\frac{
		m 
		|w_{\rm S}^{(x)}|
		w_{\rm A}^{(x)} 
		w_{\rm A}^{(y)}
	}{
		\pi\gamma w_{S}^{(y)}
	}
	=
	\frac{ 
		\sqrt{|\kappa_{\rm S}^{(x)}|\kappa_{\rm A}^{(x)} \kappa_{\rm A}^{(y)}}
	}{
		\pi\gamma \sqrt{\kappa_{\rm S}^{(y)}}
	},
\end{equation}
where the superscripts $(x)$ and $(y)$ indicate the parallel and perpendicular directions to the axis joining the stable and unstable critical points, respectively.

\begin{figure}[h!]
	\begin{center}
	\includegraphics[width=12cm]{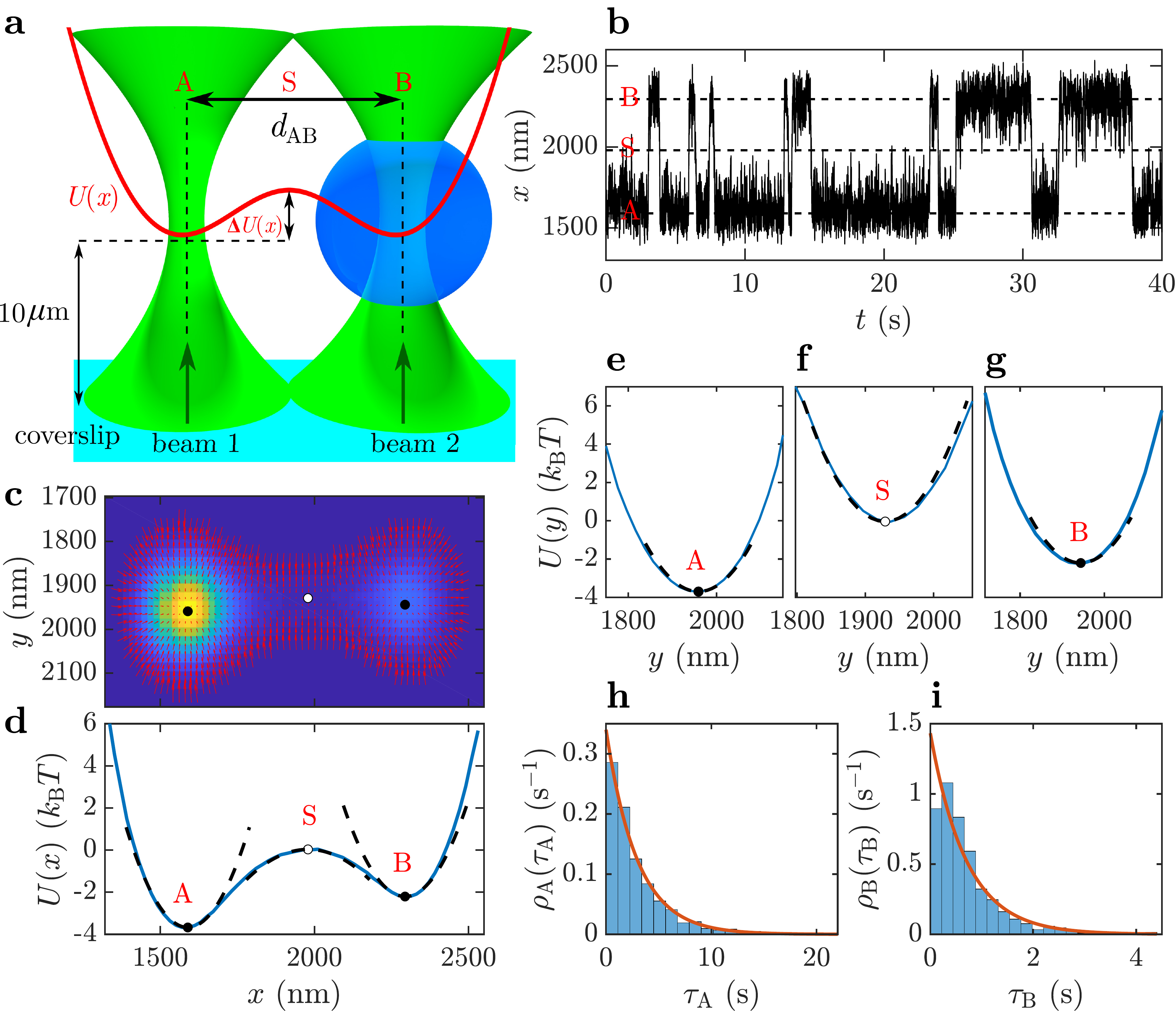}
	\caption{
	{\bf Kramers' rates.}
	(a) Schematic of a bistable potential generated with a double-beam optical tweezers. A silica bead is optically trapped with two-orthogonally-polarized and tightly-focused Gaussian beams separated by a distance $d=0.8\,{\rm \upmu m}$ along the $x$-direction and focused by a water-immersion objective (${\rm NA}=1.2$) using an inverted microscope.
	(b) Trajectory of a Brownian particle (radius $r_{\rm p}= 0.48\,{\rm \upmu m}$) in an aqueous solution (temperature $T=22^\circ{\rm C}$) in the bistable potential. The particle spends most of the time near the equilibrium points ({\bf A} and {\bf B}, whose separation is measured to be $d_{\rm AB}=0.71\,{\rm \upmu m}$), but from time to time it is thermally activated to the saddle point {(\bf S}) with a mean rate defined by equations~\eqref{eq:Kramers} and \eqref{eq:2drate}.
	(c) Reconstruction of the force field measured with FORMA (arrows, section~\ref{sec:3.8:forma}) and of the potential energy measured with the potential analysis (background color, section~\ref{sec:3.2:pot}). 
	(d-g) Optical potential (d) along the $x$-direction and (e-f) along the $y$-direction measured using FORMA.
	(h-i) Residence time probability in wells (h) {\bf A} and (i) {\bf B}. The red curves show the exponential probability function defined by equation~\eqref{eq:probkramers}, with the mean residence times estimated from the experimental data $\overline{\tau}_{\rm A}=3.03\,{\rm s}$ and $\overline{\tau}_{\rm B}=0.67\,{\rm s}$.
}
	\label{fig:27:schemedw}
	\end{center}
\end{figure}

Fig.~\ref{fig:27:schemedw}(a) shows a schematic of the configuration of the optical tweezers to generate a double-well potential.\footnote{Alternative experimental configurations have also been used to realize a bistable optical potential. For example, using two nanoholes in a metallic film \cite{zehtabi2013double}, sculpting the bistable potential with a spatial light modulator \cite{curran2012partial}, or positioning time-shared optical traps in two closely-separated positions using acousto-optical deflectors \cite{seol2009phase}.}
A silica bead\footnote{Interestingly, also $1\,{\rm \upmu m}$ polystyrene particles can be trapped with this setup, but with these particles we have not been able to observe any bistable behavior. We believe this is because the higher weight and lower refractive index of the silica beads make them more stably trapped along the axial direction of the beams allowing us to explore relatively larger distances between optical tweezers in a range where the bistable potential is neatly observed.} of diameter $d_{\rm p}=0.96\,{\rm \upmu m}$ is confined in a bistable potential generated by two parallel optical tweezers with orthogonal polarizations and separated by a distance $d=0.8\,{\rm \upmu m}$, which are generated by focusing two Gaussian beams with a water immersion objective with ${\rm NA}=1.2$. These two beams are generated with two polarizing beam splitters according to the scheme shown in Fig.~\ref{fig:5:polarizationsplitting} (section~\ref{sec:2.3.2:PS}). The mirrors in the beam splitter are controlled with piezo actuators with $30\,{\rm nm}$ step resolution leading to a resolution of the position of the optical traps in the sample cell below $20\,{\rm nm}$. The trapping region is set approximately $10\,{\rm \upmu m}$ above the bottom coverslip of the sample cell and $90\,{\rm \upmu m}$ from the top one, so that the hydrodynamic interaction with the walls of the cell is negligible. 

\begin{table}[h!]
	\begin{center}
	\begin{tabular}{c|c|c|c|c|c|c}
		\textbf{Point}
		&
		$\kappa_x\,({\rm pN\,\upmu m^{-1}})$
		&
		$\kappa_y\,({\rm pN\,\upmu m^{-1}})$
		&
		$\Delta U\,(k_{\rm B}T)$
		&
		${1 \over \tau_0}\,({\rm s^{-1}})$
		&
		${1 \over \overline{\tau}^{({\rm th})}}\,({\rm s^{-1}})$
		&
		${1 \over \overline{\tau}^{({\rm ex})}}({\rm s^{-1}})$
		\\
	\hline
		{\bf A}
		&
 		0.98
 		&
 		1.43
 		&
 		3.67
 		&
 		12.3
 		&
 		0.31
 		&
 		0.33
		\\
	\hline
		{\bf S}
		&
		-0.29
		&
		3.47
		&
		-
		&
		-
		&
		-
		&
		-
		\\
	\hline
		{\bf B}
		&
		0.86
		&
		1.36
		&
		2.23
		&
		11.4
		&
		1.22
		&
		1.49
		\\
	\hline
	\end{tabular}
	\caption{
	{\bf Characterization of equilibrium points in a bistable potential.} 
	The theoretical mean crossing rates ($1/\overline{\tau}^{({\rm th})}$) obtained using equations~\eqref{eq:Kramers} and \eqref{eq:2drate} and the fitted potential parameters are in very good agreement with those directly measured from the escape time of the trajectories trajectories ($1/\overline{\tau}^{({\rm ex})}$). 
	}
	\label{tab:10:bistabledata}
	\end{center}
\end{table}

A typical trajectory of the Brownian particle in this double-well potential is shown in Fig.~\ref{fig:27:schemedw}(b). Along the $x$-direction joining the equilibrium points, the particle spends most of the time near the equilibrium points {\bf A} and {\bf B}, occasionally crossing the saddle point {\bf S}. Along the perpendicular $y$-direction, the motion does not appear to be affected by the presence of two neighboring traps.
Knowing the friction coefficient of the particle, 
this trajectory permits us to determine the Kramers' time using equations~\eqref{eq:Kramers} and \eqref{eq:2drate}, where the characteristic frequencies (or equivalently their stiffnesses) at the critical points can be straightforwardly obtained using FORMA (section~\ref{sec:3.8:forma}).
Fig.~\ref{fig:27:schemedw}(c) shows the force field in the $xy$-plane reconstructed using FORMA, from which the equilibrium points are easily identified as the positions where the magnitude of the force field has local minima. 
The potential profile along the $x$-direction is shown in Figs.~\ref{fig:27:schemedw}(d) and those along the $y$-direction at the critical points in Figs.~\ref{fig:27:schemedw}(e-g).
Table~\ref{tab:10:bistabledata} provides the stiffnesses of the equilibrium points ($\kappa_x$ and $\kappa_y$), the energy barriers ($\Delta U$), and the overall relaxation times ($1/ \tau_0$), from which we estimate the mean crossing rates ($1/\overline{\tau}^{({\rm th})}$) using equations~\eqref{eq:Kramers} and \eqref{eq:2drate}.

The average crossing rate can be straightforwardly obtained from the trajectory of the bead by detecting the times at which the particle crosses the saddle point, shown in Fig.~\ref{fig:27:schemedw}(f).
For the validity of the approximations within Kramers' theory, these crossing events have to be rare, i.e., the time spent by the particle near the stable positions must be always longer than the relaxation time $\tau_0$ defined by the stiffnesses of the critical points, limiting the ideal situations to the cases where the energy barrier is very high in comparison to $k_{\rm B}T$. 
This is a homogeneous Poisson process, where each escape event occurs at random with small probability \cite{simon1992escape}. 
Therefore, we expect the escape times to be exponentially distributed (red curves in Fig.~\ref{fig:27:schemedw}(f)):
\begin{equation}\label{eq:probkramers}
	p(\tau)
	=
	p_0 e^{-\tau/\overline{\tau}},
\end{equation}
where $p_0=1/\overline{\tau}$ is the mean crossing rate and $\overline{\tau}$ is the mean residence time. 
The resulting experimental mean rates are reported in Table~\ref{tab:10:bistabledata} ($1 / \overline{\tau}^{({\rm ex})}$), showing very good agreement with data obtained by means of Kramers' assumptions ($1 / \overline{\tau}^{({\rm th})}$).

Bistable potentials are often employed to study non-linear dynamics. 
For example, a bistable potential such as that described in this section has been used to demonstrate the emergence of stochastic resonance when the optical tweezers are modulated in time harmonically, but asymmetrically (i.e., one potential well gets deeper while the other one gets shallower) \cite{simon1992escape}: as the modulation time approaches the Kramers escape time, the escape probability and the activation rate increases. 
A different process of synchronization between two bistable systems was also demonstrated when two optically generated bistable systems with one bead each in water are set very close to let the particles to interact \cite{curran2012partial}: when the two bistable systems have slightly different escape rates, the particles synchronize their jumps and lock their rates to an intermediate value between those featured by the two non-interacting systems.         

\subsubsection{Non-equilibrium fluctuation-dissipation relations}\label{sec:4.5.2:ness}

In its simplest form the fluctuation-dissipation theorem (FDT) \cite{callen1951irreversibility} can be expressed as $k_{\rm B} T \mu = D$ (Einstein-Smoluchowski equation~\eqref{eq:einstein}): it provides a direct relationship between the mobility $\mu = 1/\gamma$, which measures the response of its velocity to an external force, and its diffusion coefficient $D$. 

More broadly, the FDT states that for a system in contact with a heat bath at temperature $T$, upon applying a small external perturbation $h$ which changes the energy $U$ of the system as $U \rightarrow U -  h V$, where $V$ is the variable conjugate to $h$ with respect to the energy, the following relation holds for any observable $Q$ \cite{marconi2008fluctuation}:
\begin{equation}\label{eq:FDT}
	k_{\rm B} T 
	R(t - s) 
	= 
	\partial_s 
	C(t - s) 
	\quad 
	\mbox{for} 
	\quad 
	t \ge s,
\end{equation}
where $R(t - s) = \left. \frac{\delta \langle Q (t) \rangle_h }{\delta h(s)} \right|_{h = 0}$ is the linear response function of $Q$ at time $t$ to the perturbation $h$ at time $s$, $\frac{\delta}{\delta h}$ denotes a functional derivative, i.e.
\begin{equation}\label{eq:resp}
	\langle Q(t) \rangle_h 
	= \langle Q(t) \rangle_0 
	+ 
	\int_{-\infty}^t ds\,R(t-s) h(s),
\end{equation}
and $C(t-s) = \langle Q(t) V(s) \rangle_0$ is the two-time correlation function between the observable $Q$ and $V$. The brackets $\langle \ldots \rangle_h$ denote a non-equilibrium ensemble average in presence of the perturbation $h$, whereas $\langle \ldots \rangle_0$ represent an equilibrium average for $h = 0$. The expression~\eqref{eq:FDT} of the FDT can be re-written in the integral form as
\begin{equation}\label{eq:intFDT}
	k_{\rm B} T 
	\chi(t) 
	= 
	C(0) - C(t) 
	\quad 
	\mbox{for} 
	\quad 
	t \ge 0,
\end{equation}
where $\chi(t) = \int_0^t R(t-s) ds$ is the integrated response function, which is usually much easier to implement and measure than $R(t)$. 

The importance of equations~\eqref{eq:FDT} and \eqref{eq:intFDT} for experiments lies on the fact that linear response functions, which characterize states slightly away from equilibrium, and equilibrium correlation functions can be interchangeably determined by an appropriate choice of the varibles $h$, $Q$ and $V$, depending on the specific conditions. 
For example, for a colloidal particle suspended in water at constant temperature $T$ and trapped by optical tweezers with stiffness $\kappa$, the integrated response function of the particle position $x$ with respect to the trap center $x_{\mathrm{OT}}$ to an external time-dependent force can be determined by applying a Heaviside-like perturbative displacement of the trap position $x_{\mathrm{OT}}(s) = x_0 \Theta(s)$ and then by measuring the resulting time evolution of the ratio $\frac{\langle x(t) \rangle_h}{\kappa x_0}$ at $t \ge 0$. Then, it follows that $R(t) = \frac{1}{\kappa x_0}\frac{d\langle x(t) \rangle_h}{dt}$. 
Equivalently, it can be indirectly computed by means of the equilibrium quantity $\frac{1}{k_{\rm B} T } \partial_s C(t-s)$, where $t \ge s$, $Q = x$ is the observable of interest, $V = x $ is the variable conjugate to the perturbative force $h = \kappa x_{\mathrm{OT}}$, and $C(t - s) = \langle x(t) x(s) \rangle_0$, for which no external force is needed. Using the explicit expression for the positional correlation function in thermal equilibrium (section~\ref{sec:3.5:acf}), $ \langle x(t) x(s) \rangle_0 = \frac{k_{\rm B} T }{\kappa} \exp\left( - \frac{\kappa |t - s|}{\gamma} \right)$, one finds $R(t) = \frac{1}{\gamma}\exp\left( - \frac{\kappa t }{\gamma}  \right)$, $t \ge 0$, from which the time evolution of $\langle x(t) \rangle_h$  in response to \emph{any} external force $h$ can be determined by means of equation~\eqref{eq:resp}.

\paragraph{Fluctuation-dissipation relation for non-equilibrium steady states (NESS).}

Optical tweezers can be used to explore some generalizations of equation~\eqref{eq:FDT} around \emph{non-equilibrium steady states} (NESS) for mesoscopic systems such as colloidal particles, molecular motors, and biomolecules \cite{ciliberto2017experiments}. 
In particular, a 1D model system that has been extensively studied within the context of non-equilibrium statistical physics consists of an overdamped particle with friction coefficient $\gamma$, moving on a circle of radius $R_{\rm circle}$ and polar coordinate $\theta$, across a periodic potential $U(\theta) = U(\theta + 2\pi)$ under the action of a constant force $f_0$ and a thermally fluctuating force $\xi$ \cite{seifert2012stochastic}.
The dynamics of the polar coordinate $\theta$ is described by the Langevin equation
\begin{equation}\label{eq:LE}
\frac{d\theta}{dt} = - \partial_{\theta}[A\phi(\theta)] + F + \zeta,
\end{equation}
where $\phi(\theta) = \frac{U(\theta)}{\max\{ U(\theta) \}}$ is the normalized potential profile,  $A = \frac{\max\{ U(\theta) \}}{\gamma a^2}$ is its amplitude, $F = \frac{1}{\gamma a}f_0$ is a non-conservative force term,\footnote{On a circle, any constant force $f_0 \ne 0$ is non-conservative, because it cannot be expressed as the derivative of a potential due to the periodic boundary conditions.} and $\zeta = \frac{1}{\gamma a} \xi$ is a Gaussian white noise of zero mean, i.e., $\langle \zeta \rangle = 0$, and autocorrelation $\langle \zeta(t) \zeta(s) \rangle =2 D_{\theta}\delta(t - s)$, where the angular diffusion coefficient along the circle is given by $D_{\theta} = \frac{D}{a^2} = \frac{k_{\rm B} T}{\gamma a^2}$. Due to the non-conservative force, the system lacks detailed balance: for constant values of $F > 0$, $A > 0$ and $D_{\theta} > 0$, and in absence of time-dependent perturbations,  $\theta$ reaches a NESS, characterized by a distribution $\rho_{\rm NESS}(\theta)$, which is different from the equilibrium Boltzmann distribution $\rho_{\rm eq}(\theta) \propto \exp \left(  -\frac{A \phi(\theta)}{D_{\theta}}\right)$. Under such conditions, the particle is able to go beyond the potential barrier and explore the whole circle due to the combined effects of thermal fluctuations and the non-conservative force, thereby developing a constant non-vanishing probability current given by
\begin{equation} \label{eq:NESScurrent}
	j = \{ F - \partial_{\theta}[A\phi(\theta)] \} \rho_{\rm NESS}(\theta) - D_{\theta}\partial_{\theta} \rho_{\rm NESS}(\theta) > 0.
\end{equation}
Around this NESS, the expression \eqref{eq:FDT} of the FDT can be generalized by the following relation \cite{chetrite2008fluctuation}
\begin{equation}\label{eq:GFDT}
	k_{\rm B} T R(t - s) = \partial_s C(t - s) - b(t-s), \,\,\, t \ge s,
\end{equation}
where $\langle \ldots \rangle_0$ and $\langle \ldots \rangle_h$ represent now a non-equilibrium ensemble average in the NESS ($h=0$) and around the perturbed NESS ($h \neq 0$), respectively,   while the additional term
\begin{equation}\label{eq:corrFDT}
	b(t - s) = \langle Q(t) v_0(t) \partial_{\theta} V(s)  \rangle_0
\end{equation}
takes into account the  extent of the ``violation'' of the conventional FDT due to the non-vanishing current $j$. Here, the observables $Q$ and $V$ as well as $v_0$ are functions of $\theta$, while $v_0$ is called the \emph{mean local velocity}, and is given by $v_0(\theta) = j / \rho_{\rm NESS}(\theta)$. Note that the generalized expression \eqref{eq:GFDT} around a NESS reduces to the conventional FDT \eqref{eq:FDT} around thermal equilibrium, when $j = 0$. The expression~\eqref{eq:GFDT} of the generalized FDT can also be written in the integral form
\begin{equation}\label{eq:intGFDT}
	k_{\rm B} T \chi(t) = C(0) - C(t) - B(t) \quad \mbox{for} \quad t \ge 0,
\end{equation}
where $B(t) = \int_0^t b(t -s) ds$, and represents the generalization of equation~\eqref{eq:intFDT} for NESS.
Similar experimental approaches based on scanning optical tweezers have been followed to investigate fluctuation-response relations for other kinds of non-equilibrium states \cite{blickle2007einstein, toyabe2007experimental, toyabe2008energy, gomez2012fluctuations, bohec2013probing}.

\begin{figure}
	\begin{center}
	\includegraphics[width=12cm]{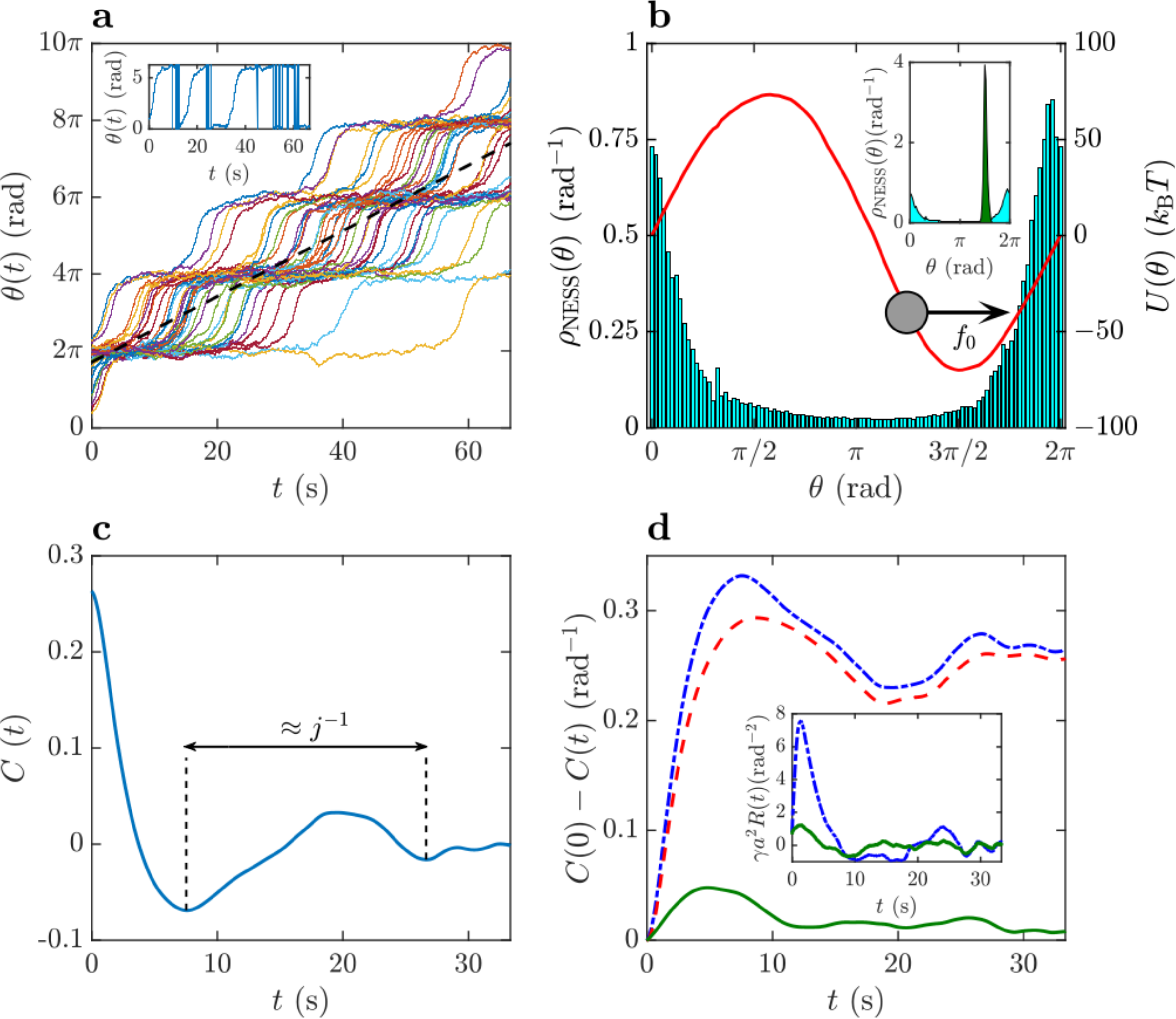}
	\caption{
	{\bf Experimental realization of a NESS}
	(a) Typical unwrapped (i.e., defined on the interval $[0,\infty)$) trajectories $\theta(t)$ of a colloidal particle in a NESS. The dashed line represents the mean drift due to the non-zero probability current $j$ induced by the non-conservative term $F$. Inset: example of wrapped (i.e., defined over $[0, 2\pi)$) trajectory $\theta(t)$.
	(b) NESS probability density function of $\theta$, defined over $[0, 2\pi)$ (bars) and reconstructed potential energy $U(\theta)$ (solid line). The arrow indicates the direction of the non-conservative force $f_0$, which shifts the maximum of  $\rho_{\rm NESS}(\theta)$ to the right relative to the minimum of $U(\theta)$. Inset: comparison between the NESS distribution  $\rho_{\rm NESS}(\theta)$  and the equilibrium one $\rho_{eq}(\theta)$ ($F = 0$, sharp peak).
	(c) Correlation function between the observable $Q(\theta) = \sin \theta$ and the variable $V(\theta) = \sin \theta$.
	(d) NESS correlation functions involved in the integral form of the generalized FDT~\eqref{eq:intGFDT}: $C(0) - C(t)$ (dotted-dashed line), $B(t)$ (dashed line) and $C(0) - C(t) -  B(t)$ (solid line). Inset: estimate of the response function $R(t)$ by means of the time derivative of $[C(0) - C(t)] / (k_{\rm B} T)$ (dotted-dashed line), and taking into account the corrective term,  $[C(0) - C(t) -  B(t)] / (k_{\rm B} T)$ (solid line).
	}
	\label{fig:28:sketchNESS}
	\end{center}
\end{figure}

\paragraph{Experimental implementation of a NESS.}

We illustrate the generalization of the FDT around a NESS with optical tweezers. To experimentally realize an overdamped system whose stochastic motion is well-modelled by equation~\eqref{eq:LE}, we consider a spherical silica particle (radius $a = 1\,{\rm \upmu m}$) suspended in water at constant temperature ($T = 20^{\circ}{\rm C}$, $\gamma = 1.89 \cdot 10^{-8}\,{\rm kg\,s^{-1}}$). 
The particle is subject to a \emph{toroidal} optical trap, which can be created by a conventional single trap, whose position is deflected on the plane transverse to the beam axis either by two perpendicular galvanometric mirrors~\cite{faucheux1995, blickle2007einstein}, or by a pair of perpendicular acousto-optic deflectors (AOD) \cite{gomez2009experimental}. 
For the latter approach, one applies simultaneously sinusoidal frequency modulations to the acoustic waves created inside each AOD X and Y, $\Delta F_{{\rm X},{\rm Y}}(t) = \Delta F \sin(2\pi f_{\rm R} t + \alpha_{{\rm X},{\rm Y}})$, with amplitude $\Delta F\sim{\rm MHz}$, scanning frequency $f_{\rm R}$, and phase difference $ |\alpha_{\rm X} -  \alpha_{\rm Y}|= \frac{\pi}{2}$. When $f_{\rm R}$ is kept fixed, the incident laser beam is deflected in a periodic manner. In this way the beam focus describes a circular trajectory inside the sample cell, $\theta_{\mathrm{OT}}(t) = 2\pi f_{\rm R} t + \alpha$, of constant radius $R_{\rm circle}$ at frequency $f_{\rm R}$ on the plane perpendicular to the beam propagation.  To achieve a periodic force landscape as described by the Langevin model~\eqref{eq:LE}, the scanning frequency of the laser beam must be on the order of $f_{\rm R} \sim  10^2\,{\rm Hz}$. For a particle with radius $a = 1\,{\rm \upmu m}$, the optical setup must be built in such away that the final radius of the circular path described by the tweezers is $R_{\rm circle} \approx 2 - 5\,{\rm \upmu m}$. Smaller values of $R_{\rm circle}$ lead to a Mexican-hat potential $U(\theta)$ in which the particle can spontaneously jump from $\theta$ to $\theta + \pi$, similar to a Kramer process in a double-well potential (section~\ref{sec:4.5.1:kramers}), while larger values of $R_{\rm circle}$ yield very slow dynamics for which it becomes experimentally difficult to sample $\theta$ with good statistics. For $R_{\rm circle} = 4.12\,{\rm \upmu m}$, $f_{\rm R} = 200\,{\rm Hz}$, and an infrared laser ($\lambda = 1064\,{\rm nm}$) with incident power of $30\,{\rm mW}$, the resulting scanning speed of the focused beam is $v_{\rm R} = 2\pi R_{\rm circle} f_{\rm R} \approx 5\,{\rm mm\,s^{-1}}$, which is so high that it is not able to trap and drag continuously the particle through the fluid because the viscous frictional force due to the surrounding water quickly exceeds the optical trapping force.  Consequently, the beam only kicks the particle a small distance $\delta s \lesssim \frac{\gamma v_{\rm R}}{\kappa}$ along the polar direction $\theta$ at each rotation. During the absence of the beam ($\approx 1/f_{\rm R} = 5\,{\rm ms}$) the particle undergoes free diffusion over a length scale $l_{\rm D} = \sqrt{\frac{D}{f_{\rm R}}}\lesssim 40\, \mathrm{nm}\ll R_{\rm circle}$ in the directions radial and perpendicular to the circle. The beam kicks the particle again during the next arrival, thus preventing its escape  by diffusion away from the region scanned by the beam focus. Therefore, the particle motion is effectively confined to a torus of major radius $R_{\rm circle}$ and minor radius $\sim l_{\rm D} / 2$, where the angular position $\theta$  is the only relevant dynamical degree of freedom. In addition, a periodic  intensity profile can be created along the main circle by sinusoidally modulating in time the laser power $P(t)$ with an amplitude $\Delta P = 2.1\,{\rm mW}$ around the mean value of $P_0 = 30\,{\rm mW}$, at the same frequency as the scanning frequency $f_{\rm R}$, i.e., $P(t) = P_0 + \Delta P \sin(2\pi f_{\rm R} t + \alpha_{\rm X})$. These optical conditions result in the desired force landscape:  a constant non-conservative force $f_0$ associated to the mean kick and a static periodic potential $U(\theta)$ due to the conservative force exerted by the sinusoidal light intensity profile \cite{gomez2009experimental, gomez2011fluctuations}. 

The instantaneous angular coordinate $\theta$ of the optically driven particle can be obtained from its 2D trajectories $(x,y)$, by $\theta = \arctan(y/x)$ (Fig.~\ref{fig:28:sketchNESS}(a)). 
The effective parameters $F$, $A$ and the profile $\phi(\theta)$ must be determined from an ensemble of independent NESS trajectories $\theta(t)$: at least 100 trajectories, each with at least a full rotation on the circle are needed to find the probability current $j$ (dashed line in Fig.~\ref{fig:28:sketchNESS}(a)) and the corresponding nonequilibrium distristriution $\rho_{\rm NESS}(\theta)$ (histogram in Fig.~\ref{fig:28:sketchNESS}(b)). 

Depending on the quantity of interest, $\theta$ must be defined either on the interval $[0, 2\pi)$ or $[0,\infty)$. For example, the probability current must be computed as \cite{gomez2009experimental}
\begin{equation}\label{eq:expNesscurrent}
	j = \frac{1}{2\pi} \frac{d\langle \theta(t) \rangle_0}{dt}, 
\end{equation}
where $\langle {\theta}(t) \rangle_0$ is the average of an ensemble of independent NESS trajectories $\theta(t)$, defined on $[0,\infty)$ (Fig.~\ref{fig:28:sketchNESS}(a)). 
This permits us to verify that $\langle {\theta}(t) \rangle_0$ is linear in $t$, thus leading to a constant $j > 0$. 
The quantity $j^{-1}$ represents the typical time the particle needs to perform a full cycle around the circle, which corresponds to the longest relaxation time of the NESS. For example, for the experiment presented here $j^{-1} = 26\,{\rm s}$, which is much larger than the typical equilibrium relaxation time of a particle in water trapped by optical tweezers ($\sim\,{\rm ms}$). 
Therefore, it is expected that the particle motion in this NESS has very long-lived correlations. 
In Fig.~\ref{fig:28:sketchNESS}(b), we plot the experimental profile of the NESS distribution $\rho_{\rm NESS}(\theta)$, computed from all the data points of the trajectories on $[0, 2\pi)$. 

From the experimental profile of $\rho_{\rm NESS}(\theta)$  and the value of $j$, the full force landscape of the particle can be reconstructed \cite{blickle2007characterizing}. 
First, taking into account the periodic boundary conditions of the NESS distribution, i.e. $\rho_{\rm NESS}(0) = \rho_{\rm NESS}(2\pi)$, from equation~\eqref{eq:NESScurrent} we get the non-conservative term 
\begin{equation}\label{eq:expF}
	F 
	=
	{j \over 2\pi} 
	\int_0^{2\pi} d\theta {1 \over \rho_{\rm NESS}(\theta)},
\end{equation}
whereas the potential is given by
\begin{equation}\label{eq:expPot}
	A\phi(\theta) 
	=
	-D_{\theta} 
	\log \rho_{\rm NESS}(\theta) 
	+  
	\int_0^{\theta} d\theta' 
		\left[ 
			F 
			-  
			{j \over \rho_{\rm NESS}(\theta')} 
		\right].
\end{equation}
The experimental values of the parameters considered here yield $F = 0.85\,{\rm rad\,s^{-1}}$ and $A = 0.87\,{\rm rad^2 s^{-1}}$, which correspond to a non-conservative force $f_0 = \gamma a F = 66\,{\rm fN}$ and an energy potential amplitude $\max\{U(\theta)\} = \gamma a^2 A = 68.8\,k_{\rm B} T$, while the normalized potential profile is sinusoidal, $\phi(\theta) = \sin \theta$ (red solid line in Fig.~\ref{fig:28:sketchNESS}(b)). 
In thermal equilibrium ($F = 0$), the particle motion would be tightly confined around the minimum of such a deep non-linear potential at $\phi_m = 3\pi/2$, with a width $\Delta \theta \approx \sqrt{\frac{k_{\rm B} T}{\gamma a^2 A}} = 0.12\,{\rm rad}$ and a vanishingly small probability to cross the potential barrier of height $2A = 137.6\,k_{\rm B} T$ (inset in Fig.~\ref{fig:28:sketchNESS}(b)). 
However, under the action of the non-conservative force, the distribution of $\theta$ is much broader, with a maximum shifted in the direction of the resulting current $j > 0$. 
This reflects the highly non-equilibrium conditions of the system. 
Indeed, they can be better assessed by means of the generalized FDT relation~\eqref{eq:intGFDT}. We focus on the observable $Q(\theta)  = \sin \theta$, which is periodic on the circle and corresponds to the dimensionless potential energy of the particle $\phi(\theta) = U(\theta) / \max\{U(\theta)\}$.\footnote{In general, $\phi(\theta)$ and $Q(\theta)$ must not necessarily be the same: while $\phi(\theta)$ represents the normalized potential energy of the particle, $Q(\theta$) is an arbitrary observable that is periodic with respect to $theta$. Since $phi(\theta) $ fullfils in particular the required periodicity, in the example described in the main text we choose $Q = phi$.} A variation in the potential amplitude, $A \rightarrow A + \delta A$ translates into the following change in the potential energy: $ \gamma a^2 A\phi(\theta) \rightarrow \gamma a^2 A\phi(\theta) - (-\gamma a^2 \delta A) \phi(\theta)$. Therefore, by noting that a variation $h$ in a control parameter leads to a change in the potential energy, i.e., $U \rightarrow U  - hV$, we conclude that in this specific example the integrated response function of $Q(\theta)$ to the perturbation $h = -\gamma a^2 \delta A$ is given by equation~\eqref{eq:intGFDT}, where the variable conjugate to the perturbation is $V(\theta) = \sin \theta$,
whereas $\partial_{\theta} V(\theta) = \cos \theta$ in equation~\eqref{eq:corrFDT}. In Fig.~\ref{fig:28:sketchNESS}(c), we plot the NESS correlation function $C(t) = \langle \sin \theta(t) \sin \theta(0) \rangle_0$, revealing an oscillatory behavior with a typical time-scale $\approx 25\,{\rm s}$, which corresponds to $j^{-1}$.  Here, to improve the statistics from the available data, the NESS ensemble average $\langle \sin \theta(t) \sin \theta(0)  \rangle_0$ is computed first by taking a time average 
\begin{equation}\label{eq:timeaverage}
	C_{t_{\mathrm{max}}}(t) = \frac{1}{t_{\mathrm{max}}-t}\int_0^{t_{\mathrm{max}}-t} du \,\sin \theta(u) \sin \theta(u + t),
\end{equation}
along a given realization of $\theta(t)$, where $t_{\mathrm{max}} > j^{-1}$ to span a full cycle of the probability current, and then by performing an additional ensemble average $\langle C_{t_{\mathrm{max}}}(t) \rangle$ over different independent NESS trajectories $\theta(t)$. The same procedure must be carried out for the correlation $B(t)$ defined through equation~\eqref{eq:corrFDT}. In Fig.~\ref{fig:28:sketchNESS}(d), we plot both the NESS terms $C(0) - C(t)$ and $B(t)$ involved in the integral generalized FDT~\eqref{eq:intGFDT}. We observe that both terms are of the same order of magnitude, thus confirming that the experimental conditions of the system are far from thermal equilibrium. Note that, without the corrective term $b$ defined in equation~\eqref{eq:corrFDT}, a determination of the integrated response function $\chi(t)$ by means of the equilibrium FDT~\eqref{eq:intFDT}, and consequently $R(t)$, would be wrong. This is shown in the inset of Fig.~\ref{fig:28:sketchNESS}(d) , where we compare the numerical time derivative of $\frac{1}{k_{\rm B}T} [C(0) - C(t)]$ with that of $\frac{1}{k_{\rm B}T} [C(0) - C(t)  - B(t)]$, which shows that without taking into account the corrective term $B$, the response function $R(t)$ would be highly overestimated around a NESS. 

\subsubsection{Stochastic Thermodynamics} \label{sec:4.5.3:st}

Stochastic thermodynamics is an emerging branch of statistical physics that extends  concepts of classical thermodynamics such as heat, work and entropy production to the level of single stochastic trajectories for systems  in contact with a heat bath and driven arbitrarily far from thermal equilibrium by means of a well-specified experimental protocol \cite{seifert2012stochastic}.  It provides a number of statistical relationships involving these fluctuating quantities, which generalize the macroscopic laws of thermodynamics and describe the non-equilibrium energetics of many mesoscopic systems of interest in soft matter and biophysics, such as colloidal particles, polymers, biomolecules, molecular motors, biochemical networks, and microelectromechanical systems \cite{ciliberto2013fluctuations, ciliberto2017experiments}. Thus, stochastic thermodynamics represents nowadays a solid theoretical framework for applications ranging from single-molecule biomechanics (section~\ref{sec:4.1:molecule}) to optimization protocols and the efficiency of microscopic heat engines (section~\ref{sec:4.5.4:engines}) .

Here, we present some general notions of stochastic thermodynamics using a paradigmatic system: a colloidal particle embedded in a viscous liquid (friction coefficient $\gamma$), in contact with a thermostat at constant temperature $T$ and subject to a possibly time-dependent force. The 1D motion of a single coordinate $x$ of the particle can be described by the Langevin equation
\begin{equation}\label{eq:1DLE}
	\dot{x} =  \mu \left[ -\frac{\partial U(x,\lambda)}{\partial x} + f(x,\lambda) \right] + \xi,
\end{equation}
where $U(x,\lambda)$ is the potential energy which leads to a conservative part $-\partial U(x,\lambda) / \partial x$ in the total force, whereas $ f(x,\lambda)$ represents a non-conservative force, as that described in section~\ref{sec:4.5.2:ness}. Both forces can be time-dependent through the control  paramenter $\lambda$, which can be externally varied according to a prescribed protocol from $\lambda(0)$ to $\lambda(t)$ during the time interval $[0,t]$. In addition, in equation~\eqref{eq:1DLE} $\mu = 1/\gamma$ is the particle mobility, whereas $\xi$ represents a Gaussian white noise with zero-mean and autocorrelation function $\langle \xi(t) \xi(s) \rangle = 2 D \delta(t-s)$, which accounts for the thermal fluctuations of the particle velocity. Stochastic thermodynamics requires that the properties of the heat bath are not affected by the presence of external driving forces, thus implying that $D = k_{\rm B} T \mu$ regardless of the strength of $-\partial U(x,\lambda)/\partial x$ and $f(x,\lambda)$. Note that a statistically equivalent description of the system modelled by equation~\eqref{eq:1DLE} can be given by means of the Fokker-Planck equation for the probability density function of $x$ at time $t$
\begin{equation}\label{eq:1DFP}
	\frac{\partial \rho(x,t)}{\partial t}=  -\frac{\partial j(x,t)}{\partial x},
\end{equation}
where the probability current is defined as
\begin{equation}\label{eq:1Dcurrent}
	j(x,t)=   \upmu \left[ -\frac{\partial U(x,\lambda(t))}{\partial x} + f(x,\lambda(t)) \right]  \rho(x,t) -  D \frac{\partial \rho(x,t)}{\partial x},
\end{equation}
supplemented with the initial condition $\rho(x,t = 0) = \rho_0(x)$. Both $\rho(x,t)$ and $j(x,t)$ allow to directly assess how far the system is from  thermal equilibrium conditions, for which $j= 0$ and $\rho \propto \exp\left( -\frac{U(x)}{k_{\rm B} T}\right)$.

\begin{figure}
	\begin{center}
	\includegraphics[width=12cm]{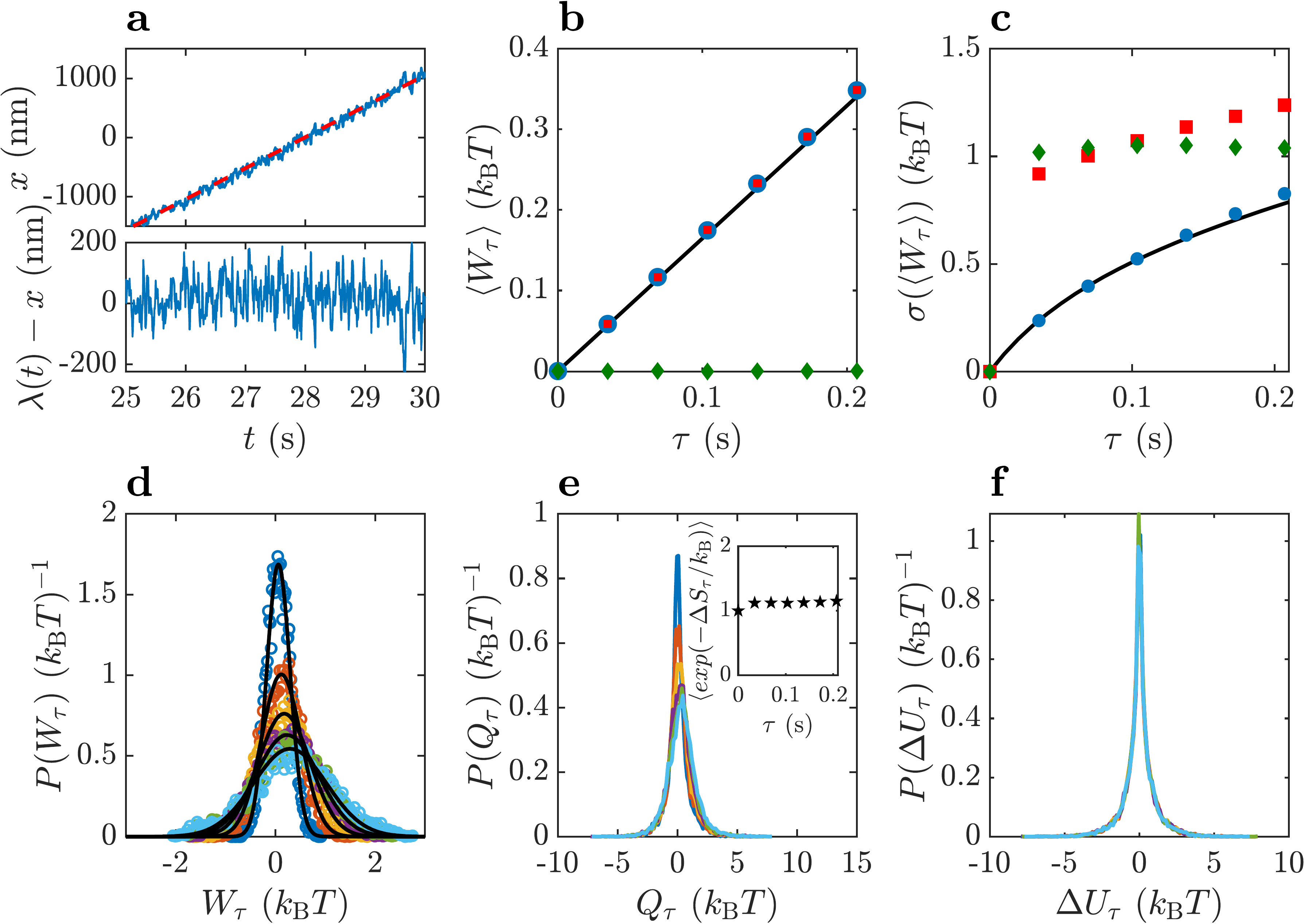}
	\caption{
	{\bf First Law in stochastic thermodynamics.}
	(a) Upper panel: Portion of the trajectory $x(t)$ (blue solid line) of a colloidal bead  ($2a = 2.73\,{\rm \upmu m}$), dragged by moving an optical tweezers ($k = 1.15\,{\rm pN \, \upmu m^{-1}}$, $v = 0.520\,{\rm \upmu m\,s^{-1}}$) through water ($T = 295\,{\rm K}$). The dashed line represents the time evolution of the minimum of the trapping potential: $\lambda(t) = vt + \lambda(0)$. 
	Lower panel: time evolution of $\lambda(t) - x(t)$.  
	(b) Dependence of the mean work $\langle W_{\tau} \rangle$ ($\circ$), mean heat dissipated into the aqueous medium $\langle Q_{\tau} \rangle$ ($\square$), and mean variation of potential energy $\langle \Delta U_{\tau} \rangle$ ($\diamond$) as a function of the measurement time $\tau$. The solid line represents the values given by equation~\eqref{eq:meanwork}.  
	(c) Dependence of the standard deviation of the work ($\circ$), of the heat dissipated into the aqueous medium ($\square$), and of the variation of potential energy ($\diamond$) as a function of $\tau$. The solid line represents the values given by the square roots of equation~\eqref{eq:varwork}. 
	(d) Distribution of the work (in units of $k_{\rm B} T$) for different values of the measurement time $\tau$. From inner to outer curves: $\tau = 0.034\,{\rm s}$, $0.069\,{\rm s}$, $0.103\,{\rm s}$, $0.138\,{\rm s}$, $0.172\,{\rm s}$, and $0.207\,{\rm s}$. 
	(e) Distribution of the dissipated heat  (in units of $k_{\rm B} T$) for different values $\tau$. Same color code as in (d). 
	Inset: verification of the integral fluctuation theorem~\eqref{eq:IFT} for the total entropy production. 
	(f) Distribution of the variation of potential energy  (in units of $k_{\rm B} T$) for different values $\tau$. Same color code as in (d). }
	\label{fig:29:StochThermo}
	\end{center}
\end{figure}

\paragraph{First Law.}

To generalize classical thermodynamic concepts to the level of an individual stochastic realization of the process $x(t)$, we first note that either a temporal variation of the control parameter $\lambda$ during the infinitesimal time duration $dt$ or a non-zero non-conservative force, $f \neq 0$, can be identified as an infinitesimal work increment done on the system:
\begin{equation}\label{eq:dwork}
	 \delta W = \frac{\partial U}{\partial \lambda} \dot{\lambda}dt + fdx.
\end{equation}
Then, from equation~\eqref{eq:dwork} an inifinitesimal variation $dU$ in the potential energy during $dt$ can be written as
\begin{equation}\label{eq:dpotential}
	dU
	= \frac{\partial U}{\partial \lambda}\dot{\lambda}dt  + \frac{\partial U}{\partial x} dx,\nonumber 
	= \delta W - \left( f - \frac{\partial U}{\partial x} \right)dx.
\end{equation}
Thus, by defining the infinitesimal increment of the heat dissipated into the medium as the product of the total force acting on the particle times the infinitesimal displacement $dx$,
\begin{equation}\label{eq:dheat}
	\delta Q = \left( f - \frac{\partial U}{\partial x} \right)dx,
\end{equation}
the expression~\eqref{eq:dpotential} can be regarded as a generalization of the \emph{First Law of thermodynamics} \cite{sekimoto1998langevin}
\begin{equation}\label{eq:1lawST}
	dU = \delta W - \delta Q
\end{equation}
on the level of the stochastic process $x(t)$. From the Langevin equation~\eqref{eq:1DLE}, the work done on the system and the heat dissipated into the medium over the finite time interval $[0,\tau]$, are
\begin{equation}\label{eq:stochwork}
	W_{\tau}[x(t)] = \int_0^{\tau} dt \left[  \frac{\partial U(x(t),\lambda(t))}{\partial \lambda} \dot{\lambda}(t) + f  \dot{x}(t) \right],
\end{equation}
\begin{eqnarray}\label{eq:stochheat}
	Q_{\tau}[x(t)] & = & \int_0^{\tau} dt \left[ f - \frac{\partial U(x(t), \lambda(t))}{\partial x} \right]\dot{x}(t),\nonumber\\
		& = &  \frac{1}{\mu}\int_0^{\tau} dt  \dot{x}(t)^2  - \frac{1}{\mu}\int_0^{\tau} dt  \xi(t) \dot{x}(t) ,
\end{eqnarray}
respectively, where the notation $[x(t)]$ highlights the fact that both quantities fluctuate over time and depend on the specific stochastic realization of $x(t)$ up to time $\tau$. 
The terms which involve $dx(t) = dt \dot{x}(t)$ in the integrals of equations~\eqref{eq:stochwork} and \eqref{eq:stochheat} must be interpreted in the Stratonovich sense \cite{sekimoto2010stochastic}. 
Note that the definition of the heat given by equations~\eqref{eq:dheat} and \eqref{eq:stochheat} is physically meaningful, since the first term on the right-hand side of equation~\eqref{eq:stochheat} is positive definite and represents the energy delivered to the thermostat, while the second term can be either positive or negative and corresponds to the thermal fluctuating energy injected by the medium. Moreover, for systems in thermal equilibrium (i.e., $f = 0$ and $\dot{\lambda} = 0$), equation~\eqref{eq:1lawST} consistently reduces to $ dU = -  \delta Q$, where any increment (decrease) in the potential energy of the particle is only due to a fluctuation of the heat taken up from (delivered to) the thermostat, being both on average equal to zero. 

We analyze the previous concepts in the specific case of a colloidal particle  (diameter $2 a = 2.73\,{\rm \upmu m}$) suspended in water  at constant $T = 295\,{\rm K}$, for which the friction coefficient is $\gamma = 2.44 \cdot 10^{-8}\,{\rm kg\,s^{-1}}$. The particle is subject to the harmonic potential of an optical trap, where the position of its minimum plays the role of the control parameter $\lambda$ in equation~\eqref{eq:1DLE}. In the case where the optical trap is uniformly translated, $\lambda$ varies linearly over time at constant speed $v$ and, under the initial condition $\lambda(t = 0) = 0$, $U(x,\lambda(t)) = {1\over2} k \left( x - vt \right)^2$, $\lambda(t) = vt$, and $\dot{\lambda}(t) = v$. Without loss of generality, we consider the simplest situation where all the forces acting on the system are conservative (i.e., $f = 0$), thus being the temporal variation of $\lambda$ the source of non-equilibrium work and dissipated heat. This well-defined protocol can be implemented by means of the deflection of the trapping beam using galvanometric mirrors or acousto-optics deflectors (section~\ref{sec:2.3.1:beamsteering}). 
Although with a different thermodyamic interpretation,\footnote{When moving the sample stage at speed $v$, an external flow is applied to the trapped particle, which results in an extra term in the viscous drag force acting on the particle, i.e. $\gamma (\dot{x} - v)$. On the other hand, the potential energy remains unaffected by $v$, $U(x) = \kappa x^2/2$. This leads to expressions for the work and heat which are different from those obtained when performing a motion of the optical trap.}
a closely related protocol can be carried out by moving the sample cell at constant speed $v$ by means of a piezoelectric stage (section~\ref{sec:3.11:active}) \cite{wang2002experimental}. 
Fig.~\ref{fig:29:StochThermo}(a) shows a typical experimental trajectory $x(t)$ of such a colloidal particle, as well as the time evolution of the position of the minimum of the optical trap, $\lambda(t)$. 
For times much longer than the trap relaxation time ($\approx 21\,{\rm ms}$), the dragged particle reaches a steady state, whose statistical properties do not depend on time. 
In this case, from the definition~\eqref{eq:stochwork}, one can find the work done by the optical tweezers on the trapped particle over a time interval of duration $\tau$ along a given stochastic realization of $x(t)$:
\begin{eqnarray}\label{eq:stochworkvel}
	W_{\tau} & = & - kv \int_0^{\tau} dt\left[ x(t) - vt \right],\nonumber\\
	& = & {1\over2}kv^2 \tau^2 - kv \int_0^{\tau} dt x(t),
\end{eqnarray}
where we have dropped the argument $[x(t)]$ for the sake of simplicity. 
Since for this control parameter and a harmonic potential, equation~\eqref{eq:1DLE} is linear and $\xi$ is Gaussian, $x$ must also be Gaussian and, thus, $W_{\tau} $ is Gaussian as well, with mean
\begin{equation}\label{eq:meanwork}
	\langle W_{\tau} \rangle = \gamma v^2 \tau
\end{equation}
and variance 
\begin{equation}\label{eq:varwork}
	\langle [W_{\tau} - \langle W_{\tau} \rangle]^2 \rangle = 2k_{\rm B} T \gamma v^2 \left[ \tau +\frac{\gamma}{k}\exp\left( -\frac{k\tau}{\gamma}\right) - \frac{\gamma}{k}   \right].
\end{equation}
Indeed, using the experimental trajectory $x(t)$ (Fig.~\ref{fig:29:StochThermo}(a)), we compute numerically $W_{\tau}$ by means of equation~\eqref{eq:stochworkvel} and check that its mean and standard deviation satisfy equations~\eqref{eq:meanwork} and \eqref{eq:varwork}, as shown by the blue circles in Figs.~\ref{fig:29:StochThermo}(b) and \ref{fig:29:StochThermo}(c), respectively. 
The distribution of $W_{\tau}$ for different values of $\tau$ is plotted in Fig.~\ref{fig:29:StochThermo}(d), where we verify that $P(W_{\tau})$ is Gaussian. Note that, while  $\langle W_{\tau} \rangle$  is always positive and increases linearly with the observation time $\tau$ as the trap drags the particle through the fluid, there can be negative work fluctuations, where the system loses mechanical energy \cite{wang2002experimental}. 
These rare events, which are at first glance in contradiction with macroscopic intuition, are due to thermal fluctuations, which are not taken into account by classical thermodynamics. 
The role of such fluctuations, quantified by the ratio $\sqrt{\langle [W_{\tau} - \langle W_{\tau} \rangle]^2 \rangle} / \langle W_{\tau} \rangle$, is very prominent at short time-scales and become negligible as $\tau$ approaches the limit of a macroscopic observation time, i.e., $\tau \gg \frac{\gamma}{k}$.

The heat dissipated into the medium due to the motion at constant speed of the trap can be derived from equation~\eqref{eq:stochheat} as
\begin{eqnarray}
	Q_{\tau}&= & - k \int_0^{\tau} dt\left[ x(t) - vt \right] \dot{x}(t)\nonumber\\
	\label{eq:135}
	              &= & - k \int_0^{\tau}\left[ x(t) - vt \right] d{x}(t).
\end{eqnarray}
In this case, some care is needed to compute the stochastic integral~\eqref{eq:stochworkvel} from the experimental values of $x(t)$. 
As previously mentioned, Stranotovich integration must be used to get physically meaningful results.\footnote{While equation~\eqref{eq:stochworkvel} for the work simply requires the product of  the deterministic time increment $dt$ with the stochastic particle position relative to the position of the trap, $x(t) - vt$, the expression for the heat (equation~\eqref{eq:135}) involves the stochastic increment of the particle position, $dx$ times $x(t) - vt$.}
Specifically, from the collection of $N$ experimental data points ${x(t_j)}$, defined at each time $t_j = \frac{j-1}{N-1}\tau$ over $[0,\tau]$, where $j = 1, \ldots , N$, a mid-point discretization of the force $-k[x(t) - vt]$ with respect to the stochastic increment $d{x}(t)$ must be performed 
\begin{equation}\label{eq:discreteheat}
	Q_{\tau} =-\frac{k}{2}\sum_{j = 1}^{N-1} \left[ x(t_{j+1})+x(t_j) - (t_{j+1} + t_j) v \right][x(t_{j+1}) - x(t_j)].
\end{equation}
In Fig.~\ref{fig:29:StochThermo}(e) we show the distribution of $Q_{\tau}$ for different values of  $\tau$, computed by means of equation~\eqref{eq:discreteheat}. In this case, all discrete time steps have a constant value fixed by the inverse of the experimental sampling frequency $f_{\rm s} = 145\,{\rm Hz}$, i.e., $t_{j+1}- t_j = f_{\rm s}^{-1} = 0.007\,{\rm s}$. It should be noted that, although the mean value of $Q_{\tau}$ coincides with that of $W_{\tau}$ for all $\tau \ge 0$ in a steady state, as shown in Fig~\ref{fig:29:StochThermo}(b), their distributions are quite different, see Fig.\ref{fig:29:StochThermo}(e). In this case, the distribution of the heat, $P(Q_{\tau})$ has non-Gaussian tails, with standard deviation larger than the variance of the work fluctuations (Fig.~\ref{fig:29:StochThermo}(c)). In addition, similar to the behavior of the stochastic work, rare events where $Q_{\tau}$ is negative are also possible for small values of $\tau$ at which thermal fluctuations are significant, i.e., the heat spontaneously flows from the surrounding medium to the particle even if mechanical work is done on it.

Finally, in Fig.~\ref{fig:29:StochThermo}(f) we plot the distribution $P(\Delta U_{\tau})$ of the variation of the potential energy over a time interval of duration $\tau$:
\begin{eqnarray}\label{eq:discreteenergy}
	\Delta U_{\tau} & = & U(x(\tau),\lambda(0)) - U(x(0), \lambda(0)) \nonumber\\
				& = & \frac{k}{2} \left[ [x(\tau) - v\tau]^2 - x(0)^2  \right].
\end{eqnarray}
In this case, while the mean value of $\Delta U_{\tau}$ is equal to 0 for all values of $\tau$ (Fig.~\ref{fig:29:StochThermo}(b)), consistent with the fact that in a steady state the potential energy of the system remains constant on average, its distribution is strongly non-Gaussian, due to the non-linear dependence of $\Delta U_{\tau}$ on $x(t)$, with standard deviation which quickly saturates to a constant value, as observed in Fig.~\ref{fig:29:StochThermo}(c).

The statistical properties of $W_{\tau}$,  $Q_{\tau}$ and  $\Delta U_{\tau}$ for this simple experiment illustrate the main essence of stochastic thermodynamics. Even though the average values of such quantities satisfy the expected thermodynamics behavior, the probability of observing \emph{rare} events, which are not described by the classical thermodynamics, is not negligible for mesoscopic systems at sufficiently short time scales.

\paragraph{Second Law and entropy.}

Stochastic thermodynamics also offers the possibility to extend the concept of entropy to stochastic trajectories of systems in non-equilibrium states, and provides quantitative relationships that generalize the macroscopic Second Law of thermodynamics. In the case of a system described by equation~\eqref{eq:1DLE}, the variation of the \emph{total} entropy is defined over a given time interval $[0,\tau]$ as
\begin{equation}\label{eq:totalentropy}
	\Delta S_{\tau} = \frac{Q_{\tau}[x(t)]}{T} - k_{\rm B} \ln \frac{\rho(x(\tau),t)}{\rho(x(0),0)}.
\end{equation}
The first term on the right-hand side of equation~\eqref{eq:totalentropy} can be identified as a change in entropy due to the heat dissipated into the medium, defined in equation~\eqref{eq:stochheat}, while the second term is called \emph{stochastic} entropy, and must be determined by solving first the corresponding Fokker-Planck equation~\eqref{eq:1DFP} and then evaluating the solution $\rho(x,t)$ along the specific stochastic trajectory $x(t)$ with initial condition $x(0)$ at $t=0$. Therefore, unlike work, heat and potential energy, the stochastic entropy does not only depend on the particular realization of $x(t)$ but on the ensemble through the quantity $\rho(x,t)$. For example, for the linear motion ($\lambda(t) = vt$) of the position of the optical trap considered in the experiment above, the solution of the Fokker-Planck equation~\eqref{eq:1DFP} is  $\rho(x,t) \propto \exp \left[  - \frac{k(x - vt + \frac{\gamma v}{k})^2}{2k_{\rm B} T} \right]$ under the condition that the system has already achieved a steady state in the comoving frame of the trap. In this case, the variation of the stochastic entropy over the time interval $[0,\tau]$ is given by $\frac{k}{2T}\left[  \left( x(\tau) - v\tau + \frac{\gamma v}{k} \right)^2 - \left( x(0) + \frac{\gamma v}{k} \right)^2\right]$, from which the corresponding change of the total entropy can be computed using equations~\eqref{eq:stochheat} and \eqref{eq:totalentropy}.

Remarkably, the change in the total entropy defined by equation~\eqref{eq:totalentropy} satisfies the \emph{integral fluctuation theorem} (IFT)  \cite{seifert2005entropy}
\begin{equation}\label{eq:IFT}
	 \left\langle \exp \left( -\frac{1}{k_{\rm B}}\Delta S_{\tau}  \right) \right\rangle= 1
\end{equation}
for all measurement times $\tau$ and any kind of protocol involving a non-equilibrium temporal variation of the control parameter $\lambda$ from $\lambda(0)$ to $\lambda(\tau)$, and possibly including non-conservative forces $f \neq 0$ that break detailed balance. Using the Jensen's inequality, $\langle e^{a} \rangle \ge e^{\langle a \rangle}$, equation~\eqref{eq:IFT} implies that $\langle \Delta S_{\tau}  \rangle \ge 0$, which corresponds to one of the expressions of the Second Law of thermodynamics. Therefore, the IFT represents a refinement of the Second Law of classical thermodynamics. For the experiment above, we verify the IFT for different values of $\tau$  in the inset of Figure \ref{fig:29:StochThermo}(e), where the left-hand side of equation~\eqref{eq:IFT} is numerically determined using the computed values of $\Delta S_{\tau}$.

Depending on the specific protocol for the variation of the control parameter $\lambda$ and on the state of the system (e.g., transient or steady), alternative versions of the Second Law can be derived within the context of stochastic thermodynamics. 
For example, when driving the system  by changing the control parameter from $\lambda(0)$ to $\lambda(\tau)$, the non-equilibrium work $W_{\tau}$ done on the system, defined by equation~\eqref{eq:stochwork}, and the corresponding difference in free energy  $\Delta F = F(\lambda(\tau)) - F(\lambda(0))$, satisfies the relation \cite{jarzynski1997nonequilibrium}
\begin{equation}\label{eq:Jarzynski}
	 \left\langle \exp \left( -\frac{W_{\tau}}{k_{\rm B} T}  \right) \right\rangle= \exp\left(-\frac{\Delta F}{k_{\rm B} T}\right)
\end{equation}
for any arbitrary non-equilibrium protocol $\lambda(t)$, $0 \le t \le \tau$ , provided that the system is initally in thermal equilibrium at $t=0$, and all the forces acting on the system are conservative (i.e., $f=0$ in equation~\eqref{eq:1DLE}). Equation~\eqref{eq:Jarzynski} is called the \emph{Jarzynski relation} \cite{jarzynski1997nonequilibrium}, and derives directly from the IFT~\eqref{eq:IFT}, in which case the total entropy takes the particular form $\Delta S_{\tau} = (W_{\tau} - \Delta F)/T$, where $W_{\tau} - \Delta F$ is known as dissipated work.
A closely related expression, from which the Jarzynski relation~\eqref{eq:Jarzynski} can alternatively be derived is the so-called \emph{Crooks fluctuation theorem} \cite{crooks1999entropy}. It quantifies the ratio between the probability density of the work spent in the \emph{forward} process, $\rho(W_{\tau})$, when driving the system from thermal equilibrium at $\lambda(0)$ through an arbitrary non-equilibrium protocol up to $\lambda(\tau)$, and the distribution of the work done in the reversed process, $\tilde{\rho}(W_{\tau})$,  where the control parameter is varied as $\tilde{\lambda}(t) = \lambda(\tau - t)$, starting in thermal equilibrium at $\tilde{\lambda}(0) = \lambda(\tau)$. The expression of the Crooks fluctuations theorem is
\begin{equation}\label{eq:Crooks}
	\frac{\tilde{\rho}(W_{\tau} = -W)}{\rho(W_{\tau} = W)}= \exp\left(-\frac{(W - \Delta F)}{k_{\rm B} T}\right).
\end{equation}
The practical importance of the relations~\eqref{eq:Jarzynski} and \eqref{eq:Crooks} is that, unlike quasistatic experimental protocols based on classical thermodynamics, they allow to determine  free energy differences between two equilibrium states from non-equilibrium measurements of the work performed at an arbitrarily fast rate, for example when studying biomolecules (section~\ref{sec:4.1:molecule}).

\subsubsection{Carnot cycles and Maxwell's demons}\label{sec:4.5.4:engines}

In this section, we will explore how optical tweezers can be used to get a deeper understanding of Carnot cycles, work efficiency and Maxwell engines. Although the results presented in this section have not been implemented by us in the laboratory directly, we provide sufficient instructions for the readers to be able to readily realize these experiments.

\paragraph{Carnot cycles and microscopic engines}

In 1824 Sadi Carnot tried to answer  basic questions about the efficiency, construction and operation of macroscopic heat engines \cite{carnot1824reflexions}. This seminal work gave birth to thermodynamics, for which Carnot is considered one of its fathers. It also introduced concepts that later were used by Clausius and Kelvin to define the concept of entropy and introduce the Second Law of thermodynamics. Almost two centuries later, similar questions posed by Carnot on big, chunky, greasy, macroscopic heat engines, are finally being explored in the microscopic realm \cite{martinez2017colloidal}.

\begin{figure}[ht]
	\begin{center}
	\includegraphics[width=12cm]{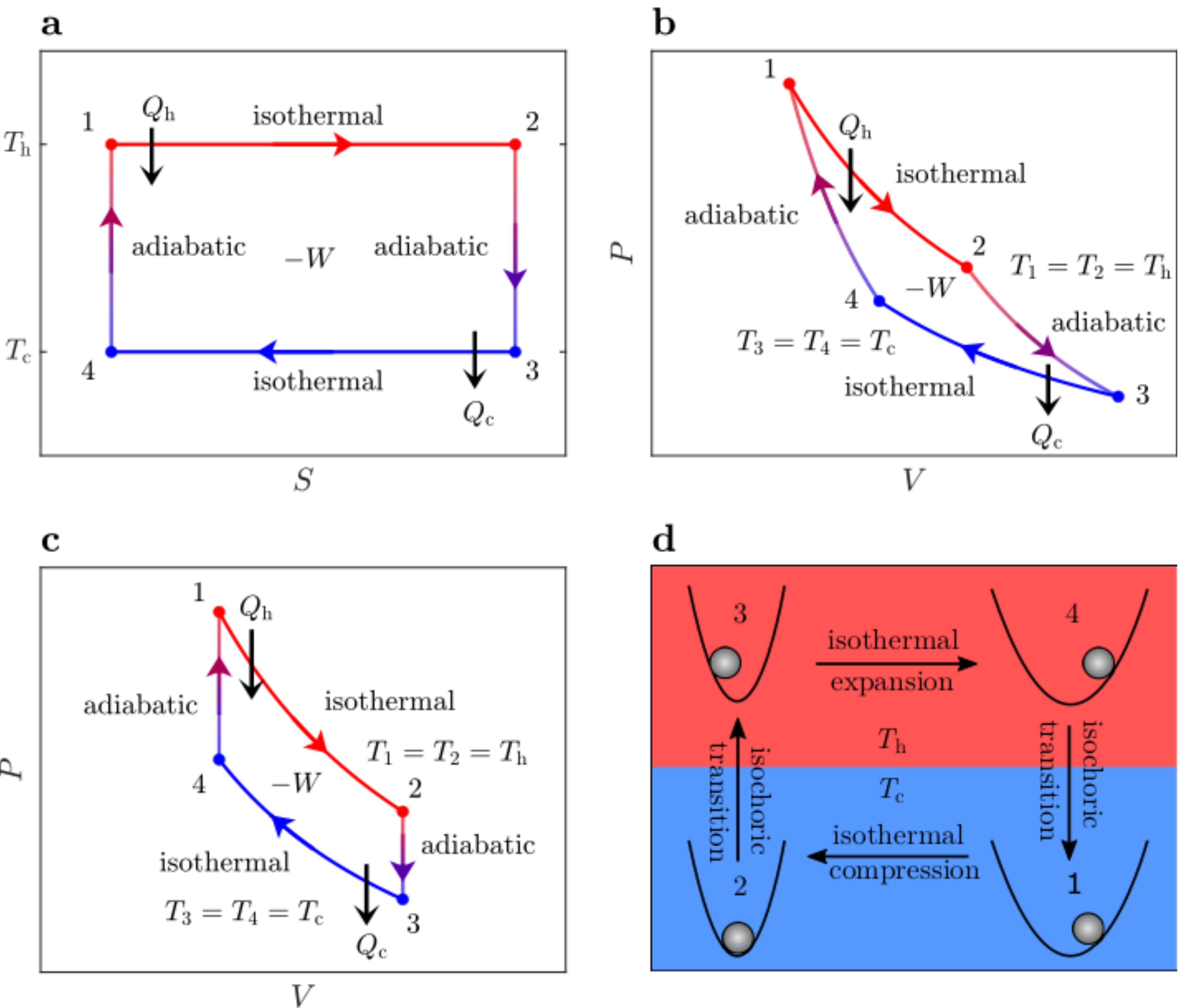}
	\caption{
	{\bf Microscopic engines.}
	(a) A sketch of Carnot's cycle in the $(S,T)$-plane and 
	(b) in the $(V,P)$-plane.
	(c) The Stirling cycle in the $(V,P)$-plane. 
	(d) Implementation of the Stirling cycle using a colloidal particle in an optical trap.
	}
	\label{fig:30:carnot}
	\end{center}
\end{figure}

Colloquially \emph{Carnot's cycle} refers to a process in which one extracts work from a system coupled to two reservoirs at different temperatures $T_{\rm c}< T_{\rm h}$. More precisely, given a system interacting with two thermal baths, Carnot's cycle consists of four steps:
\begin{enumerate}
\item An isothermal expansion at temperature $T_{\rm h}$, during which an amount of heat $Q_{\rm h}$ is absorbed by the system
\item An adiabatic expansion from $T_{\rm h}$ to $T_{\rm c}$.
\item An isothermal contraction at temperature $T_{\rm c}$, during which an amount of heat $Q_{\rm c}$ is released from the system.
\item An adiabatic contraction  from $T_{\rm c}$ to $T_{\rm h}$, returning the system back to its original state.
\end{enumerate}
The four steps of this cycle are depicted in Fig.~\ref{fig:30:carnot}(a) in the $(S,T)$-plane, where $S$ refers to the entropy. For a generic fluid with volume $V$ and pressure $P$, Carnot's cycle in the $(V,P)$-plane is shown in Fig.~\ref{fig:30:carnot}(b).

During one cycle, the extracted work is given by the First Law of thermodynamics
\begin{equation}
	-W=Q_{h}+Q_{\rm c}
\end{equation}
and the entropy production is
\begin{equation}
	\Delta S_{\text{tot}}=-\frac{Q_{\rm h}}{T_{\rm h}}-\frac{Q_{\rm c}}{T_{\rm c}}.
\end{equation}
This implies that for a reversible operation (i.e., for a process for which $\Delta S_{\text{tot}}=0$) we obtain the following relationship between the heat exchange and the temperature of both baths:
\begin{equation}
\frac{Q_{\rm c}}{Q_{\rm h}}=-\frac{T_{\rm c}}{T_{\rm h}}.
\end{equation}
Carnot showed that, for a reversible process, the efficiency of Carnot's cycle is
\begin{equation}
\eta_{\rm C}\equiv\frac{-W}{Q_{\rm h}}=\frac{Q_{\rm h}+Q_{\rm c}}{Q_{\rm h}} =1-\frac{T_{\rm c}}{T_{\rm h}},
\end{equation}
where the efficiency $\eta = -W/Q_{\rm h}$ is defined as the ratio between the extracted work and the absorbed heat.
Ideally, for the process to be reversible, it must be performed infinitely slowly or, more precisely in a time $\tau$ much slower than any characteristic relaxation time of the system. This automatically means that the output power, defined as  $P\equiv -W/\tau$ is zero for an  ideal reversible heat engine.

In reality, Carnot's cycle must be performed in a finite time $\tau$, which implies a trade-off between output power and efficiency. 
Due to the Second Law of thermodynamics, , which imposes $\Delta S_{\rm tot} > 0$, we have that $\eta < \eta_{\rm C}$.

Moreover, practical heat engines have additioanl limitations that further reduce their efficiency \cite{vining2009inconvenient}. 
For example, accounting for heat transfer yields a maximum power output given by the \emph{Novikov-Curzon-Ahlborn formula} \cite{novikov1958efficiency, curzon1975efficiency}
\begin{equation}
\eta_{\rm NCA}=1-\sqrt{\frac{T_{\rm c}}{T_{\rm h}}},
\end{equation}
which has been confirmed experimentally with different protocols \cite{calvo2014heat}.

Another closed thermodynamic cycle, the \emph{Stirling cycle}, is often used to describe engines that work by compressing and expanding a working fluid at different temperatures.\footnote{The Stirling's cycle has the same efficiency as Carnot's cycle.} This is also composed of four steps, namely:
\begin{enumerate}
\item An isothermal compression during which heat is extracted towards the cold bath $T_{\rm c}$.
\item  An isochoric heat-addition. 
\item An isothermal expansion during which the system absorbs heat from the hot bath $T_{\rm h}$.
\item  An isochoric heat-removal. 
\end{enumerate}
A sketch of Stirling's cycle can be seen in Fig.~\ref{fig:30:carnot}(c).

In the last decade, various research groups have probed the concept of Carnot efficiency in microscopic heat engines, particularly focusing on single-particle or colloidal heat engines. 
The first of these colloidal heat engines implemented a Stirling engine by using an optically trapped particle \cite{blickle2012realization}, as shown in Fig.~\ref{fig:30:carnot}(d). In this design, a colloidal particle plays the role of the working fluid, while the optical trap can be understood as the engine cylinder where the compression and expansion occur. The cycle includes the following phases:
\begin{enumerate}
\item The isothermal compression by increasing the trap stiffness (i.e., increasing the laser output power).
\item The isochoric head-addition by heating up the colloidal suspension from $T_{\rm c}$ to $T_{\rm h}$. 
\item The isothermal expansion by decreasing the trap stiffness (i.e., decreasing the laser power). 
\item The  isochoric heat-removal by cooling down the colloidal suspension from $T_{\rm h}$ to $T_{\rm c}$. 
\end{enumerate}
Being a microscopic engine, the associated extracted work is a stochastic variable that depends strongly on the cycle duration $\tau$. The first experimental realization of this microscopic Stirling cycle \cite{blickle2012realization} operated between two baths at $T_{\rm c}=22^{\circ}{\rm C}$ and $T_{\rm h}=90^{\circ}{\rm C}$ with the maximum mean power $-W/\tau$ obtained for $\tau=7\,{\rm s}$. In these conditions, the average work extracted per cycle was about $ 4.5 \cdot 10^{-22}\,{\rm J}$, 

In 2014, a micrometer-sized equivalent of a piston steam engine was realized in an optical tweezer, where a single colloidal microparticle plays the role of the piston. The particle absorbs a small amount of light enabling to locally raise the temperature when the particle is exposed to light. At the beginning of the cycle, the gradient force of an optical tweezers pulls the particle towards the laser focus, giving rise to a dramatic increase in temperature and, consequently, the explosive vaporization of the surrounding liquid. This causes a bubble, which pushes the particle out of the focal region. As a consequence, the particle cools down and is pulled back again by the optical tweezers \cite{quinto2014microscopic}. To achieve the sudden bubble formation, this experiment exploited a superheated liquid, which becomes unstable and relaxes back towards equilibrium via the nucleation of a vapor cavitation bubble.
The mean extracted work per cycle exceeded that found in Ref.~\cite{blickle2012realization} by a striking six to eight orders of magnitude, because both the cycle frequency and maximum displacements were around two orders of magnitude larger. 

These two approaches to microscopic heat engines rely on laser heating of the solution where the particle is immersed. The heat exchange rates of the heating and cooling processes effectively limit the range of accessible temperatures and the timescales of the experiments. 
This technical difficulty can be overcome by adding a synthetic noise to the Brownian motion \cite{martinez2013effective}. This permits one to heat without having to raise or lower the physical temperature of the colloidal suspension, allowing fast heating and cooling as well as removing temperature limits, at least in principle.
This method is called \textit{white-noise technique}. Mathematically, the idea works as follows. Consider the underdamped one-dimensional Brownian particle:
 \begin{equation}
 m\frac{d^2 x(t)}{dt^2}=-\gamma\frac{dx(t)}{dt}-\kappa(t)x+\xi(t)+\eta(t)\,,
 \end{equation}
 where $\xi(t)$ corresponds to the usual thermal noise with
 \begin{equation}
 \langle \xi(t)\rangle=0\,,\quad\quad  \langle \xi(t)\xi(t')\rangle=2 k_{\rm B} T\delta(t-t')\,,
 \end{equation}
 while $\eta(t)$ is, for example, an electrostatic noise such that
 \begin{equation}
 \langle \eta(t)\rangle=0\,,\quad\quad  \langle \eta(t)\eta(t')\rangle=\sigma^2_{\rm noise}\delta(t-t')\,.
 \end{equation}
Consider, in addittion, that these two noise sources are independent. This model is statistically equivalent to an effective model for stochastic thermodynamics,
 \begin{equation}
 m\frac{d^2 x(t)}{dt^2}=-\gamma\frac{dx(t)}{dt}-\kappa(t)x+\xi_{\text{eff}}(t)\,,
 \end{equation}
where $\xi_{\text{eff}}(t)$ can be understood as thermal noise with effective temperature\footnote{See also section~\ref{sec:5.7:feedback}, where we discuss feedback cooling of trapped particles in vacuum, which can also be discussed in terms of an effective temperature.}
\begin{equation}
T_{\text{eff}}=T+\frac{\sigma^2_{\rm noise}}{2 k_{\rm B}\gamma}\,.
\end{equation}
The white-noise techniques allow to reach effective temperatures as high as $T=3300\,{\rm K}$. 
The synthetic noise can be generated by placing the colloidal suspension in an electrophoretic fluid chamber with two electrodes connected to a computer-controlled electric generator that produces a sequence of noisy random electric fields, whose statistical properties closely resemble that of white Gaussian noise \cite{martinez2013effective}. 
A similar effect of adding a random force applied to the particle can be achieved by modulating the position of the trap  by using an acousto-optic deflector \cite{gomez2010steady, berut2014energy}.

A remarkable step forward was achieved in a series of works where the concept of microadiabaticity was introduced \cite{martinez2015adiabatic},  and later used for the creation of the first Brownian Carnot engine by concatenating isothermal and microadiabatic protocols \cite{martinez2016brownian}. 
The basic idea is as follows. Recall that, macroscopically,  an adiabatic  process occurs without  exchange of heat (or mass) between the system and its surroundings (i.e., $Q(t)=0$). 
For microscopic systems, the amount of heat is a random variable, so the simplest generalization is to consider  those processes for which the average amount of exchanged heat is zero (i.e., $\langle  Q(t)\rangle=0$). Moreover, if we further assume that the process is performed quasi-statically, we also have that the average value of the variation of the total entropy is zero, $\langle  \Delta S_{\text{tot}}(t)\rangle=0$, and, consequently, the process is isentropic, $\langle  \Delta S_{\text{sys}}(t)\rangle=0$. In this scenario, a trick to implement a microadiabatic process consists in using protocols which keep the system's entropy constant. Importantly, this process can be  quantified fairly precisely. Indeed, since the system's entropy is given by
\begin{equation}
S_{\text{sys}}(t)=-k_{\rm B}\log P_t(x(t),v(t))
\end{equation}
and assuming that the system is close to equilibrium, i.e.,
\begin{equation}
P_t(x(t),v(t))=\frac{1}{Z}e^{-\frac{H(t)}{k_{\rm B} T(t)}},
\end{equation}
where $H(t)$ is the Hermitian of the system,
we obtain that
\begin{equation}
\langle  S_{\text{sys}}(t)\rangle \propto \langle x^2(t)\rangle \langle v^2(t)\rangle\propto \frac{T^{2}(t)}{\kappa(t)}.
\end{equation}
Thus,  a pseudo-adiabatic process corresponds to keeping $T^2(t)/\kappa(t)$  constant. Notice that the microscopic Carnot engine was also explored in the work in Ref.~\cite{schmiedl2007efficiency}, but only considering the configurational space and not the phase space. 

\paragraph{Maxwell's demons and Szilard engines.}

The thought experiment known as \emph{Maxwell's demon} \cite{maxwell1871theory} addressed the issue that the Second Law of thermodynamics might be violated by the inclusion of ``intelligent beings'' capable of processing information. In its original version, an ideal gas at temperature $T$ is enclosed in an isolated container divided into two equal parts by a fixed wall. In this wall, there is a trap door operated by a ``demon'' who will open it if an incoming particle from the left (right) is faster (slower) than the average. After a while, the right compartment will have the fastest particles and, consequently, a higher temperature. Therefore, the effect of the demon is to generate a temperature gradient, which can be used to obtain work from the system, without the corresponding energy expenditure, thus violating the Second Law.

To better understand the origin of such violation, Leo Szilard proposed his  eponymous simplified model. The \emph{Szilard engine} consists of a single particle within a container divided into two equal partitions separated by a movable, frictionless wall \cite{szilard1929uber}.  When the container is put into contact with a single thermal reservoir at temperature $T$, the movable wall may then be used to extract work (e.g., by lifting a weight), as the particle transfers kinetic energy in successive elastic collisions.  The isothermal volume expansion results in a work extraction $W_{\rm ext}= k_{\rm B} T \ln(V_f/V_0)$, where $V_0$ and $V_f$ are the initial and final volumes of the partition that contains the particle. Since the final volume $V_f$ is that of the total box, then $V_0/V_f=1/2$ is the fraction of the volume that contains the particle.  This implies, in turn, that the expected work extracted from the engine is simply $k_{\rm B} T \ln 2$, which is  proportional to the \emph{Shannon entropy} measuring the uncertainty on the initial position of the particle in a symmetric partition.

Leon Brillouin suggested that the work extraction of the Szilard engine is precisely compensated by the work of measurement and, therefore, no net extraction of work is possible \cite{brillouin1952maxwell}.  After a decade, Rolf Landauer introduced a concept of logical irreversibility \cite{landauer1961irreversibility} and Charles H. Bennett explained that the work extraction is precisely compensated by the erasing process of the memory apparatus \cite{bennett1982thermodynamics}. Recently, a modified version of the Second Law was derived \cite{sagawa2008second, sagawa2012thermodynamics}, the \emph{Jarzynski-Sagawa-Ueda relation}, which has also been confirmed experimentally \cite{toyabe2010experimental}. This latter result claims that the work extraction is compensated by the mutual information between the system and the measurement apparatus which is involved in both the measurement and erasing processes \cite{sagawa2009minimal}.
In a relative recent experiment \cite{roldan2014universal}, the authors used a colloidal suspension to implement a Szilard engine to explore how the entropy decreases due to the spontaneous symmetry breaking that is the cornerstone of the Szilard engine and Landauer's principle, and to measure directly this entropic change for a Brownian particle  in a bistable potential implemented by two optical traps.
Very recently, a continuous versions of the Maxwell's demon, which is  capable of extracting arbitrarily large amounts of work per cycle by repeated measurements of the state of a system, was proposed and was  experimentally implemented in a single DNA hairpin pulling experiment \cite{ribezzi2019large}.

\subsubsection{Other applications to statistical physics and outlook}

In this section, we have presented some well-known  applications of optical tweezers to investigate both equilibrium and non-equilibrium properties of model systems in statistical physics.
Here, we briefly mention some emerging areas where optical tweezers also represent a powerful tool to study statistical quantities of microscopic systems.
\begin{description}

\item[Synchronization.] Coordinated motion of self-sustained oscillators mediated by means of hydrodynamic interactions is a fundamental behavior in many biological systems, such as cilia and flagella. Using optically trapped microparticles, it has been possible to elucidate and directly visualize different aspects of the hydrodynamic synchronization of oscillators and rotors. There are different approaches to define these oscillators, for example with light-propelled chiral particles \cite{dileonardo2012hydrodynamic}, using angular momentum in optical tweezers to propel birefringent or asymmetric particles \cite{arzola2014rotation}, by means of rotating energy landscapes \cite{lhermerout2012collective, koumakis2013stochastic, kotar2013optimal} or, as we already mentioned in section~\ref{sec:4.5.1:kramers}, by means of bistable potentials sculpted with light \cite{curran2012partial}. In all these cases, when two of these self-sustained oscillators are close enough, the hydrodynamic interactions perturb and tend to synchronize their phases and their frequencies. Since all of these experiments have been realized at microscopic scale, thermal fluctuations are not negligible and escape transitions play an important role in their dynamics.  

\item[Active matter.] Active particle systems, such as motile microorganisms, are non-equilibrium systems capable of converting autonomously the energy from their surroundings into directed motion. In recent years, also the possibility of developing artificial active matter systems has been extensively explored \cite{bechinger2016active, cichos2020machine}. Optical tweezers can be used to probe such phenomena and their connection to non-equilibrium relations \cite{argun2016non}.
	
\item[Phase transitions.] Optical tweezers can be used in a variety of situations to induce phase transitions and to probe critical phenomena \cite{paladugu2016nonadditivity}. One can, for example, use optical lattices to control colloidal phase transitions via tunable interactions \cite{mangold2003phase, yethiraj2007tunable, zaidouny2013light}.

\end{description}
Further advances in optical trapping could also enable the possibility to address more fundamental questions on statistical aspects of small systems, such as memory effects in  
non-equilibrium systems. Moreover, most of the studies discussed in this section deal with the behavior of individual particles, but it is natural to wonder how the properties of these systems change in many-body systems. Equally interesting would be to explore how microengines or active baths can be used to perform practical tasks, such as energy harvesting at the microscale.

\subsection{Transport in extended potential landscapes}
\label{sec:4.6:transport}

Microscopic particles in a fluid are constantly diffusing: the smaller the particle, the faster it diffuses, following a power law dependence $D\propto 1/a$, where $a$ is the particle radius. As $D$ is a tiny quantity in typical conditions (of the order of ${\rm \upmu m^2\,s^{-1}}$), diffusive transport is very slow. For example, on average a $1\,{\rm \upmu m}$ particle in water takes approximately 26 days to move by about $1\,{\rm mm}$ away from its original position.

Transport can be greatly enhanced by external forces. These forces generate directed motion, where diffusion generally plays a pernicious role, since the particle will be gradually dispersed away from the path defined otherwise by the force. In Nature, several microorganisms have developed different strategies for exploring their environment in a more efficient and precise way than that of simple diffusion, for example using flagella or cilia.

Microscopic objects are often subject to the influence of structured force landscapes, which may either boost or inhibit their motion. For example, some molecular motors inside eukaryotic cells move along the paths defined by microtubules that exhibit a periodic structure of two proteins, acting as a periodic potential. 
A paradigmatic model that serves as a starting point for studying microscopic dynamics is the motion of a Brownian particle in a corrugated potential landscape under the action of an external force. 
Being able to study and understand these transport mechanisms is not only important from the fundamental science perspective, but it also opens up the possibility to reproduce them and develop new techniques to perform tasks, such as particle sorting and delivery in lab-on-a-chip platforms. 

As we will see in detail in this section, colloids are ideal prototypes for the study of microscopic transport phenomena, and extended light patterns provide a powerful and versatile way to create reconfigurable potential energy landscapes in optical micromanipulation systems. We will be primarily concerned with conservative optical forces in this section, but there are also interesting light-induced transport phenomena that emerge in the presence of non-conservative forces (e.g., scattering forces), as we will briefly comment on.   

\subsubsection{Optical forces and potentials in extended light patterns}

Table~\ref{tab:11:patterns} presents some examples of extended light patterns used in optical manipulation: the images on the left illustrate the optical field of interest generated at the sample (i.e., at the focal plane of the microscope objective or the last lens of the optical trapping setup); the images on the right provide information about how these patterns can be experimentally generated.
As explained in section~\ref{sec:2.3:alternatives}, there are different ways for creating extended light patterns, but here we will focus our attention on the two most common approaches using interference or diffractive optical elements. 

\begin{table}[ht!]
	\centering
	\begin{small}
	\begin{tabular}{c b{7.8cm}}
	\hline
		\includegraphics[height=21mm]{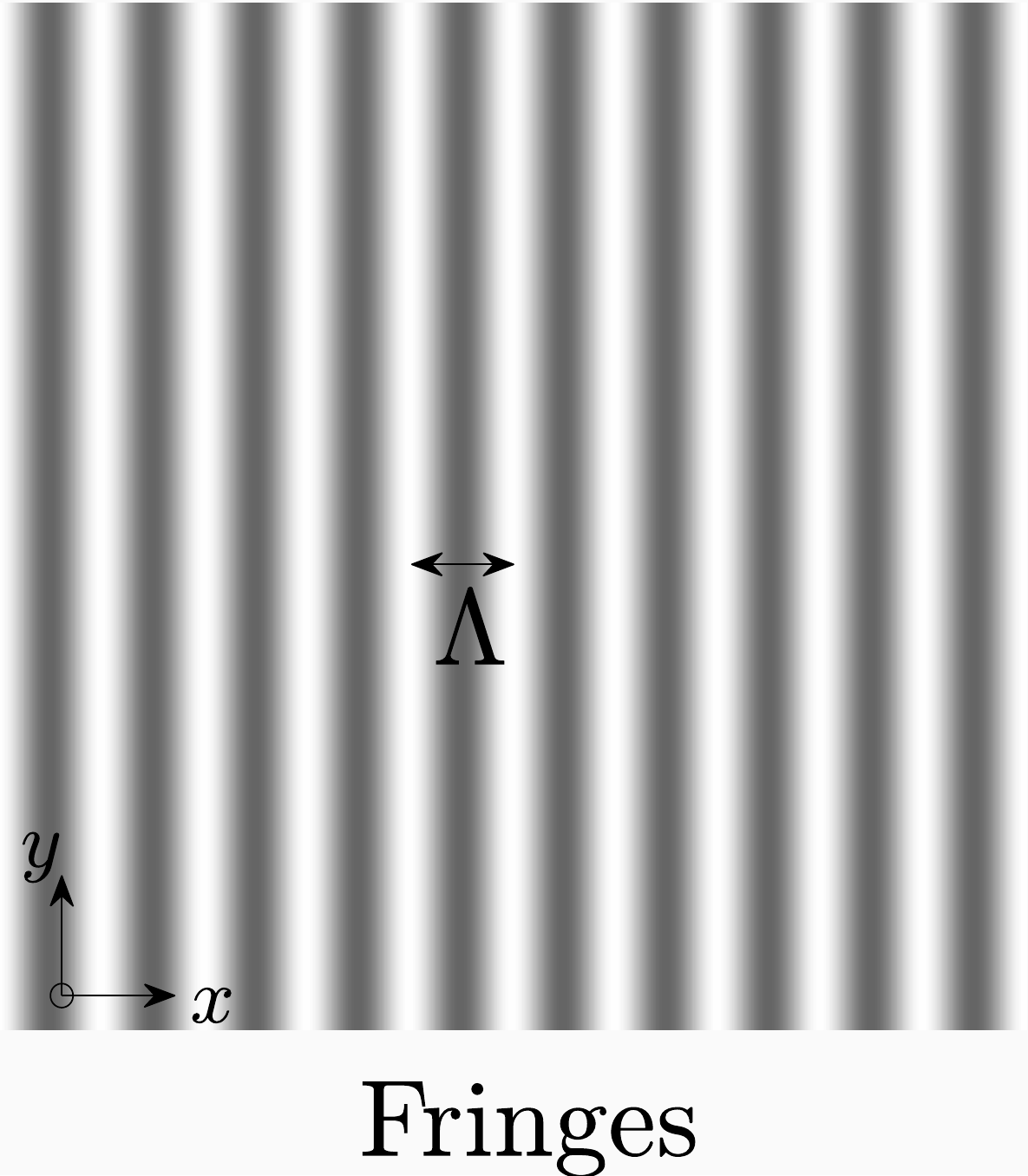}
		\includegraphics[ height=21mm]{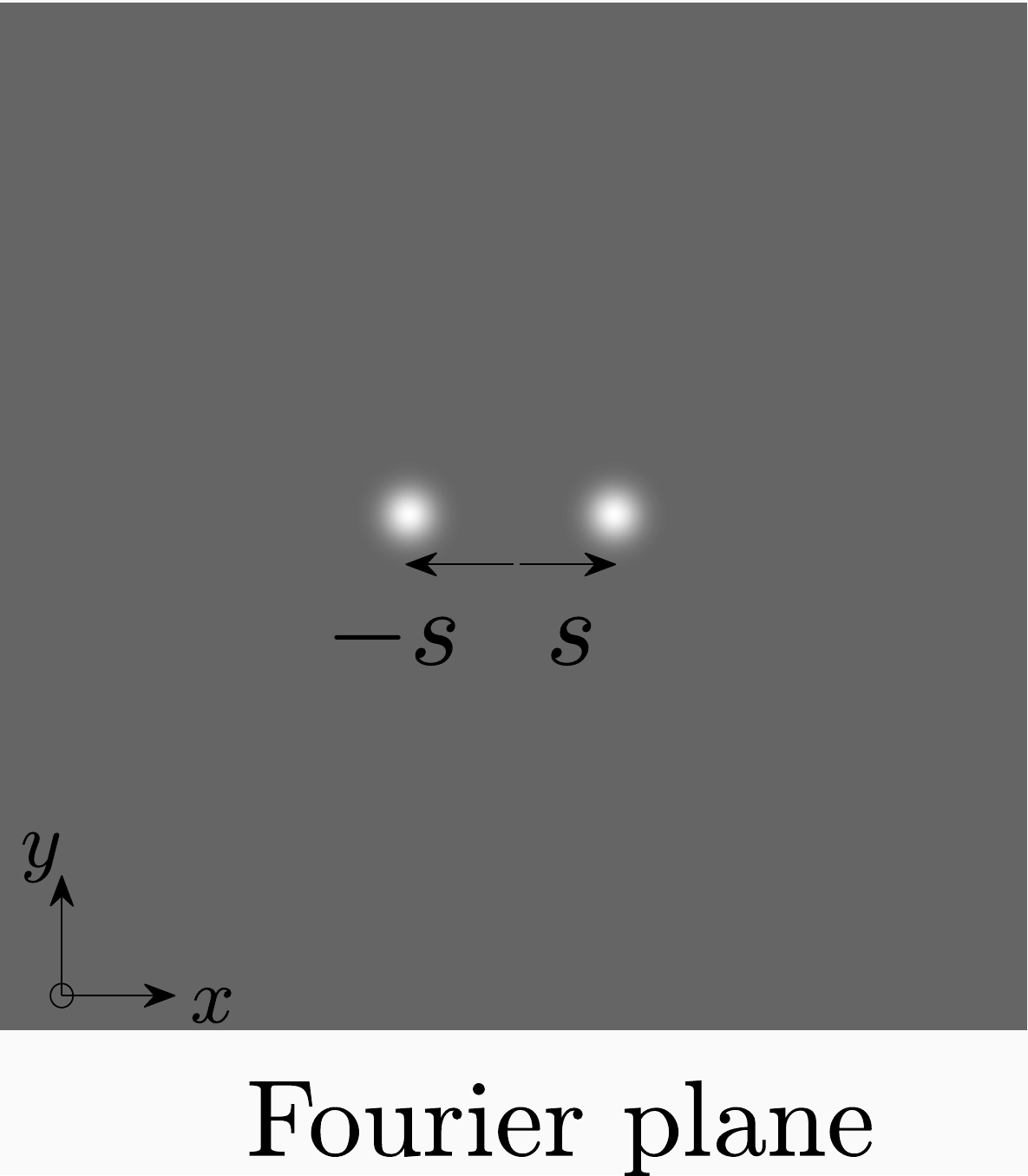} 
		& 
		\textbf{Sinusoidal pattern (periodic in 1D) produced by the interference of two beams (left) and its Fourier spectrum (right).} 
Any interferometer is suitable to produce this pattern, such as Fresnel biprisms \cite{rubinov2003physical} or a Mach-Zehnder interferometer \cite{macdonald2001trapping, ricardez2006modulated}. A diffraction grating generated with a SLM can also be used to split the beam \cite{jakl2014optical}.  
\vspace{0.1pt}
		\\
	\hline
		\includegraphics[height=21mm]{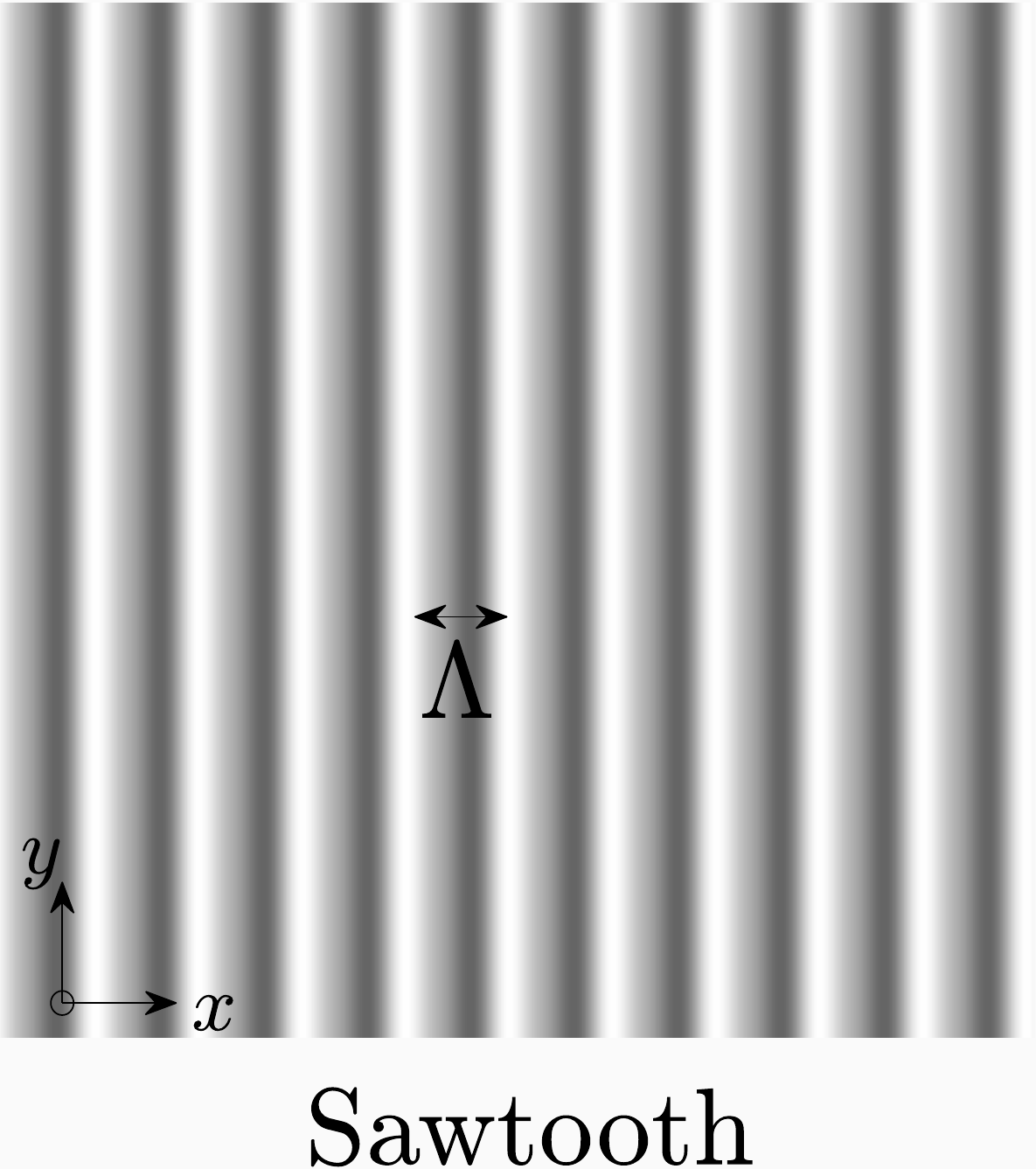} 
		\includegraphics[height=21mm]{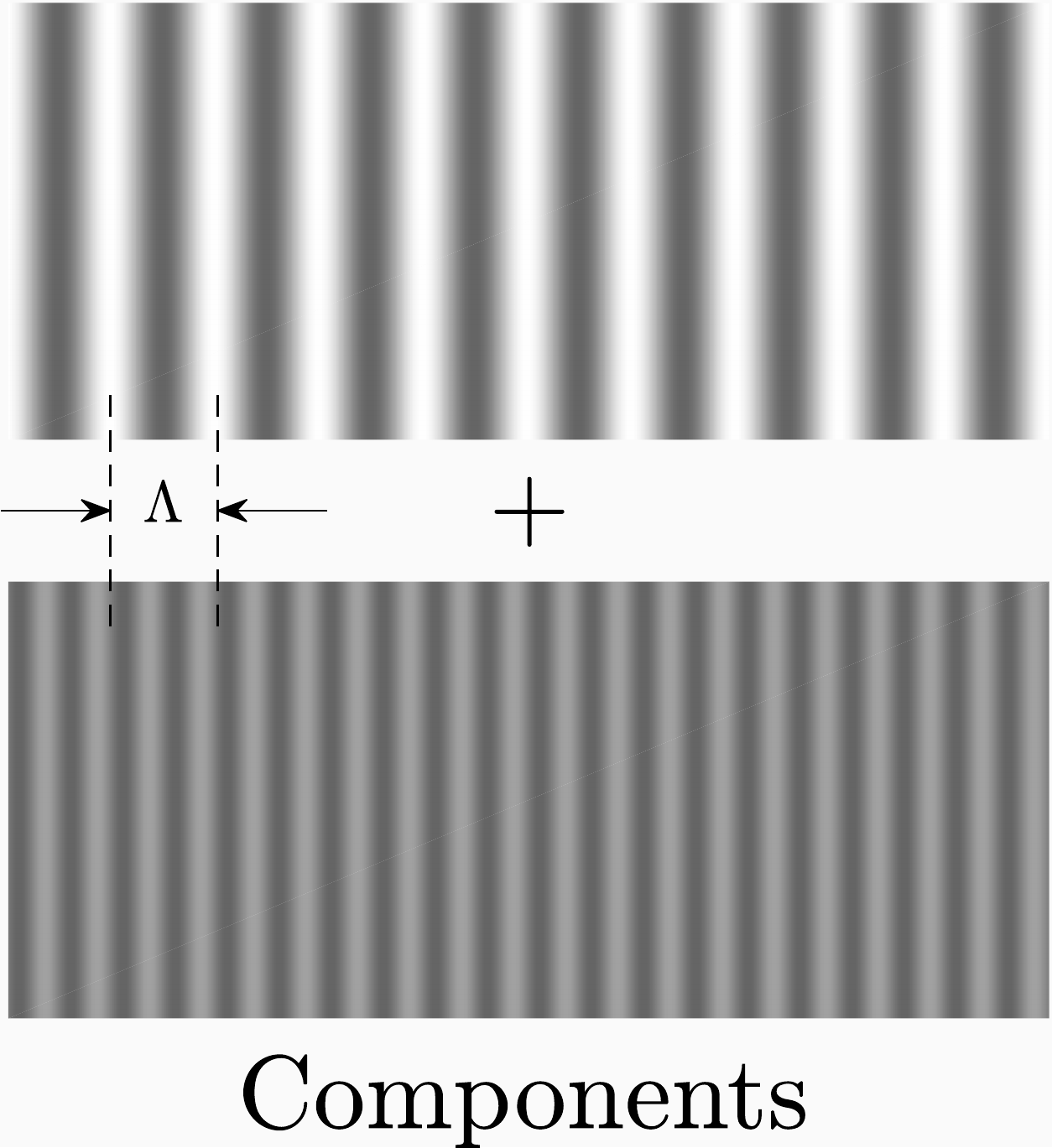} 
		&
		\textbf{Sawtooth pattern (periodic in 1D) (left) generated by the superposition of two orthogonally polarized sinusoidal patterns (right).}
A three-armed Mach-Zehnder interferometer can produce two patterns of fringes, one with twice the spatial period of the other, by controlling the polarization of the input beams. The phase difference and intensity ratio determines the asymmetry \cite{arzola2011experimental, arzola2013dynamical}.
		\\
	\hline
		\includegraphics[height=21mm]{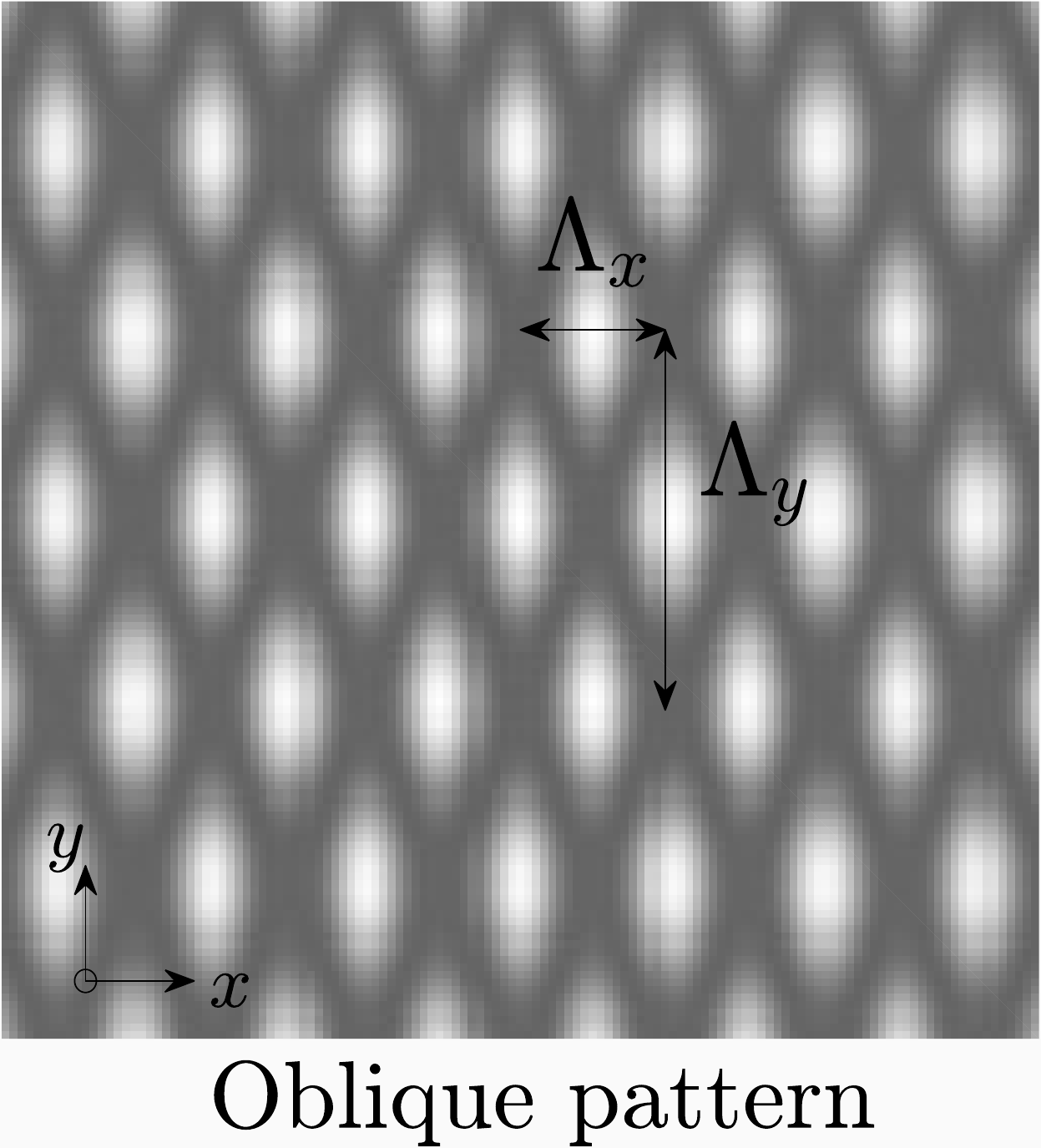}
		\includegraphics[ height=21mm]{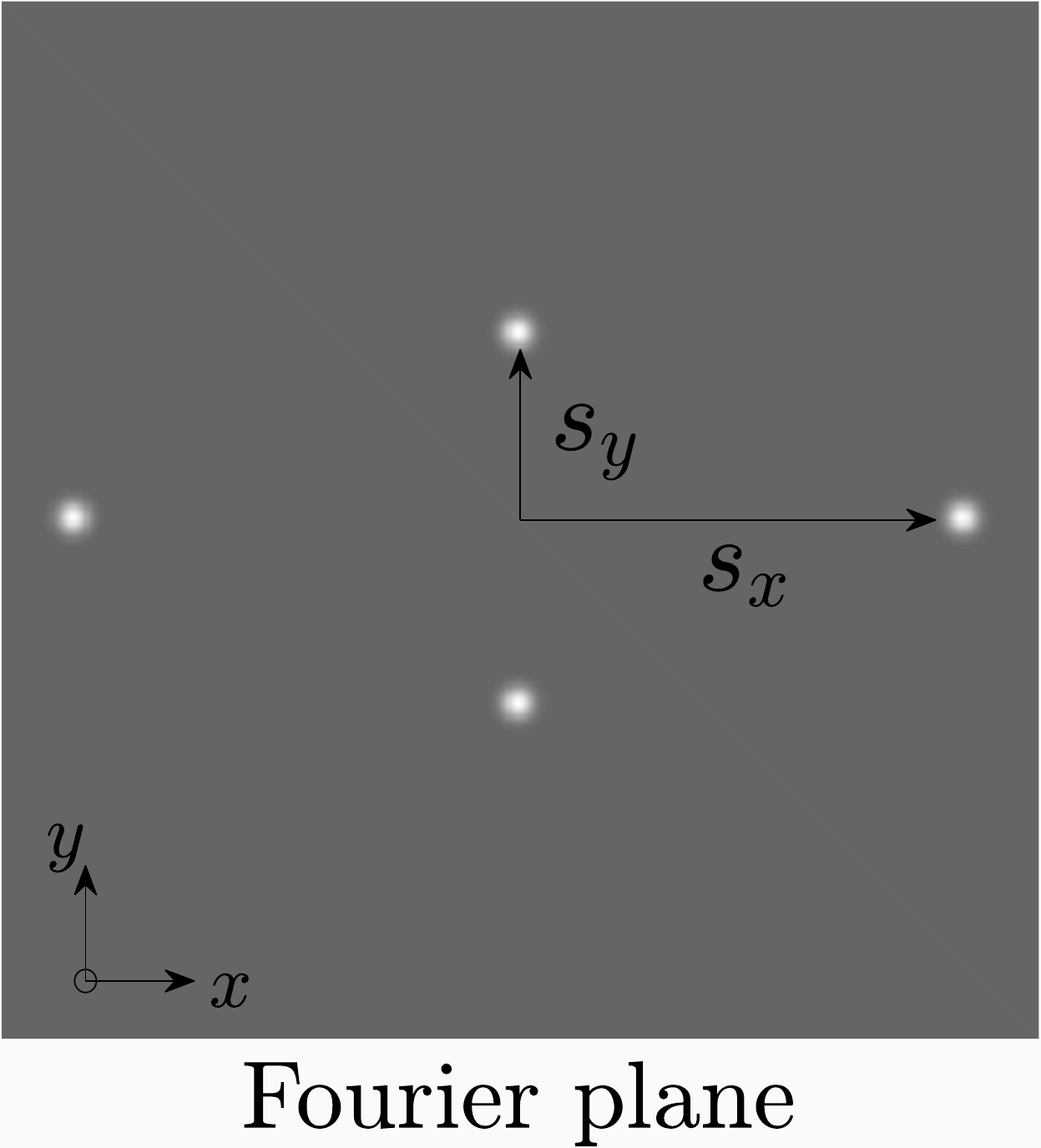} 
		& 
		\textbf{Oblique lattice (periodic in 2D): Interferometric pattern of four beams (left) and its Fourier spectrum (right).}
These patterns can be generated using a tandem of two interferometers or using diffraction gratings designed to split the input beam into multiple beams travelling along the desired directions \cite{macdonald2001trapping}.
		\\
	\hline
		\includegraphics[height=21mm]{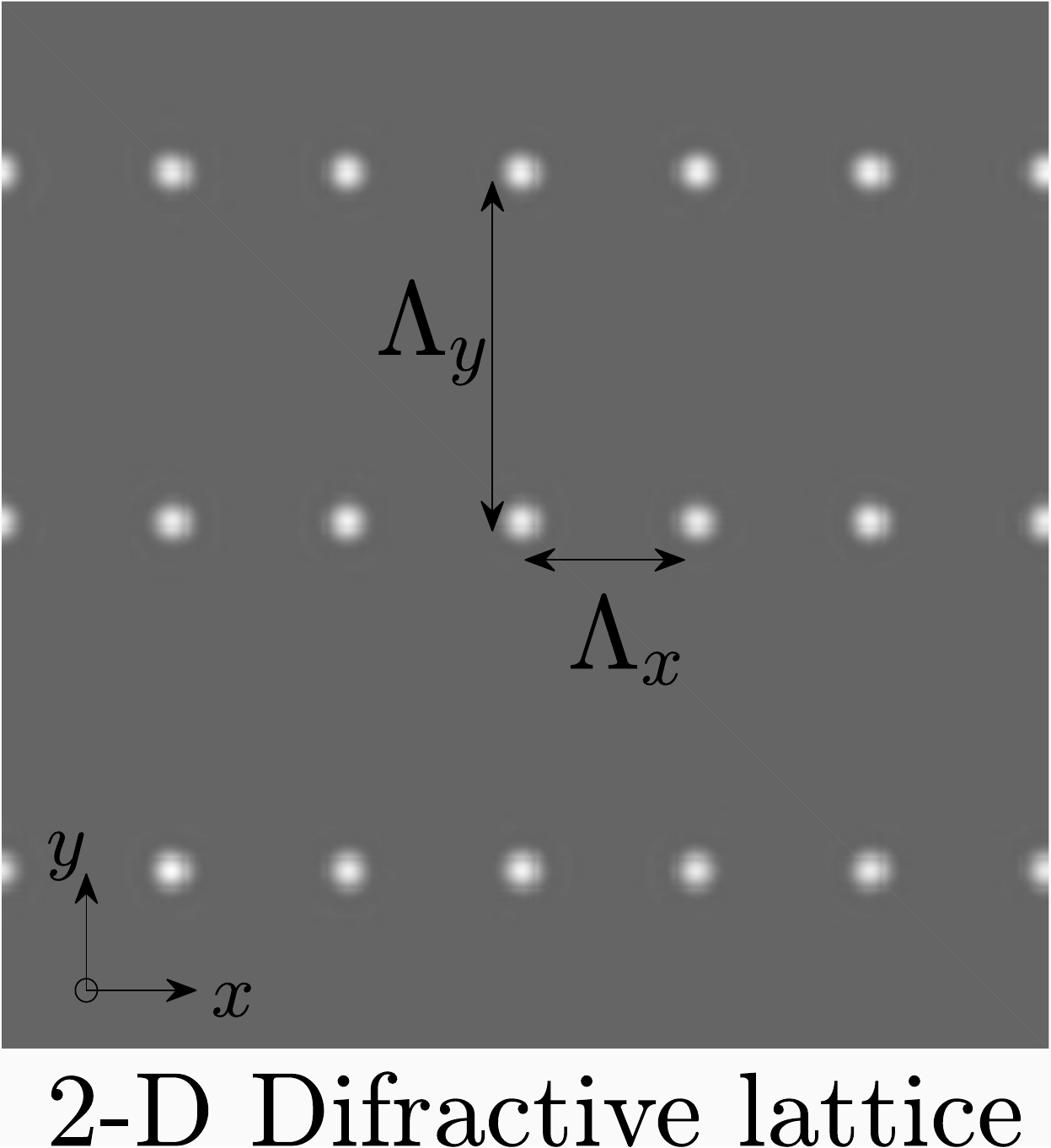} 
		\includegraphics[ height=21mm]{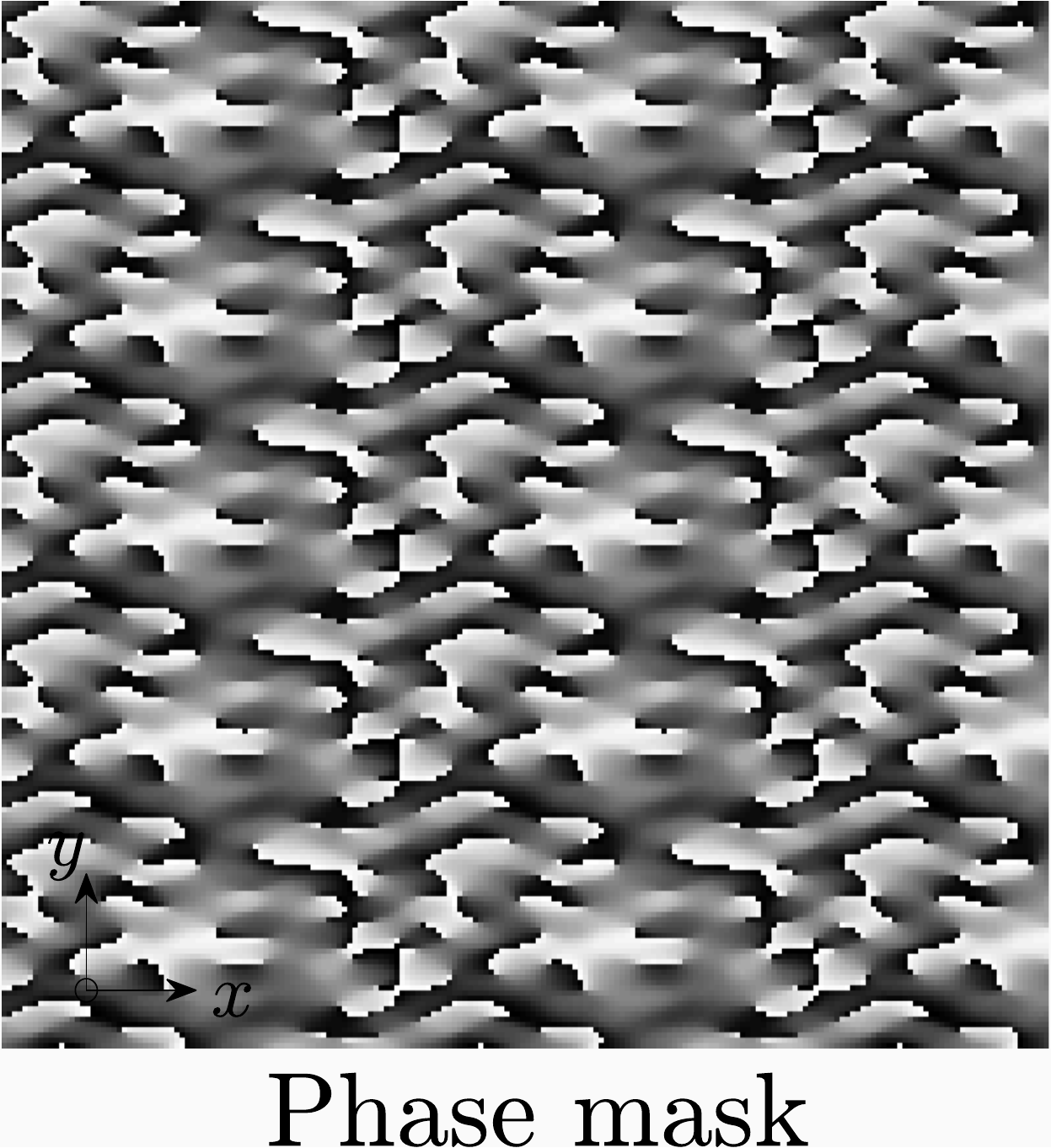}  
		& 
		\textbf{Rectangular lattice (periodic in 2D): Diffractive pattern (left) generated with a phase hologram (right).}
Holographic phase masks can be designed to produce the target distributions of individual light spots using iterative optimization algorithms \cite{jones2015optical}. These can be displayed on a SLM and dynamically reconfigured.
		\\
	\hline
		\includegraphics[height=21mm]{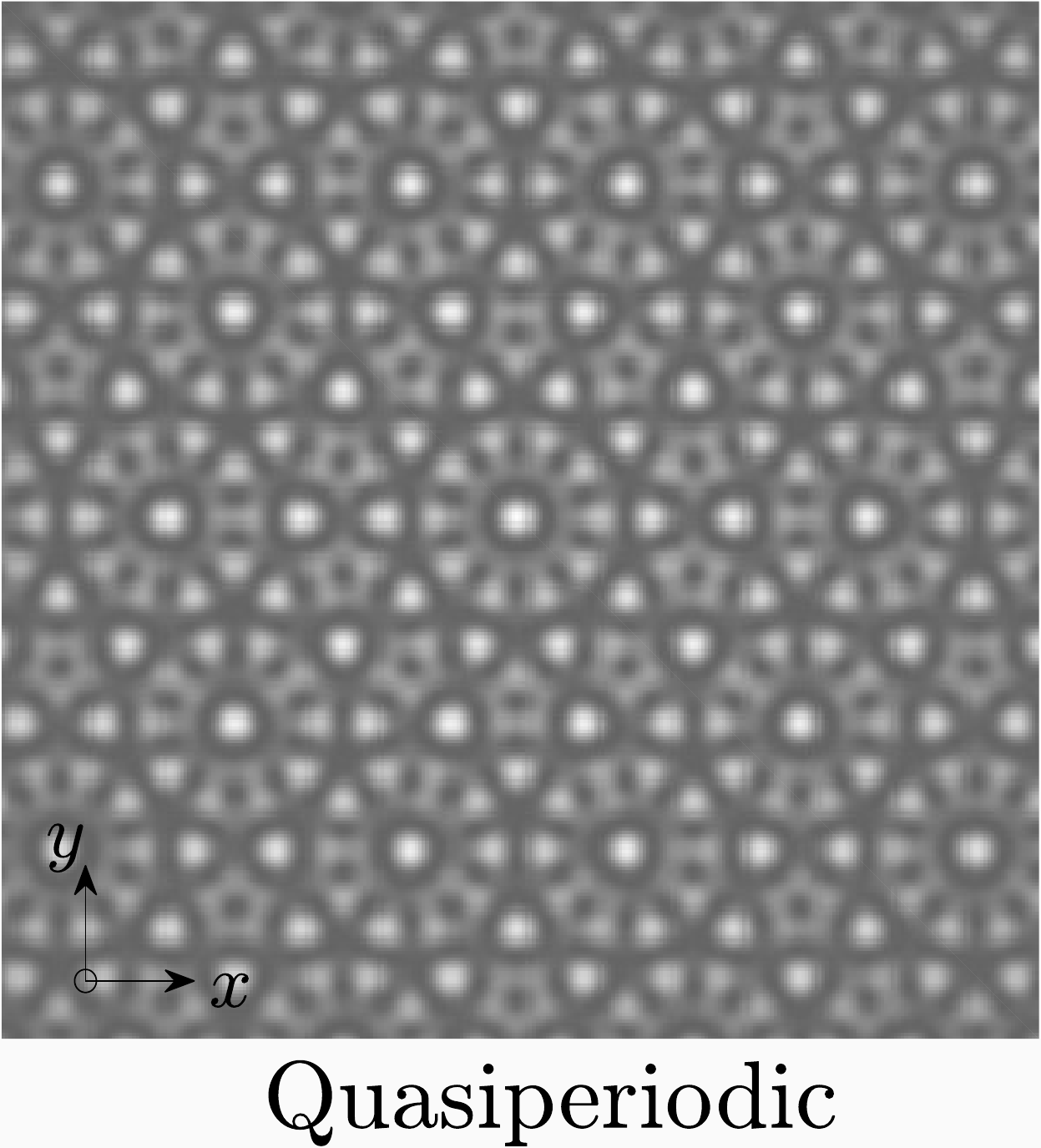}   
		\includegraphics[ height=21mm]{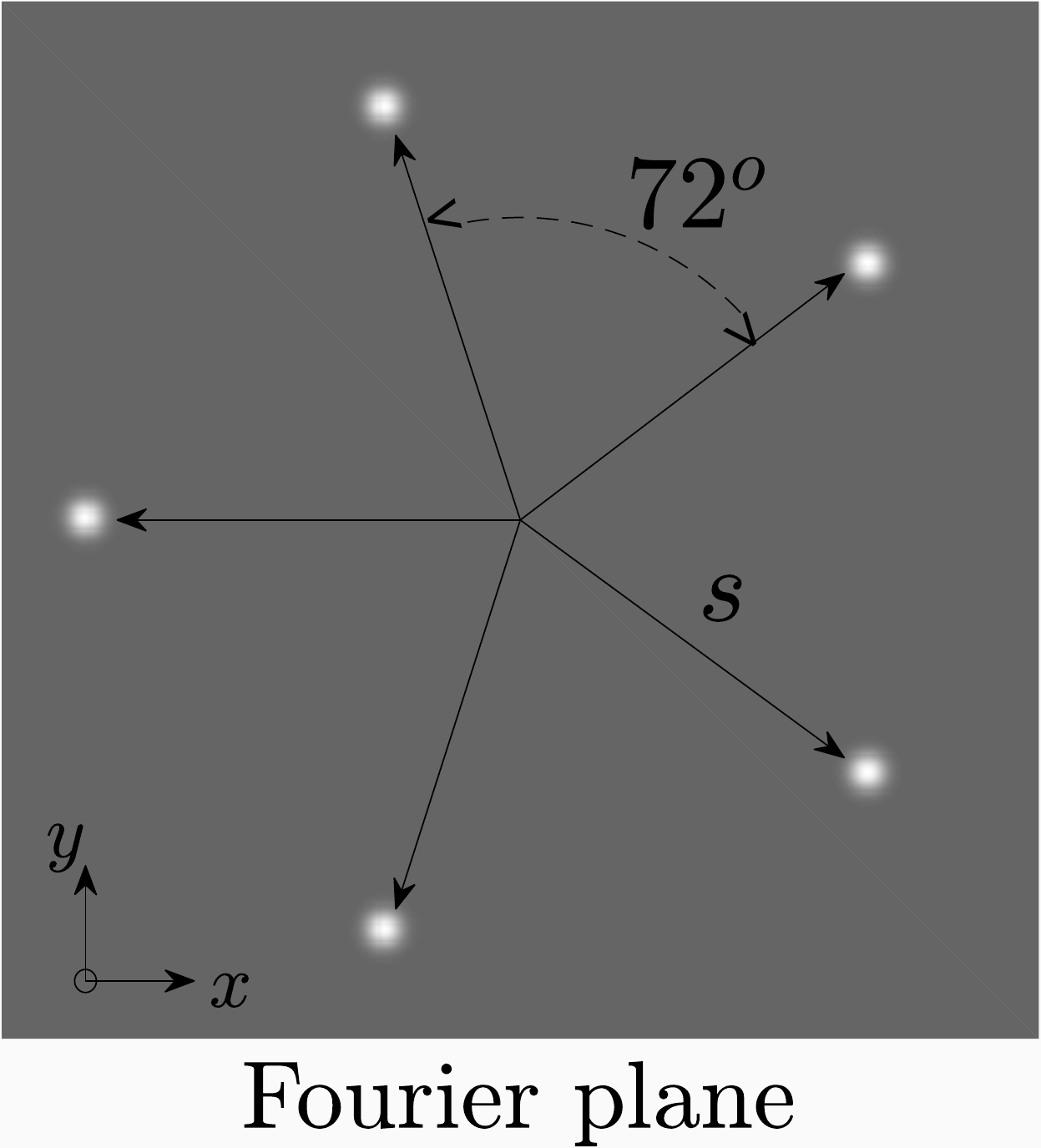} 
		&        
		\textbf{Quasi-periodic lattice: Interferometric pattern of five equidistant beams (left) and its Fourier spectrum (right). }
A pentagon-shaped diffraction grating can be displayed on a phase-only SLM to split the input beam into five beams travelling at the desired angle. 
\vspace{0.4pt}
		\\
	\hline
		\includegraphics[height=21mm]{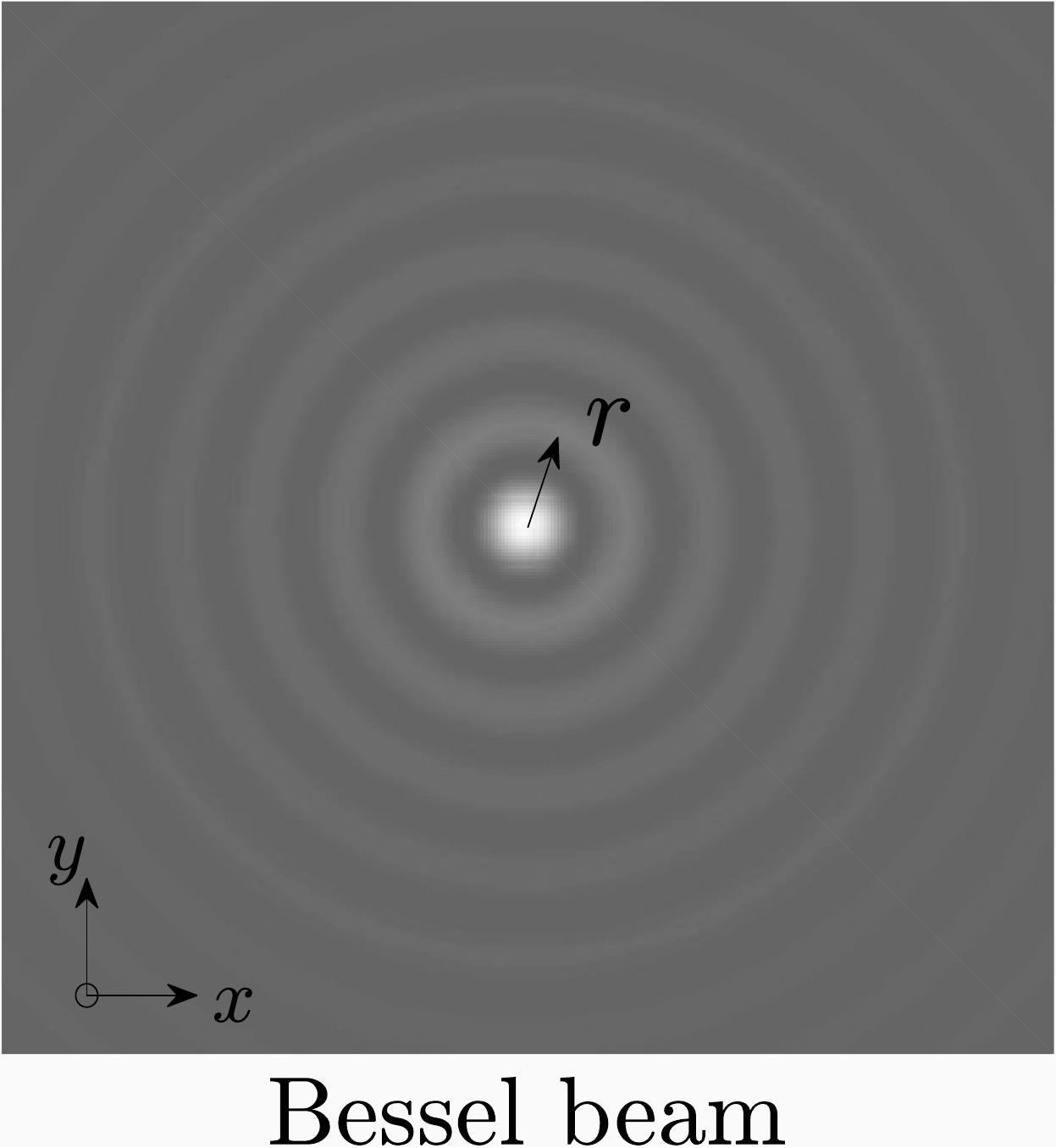}   
		\includegraphics[ height=21mm]{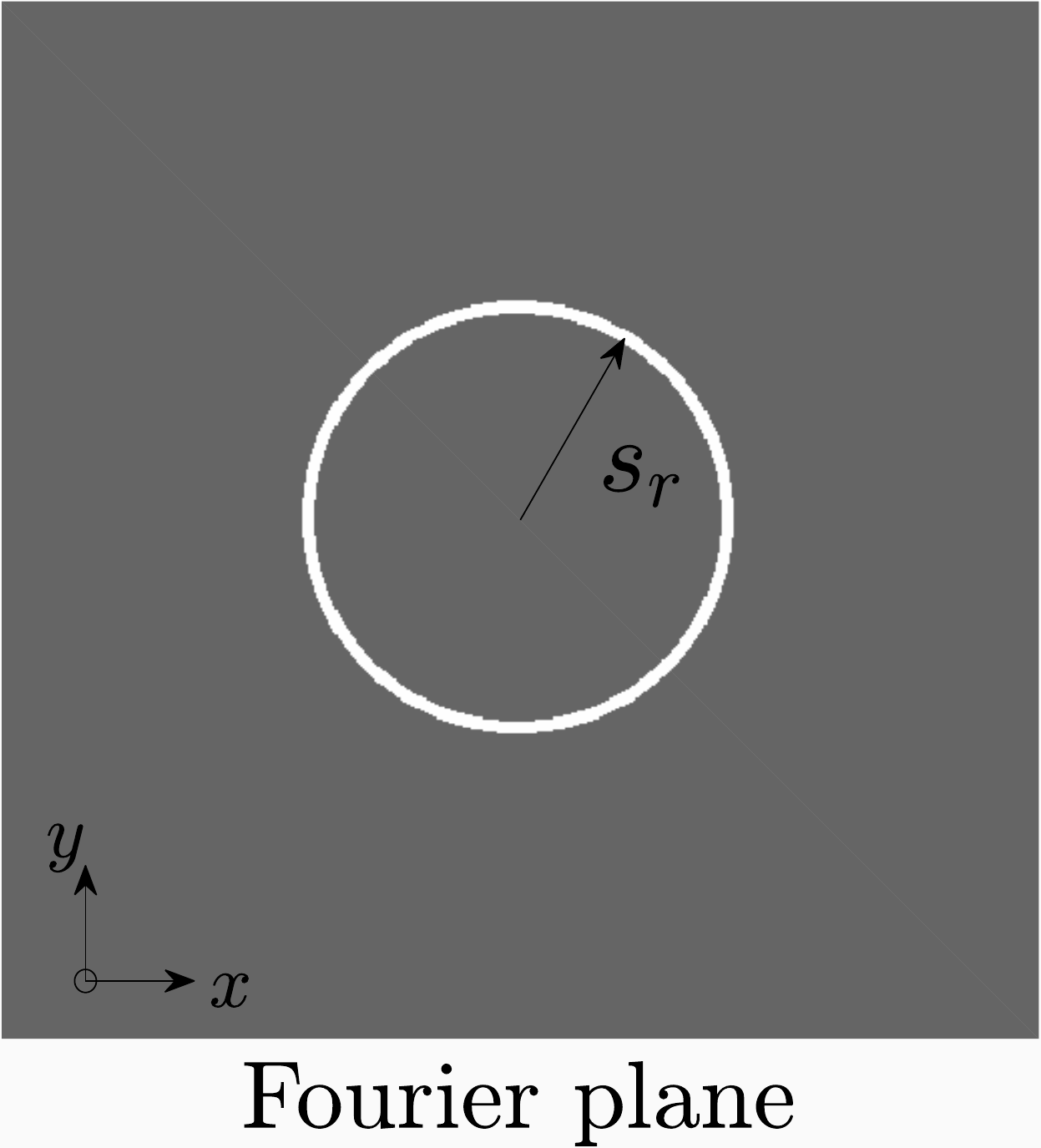} 
		& 
		\textbf{Zeroth-order Bessel beam (left) and its annular Fourier spectrum (right).}
Bessel beams can be obtained efficiently with a conical lens or axicon. Any nondiffracting beam can also be generated by means of a properly encoded phase mask displayed on a SLM \cite{arrizon2009efficient, hernandez2010experimental}.
		\\
	\hline
		\includegraphics[ height=21mm]{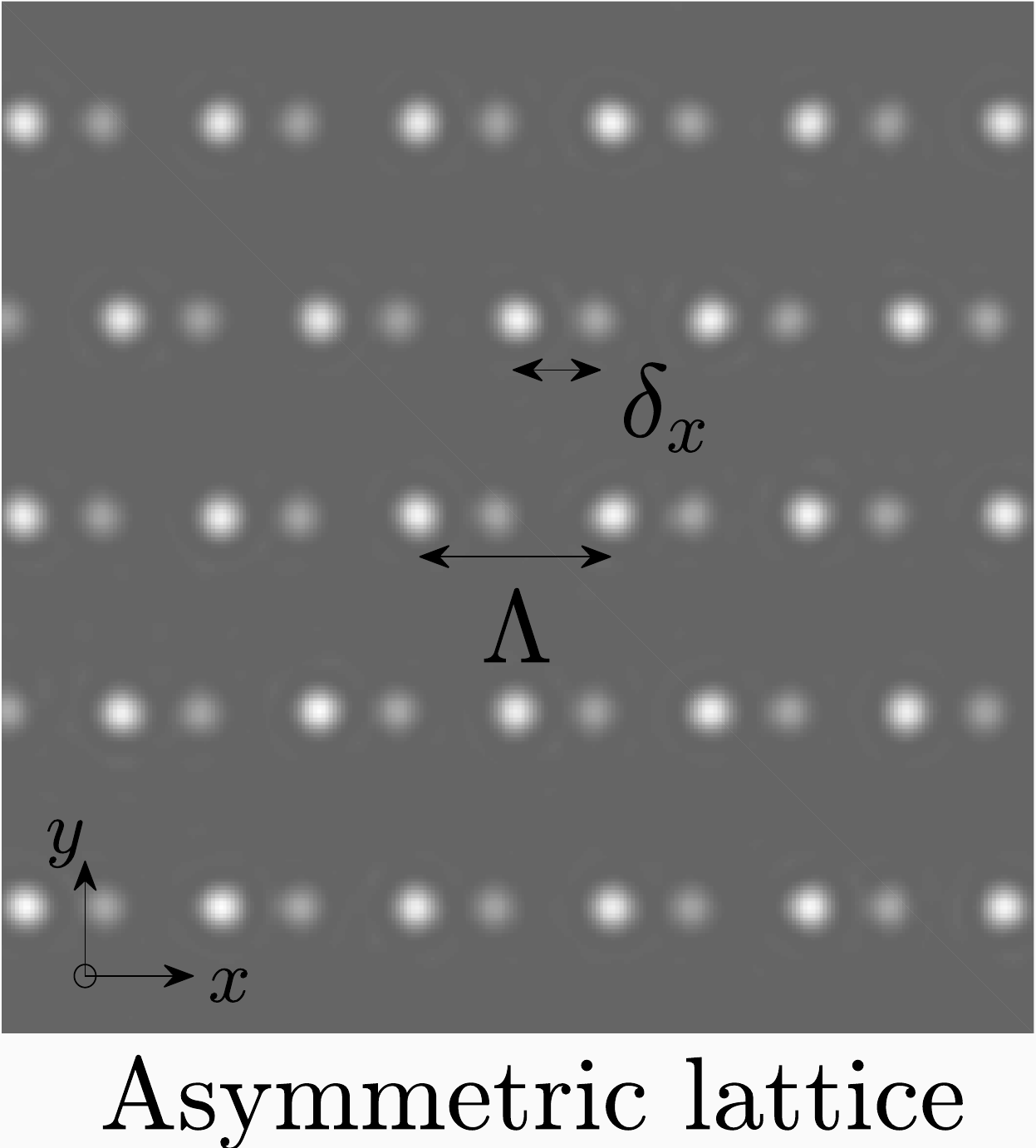} 
		\includegraphics[ height=21mm]{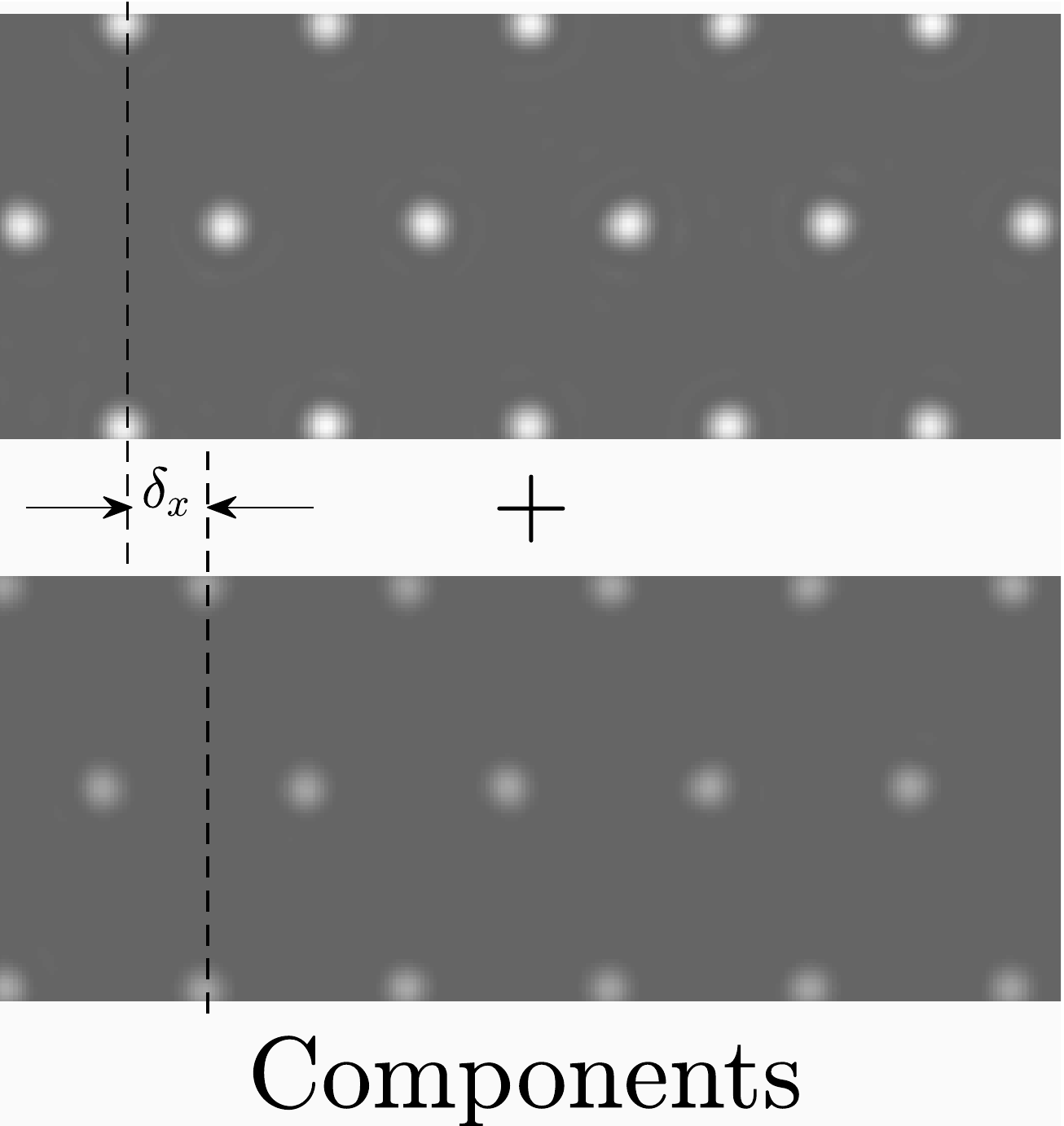} 
		& 
		\textbf{Lattice of asymmetric traps in 2D (left) generated by the superposition of orthogonally polarized identical patterns.}
A light lattice generated with a phase hologram can be introduced into a polarization splitting Mach-Zehnder device to produce two identical patterns, whose relative positions and intensity ratio  determine the asymmetry of the resulting pattern \cite{arzola2017omnidirectional}.
		\\
	\hline
		\includegraphics[ height=21mm]{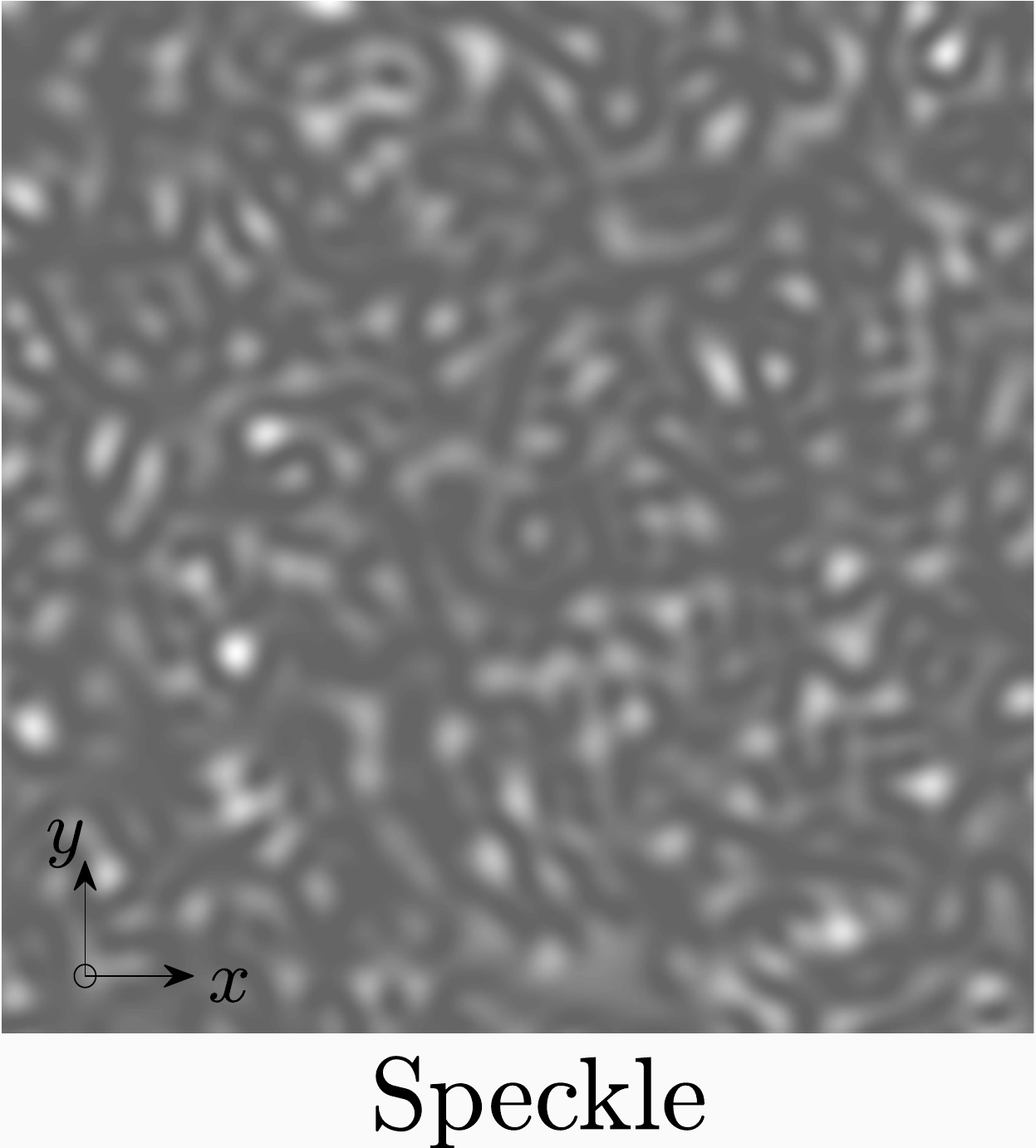} 
		\includegraphics[ height=21mm]{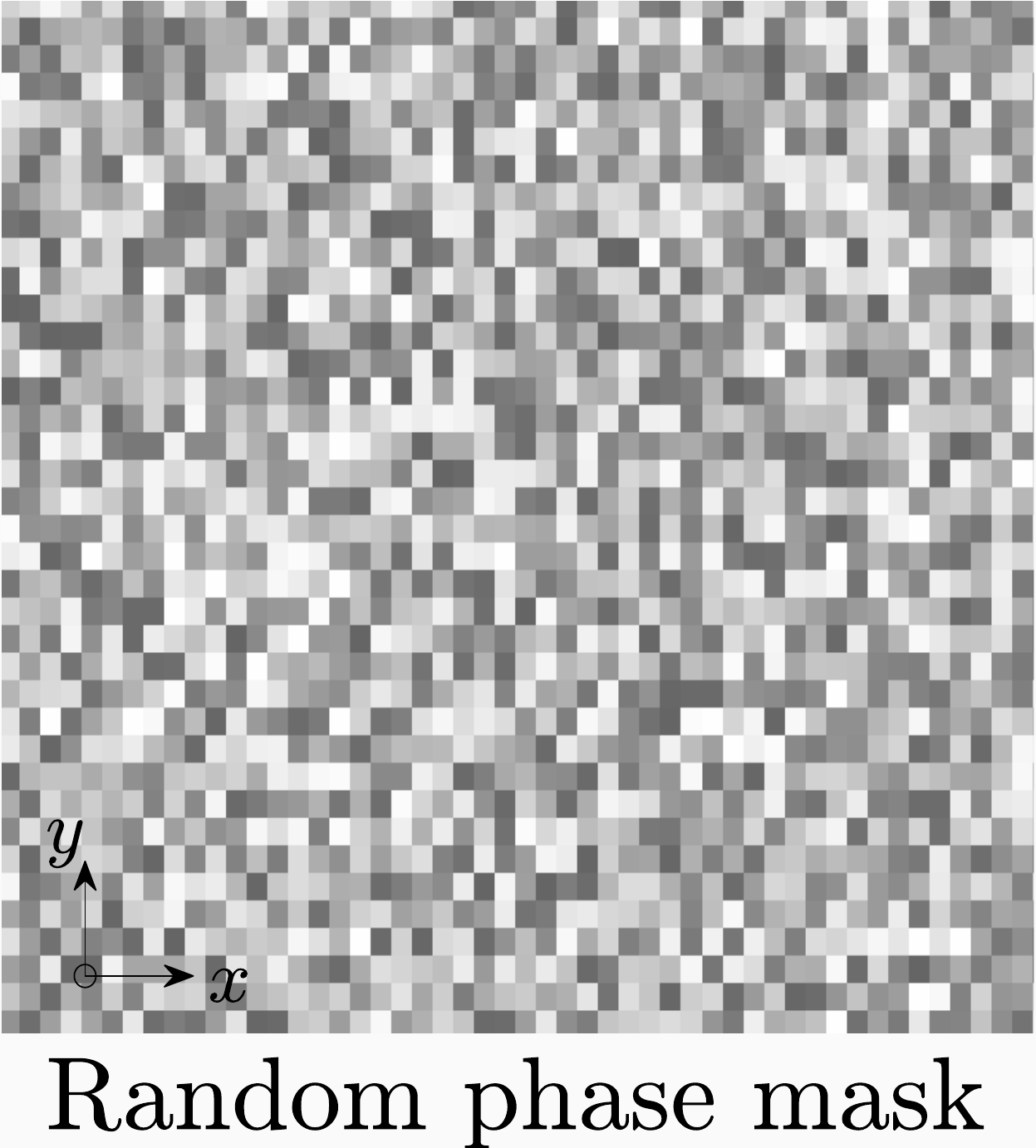} 
		&
		\textbf{Speckle pattern (left) generated with a random phase mask (right).}
Speckle can be generated by passing a laser beam through a diffuser plate \cite{shvedov2010selective}, by coupling the laser into a multimode optical fiber \cite{volpe2014brownian, bianchi2016active}, or with a random phase mask displayed on a SLM \cite{perez2018high}. 
\vspace{0.1pt}
		\\
		\hline
	\end{tabular}
	\caption{
	{\bf Extended light patterns.}
	}
	\label{tab:11:patterns}
	\end{small}
\end{table}

A subwavelength particle is pulled by the conservative optical gradient forces towards the brightest regions of an extended light pattern. In this case, the particle moves in an extended potential energy landscape mimicking the shape of the intensity pattern, with valleys at maxima and hills at minima of intensity. In this regime, the magnitude of the forces, or the depth of the valleys, increases in proportion to the cube of the particle radius and linearly with optical power. 
However, for particles whose size is comparable or larger than the wavelength, the shape of the energy landscape might be remarkably different from the intensity distribution. 

Importantly, when working with extended patterns of light, one should keep in mind that the larger the area covered by the pattern, the lower the intensity for a given optical power and, therefore, the magnitude of the optical force decreases. Hence, it is important to estimate the appropriate area of the pattern so that the average intensity is at least of the order of $0.1\,{\rm mW\,\upmu m^{-2}}$.

\paragraph{One-dimensional potentials}

Let us analyze the simplest case: an extended pattern of fringes obtained from the interference of two collimated beams, as that in the first row of Table~\ref{tab:11:patterns}. At the back aperture of the microscope objective, the Fourier transform of the two beams corresponds to a pair of focused spots, whose separation ($2s$) determines, together with the focal length of the microscope objective ($f$), the period of the interference pattern at the sample plane: $\Lambda=\frac{\lambda f}{2s}$, where $\lambda$ is the wavelength of the laser in the background medium. The corresponding intensity distribution is given by
\begin{equation}
I(x,y)=2I_0\left[1+\cos\left(\frac{2\pi x}{\Lambda}\right)\right]=4I_0\cos^2\left(\frac{\pi x}{\Lambda}\right),
\end{equation}
where we have assumed that both interfering beams are plane waves with equal intensity, $I_0$. In reality, there is always an envelope function modulating the intensity, often with a Gaussian shape $I_0(x,y)=A\exp[-2(x^2+y^2)^2/w_0^2]$, but as long as $w_0\gg \Lambda$, the plane-wave assumption works reasonably well in the central region of the envelope.

There are several approaches with different levels of complexity for calculating the optical force exerted by this pattern on a dielectric particle of diameter $D_{\rm p}>\lambda$. 
In all cases, the force is also periodic with the same periodicity as the intensity profile, so it can be written as
\begin{equation}
F(x)=A\sin(2\pi x/\Lambda).    
\end{equation}
Notice that there is a $\pi/2$-phase difference between the expressions for the intensity and that for the force, since both maxima and minima of the intensity function correspond to zeroes of the force, representing either stable or unstable equilibrium positions. The amplitude factor $A$ is proportional to the intensity of light and, more importantly, it depends on the size of the particle and its refractive index, as well as on the period $\Lambda$ of the pattern \cite{cizmar2006optical_sorting, ricardez2006modulated, arzola2009force}, i.e. $A=A(D_{\rm p},n_{\rm p},\Lambda)$.  

\begin{figure}[h]
	\centering
	\includegraphics[width=12cm]{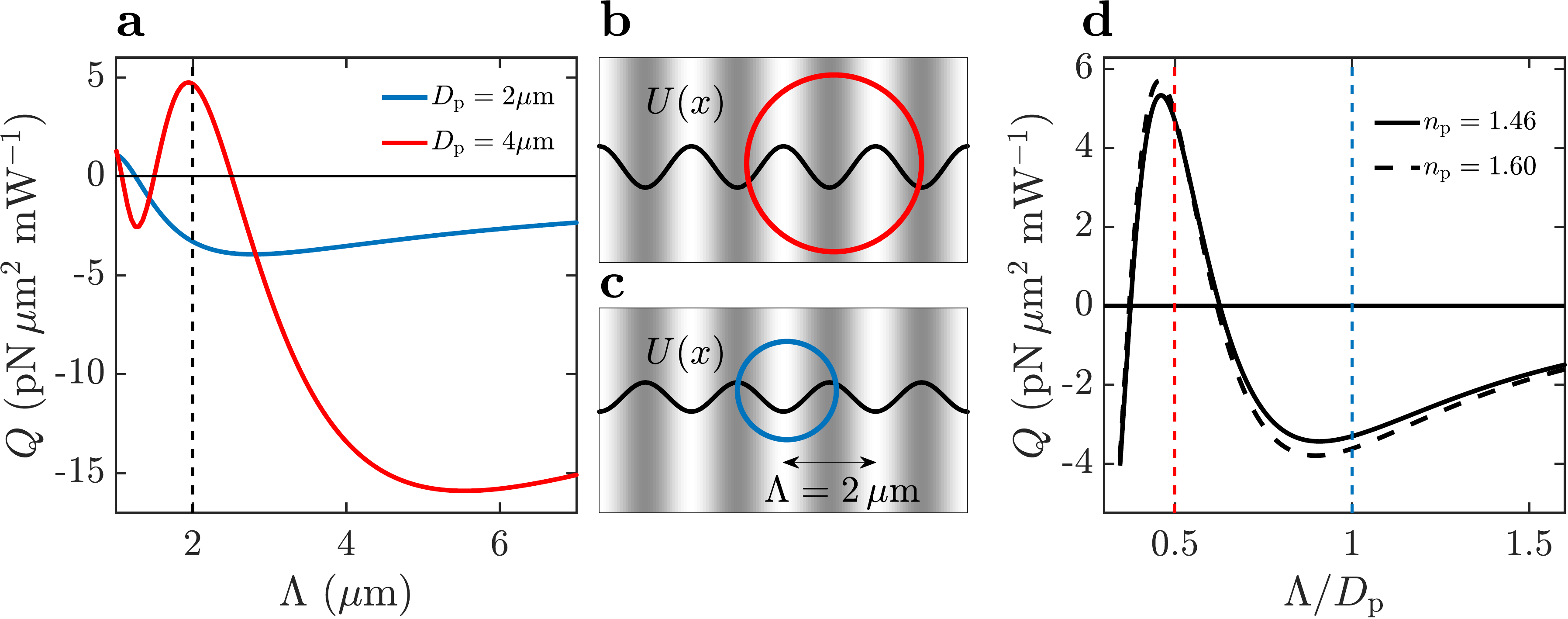}
	\caption{
	{\bf Energy potentials and trapping positions in a pattern of fringes.}  
	The force in a pattern of fringes has a sinusoidal shape with the same period as the fringe pattern. 
	(a) Force amplitude normalized to the average intensity, $Q=A/2I_0$, as a function of the period of the fringes, $\Lambda$, for two silica particle of different diameters: $D_{\rm p}=2\,\upmu \textrm{m}$ (blue) and $4\,\upmu \textrm{m}$ (red). Negative (positive) values of $Q$ give rise to a potential with energy minima at the brightest (dark) regions of the pattern. 
	(b-c) Optical potentials $U(x)$ for the particles with (b) $D_{\rm p}=2\,\upmu \textrm{m}$ (blue) and (c) $D_{\rm p}=4\,\upmu \textrm{m}$ for a pattern with period $\Lambda_x=2\,{\rm \upmu m}$ (corresponding to the dashed vertical line in (a)). 
	(d) Variation of $Q$ as a function of the ratio $\Lambda_x/D_{\rm p}$ for microscopic beads made of silica and polystyrene, with respective refractive indices of $n_{\rm p}=1.46$ and $n_{\rm p}=1.60$ (values defined at $\lambda=532\,{\rm nm}$). The red and blue vertical dashed lines correspond to the values of the ratio $\Lambda_x/D_{\rm p}$ for the examples shown in (b) and (c), respectively. The optical forces and potentials were computed using the ray optics model.
	}
    	\label{fig:31:fringes}
\end{figure}

To illustrate this fact, Fig.~\ref{fig:31:fringes}(a) shows the dependence of the normalized amplitude of the force ($Q=A/2I_0$) as a function of the period, for two silica spheres (refractive index $n_{\rm p}=1.46$) with diameters $D_{\rm p}=2\,{\rm \upmu m}$ and $4\,{\rm \upmu m}$, respectively, immersed in water (refractive index $n_{\rm m}=1.33$). If $\Lambda$ is larger than the bead size, the particle feels a potential with equilibrium points at the intensity maxima, corresponding to negative values of $Q$, similar to what happens with subwavelength particles. There is a global minimum in each curve of Fig.~\ref{fig:31:fringes}(a), which represents the optimum period, slightly larger than the diameter of the sphere, that maximizes the magnitude of the force. Now, as the period is progressively reduced, the magnitude of the force decreases until it vanishes, meaning that a light pattern with such a spatial period will have no effect on the sphere. If $\Lambda$ becomes even shorter, the force changes its sign ($Q>0$) and the optical potential reverses, having stable equilibrium points in places where the intensity is minimum. This is schematically illustrated in Figs.~\ref{fig:31:fringes}(b-c), showing each particle at one of its respective equilibrium positions in the same intensity pattern (background), which has the period $\Lambda=2\,{\rm \upmu m}$ (marked by the vertical dashed line in Fig.~\ref{fig:31:fringes}(a)). The resulting energy potentials for these particles, $U(x)$, are also plotted, exhibiting a sign reverse with respect to one another.  

In general, the most important aspect for determining the trapping positions in an extended pattern is the relative size of the particle with respect to the characteristic size of the pattern. In that sense, it is convenient to plot $Q$ in terms of the ratio $\Lambda/D_{\rm p}$. Accordingly, Fig.~\ref{fig:31:fringes}(d) depicts the normalized force amplitude as a function of $\Lambda/D_{\rm p}$ for spheres made of silica $n_{\rm p}=1.46$ (solid curve) and polystyrene $n_{\rm p}=1.6$ (dashed curve). The red and blue vertical dashed lines indicate the values of the parameter $\Lambda/D_{\rm p}$ corresponding to the examples in Figs.~\ref{fig:31:fringes}(b) and \ref{fig:31:fringes}(c), respectively. This kind of pattern has been widely used for optical manipulation of multiple particles and optical sorting \cite{chiou1997interferometric, macdonald2001trapping, arlt2002moving, cizmar2006optical_sorting, ricardez2006modulated, siler2008surface, jakl2014optical}.

\begin{figure}[h]
	\centering
	\includegraphics[width=6cm]{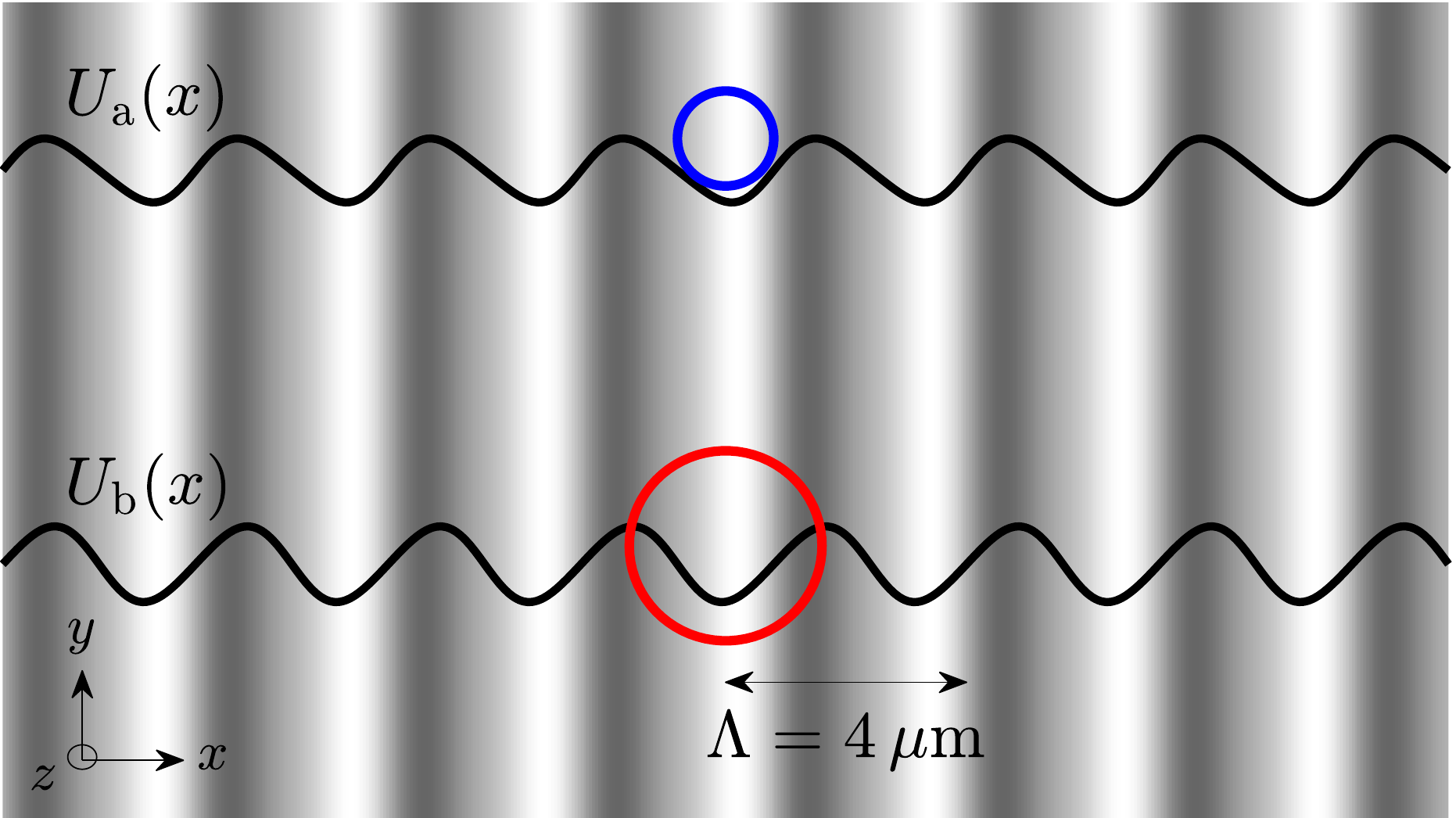}
	\caption{
	{\bf Particle-dependent potential or \textit{size effect} in asymmetric patterns.} 
	Asymmetric pattern of light with period $\Lambda=4\,{\rm \upmu m}$ (background) and the resulting optical potentials, $U_{\rm a}(x)$ and $U_{\rm b}(x)$, for two silica beads with diameters $D_{\rm a}=2\,{\rm \upmu m}$ and $D_{\rm b}=4\,{\rm \upmu m}$, respectively. The shape of the optical potential for a given pattern of light can be very different depending on the size of the particle. The potential was computed using the ray optics model.
	}
	\label{fig:32:asymtwoparticles}
\end{figure}

The particle-dependent response to the structure of a light pattern is sometimes referred to as the \textit{size effect} of the optical forces and takes place regardless of the complexity of the pattern. As the optical landscape becomes more intricate, nevertheless, not only the magnitude and the sign of the optical force may change like in our simple example, but also the shape of the potential may be drastically altered depending on the size of the particle. 
For example, Fig.~\ref{fig:32:asymtwoparticles} shows the case of a sawtooth pattern of fringes, also shown in the second row of Table \ref{tab:11:patterns}. This pattern can be generated by superimposing three beams with linear polarization oriented along the horizontal, vertical and diagonal directions, so that the latter partially interferes with the former two, creating two patterns of fringes orthogonally polarized, one with twice the period of the other \cite{arzola2011experimental}. Given the asymmetric intensity in the fringes, a consequent asymmetry could be expected in the potential, and this is true for the smaller particle ($D_{\rm a}=2\,{\rm \upmu m}$) shown in Fig.~\ref{fig:32:asymtwoparticles}. However, a larger particle ($D_{\rm b}=4\,{\rm \upmu m}$) experiences a completely different potential, with a slight asymmetry in the opposite direction. The modification of the potential can be even more radical for other particle sizes. The application of this effect for the transport of particles was analyzed in detail in Refs.~\cite{arzola2011experimental, arzola2013dynamical}. 

\paragraph{Two-dimensional potentials}

When the periodicity of the pattern is in 2D the size effect might give rise to different behaviors along each symmetry axis, and thus the equilibrium positions for a given particle might be unpredictable without performing an explicit calculation of the optical potential. Different kinds of light lattices in 2D have been used to study optical binding and colloidal crystallization \cite{burns1989optical, burns1990optical}, to explore structural phase transitions in colloidal monolayers \cite{bechinger2001phase, brunner2002phase, mangold2003phase},
to tailor bandgap materials \cite{baumgartl2007tailoring}, and to optically sort particles in combination with microfluidics \cite{macdonald2001trapping}. 
Periodic light lattices in 2D can be produced by the interference of at least three beams. For example, the coherent superposition of three waves, depending on the directions of their wave vectors, can lead to centered-rectangular or hexagonal lattices, whereas the interference of four waves can produce square, oblique or rectangular lattices. The example in the third row of Table~\ref{tab:11:patterns} corresponds to an oblique or rhombic lattice. 
An alternative approach to generate 2D-periodic patterns is by means of phase holograms \cite{dufresne1998optical}, such as the rectangular array of individual spots illustrated in the fourth row of Table~\ref{tab:11:patterns}. In contrast with the multiple interference fields, diffractive patterns can act as a collection of individual traps for 3D confinement, overcoming the size effect and hence facilitating the control of the equilibrium positions of individual particles. 
A SLM can be used as well to display diffractive gratings and produce multiple interfering beams \cite{jakl2014optical}, whilst the 2D-trapping capabilities can be extended to manipulate ordered structures in 3D by taking advantage of the self-imaging Talbot effect \cite{schonbrun20053d}. Another interesting case is that of a quasi-periodic pattern arising from the superposition of five waves, which is shown in the fifth row of Table~\ref{tab:11:patterns}. This has been used to study the formation of quasi crystalline structures in colloidal monolayers \cite{mikhael2008archimedean, mikhael2010proliferation}. 

A type of extended patterns with circular, elliptical or parabolic symmetry in the transverse plane correspond to Bessel, Mathieu and parabolic nondiffracting beams, respectively \cite{durnin1987exact, gutierrez2000alternative, bandres2004parabolic}. Their Fourier spectrum at the back aperture of the microscope objective lies on a thin annulus, where the particular beam geometry is determined only by the intensity and phase modulation around the annulus. The optical forces in a Bessel beam, like the pattern depicted in the sixth row of Table \ref{tab:11:patterns}, exhibit very interesting features due to the size effect \cite{volke2004three, cizmar2005optical, milne2007transverse}, and similar attributes can be expected in the other types of beams. Namely, the multi-ringed structure of the intensity pattern has an envelope that decreases radially outwards. Depending on its size with respect to the pattern, a particle might respond predominantly to the fine structure or to the envelope. This can be used as a passive mechanism to sort particles \cite{paterson2005light} or to build what is known as a washboard potential \cite{tatarkova2003brownian}. 

Another version of an asymmetric periodic pattern, but now in 2D, is illustrated in the seventh row of Table \ref{tab:11:patterns}. This is generated by superimposing two identical light lattices, orthogonally polarized, produced with a single phase hologram  \cite{arzola2017omnidirectional}. By controlling the relative position and the intensity ratio between the two patterns, the asymmetry of the light lattice can be tailored. In general, periodic and asymmetric patterns are very important in the study of the {\em ratchet effect} \cite{faucheux1995optical, lee2005observation, arzola2011experimental, arzola2017omnidirectional}, which is the emergence of directed transport in the presence of unbiased external forces due to a spatiotemporal symmetry breaking and, furthermore, is thought to be at the core of the operation mechanism of some biological engines \cite{hanggi2009artificial}. 

Random light patterns like speckle (bottom row of Table \ref{tab:11:patterns}) are also very important in the context of particle diffusion and transport. In practice, speckle can be generated by transmitting a laser beam through a glass diffuser, by coupling the laser into a multimode optical fiber, or by encoding a random phase mask on a SLM. A very interesting aspect of the interaction of Brownian particles with a random potential landscape is its major role in diffusion, and how it depends on the several parameters of the potential \cite{volpe2014brownian, bianchi2016active, shvedov2010selective, bewerunge2016experimental}. 

\subsubsection{Transport mechanisms}

So far, we have examined optical forces and potentials in a static situation. Now we will examine the conditions for inducing directed transport in the system, i.e., a net current of particles in a given direction through their interaction with light and, possibly, other external forces. To address this subject, we will start by looking at the general equation of motion for a single particle interacting with an optical potential in the overdamped regime, which is a good description of the micromanipulation systems under consideration. For the sake of simplicity, only one dimension will be examined, this is:
\begin{equation}\label{eq:langevin}
   \gamma\Dot{x}=-\frac{d}{dx}U(x,t)+F(t)+\xi(t),
\end{equation}
where $\gamma$ represents an effective drag coefficient, $U(x,t)$ is the optical potential, $F(t)$ denotes an external force, and $\xi(t)$ is the thermal noise, having a correlation function $\langle \xi(t) \xi(t')\rangle=2D\delta(t-t')$, with $D=k_{\rm B}T/\gamma$ being the diffusion coefficient. Although noise usually plays an important role in microscopic-scale dynamics and some transport mechanisms even rely on it, this can be neglected under certain conditions, giving rise to a deterministic dynamics, $\xi(t)\approx 0$. For example, Brownian motion of silica or latex spheres immersed in water at room temperature is very subtle if the diameter is $D\gtrsim5\,{\rm \upmu m}$, and it becomes practically negligible for $D\gtrsim10\,{\rm \upmu m}$.  

In general terms, we can distinguish between two main transport mechanisms: a dynamic potential and a static potential in the presence of an external driving force. A very simple realization of the former case corresponds to a light pattern of fringes that moves with a constant speed $v$ with respect to the static sample. This corresponds to a travelling periodic potential: $U(x,t)=U(x-vt)$. The dynamics of the particle can be analyzed in a reference frame that moves along with the potential by making the change of variable $u=x-vt$ in equation~\eqref{eq:langevin}, leading to 
\begin{equation}\label{eq:langevin_drag}
    \gamma\Dot{u}=-\frac{d}{d u}U(u)-\gamma v +\xi(t).  
\end{equation}
Notice that this is equivalent to adding a constant drag force to a static pattern and thus the following discussion will be valid for the case of a constant external driving force. 
Whether the particle will follow the pattern or not depends on the relative magnitude between the optical force, $F_{\rm{op}}(u)=-dU/du$, and the drag force, $F_{\rm d}=-\gamma v$, in the moving reference frame. The constant force has the effect of tilting the potential energy landscape, giving rise to what is sometimes referred to as a washboard potential. If $|F_{\rm{op}}|>|F_{\rm d}|$, the particle will not be able to escape from the well and, therefore, it will be locked-in to the pattern, moving with the same velocity as the pattern with respect to a fixed reference frame. A particle with this behavior has been called a {\em Brownian surfer} \cite{reimann2002brownian}. If $|F_{\rm{op}}|<|F_{\rm d}|$, the particle will be pulled out from the potential well by the drag force in the moving reference frame or, in the fixed frame, it will not follow the pattern but it will diffuse around its initial position. This is known as a {\em Brownian swimmer} \cite{reimann2002brownian}. Depending on the thermal noise level, there might be intermediate situations, in which the particle moves in the same direction as the pattern but with a reduced speed $v_{\rm p}<v$. This happens when $|F_{\rm{op}}|\gtrsim |F_{\rm d}|$ or, in terms of the potential energy barrier between neighboring wells, when $\Delta U\gtrsim k_{\rm B}T$, so that thermal energy may sometimes help the particle to surmount the potential barrier between consecutive wells. 

Due to the size effect discussed above, in a polydisperse mixture of particles, some of them may follow the pattern while others may not, depending on their size and refractive index, or even if two types of particles move along with the pattern, their average speeds may be different. Therefore, a moving potential can be used as a mechanism to separate particles of different characteristics, i.e. as an optical sorting device.

\begin{figure}[h] 
	\centering
	\includegraphics[width=8cm]{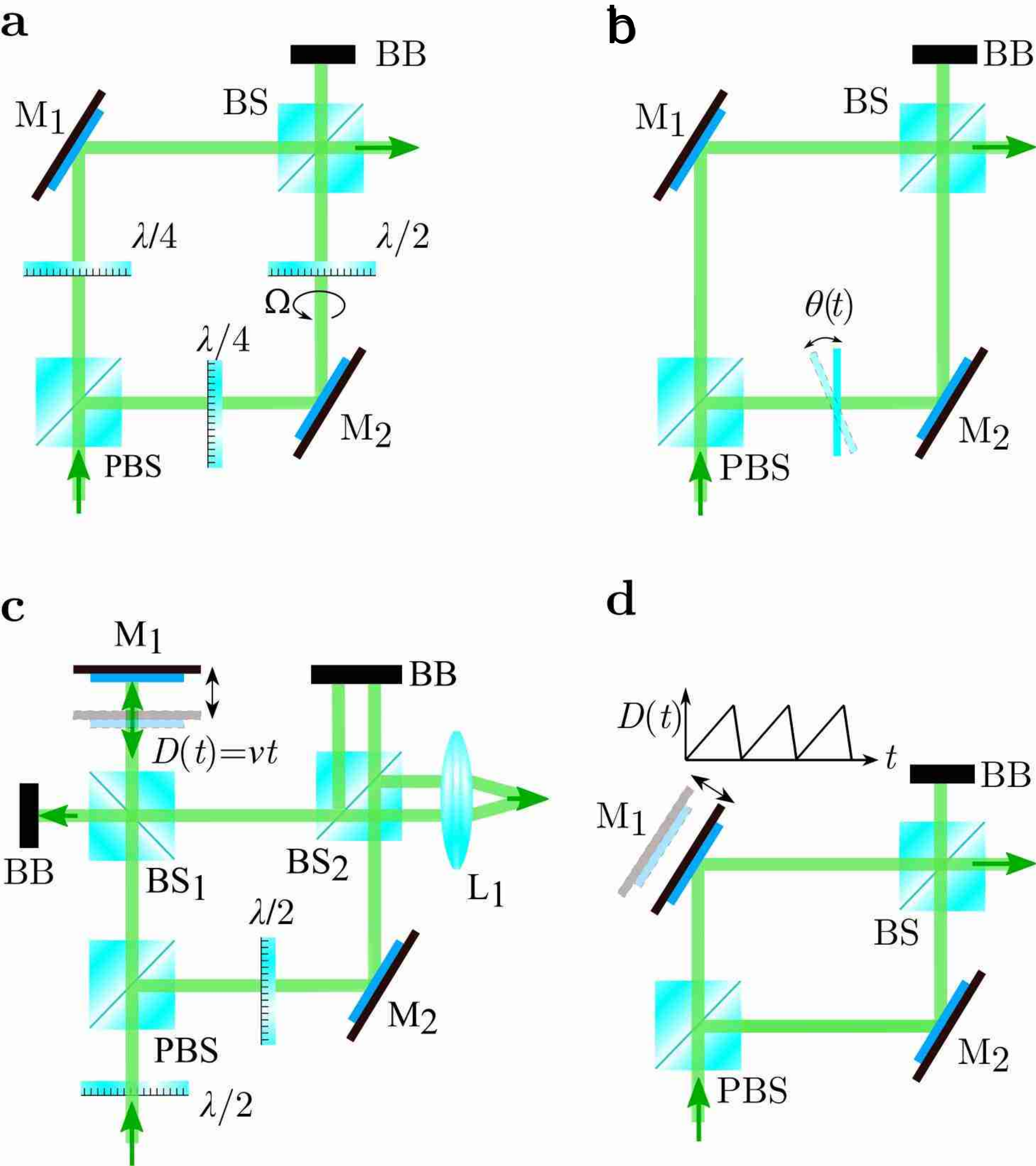}
	\caption{
	{\bf Schemes for the generation of a moving pattern of fringes using a Mach-Zehnder interferometer}. 
	(a) By exploiting the angular Doppler effect with a $\lambda/2$ plate rotating at a constant angular speed $\Omega$. 
	(b) By gradually tilting a glass plate a small angle $\theta(t)$. 
	(c) By means of a linear displacement $D(t)=vt$ of a mirror. 
	(d) By a periodic and asymmetric displacement of a mirror, according to a sawtooth function $D(t)$. 
	PBS: polarizing beam splitter. M: mirror. BB: beam block. BS: non-polarizing beam splitter. L: lens. }
	\label{fig:33:moving}
\end{figure}

In practice, there are several ways to move an interference pattern of fringes. Let us consider, for example, an optical micromanipulation setup with a Mach-Zehnder interferometer. We can express the resulting interference pattern at the sample plane as $I(x,y)=4I_0\cos^2\left(\pi x/\Lambda+\varphi(t)/2\right)$, where $\varphi(t)$ represents a time-dependent phase difference introduced in one arm of the interferometer. To generate a travelling potential, $\varphi(t)$ should be a linear function of time. 
An elegant way to achieve this is by using the angular Doppler effect \cite{garetz1979variable, garetz1981angular}, arising when a circularly polarized beam passes through a rotating $\lambda/2$-plate, which gives rise to a frequency shift $\Delta\omega=\pm 2\Omega$, where $\Omega$ is the rotation speed of the $\lambda/2$-plate. This can be easily implemented in the Mach-Zehnder interferometer \cite{arlt2002moving}, as illustrated in Fig.~\ref{fig:33:moving}(a). A polarizing beam splitter (PBS) is followed by a $\lambda/4$-plate in each arm of the interferometer, producing orthogonal circular polarization states. The rotating $\lambda/2$-plate is inserted in one arm, such that it inverts the handedness of the circular polarization at the same time that it introduces an angular Doppler frequency shift. A non-polarizing beam splitter is placed at the output of the interferometer to superimpose the beams. The resulting phase difference between the two beams is $\varphi(t)=(\Delta\omega) t=\pm2\Omega t$, where the sign depends on the rotation direction of the  $\lambda/2$-plate with respect to the polarization handedness \cite{garetz1981angular}. At the sample plane, this will define the direction of motion of the fringes ($\pm x$), whereas the speed of motion is $v=\Omega/k$.

There are other alternatives for producing motion of the fringes. For example, a very simple option, which can be easily implemented in different types of interferometers, is to introduce a thin glass plate in the path of one of the interfering beams \cite{rubinov2003physical}. As illustrated in Fig.~\ref{fig:33:moving}(b) for the particular case of a Mach-Zehnder interferometer, by gradually tilting the glass plate a small angle $\theta(t)$, the optical path length of the corresponding beam is modified, resulting in a relative phase shift between the two beams and the consequent motion of the pattern. A straightforward analysis using geometrical optics yields that, if $\theta\ll 1$, then the change in the optical path length is given by $\Delta(\theta)\approx T_g/2(1-1/n)^2\theta^2$, where $T_g$ is the thickness of the glass plate and $n$ its refractive index. Therefore, as $\Delta(\theta)$ is not linear, the phase shift is not linear either: $\varphi(t)=k\Delta(\theta(t))$. Although this complicates the theoretical description of the system, it is a cheap and accessible choice if the aim is just to make a demonstration of particle transport.

Another configuration that provides full control of the phase shift $\varphi(t)$ makes use of a displacement of one of the mirrors in a Mach-Zehnder interferometer. For example, in the first interferometric optical tweezers setup \cite{chiou1997interferometric}, depicted in Fig.~\ref{fig:33:moving}(c), an extra beam splitter is placed in one arm of the interferometer (BS1), so that mirror M1 retroreflects the beam and, mounted on a motorized translation stage, it is moved with a constant speed: $D(t)=vt$. This produces a phase shift between the interfering beams of the form $\varphi(t)=2kD(t)$, leading to a continuous motion of the fringes at the same speed as the mirror ($v$), while keeping the spatial period of the fringes unaltered. Indeed, the period of the fringes is a function of the lateral distance between the output beams and the focal length of lens L1 (see Fig.~\ref{fig:33:moving}(c)). Notice, however, that the presence of the beam splitter BS1 yields an additional intensity loss. To equalize the intensity in both arms of the interferometer, two $\lambda/2$-plates should be used, one at the entrance of the setup in combination with a polarizing beam splitter (PBS) to adjust the amount of light in each arm, and the other one to set the same polarization in the two beams.

\begin{figure}[h]
	\centering
	\includegraphics[width=12cm]{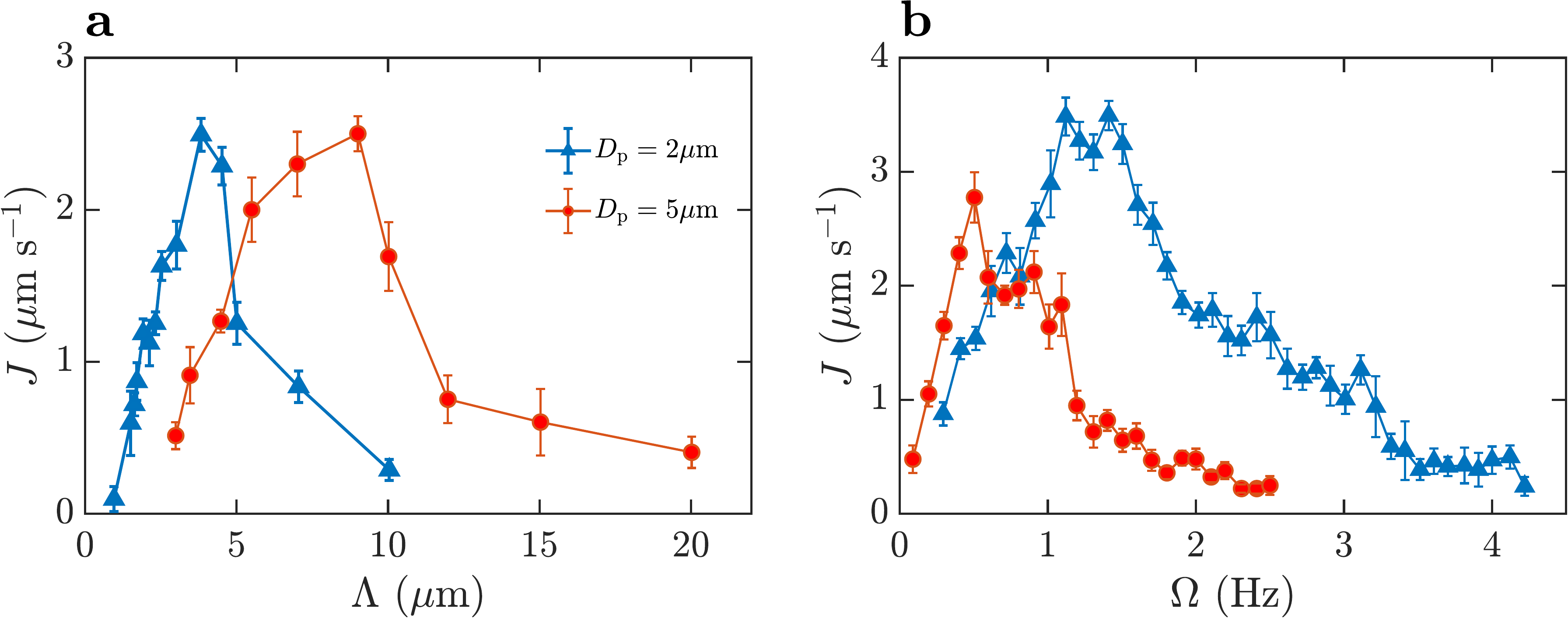}
	\caption{
	{\bf Current in an optical sieve}. 
	(a) Mean current as a function of the period of the fringes $\Lambda$ for particles with diameter $D_{\rm p}=2\,{\rm \upmu m}$ and $5\,{\rm \upmu m}$ for a fixed optical power $P=130\,{\rm mW}$ and frequency $f=1\,{\rm Hz}$. 
	(b) Mean current as a function of the angular frequency $\Omega$ of the sawtooth function for the optimum spatial period in each case and fixed power $P=130\,{\rm mW}$. 
	The two particle kinds feature different optimal currents as a function of $\Lambda$ and $\Omega$, enabling to classify them with high throughput. 
	Data from Ref.~\cite{ricardez2006modulated}.
	}
	\label{fig:34:sieve}
\end{figure}

A variant of this idea, which avoids the extra beam splitter, is shown in Fig.~\ref{fig:33:moving}(d). A periodic and asymmetric displacement of the mirror M1 might be equivalent to a continuous travelling pattern under certain conditions. This can be done by mounting the mirror on a piezo-stage\footnote{Any programmable motorized translation stage would be useful and, for example, stepper motors are cheaper, at the cost of loosing precision and speed in the motion control.} driven with a sawtooth function \cite{ricardez2006modulated}:
\begin{equation}\label{eq_shift}
D(t)=
\begin{cases}
v_1t & \text{if}\ \ 0< t\leq \tau\\
-v_2t & \text{if}\ \ \tau< t\leq \tau+\delta t\\
\end{cases},
\end{equation}
where $\tau\gg \delta t$, $v_1=D_{\rm max}/\tau$ and $v_2=D_{\rm max}/\delta t$, with $D_{\rm max}$ denoting the maximum displacement of the mirror, which should be small to prevent distortion in the period of the fringes. The frequency of $D(t)$ is $f=1/T$, with $T=\tau + \delta t$ the period of the function: $D(t+T)=D(t)$. The motion of the mirror is related to the phase shift of the pattern as $\varphi(t)=2\sqrt{2}kD(t)$, implying that the fringes at the sample mimic the sawtooth-like displacement, with speeds given by $u_i=2\sqrt{2}\Lambda v_i/\lambda$, where $i=1,2$. The main idea is that the particle is able to follow the pattern in the direction of the slower motion ($u_1$), but not during the fast motion ($u_2$) in the opposite direction, resulting in a unidirectional transport. Of course, the selection of the parameters\footnote{There are alternative choices, but in any case, the dynamic phase $\varphi(t)$ is defined by a set of three independent parameters, set through the motion function of the mirror.} $D_{\rm max}$, $f$, and $\tau$, along with the spatial period of the fringes $\Lambda$, is crucial to optimize the current of particles. For example, if $D_{\rm max}=m\lambda/(2\sqrt{2})$, with $m$ a positive integer, the particle will advance a distance $m\Lambda$ in one direction and will end up at another minimum of the potential after an entire cycle, provided that $\Lambda$ is such that optimizes the optical force for the given particle size. Figs.~\ref{fig:34:sieve}(a-b) illustrate the resulting particle current for two sizes of particles as a function of $\Lambda$ and $\Omega=2\pi f$, respectively. An optimum can be clearly identified in each case. This device was used as an optical sieve to separate particles according to their size and it can also be used to sort particles according to their refractive indices \cite{ricardez2006modulated}. In the latter case, the strategy is to start with a low intensity and gradually increase the optical power up to a value for which only the particles with the highest refractive index contrast with respect to the medium will respond.

A related approach consists in generating phase gratings with a SLM aimed at splitting an incident beam to produce interference. The advantage is that the angular separation between the beams can be varied by changing the grating, so that the spatial period of the fringes can be tuned and its orientation can be modified as well. Furthermore, the motion of the pattern can be also controlled with the SLM, by adding a continuous phase shift to one of the beams. Such a device has been used to move and sort living objects and particles of different shapes \cite{jakl2014optical}.

It is worth mentioning that the techniques discussed above can be extended to more complex interference patterns, such as rotating beams, allowing circular motion of beads and structures \cite{macdonald2002creation, paterson2001controlled}. Similar ideas have been applied as well to build optical conveyors that allow bidirectional transport in 3D. These kinds of devices use coaxial beams, either counter-propagating to generate axial standing waves or co-propagating, and not just Gaussian beams but also other structured fields, like Laguerre-Gaussian and Bessel beams \cite{cizmar2005optical, cizmar2006optical_sorting, cizmar2006optical_nanotrap, cizmar2010holographic, ruffner2012conveyor}, including evanescent fields \cite{siler2008surface}. The operation principle is basically the same as before: to manage the motion of an interference field by adding a controlled phase shift between the beams.  

There have also been different approaches to yield moving patterns based on digital holography. Among the first schemes, we can mention a mechanism called optical peristalsis, in which a lattice of optical traps is created and controlled with a set of phase holograms displayed on a SLM \cite{koss2003optical}. Following a periodic sequence of three states, the pattern is consecutively displaced by one-third of the lattice constant, reaching the initial state again after an entire cycle. A particle trapped in a given potential well is released when that pattern is turned off and it is immediately attracted towards the nearest shifted well once the next-state pattern is turned on. By repeating this process, the particle is transferred from one well to the neighboring trap after each cycle, experiencing a constant motion in one direction, resembling the operation of a peristaltic pump. The geometry of the lattice can be varied to achieve, for example, a radial current in circular patterns. If the particle's diffusion during the time between consecutive states is negligible, the process is deterministic and the current is given by $v=L/T$, where $L$ and $T$ are the lattice constant and the temporal period of the sequence, respectively.

\subsubsection{Other transport mechanisms and applications}

As we have already mentioned, a travelling potential is dynamically equivalent to having a particle in a fixed light pattern subject to a constant drag or, more generally, a driving force. The drag force can be provided, for example, by tilting the sample \cite{arzola2009force}, by moving the sample with respect to the pattern \cite{roichman2007colloidal,arzola2011experimental}, or by means of a controlled microfluidic flow. The latter option has the important advantage of being suitable to integrate into lab-on-chip platforms \cite{wang2011enhanced, dholakia2007optical}. In addition, sorting capabilities can be boosted by using 2D light lattices instead of 1D-periodic patterns, offering new control parameters and thus more versatility. Indeed, biological or colloidal assays flowing through two-dimensional patterns may be deflected from the body of the flowing fluid depending on the orientation of the lattice with respect to the flow and on the strength of the optical force, giving rise to discrete angle deviations or fractionation \cite{korda2002kinetically, macdonald2001trapping, ladavac2004sorting}. Using this approach, a high throughput of sorted samples can be practically and efficiently achieved. Also, microfluidic techniques can be combined with other schemes of optomechanical control, like dynamic or blinking potentials \cite{dasgupta2012microfluidic}.

A fundamentally different and intriguing mechanism to selectively move objects at the microscopic scale is based on the ratchet effect. In contrast to previous examples where there exist a biased external field to induce a net current of particles, the ratchet effect makes use of non-biased time-varying forces to induce transport. Since these forces are of zero-mean in time and in space, a non-trivial symmetry has to be broken  to observe a net current \cite{hanggi2009artificial}. The light patterns represented in the second and seventh rows of Table~\ref{tab:11:patterns} are examples of ratchet potentials with a space asymmetry, yet the symmetry breaking may be much more subtle, including variations in time. Actually, the spatial structure and time-modulation of the potential along with the time-dependence of the external force define many different types of ratchets \cite{faucheux1995optical, libal2006dynamics, arzola2011experimental, arzola2017omnidirectional, lee2005one, lee2005observation, xiao2011two, leon2017noise}. Moreover, although most of these systems are noise-assisted when particle diffusion is non-negligible, there are also deterministic ratchets \cite{arzola2011experimental}. The rich dynamics arising in these systems is due to the delicate interplay among the whole set of encompassed parameters and manifests itself in the diverse phenomena that can be observed, including bidirectional transport depending on size, currents in different directions with respect to the driving, current reversals and chaotic behavior. Interestingly, the ratchet mechanism is ubiquitous, as it appears in many biological, physical and chemical systems, and its study has paved the way to broaden our understanding of transport processes at the micro- and nanoscales \cite{hanggi2009artificial}.

So far we have looked mainly at ordered patterns or light lattices, but random patterns have also shown an excellent selectivity for sorting particles  \cite{shvedov2010selective, volpe2014brownian}, with the additional advantage of making feasible to sort and select a continuous throughput of chosen constituents in turbid media, common in many biological assays. 

Although we did not deal with scattering forces in our previous discussion, it is important to remind that the use of scattering rather than gradient forces was really the first means to transport and separate particles. Indeed, since his earliest experiments in 1970, Ashkin noted that radiation pressure acting on particles of different sizes along the propagation axis of a moderately focused beam was proportional to the size of the particles, so that the beads were guided along the beam axis with different velocities \cite{ashkin1970acceleration}. This idea is the basis of the so-called optical chromatography, in which a Gaussian light beam opposes a fluid flow, causing that particles of various sizes and/or refractive indices find different equilibrium positions along the beam axis where the scattering force is balanced by the force due to the fluid flow \cite{imasaka1995optical, hart2003refractive}. Other schemes combine the action of scattering and gradient forces. Consider, for example, the interference pattern of fringes discussed above. In addition to the transverse gradient force, there is usually a scattering force along the propagation direction, which may push the particles selectively according to their relative size with respect to the fringes, so that small particles, attracted to the bright fringes, will be pushed much stronger than larger particles attracted to the dark fringes. Using this interplay between gradient forces and scattering forces, it is possible to activate transport with an all-optical static configuration \cite{jakl2008static}. These kinds of ideas have been extended further to include the use of evanescent waves generated on the surface of prisms (Kretschmann configuration) or with total internal reflection (TIR) microscope objectives \cite{marchington2008optical, siler2006optical, jonavs2008light}. An advantage is that the use of gradient and scattering forces arising from evanescent waves in near-field photonic configurations allows one to scale down to tens of nanometers the size of the manipulated objects \cite{jonavs2008light, erickson2011nanomanipulation}. 

Another puzzling transport mechanism appears in the so-called \emph{tractor beam}. This is a counterintuitive effect, in which particles move opposite to the direction of the beam propagation due to a pulling scattering force \cite{brzobohaty2013experimental}.
In the case of a highly nonparaxial wave field, whose angular spectrum components subtend large angles with respect to the main propagation axis, part of the diffracted radiation may indeed have larger wave vector components along the propagation axis than those of the incident field, giving rise to strong forward scattering. The net effect might be a negative radiation pressure (pulling force), depending on the shape of the object, its size, and its interaction with the wave field. It is important to stress though, that this effect is not based on intensity variations and it is not derived from a potential, in contrast with the conservative gradient force acting in conventional optical tweezers. Remarkably, this phenomenon can happen also in rotational dynamics, where the spin angular momentum of light can exert transverse non-conservative forces in the opposite direction than expected in a chiral object \cite{tkachenko2014optofluidic, hayat2015lateral}.  

The simplest description of the dynamical mechanisms previously discussed considers diluted samples, and this is usually a good approximation even if the sample is not well diluted and some interactions exist. When the interactions are strong, a more complete description becomes necessary to fully understand the phenomenology. The origin of the interactions can be manifold, like hydrodynamic, electrical or optical (optical binding), and they may promote collective behaviors such as dynamical ordering, phase transitions or cluster formation \cite{brzobohaty2013experimental, jakl2014optical, cao2019orientational, brazda2018experimental,zaidouny2013light, bohlein2012experimental,bohlein2012observation}. 

Understanding and controlling transport from the mesoscale down to the nanoscale is still a key topic in biophysics, nanomedicine and nanotechnology. There is still plenty of room for basic and applied research on the many aspects involved in transport of microscopic objects, including studies on self-propelled particles \cite{bechinger2016active}. 
Especially in biological environments, the role of light-induced forces has been barely studied so far.
On the other hand, many of the ideas and techniques discussed here are being extended to other realms where force fields can be used for contactless manipulation, either light, like in plasmonic \cite{lehmuskero2015nanometal, righini2007parallel, righini2008surface, juan2011plasmon, marago2013optical, shoji2014plasmonic}, optoelectronic \cite{chiou2005massively} and some recent thermophoretic devices\cite{lin2017thermophoretic, li2018opto}, or without light, like in acoustic manipulation \cite{shi2009acoustic, friend2011microscale, baresch2016observation, marzo2019holographic}. This range of techniques has broaden the trapping conditions and the accessible range of particle sizes and force magnitudes, going down to femtonewtons for nanoparticle maneuvering and up to micronewtons to trap millimiter-sized beads, and possibly beyond in both directions in the near future. Also, there is a great effort invested in turning the basic concepts into real applications, specially in the development of lab-on-chip platforms, where many modern areas can converge to perform practical tasks, including optomechanics, microfluidics, acoustofluidics, nanoplasmonics, thermoplasmonics, optoelectronics, near-field photonics, spectroscopy, to name but a few.  
A more extended discussion on light-induced mechanisms of controlled transport can be found in a very recent and comprehensive review \cite{zemanek2019perspective}.

\section{Optical tweezers in vacuum}
\label{sec:5:vacuum}

Even though Ashkin conducted his pioneering experiments on optical forces and radiation pressure  in a gas or vacuum environment \cite{ashkin1977feedback, ashkin1976optical, ashkin1971optical, ashkin1970acceleration}, for many decades optical forces have been primarily harnessed to conduct optical tweezers experiments in liquids or for trapping and cooling of atoms \cite{chu1998nobel}. Only recently there has been a renewed interest to use optical tweezers in vacuum. The fist single-beam gradient-force optical trap in air was demonstrated in Ref.~\cite{omori1997observation} 10 years after its realization in water \cite{ashkin1986observation} and it took another 15 years to demonstrate single beam optical levitation in high vacuum \cite{gieseler2012subkelvin}.

This section will introduce the basic concepts for realizing experiments with optically levitated particles in high vacuum.  We will discuss the implementation with a single beam optical trap, even though other geometries for optical trapping such as parabolic mirrors \cite{salakhutdinov2016optical, rashid2016experimental} or counter-propagating beams \cite{li2010measurement, ranjit2015attonewton} have also been realized.
Specifically, we will focus on the practical aspects such as particle loading, sources of noise, feedback stabilization, calibration, and provide the literature references that will allow the reader to independently implement optical levitation experiments in high vacuum.
We begin in section~\ref{sec:5.1:levitation_applications} with  a concise review over the applications of levitated particles in high vacuum.

\subsection{Applications of optical levitation}\label{sec:5.1:levitation_applications}

In vacuum the motion of an optically levitated dielectric nanosphere is well isolated from the environment. This makes it a promising candidate to demonstrate quantum behavior of a macroscopic object at room temperature \cite{romero2010toward, chang2010cavity, barker2010cavity}. To observe quantum effects, the motional energy of the levitated particle has to be reduced by more than seven orders of magnitude to the level of a single quantum excitation. Ongoing efforts into this direction have already reached impressive cooling rates using feedback cooling \cite{tebbenjohanns2019cold, jain2016direct, gieseler2012subkelvin,  conangla2019optimal, tebbenjohanns2019motional} and cavity cooling schemes \cite{delic2019cavity, windey2019cavity, meyer2019resolved}. While the ultimate cooling limit, i.e., the quantum ground state, has already been reached in a range of nano and micromechanical systems \cite{teufel2011sideband, chan2011laser}, ground state cooling of a levitated\footnote{To distinguish trapping in liquid from trapping in a dilute environment, we refer to the latter as optical levitation.} nanoparticle has only be claimed recently \cite{delic2020cooling}.
The quantum mechanical ground state is an important milestone towards preparation of these massive objects in non-classical quantum states that are otherwise only known from elementary particles \cite{haroche2013nobel}, atoms, ions \cite{wineland2013nobel}, and molecules \cite{tuxen2010quantum, hornberger2012colloquium}.

Investigating quantum mechanical states of motion of massive objects allows for experimental tests of fundamental physics in unexplored parameter regimes\cite{kleckner2008creating, kaltenbaek2012macroscopic, romero2011quantum,romero2011large, bateman2014near, bassi2013models}.  Preparation of non-trivial quantum mechanical states requires either a strong optomechanical interaction \cite{yin2013optomechanics, aspelmeyer2014cavity, bowen2015quantum} to map non-classical states of light onto mechanical motion \cite{keil2011classical}, a projective quantum measurement such as photon counting \cite{galland2014heralded, cohen2015phonon, hong2017hanbury}, or a quantum nonlinear interaction, which can be provided for example by a defect inside the levitated particle such as a nitrogen-vacancy (NV) center in diamond \cite{rabl2010cooling, rabl2009strong, yin2015hybrid}.
However, the necessary coupling strengths have yet to be demonstrated experimentally and optical levitation experiments with nanodiamonds have been limited by optical absorption \cite{neukirch2015multi, neukirch2013observation, frangeskou2018pure, rahman2016burning, hoang2016electron}.

Ultimately, levitated particles in the quantum regime might benefit from their simple mechanical mode structure.
Conventional mechanical oscillators have a series of normal vibrational modes. Typically, only one mode is considered and the remaining modes are neglected. 
However, in measurement-based approaches to quantum state preparation \cite{vanner2013cooling, vanner2011pulsed}, the displacement of the resonator is measured with a light pulse that is short compared to the oscillation frequency. The light pulse measures over a large bandwidth and thus measures not only the mode of interest but it also measures the displacement due to all the other modes that fall within the bandwidth. As a consequence, these so-called spectator modes add noise, which ultimately limits the efficiency at which the mode of interest can be measured and controlled \cite{riedinger2018single}.
Inside nanoparticles, the frequencies of bulk modes are several orders of magnitude higher than those of the center-of-mass motion. As a consequence, one usually only observes three modes corresponding to the center-of-mass motion in each dimension when working with spherical particles.  With less symmetrical particles additional torsional \cite{hoang2016torsional} and precession \cite{rashid2018precession} modes have been observed.

The extreme isolation from the environment also means that the mechanical quality factor (Q-factor) is very large and, hence, the thermo-mechanical noise of these nanomechanical resonators is very low. For this reason, levitated particles can be used as nanomechanical sensors for force \cite{ranjit2016zeptonewton, hempston2017force, blakemore2019three}, torque \cite{hoang2016torsional}, and acceleration \cite{monteiro2017optical}.
Applications of these sensors are envisioned to detect exotic forces \cite{rider2016search, moore2014search}, such as non-Newtonian gravity \cite{geraci2015sensing}, 
Casimir forces \cite{canaguier2011casimir, manjavacas2017lateral, nie2013dynamics} and torques \cite{xu2017detecting}, and gravitational waves \cite{arvanitaki2013detecting}.
Thanks to the high mechanical Q-factor and tunability of the system parameters, levitated particles are also ideally suited to study nonlinear nanomechanical phenomena \cite{lifshitz2008nonlinear, gieseler2014nonlinear, ricci2017optically}, which are important for the development of future nano and micromechanical devices; in fact, levitated particles are the first systems where nonlinear effects have been witnessed already at the level of thermal fluctuations \cite{gieseler2013thermal}.

Optically levitated particles have unique properties not available in traditional nanomechanical resonators.
For example, the mechanical properties of a levitated particle can be changed rapidly \emph{in situ} by simple adjustments in the trapping light. This allows to perform free-fall experiments, which provide new opportunities in force sensing \cite{geraci2015sensing, hebestreit2018sensing} and  quantum physics \cite{stickler2018probing}.
Unlike traditional clamped mechanical resonators, a levitated particle can rotate freely \cite{arita2013laser,kuhn2015cavity, kuhn2017full, rahman2017laser}. The rotation can be accelerated to GHz frequencies, faster than any other object \cite{ahn2018optically, reimann2018ghz}, simply by changing the trapping laser polarization from linear to circular. At these fast rotations, the centrifugal stress is close to the ultimate tensile strength of the material. Hence, levitated particles can be utilized to test material limits on the nanoscale or to study friction at a fundamental level \cite{manjavacas2017lateral,zhao2012rotational}.

The high control over the system parameters, in particular the ability to tune the interaction with the environment via the gas pressure, make levitated nanoparticles also an excellent testbed for studies of single particle thermodynamics \cite{gieseler2018levitated}, a field also known as stochastic thermodynamics \cite{seifert2012stochastic}. 
As we have seen in section~\ref{sec:4.5.3:st}, many studies in this field have already been conducted in the \emph{overdamped} regime with optically trapped particles in a liquid medium.
In contrast, levitated particles can also operate in the \emph{underdamped} regime and have enabled the first observation of ballistic Brownian motion \cite{li2010measurement}, the first quantitative measurement of the Kramers turnover \cite{rondin2017direct}, experimental tests of fluctuation theorems \cite{gieseler2014dynamic, hoang2018experimental}, and fundamental tests of thermodynamic laws \cite{debiossac2019thermodynamics}.

\subsection{Dynamics in the underdamped regime}\label{sec:5.2:dynamics}

When transitioning from a viscous medium like water to a dilute gas or vacuum environment, the dynamics of a particle in an optical trap changes fundamentally because of the particle's finite inertia. Including the inertial term, the equation for the center-of-mass motion along a single spatial direction is (equation~\eqref{eq:8:langevin}) \begin{equation}\label{eq:eq_of_motion}
	\ddot{q} +\gamma_0\dot{q}+\Omega_0^2 \left(1 + \xi q^2\right) q = \mathcal{F}_L/m + F_\text{ext}/m,
\end{equation}
where $\Omega_0 = \sqrt{\kappa/m}$ is the resonance frequency, $\kappa$ is the stiffness of the optical trap, and $m$ is the mass of the particle. 
Under most experimental conditions, the fluctuating force $\mathcal{F}_{\rm L}$ and damping rate\footnote{Note that here we define $\gamma_0$ as a rate with units of Hz, in contrast to $\gamma = \gamma_0 m$ in equation~\eqref{eq:8:langevin}, which has units of [Hz kg].} $\gamma_0$ are due to collisions between the particle and residual gas molecules. We will come back to the stochastic forces and damping contributions in section \ref{sec:5.5:dissipation}.
$F_\text{ext}$ summarizes additional external forces, excluding the optical forces that generate the optical trap, which correspond to the last term on the left hand side. These can be used for example to drive the particle or to apply feedback cooling, as will be discussed in greater detail in section \ref{sec:5.7:feedback}.
The Duffing term $\xi \sim 1/ w_0^2$, where $w_0$ is the width of the optical trap, accounts for deviations from a parabolic trap \cite{gieseler2013thermal}.
The optical potential term also contains nonlinear terms that couple the three orthogonal spatial  modes nonlinearly \cite{gieseler2014nonlinear,gieseler2013thermal}.
However, these terms become only relevant for sufficiently large oscillation amplitudes and can in most situations be neglected, in particular when feedback cooling is applied.
This justifies the treatment of the three-dimensional motion of the trapped particle as three independent harmonic oscillators.
Note that, while the modes are uncoupled in a static trap, the two modes that are transverse to the propagation direction can be coupled by modulating the polarization of the trapping light in time. The coupling allows to coherently exchange energy between the modes, since this coupling can be turned on and off on demand \cite{frimmer2016cooling}.
Overall, this platform is very versatile because all system parameters can be controlled to a large degree in the experiment.

In the underdamped regime, which is defined by the condition $\gamma_0\ll \Omega_0$, the particle undergoes $\sim Q = \Omega_0 / \gamma_0$ phase coherent oscillations.
This becomes very clear in the PSD
\begin{equation}\label{eq:PSD_underdamped}
S_\text{qq}(\omega) = m^{-2}|\chi(\omega)|^2 S_{\rm ff} =  \frac{k_{\rm B} T}{m \pi}\frac{\gamma_0}{(\omega^2-\Omega_0^2)^2+\omega^2 \gamma_0^2},
\end{equation}
which in the underdamped regime is strongly peaked around the oscillation frequency $\Omega_0$. 
Here, $\chi(\omega) = 1/\left[\Omega_0^2-\omega^2+i\omega\gamma_0)\right]$ is the response function of a harmonic oscillator and we assumed thermal noise with a PSD $S_{\rm ff} = \gamma_0 m k_{\rm B} T / \pi$.
As we have already seen in section~\ref{sec:3.6:psd}, in the overdamped regime ($Q\ll 1$), equation~\eqref{eq:PSD_underdamped} can be approximated by a Lorentzian distribution
\begin{equation}
\label{eq:PSD_overdamped}
S_\text{qq}(\omega) \approx \frac{k_{\rm B} T}{2m \pi\Omega_0}\frac{\gamma_0/2}{(\omega-\Omega_0)^2+ (\gamma_0/2)^2}.
\end{equation}

\begin{figure}[ht]
	\begin{center}	
	 \includegraphics[width=12cm]{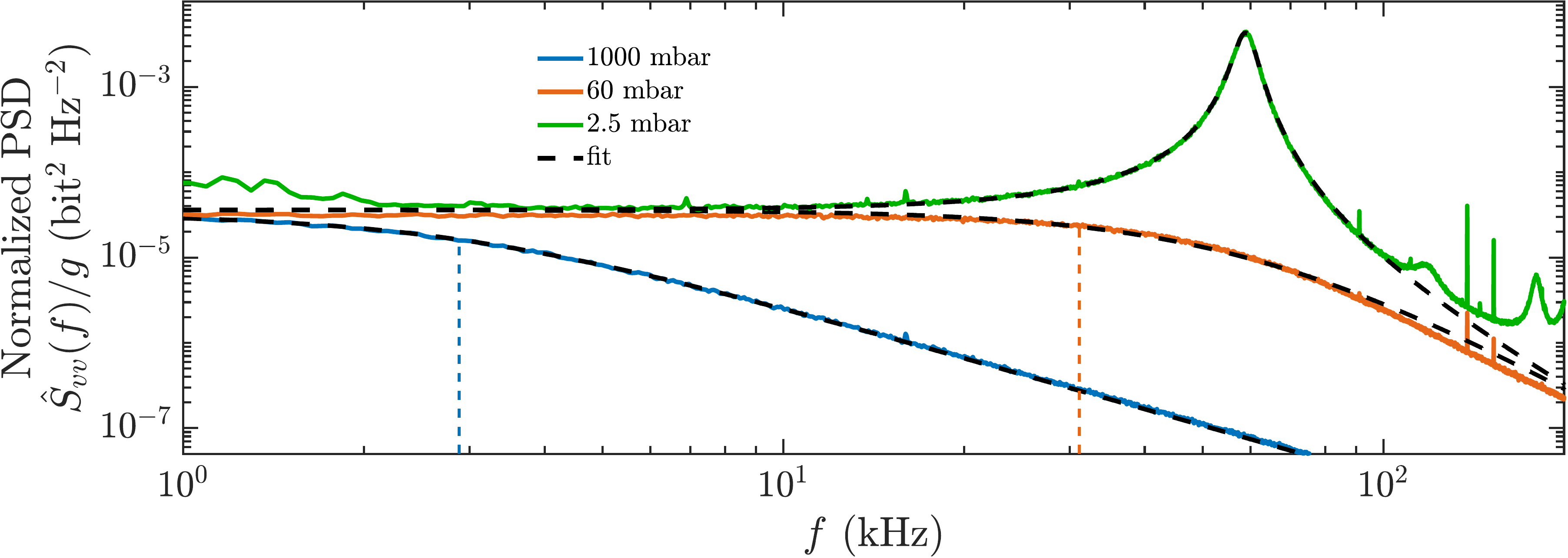}
	\caption{
	{\bf From overdamped to underdamped.}
 	We plot the measured PSD for the $z$-axis motion, normalized to the damping rate $\gamma_0/(2\pi)$, at three different pressures for a levitated particle (radius $a = 68\,{\rm nm}$, laser power $P \approx 150\,{\rm \rm mW}$) that is overdamped (blue line), critically damped (orange line), and underdamped (green line).
	The dashed black lines are least-square fits to equation~\eqref{eq:PSD_overdamped} ($1000\,{\rm mbar}$) or equation~\eqref{eq:PSD_underdamped} ($60$ and $2.5\,{\rm mbar}$).
	The colored vertical solid lines indicate the roll-off frequencies $\Omega_c/2\pi$.
	For the time traces at $60$ and $2.5\,{\rm mbar}$, the spectra contain leakage signals from the other oscillation axes above $100\,{\rm kHz}$.
	Data from Ref.~\cite{hebestreit2017thermal}.
	}
	\label{fig:35:overdamped_to_underdamped}
	\end{center}
\end{figure}

Fig.~\ref{fig:35:overdamped_to_underdamped} shows experimental data that clearly illustrate the transition from the overdamped to the underdamped regime.
In the experiment, the damping $\gamma_0$ is changed through the pressure inside the vacuum chamber.
At atmospheric pressures, the gas damps the particle motion ($\gamma_0 \gg 2\Omega_0$, blue) and we observe the spectrum familiar from optical tweezers experiments in liquids (section~\ref{sec:3.6:psd}), with the maximum at zero frequency.
At a pressure of $\sim 60\,{\rm mbar}$, the particle is critically damped ($\gamma_0 = 2\Omega_0$, orange).
At even lower pressures a strong peak evolves at $\Omega_0$, which is the signature of coherent oscillations.
The vertical lines indicate the cut-off frequency $\Omega_c = \Omega_0^2/\gamma_0$.

\subsection{Optical tweezers setup for vacuum operation} \label{sec5.3:setup}

\begin{figure}[hbt]
	\begin{center}
	\includegraphics[width=12cm]{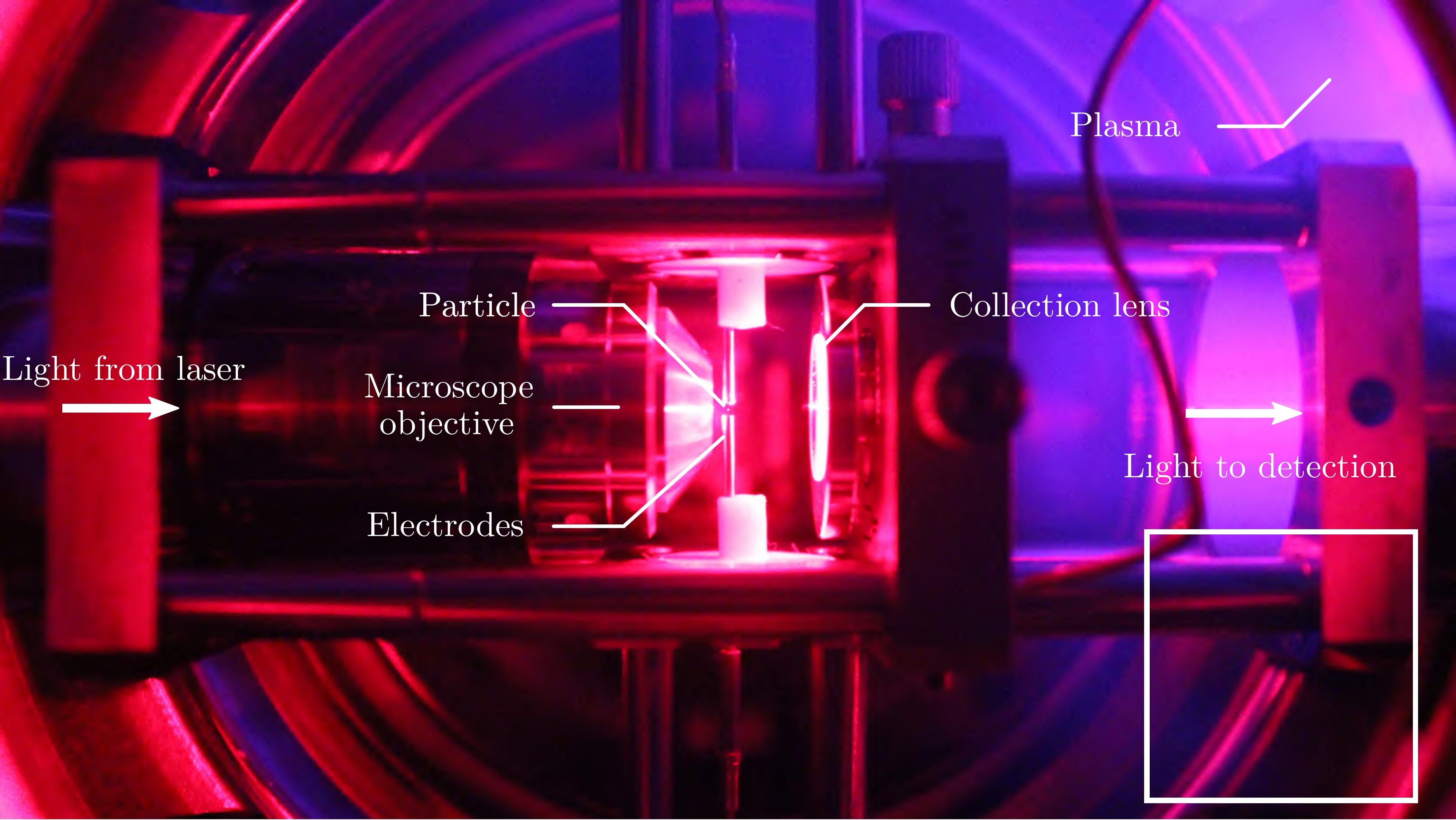}
	\caption{
	{\bf Optical trapping in vacuum.}
	Photo of the sample chamber of an optical levitation setup. An optically trapped particle is shown in the inset.
	Figure adapted from Ref.~\cite{ricci2019levitodynamics}.
	}
	\label{fig:36:Picture_Setup}
	\end{center} 
\end{figure}

\begin{figure}[hb]
	\begin{center}
	\includegraphics[width=12cm]{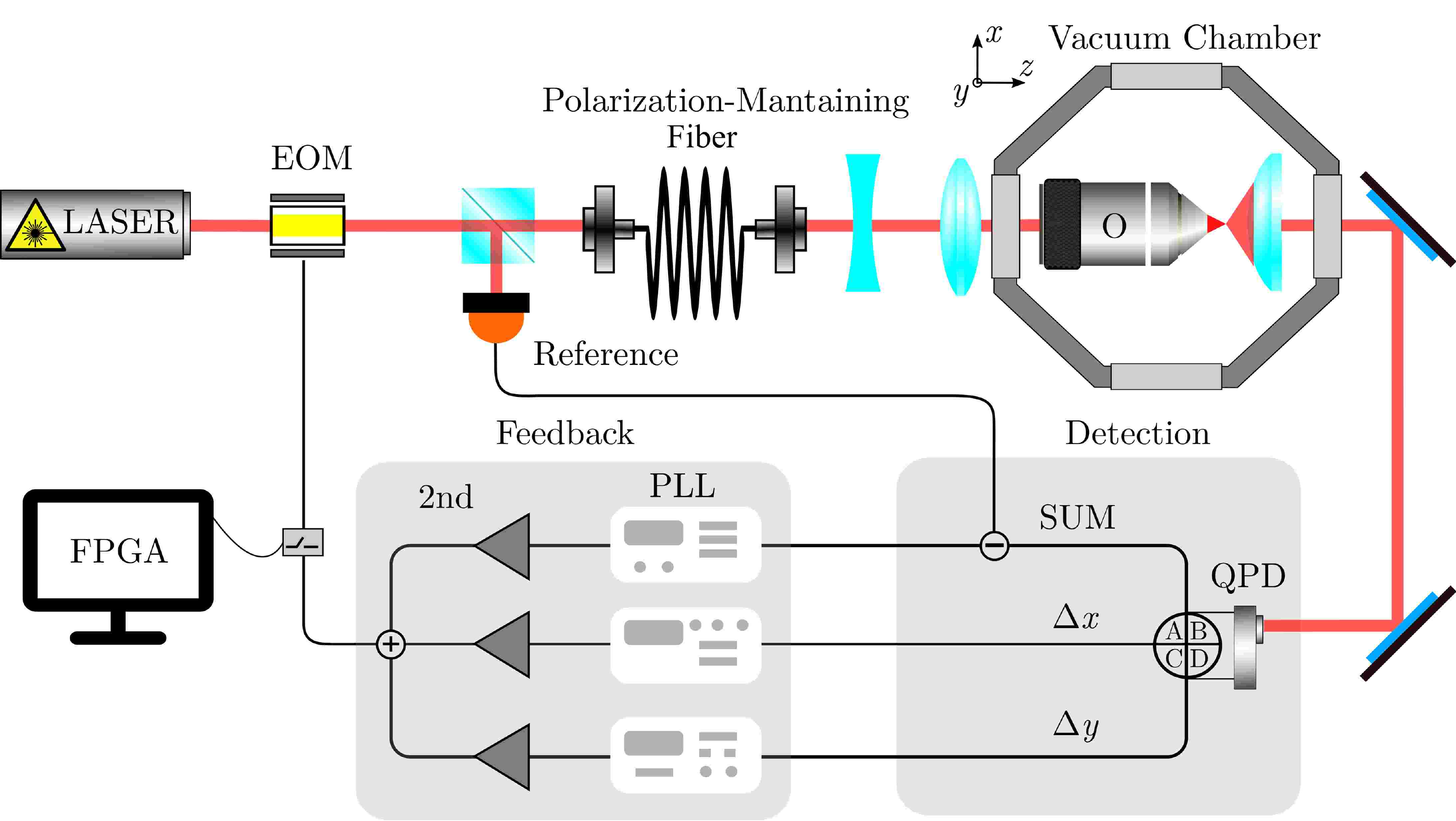}
	\caption{
	{\bf Simplified setup for optical trapping in vacuum.}
	Schematic of a setup for the trapping of particles in vacuum, based on the setup employed in Ref.~\cite{hebestreit2017thermal}.
	EOM: electro-optic modulator. O: objective. PLL: Phase-locked loop. QPD: quadrant photodetector.
	 }
	 \label{fig:37:setup}
	\end{center} 
\end{figure}

A standard setup for optical levitation in vacuum consists of a vacuum system that allows to control the pressure and thereby the friction the particle experiences, optics that strongly focus a laser beam for trapping, optical detection for real-time readout of the particle motion, and feedback control to stabilize the particle in high vacuum.
An example of a nanoparticle optically trapped in a vacuum chamber is shown in Fig.~\ref{fig:36:Picture_Setup}.
Operating an optical tweezers in vacuum introduces a range of significant changes to the experimental apparatus and also imposes limits to the type of particles that can be trapped.
Naturally, there are many ways in which such a setup can be constructed. Here, we follow largely the description given by Hebestreit \cite{hebestreit2017thermal}, a schematic picture of whose setup is shown in Fig.~\ref{fig:37:setup}.

\subsubsection{Particle types}

One of the great benefits of operating in vacuum is that the particle is very well isolated from the environment. However, this also means that the particle doesn't thermalize with the environment. As a consequence, optical absorption of the trapping laser significantly heats up the internal (bulk) temperature of the particle (section~\ref{section:internal_temperature}).
This limits vacuum trapping to particles that absorb very little of the trapping laser (e.g., silicon and silica) and to wavelengths where optical absorption is minimized (e.g., infrared, the most popular choices being $1064\,{\rm nm}$ and $1550\,{\rm nm}$).
There has also been great interest in trapping nanodiamonds, since these can host interesting optical defects such as NV-centers \cite{gruber1997scanning}; however, trapping in high vacuum has not been reported yet due to excessive optical absorption from impurities and defects in the diamond 
\cite{frangeskou2018pure}.

A liquid medium that surrounds the particle not only mitigates heating from optical absorption, it also keeps the particle afloat. Without it, gravity pulls the particles down to the bottom of the chamber, which complicates the loading of individual particles into the trap (section \ref{sec:5.4:particle_loading}).

In addition, without a high-refractive-index medium such as water ($n_{\rm m} = 1.33$), the refractive index contrast is much larger.
Consequently, the scattering force is more prominent in vacuum since the ratio of the scattering force to the gradient force scales as \cite{gieseler2014dynamics, spesyvtseva2016trapping}
\begin{equation}
  F_\text{scat} / F_\text{grad} \propto \alpha'' / \alpha'\propto (k a)^3 \Delta \epsilon
\end{equation}
where $\Delta \epsilon = (n_{\rm p}^2-n_{\rm m}^2) / (n_{\rm p}^2+2n_{\rm m}^2)$ is proportional to the relative index contrast between the particle ($n_{\rm p}$) and the surrounding medium ($n_{\rm m}$).
Therefore, trapping of spherical particles in vacuum with a single beam optical tweezers is limited to radii $a$ ranging from $\sim 30\,{\rm nm}$ to $\sim 150\,{\rm nm}$ \cite{gieseler2014dynamics}.
However, this is not a fundamental limitation, since the scattering force can be canceled, e.g., with counter-propagating beams \cite{divitt2015cancellation, li2010measurement}.

Also non-spherical particles (e.g., dumbells \cite{ahn2018optically}, clusters \cite{rashid2018precession}, and rods \cite{kuhn2017full}) have been successfully trapped in vacuum. They have quite different scattering properties and therefore experience different optical forces and torques than spherical particles.

\subsubsection{Optical trap}

The most popular choice for the trapping laser is infrared light at $1064\,{\rm nm}$ or at $1550\,{\rm nm}$.
In addition, the laser should have a small relative intensity noise (RIN), low pointing instability, and provide a power at the laser focus on the order of $P_{\rm opt} = 70\,{\rm mW}$, which requires a laser output power of $\approx 0.5\,{\rm W}$.
Possible back-reflections into the laser can destabilize the laser output. This can be avoided by introducing a Faraday isolator right after the laser output.
A low RIN is important because noise in the intensity at twice the particle oscillation frequency excites its center-of-mass motion \cite{gehm1998dynamics, gonzalez2019theory, jain2017levitated} and causes fluctuations of the oscillation frequency, since the particle frequency scales with $\Omega_0\propto \sqrt{P_{\rm opt}}$.

To parametrically drive the particle motion, the laser intensity has to be modulated at twice the oscillation frequency, i.e., at up to $400\,{\rm kHz}$ (section~\ref{sec:5.7:feedback}).  Electro-optic modulators (EOMs) or acousto-optic modulators (AOMs) provide the necessary modulation bandwidth.
While EOMs have a higher transmission (up to 95\%, in contrast to maximally 80\% for AOMs) and a higher bandwidth, they feature a much smaller contrast ratio than AOMs \cite{hebestreit2018sensing}.

The trapping laser can also be used to detect the particle motion even during parametric control since the parametric modulation is at a twice the motional frequency and can therefore be filtered out spectrally (section~\ref{sec:5.7.2:parametric_nonlinear_fb}).
However, to exclude any cross-talk, it is convenient (albeit not necessary) to have an independent measurement of the particle with a constant laser.
Interferences should be avoided, since they can lead to strong fluctuations of the optical trap.
This can be achieved with a single laser source that is divided into a high-power branch for trapping and a low-power branch for detection. A polarizing beam splitter (PBS) in combination with a half-wave plate provides a convenient way to cross-polarize and to adjust the power ratio between the two branches. In addition, one branch is frequency-shifted with an AOM, to avoid interference. A second PBS re-combines the two branches, after which they enter the microscope objective colinearly.

A stable three-dimensional trap with a single optical beam requires strong focusing. This can be achieved with a high-NA objective (${\rm NA} \geq 0.8$). Note that optical absorption in the objective can lead to small shifts of the focus \cite{hebestreit2017thermal}. This can heat the particle motion for large laser intensity modulation, for example in free-fall experiments \cite{hebestreit2018sensing}. 
Aspheric lenses provide a cheap alternative to more costly microscope objectives. In addition, they usually have a higher transmission and are available with anti-reflection coatings for a large range of wavelengths. This reduces the power requirements and absorption losses. However, they are more sensitive to misalignment than microscope objectives, which makes optical trapping with aspheric lenses much more challenging \cite{hebestreit2017thermal}.

Both AOMs and EOMs distort the laser beam as it passes through. 
To recover a nice Gaussian beam shape, the laser mode profile is cleaned with a spatial filter before it enters the microscope objective. This increases the quality of the focus and, hence, the trap stiffness.
Spatial filtering is achieved either with a pinhole of appropriate size, or more conveniently with a single-mode fiber, which has the advantage of mechanically decoupling different sections of the setup.
A polarization-maintaining single-mode fiber maintains the input polarization at the output, while a single-mode fiber with a fiber-based polarization controller allows to create an arbitrary output polarization.

A second microscope objective or aspheric lens inside the vacuum chamber collects and re-collimates the trapping laser and the measurement beam.  After the vacuum chamber, the two beams are separated again with another PBS. The trapping beam is dumped, while the probe beam is sent towards the detection setup. Efficient separation of the beams is paramount to avoid detecting the intensity modulation introduced by the EOM, that would then be deleteriously fed into the system via the feedback loop. 
In fact, mixed polarization states appear in this side of the set-up due to the tight focusing and especially to the mechanical stresses present in the glass windows of the vacuum chamber. 
To undo these changes in the  polarization state, one introduces a half waveplate and a quarter waveplate after the vacuum chamber and minimizes the unwanted trapping light  after the PBS while maximizing the probe light, which then goes to the detection setup.

\subsubsection{Detection}

In liquids, momentum relaxation occurs within a fraction of a microsecond due to the high viscosity of the surrounding medium. This is faster than what most experiments are able to resolve \cite{kheifets2014observation}.
In contrast, the viscosity of a dilute gas is orders of magnitude smaller. Hence, inertial effects on the dynamics of a particle in an optical tweezers are easily accessible \cite{li2010measurement}.
While optical tweezers experiments in liquids are mostly concerned with the diffusive motion at low frequencies on the order of the corner frequency $\omega_c / (2\pi) = k\left/(m \gamma_0)\right. \sim{\rm kHz}$, the frequency range of interest in experiments under vacuum is centered around the natural oscillation frequency $\Omega_0 / (2\pi) = \sqrt{\kappa/m}\sim 100 \rm kHz$, where $\kappa$ is the trap stiffness and $\gamma_0$ the damping coefficient.

As in the case of traditional optical tweezers, the particle motion is measured optically by extracting information contained in the field scattered by the particle.
In single-beam optical tweezers, the trapped particle are sub-wavelength scatterers.
As such, most of the information about the particle position is contained in the phase of the scattered field. This requires an interferometric measurement to read out the phase.

Forward detection provides a convenient way to access the phase, since the non-scattered transmitted light provides a stable reference field without requiring additional optical path stabilization \cite{rohrbach2002three}.
The signal-to-noise ratio of the detector signal is maximized by increasing the detection efficiency of the scattered photons and by increasing the optical power of the detection beam. Maximum power is provided by using the trapping beam itself, at the expense of introducing cross-talk as mentioned before. 
Hence, the best signal-to-noise ratio is obtained with a high-NA collection lens and by using all the scattered light to generate the detector signals.
In forward scattering, the collected power from the trapping beam amounts to $\approx 50\,{\rm mW}$.
Hence, optimal detection in forward scattering requires a quadrant photodetector that can handle at least $50\,{\rm mW}$ optical power, has a high quantum efficiency at the trapping wavelength, has a bandwidth of at least $1\,{\rm MHz}$, and is shot-noise limited at power levels above $10\,{\rm mW}$ per quadrant.
Details on a custom detector design that fulfills the above requirements can be found in Refs.~\cite{jain2017levitated, hebestreit2017thermal}.

Forward detection is attractive due to its simplicity. However, the ratio of scattered power to reference power is fixed by the scattering cross section of the particle.
Backward detection allows to collect only the scattered photons and add a reference field with variable power. 
The ultimate position sensitivity to particle motion that is transverse to the optical axis is the same for forward and backward detection. Interestingly, due to interference between incident and scattered photons this is not the case for motion along the optical axis. In the forward direction, destructive interference reduces the sensitivity, with the counterintuitive result that a higher NA collection lens can yield a lower sensitivity. In contrast, constructive interference in the backward direction allows for up to 60\% collection efficiency even with a moderate ${\rm NA}=0.8$. That is, more information about the particle position along the optical axis is contained in the backward scattered light than in the forward scattered light \cite{volpe2007backscattering, tebbenjohanns2019optimal}.

Interferometric detection provides the best signal-to-noise ratio. However, it is limited to a small range of particle motion. In contrast, video analysis allows to determine the particle position over a much larger range with a limited signal-to-noise ratio and it requires extremely fast cameras to resolve the relevant timescale $\sim {\rm \upmu s}$ in high vacuum \cite{svak2018transverse}.

\subsubsection{Feedback electronics}

Feedback cooling allows to extract energy from the center-of-mass motion and stabilizes the trap.
The feedback force can be applied either directly (e.g., with electric fields when the particle carries a charge) or parametrically (by modulation of the laser intensity).
The electronic circuit that processes the detector signal can be analog \cite{gieseler2014dynamics}, digital \cite{tebbenjohanns2019cold, conangla2019optimal}, or an off-the-shelf lock-in amplifier \cite{jain2016direct} (section~\ref{sec:5.7:feedback}).
Here, we focus on the latter following the description for parametric feedback based on a phase-locked loop (PLL) given by Hebestreit \cite{hebestreit2018sensing}.

The detector signal for each axis is fed to one of the PLLs. Each PLL detects the frequency and phase of the oscillation along one spatial dimension. The locking range of the PLLs is set to $4$ to $8\,{\rm kHz}$ around the expected center frequency to prevent the PLL from losing lock on the particle oscillation signal. The bandwidth of the loop filter is adjusted to be large enough to follow frequency changes sufficiently fast, and small enough to exclude noise from the feedback signal. In practice, this amounts to a bandwidth of $1$ to $16\,{\rm kHz}$ \cite{hebestreit2018sensing}.

The PLL synthesizes a signal that is phase-locked and at the same frequency or at a harmonic of the input signal with adjustable amplitude and phase. For parametric feedback, one uses the second harmonic and tunes the amplitude and phase to minimize the particle motion.
The independent signals for the three oscillation axes are added within the lock-in amplifiers to form the final feedback signal.
This feedback signal is applied to the EOM, which in turn modulates the power of the trapping beam.
The amplitude of the feedback signals is usually set to a few ${\rm mV}$ for each axis. For a half-wave voltage of $V_\pi = 240\,{\rm V}$ and amplifier with $400\,{\rm V/V}$ gain, the optical power is modulated by $2.5\%$ at a feedback signal amplitude of $10\,{\rm mV}$ \cite{hebestreit2018sensing}.

A switch allows to rapidly turn off the feedback signal. This allows to perform relaxation experiments \cite{gieseler2014dynamic}.
Additional signals can be added to the feedback signal.
This allows to stabilize this oscillation frequency by adjusting the power of the trapping beam with a PID (proportional-integral-derivative) control loop or to switch off the trap for short amounts of time for free-fall experiments \cite{hebestreit2018sensing}. 
A harmonic signal at twice the particle's motional frequency allows to excite the particle into the nonlinear region of the optical trap \cite{gieseler2014nonlinear, ricci2017optically}.

\subsubsection{Vacuum system}

\begin{figure}[hbt]
	\begin{center}
	\includegraphics[width=12cm]{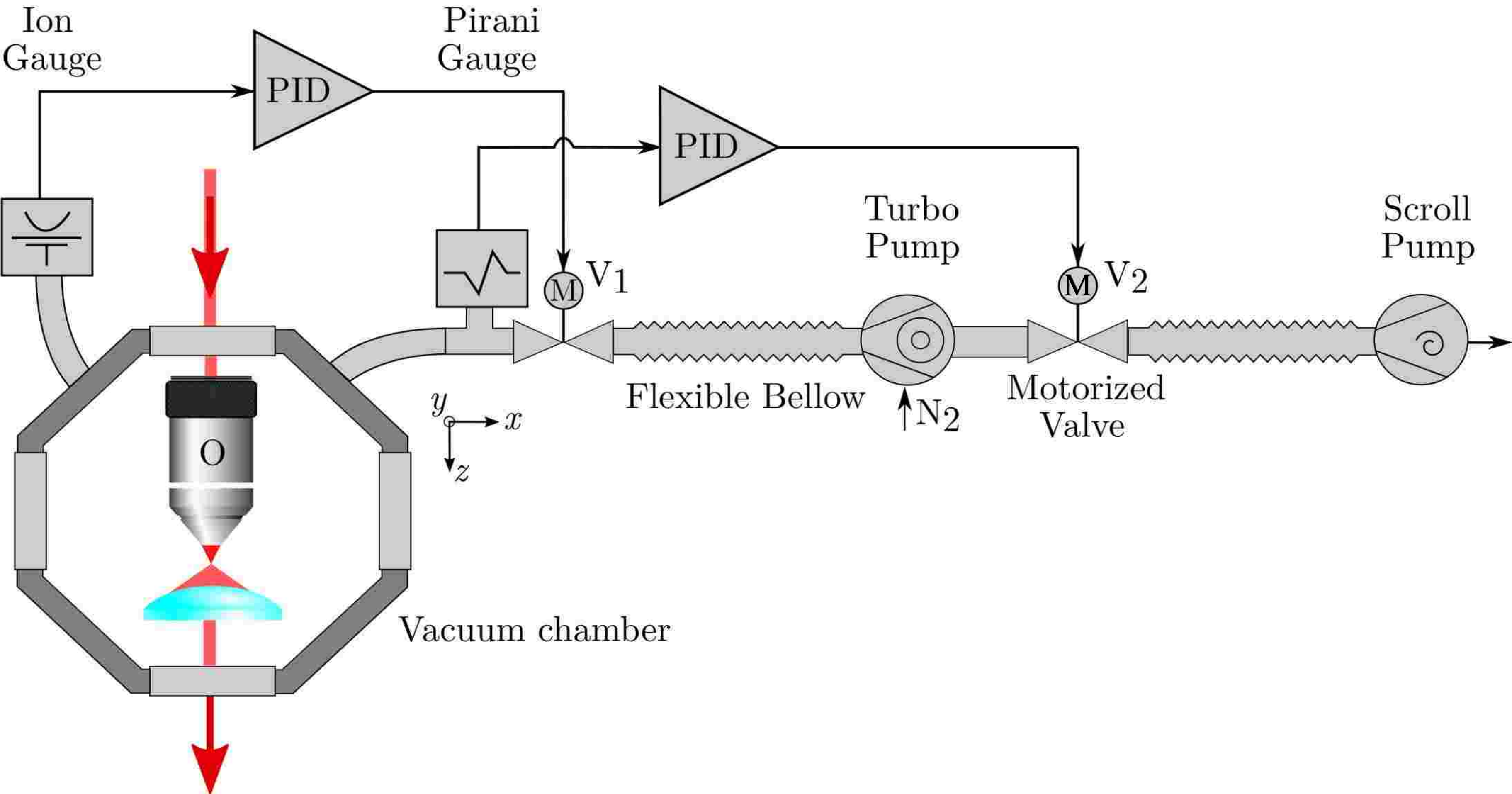}
	\caption{
	{\bf Vacuum system.}
	The vacuum system consists of a vacuum chamber with optical vacuum windows, in which the actual trapping takes place. Here, we present a system based on that employed in Ref.~\cite{hebestreit2017thermal}. The pressure is monitored using an ion and a Pirani gauge. A turbomolecular pump is connected to the chamber using a flexible bellow to suppress vibrations. A scroll pump supports the turbo pump. The pumping speeds can be regulated manually or with a PID controller using motorized valves (${\rm V_1}$ and ${\rm V_2}$). The chamber is vented with nitrogen to reduce contamination. 
	O: objective. PID: proportional integral derivative controller.
	}
	\label{fig:38:vacuum_species}
	\end{center}
\end{figure}

Optical levitation in high vacuum requires an experimental setup containing a vacuum system to control the gas pressure, optical access for the trapping and detection optics, and an access door on the top if particles are loaded at ambient pressures with a nebulizer (section~\ref{sec:5.4.1:nebuliser}).
A schematic of a vacuum system is shown in Fig.~\ref{fig:38:vacuum_species}.
The pressure in the chamber can be monitored continuously over the entire range from ambient pressure down to below $10^{-7}\,{\rm mbar}$ with a combination of gauges, e.g., a Pirani gauge for high pressure combined with a hot cathode filament gauge for low pressure.

A set of pumps and valves controls and stabilizes the pressure.
The main pump is a turbo-molecular pump. A motorized valve between the chamber and the turbo-molecular pump is used to stabilize the pressure below $10^{-3}$ mbar with a PID controller.
The turbo-molecular pump is backed by a dry scroll pump via another motorized valve that is mainly used to ensure a smooth evacuation of the chamber without sudden changes in the chamber pressure during the initial pump-down.
This is important to prevent loosing a trapped particle during pump-down from ambient pressure after
loading.
To avoid mechanical vibrations from the vacuum pumps to couple into the optical setup, the pumps are connected to the vacuum chamber through flexible bellows, which decouple the vibrations. Vibration isolation can be further improved by casting the bellows into a sand or concrete block.
At pressures below $10^{-6}$ mbar, ion pumps are sufficient to maintain a pressure without introducing vibrations.

Pressures below $10^{-6}\,{\rm mbar}$ require to bake the vacuum chamber to desorb residual water. 
Many experiments load particles at ambient pressure. Hence, the lowest pressures can be achieved if the chamber is baked \emph{in situ} while the particle is trapped.
Optical access while baking can be provided with heating bands that are wrapped around the vacuum chamber.
However, temperatures are limited to $\approx 150^\circ\rm{C}$ to avoid damaging the optical components inside the vacuum chamber and the optical coatings of the windows.
The baking process also causes thermal deformation of the optics inside the vacuum chamber.
This can then lead to particle loss during baking. 
Nonetheless, even if particles are not kept in the trap during bake out, baking the chamber in combination with purging with dry air (instead of venting with humid ambient air) reduces the overall water content in the vacuum chamber and leads to improved pump speeds and better vacuum.

\begin{figure}[hbt]
	\begin{center}
  	\includegraphics[width=12cm]{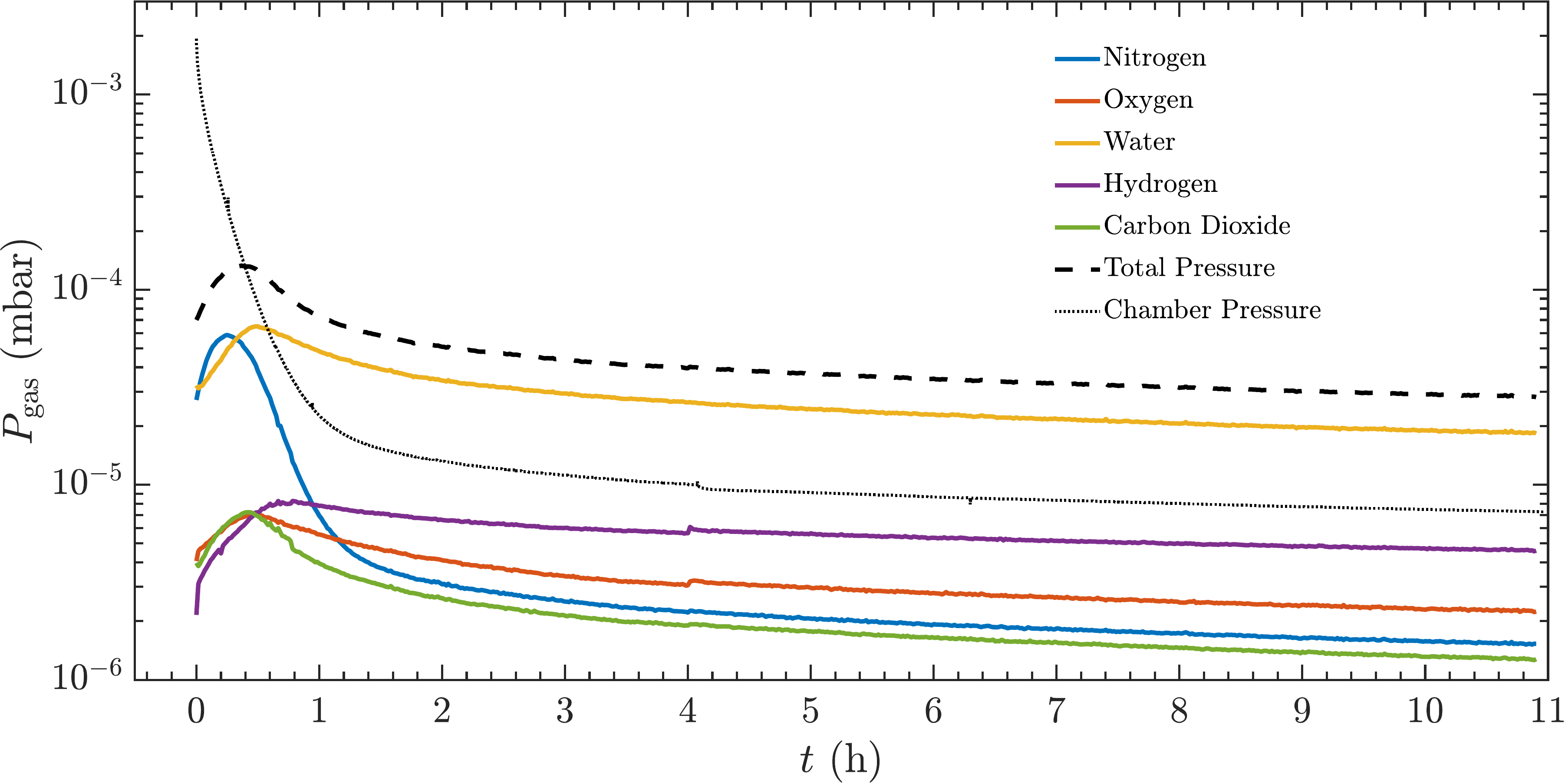}
  	\caption{
	{\bf Gas composition in a vaccum chamber.}
	Gas composition in the vacuum chamber at pressures below $10^{-3}\,{\rm mbar}$ plotted over time. 
	The solid lines show the partial pressures of different gas species measured with a residual gas analyzer (RGA). 
	The sum of all the partial pressures yields the total pressure at the RGA (dashed line).
	Due to the configuration of the vacuum system and the reduced pumping speed at the RGA, the pressure at the RGA deviates from the pressure in the main vacuum chamber (dash-dot line).
	The initial rise in partial pressures after turning on the RGA is attributed to the warm-up process and desorption of gases from the filament of the RGA. 
	Reproduced from Ref.~\cite{hebestreit2018sensing}.
	}
	\label{fig:39:vacuum_species}
	\end{center}
\end{figure}

Hebestreit \cite{hebestreit2018sensing} quantified the gas composition in the vacuum chamber during a trapping experiment with a residual gas analyzer (RGA). 
The RGA is a mass spectrometer that allows to measure the mass spectrum of a gas and thus to extract the partial pressures of the different gases species.
The measured partial pressures of the prominent gas species during a typical pump-down cycle are shown in Fig.~\ref{fig:39:vacuum_species}.
The total pressure at the RGA (the sum of all the partial pressures, dashed line) is higher than the pressure measured in the main vacuum chamber with an ion gauge (dashed line), because the pumping speed at the RGA is reduced compared to the main chamber.
Initially, the chamber is filled with nitrogen. 
Interestingly, at low pressures the composition is vastly different.
The turbo molecular-pump removes the initial nitrogen very efficiently. Thus, its partial pressure reduces quickly.
In contrast, water and hydrogen are continuously desorbed from the walls of the vacuum chamber.
As a consequence, the gas in the chamber is dominated by water vapor at pressures below $10^{-4}\,{\rm mbar}$.
Baking out the vacuum chamber accelerates the desorption process and thereby helps to get the water and hydrogen out of the chamber.
The final gas composition in the pressure range of $10^{-6}$ to $10^{-5}\,{\rm mbar}$ is mainly water vapor  ($58\pm9\%$), nitrogen ($19\pm8\%$), hydrogen ($12\pm1\%$), oxygen ($7\pm1\%$), and carbon dioxide ($4\pm1\%$).
Other gases, which contribute with less than 1\% are argon, isopropyl alcohol, and oil. The last two probably remain from the solvent of the particle solution and residual machining grease, respectively.
As a consequence, the molar mass $M\approx 20\pm11 \cdot 10^{-3}\,{\rm kg\,mol^{-1}}$ at low pressure is significantly lower than the molar mass at high pressures, the latter being equal to the molar mass of dry air $M_\text{air} =28.97\cdot 10^{-3}\,{\rm kg\,mol^{-1}}$ .

\subsection{Particle loading}\label{sec:5.4:particle_loading}

In contrast to a liquid environment, loading a particle into an optical trap in a dilute gas or vacuum is challenging because the particles will quickly fall down due to gravity.
The size of an optical trap is rather small, on the order of the focal volume $\sim \lambda^3$.
Therefore, a particle that passes by the focus at a distance larger than $\sim \lambda$ will not be captured.
Furthermore, if a particle enters the focal volume at high speed, the low damping in vacuum will not slow the particle down sufficiently  for it to get trapped.
The maximum speed for successful trapping is $v_{\rm max}\sim \lambda \gamma_0$, where $\gamma_0$ is the damping constant.
In water, the viscosity at room temperature is $\eta^{(\rm water)}=890\, {\rm \upmu Pa\,s}$ so that, using Stokes formula (equation~\eqref{eq:stokes}), $\gamma_0 = 6\pi\eta  a/m$ and, therefore, $v_{\rm max}^{(\rm water)}=344\,{\rm m\,s^{-1}}$ for a silica particle of radius $a=75\,{\rm nm}$.
In air, the viscosity is about two orders of magnitude smaller ($\eta ^{(\rm air)}=18\,{\rm \upmu Pa\, s}$) so that $v_{\rm max}^{(\rm air)}=7\,{\rm m\,s^{-1}}$.
The challenge lies, therefore, in finding a technique by which a slow single nanoparticle is brought into the focal volume \cite{gieseler2014dynamics}.

\subsubsection{Nebuliser}\label{sec:5.4.1:nebuliser}

The most common approach to particle loading is the \emph{nebulizer approach} thanks to its simplicity \cite{summers2008trapping}.
In this approach, a solution containing silica beads (e.g., Microparticles GmbH, Bangs Laboratories)  with radii ranging from $a=50$ to $100\,{\rm nm}$  is sprayed into the open vacuum chamber with a nebulizer (e.g., Omron NE-U22-W) and the particles are trapped at ambient pressure \cite{gieseler2014dynamics} or mild vacuum \cite{delic2019cavityphd}.

\begin{figure}[hbt]
	\begin{center}	
	\includegraphics[width=12cm]{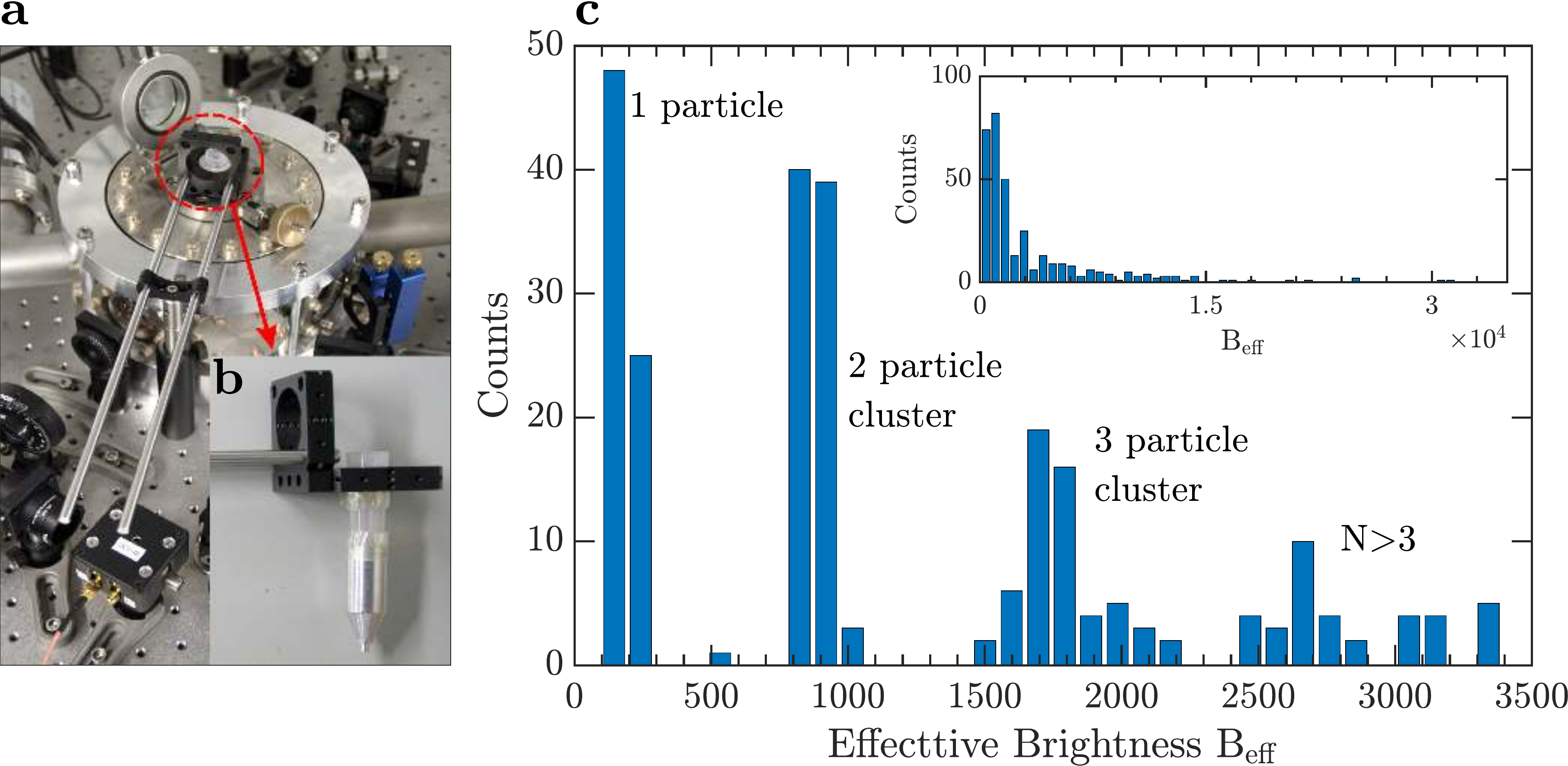}
	\caption{
	{\bf Particle loading with a nebulizer.}
	The particle is loaded by spraying a solution
of nanoparticles through a nozzle which is placed above the focus.
	(a) Positioning of the nozzle in the vacuum chamber.
	(b) Nozzle to funnel the falling particles towards the focus of the trapping laser. 
	(c) Histogram of brightness observed with a camera from the side. 
	The inset shows the brightness over a wider range.
	Reproduced from Ref.~\cite{gieseler2014dynamics} (a,b) and adapted from Ref.~\cite{ricci2019levitodynamics} (c).
	}
	\label{fig:40:nebulizer}
	\end{center}
\end{figure}

The nebulizer consists essentially of a mesh on top of a piezo element. A little bit of liquid gets into the space between the piezo and the mesh. The motion of the piezo then pushes the liquid through the mesh. The mesh breaks the liquid into little droplets of diameter $~2\,{\rm \upmu m}$ and a nozzle funnels the falling particles to a region near the laser focus (Figs.~\ref{fig:40:nebulizer}(a-b)).

Successful trapping of a particle can be easily noticed observing the laser light scattered by the particle with a camera. However, since the scattered light has a dipole radiation pattern, the polarization of the laser beam should be orthogonal to the direction of observation to see the particle with the camera.
The scattered light intensity on the camera scales with the size of the particle as $B\propto a^6$. This allows to distinguish whether the object in the trap is a single nanoparticle or a cluster of multiple particles by comparing the brightness on the camera for several trapping events. For example, in Fig.~\ref{fig:40:nebulizer}(c), the histogram of the brightness on the camera for $\sim 350$ particles shows clearly distinct peaks, which can be associated with the number of particles in the trap \cite{ricci2019levitodynamics}.

Most of the liquid in the droplet evaporates before or immediately after trapping. However, $\sim 10$ layers of water remain physically absorbed on the surface of the particle.
These physically absorbed layers evaporate as the gas pressure is reduced down to $\sim 1\,{\rm mbar}$. This dehydration process is reversible and water molecules get reabsorbed when the pressure is increased.
In the region from $0.1$ to $0.5\,{\rm mbar}$, one can often observe a sudden decrease in the mechanical frequency and power of the scattered light.
This is attributed to a sudden loss of the remaining water from the particle as its internal temperature increases to above $\sim 200^\circ{\rm C}$. This effect occurs only once and is \emph{irreversible} when cycling the pressure. 
Therefore, once the particle has lost the final water layer its water uptake is extremely slow and for all practical purposes the particle size and material composition can be considered pressure independent.
For more details on the dehydration process see Refs.~\cite{ricci2019levitodynamics, delic2019cavity, delic2019cavityphd}.

\subsubsection{Dry approaches}

Spraying solvents into a vacuum chamber introduces contaminants that are difficult to pump out, e.g., water molecules.
This motivates a dry approach, where particles are released from a substrate instead of being injected from a solution. 
However, small particles stick to surfaces due to dipole-dipole interactions known as Van der Waals-forces, which scale linearly with particle size; the magnitude of these forces is 
\begin{equation}
F_{\rm VdW} = 4\pi a \gamma_{\rm s},
\end{equation}
where $\gamma_{\rm s}$ is the effective surface energy.
For example, measurements on silica spheres on glass give $F_{\rm VdW}=176 \,{\rm nN}$ for $a=1\,{\rm \upmu m}$ \cite{heim1999adhesion}.
For comparison, the gravitational force for a particle of this size is only $\sim 0.1\,{\rm pN}$ and, therefore, much too weak to remove the particle.
There can also be additional surface forces that lead to even stronger adhesion, for example meniscus forces arising from a thin water layer that is formed on the substrate.

Therefore, even larger forces have to be applied to release the particle from the substrate.
Two common approaches, which provide dry loading methods and do not require to open the vacuum chamber as in the nebulizer approach, are based on launching particles from surfaces using piezoelectric transducers or laser-induced acoustic desorption (LIAD).

A piezoelectric transducer works really well for micrometer-sized particles \cite{ashkin1971optical, li2012fundamental} and allows even for repeated loading of the same particle \cite{park2016contact}.
When the piezoelectric transducer is driven with oscillation amplitude $q_{\rm p}$ and frequency  $\omega_{\rm p}$,  the particle feels a force
\begin{equation}
F_{\rm piezo} =  m\omega_{\rm p}^2 q_{\rm p},
\end{equation}
due to its inertia, where $m=V \rho_{\rm p}$ is the particle mass with $\rho_{\rm p}$ and $V$ the particle density and volume, respectively.
Consequently, the acceleration required to release the particle is
\begin{equation}
a_{\rm p}=\omega_{\rm p}^2 q_{\rm p}=4\pi a \gamma_{\rm s}/m\propto a^{-2}.
\end{equation}
Hence, for a given maximum acceleration the piezoelectric transducer can provide, there is a minimum particle size that can be shaken off. The smallest particles that have been released with this method have a radius of $150\,{\rm nm}$ and further details on how to implement this can be found in Ref.~\cite{atherton2015sensitive}.

The limited acceleration of the piezo approach can be overcome by generating a shockwave on the surface of the substrate in the LIAD approach. 
The shockwave can be created by a pulsed frequency-doubled Nd:YAG laser ($5\,{\rm ns}$ pulses of $3\,{\rm mJ}$ at $532\,{\rm nm}$).
When focused down to $200\,{\rm \upmu m}$ on the back side of a thin metallic foil, the laser creates an acoustic wave that launches particles from the front side, which had been previously spin coated with particles \cite{asenbaum2013cavity}.

Despite the availability of these techniques, directly loading particles into an optical trap under high vacuum conditions remains an open challenge. 
With the dry approaches outlined above, particles are still loaded into the trap at ambient pressures or moderate vacuum, where the background gas provides sufficient dissipation because capture requires an additional dissipation mechanism due to the conservative nature of gradient trapping forces and the small size of the trapping region of an optical tweezers.
Recently, it was demonstrated that high-vacuum loading into a Paul trap can be achieved \cite{bykov2019direct}. The Paul trap provides a much larger trapping volume and trap depth than an optical tweezers. This allows to turn on the trap exactly when the particle is near the center of the trap. As a consequence, the particle's total energy is given only by its kinetic energy. If the kinetic energy is less than the trap depth, the particle will be stably trapped.
Paul traps are limited to charged particles; however, this is not a strong limitation for loading since the particles generally carry several tens of elementary charges (section~\ref{sec:5.6.2:calibration_charge}).

Transferring particles from a contaminated chamber, where the particle is initially trapped, to a clean science chamber, which always remains in high-vacuum conditions, can be achieved with a mobile optical trap (MobOT) in a load-lock \cite{mestres2015cooling}. However, this solution is rather bulky and has a limited throughput at low pressures. 
Hollow-core photonic crystal fibers provide also a compact solution with high throughput, but have been limited so far to pressures above $10^{-2}\,{\rm mbar}$ \cite{bykov2015flying, grass2016optical}.

\subsection{Dissipation and noise}\label{sec:5.5:dissipation}

A particularly attractive feature of trapping in a dilute or vacuum environment is that the interaction with the environment can be adjusted over several orders of magnitude. The interactions between the particle and its environment result in stochastic forces that excite the particle motion and lead to stochastic dynamics, which are of particular interest when studying thermodynamics of individual small particles and are to be minimized when aiming for the quantum regime.

The intensity of a stochastic force is characterized by its PSD $S_{\rm ff}^{\rm noise}$.
While the stochastic forces excite the center-of-mass motion of the trapped particle, the interactions with the environment also lead to dissipation at a rate $\gamma^\text{noise}$, which damps the particle motion.
There are several sources that contribute to the total damping rate and stochastic noise intensity, which we will discuss in this section.

The total damping rate and stochastic noise intensity is simply given by the sum of the individual contributions $\gamma_0 = \sum_\text{noise}\gamma^\text{noise}$ and $S_{\rm ff} = \sum_\text{noise}S_{\rm ff}^{\rm noise}$, respectively.
Note that for most purposes, the PSD can be considered as frequency independent (white noise) over the bandwidth of the mechanical resonance of the particle, especially when the bandwidth becomes quite narrow in high vacuum.
The average energy of each center-of-mass degree of freedom evolves as 
$\dot{\langle E\rangle} = -\gamma_0 \langle E\rangle+ \Gamma$
where the heating power due to the force noise is $\Gamma = \pi S_{\rm ff}/m$, $S_{\rm ff}$ being the \emph{double sided} PSD.
After a time $\approx 1/\gamma_0$, the particle reaches an  equilibrium, where the ensemble average energy stays constant.
This allows to define the effective temperature through the fluctuation-dissipation relation:
\begin{equation}
\label{eq:temp_def}  
T_{\rm cm} =  k_{\rm B}^{-1}\frac{\Gamma}{\gamma_0} = \frac{\pi S_{\rm ff}{} }{k_\text{B} m \gamma_0 }.
\end{equation}
The effective temperature corresponds to the particle energy averaged over an ensemble of identical experiments.
Importantly, since we are concerned here with the energy of a single particle, the energy in each experiment will still fluctuate with a standard deviation $\sim T_{\rm cm}$.
Generally, one has to be careful to define a temperature for a system out of equilibrium \cite{casas2003temperature}. However, the situation we present here is somewhat simplified because it is steady-state and for many practical purposes the effective bath model that is characterized by an effective damping/temperature gives a good description. However, one can also create situations where this is no longer true. For example, by parametric feedback damping, the temperature alone is not sufficient to give a full description of the bath (section~\ref{sec:5.7:feedback}) 

For a levitated particle in a dilute gas, the main contribution to the stochastic force noise are collisions with air molecules and radiation damping at pressures below $\sim 10^{-7}\,{\rm mbar}$. 
In addition, there are technical noise sources that do not pose a fundamental limitation but can impose practical limitations. These are primarily displacement noise of the trap and laser intensity fluctuations.
Noise sources can also be introduced intentionally to emulate a different thermal environment without actually changing the real temperature of the environment \cite{mestres2014realization, martinez2013effective}. This can be a useful tool for single particle thermodynamics experiments \cite{gieseler2018levitated}.

\subsubsection{Gas damping}

From kinetic gas theory, the pressure-dependent gas damping for a spherical particle is \cite{beresnev1990motion, gieseler2014dynamics}
\begin{align}
    \frac{\gamma_{\rm gas}}{2\pi} &=
 \frac{3 a \eta_\nu}{m} \frac{0.619}{0.619+{\rm Kn}}(1+c_{\rm K}),
\end{align}
where ${\rm Kn} = \bar{l}/a$ is the Knudsen number, $c_{\rm K} = 0.31{\rm Kn}/\left(0.785+1.152{\rm Kn}+{\rm Kn}^2\right)$,  $\bar{l} = \sqrt{\frac{\pi k_{\rm B} T}{2m_{\rm gas}}}\frac{\eta_\nu}{P_{\rm gas}}$ is the mean free path, $\eta_v$ is the viscosity, and $m_{\rm gas} = M/N_{\rm A}$ the molecule mass \cite{beresnev1990motion},  $N_{\rm A}$ being Avogadro's number.  For dry air, the molar mass $M = 28.97 \cdot 10^{-3}\,{\rm kg\,mol^{-1}}$, however the actual gas composition can be different (Fig.~\ref{fig:39:vacuum_species}).
For high pressure, the interaction with the gas is so strong that the particle motion is heavily damped and its internal temperature $T_{\rm bulk}$ and centre-of-mass temperature $T_{\rm cm}$ quickly thermalize with the gas temperature $T_{\rm gas}$. In this regime, the damping does not depend on  pressure and is given by Stokes law, i.e., $\gamma_{\rm gas}/(2\pi) \approx 3 a \eta_v/m$.

For pressures below $P_{\rm gas}' \approx 54.4\,{\rm mbar}\, (a/{\rm\upmu m})^{-1}$, where the mean free path $\bar{l}$ is much larger than the radius of the particle $a$, the damping is linear in the pressure $P_{\rm gas}$ and given by 
\begin{equation}\label{eq:gas_damping_lin}
 \frac{\gamma_{\rm gas}}{2\pi}   \approx 0.354\sqrt{\frac{m_{\rm gas}}{k_\text{B} T_{\rm gas} }}\frac{P_{\rm gas}}{\rho_{\rm p} a}.
\end{equation}
In typical experiments, $\gamma_{\rm gas}/ 2\pi = c_{\rm P} P_{\rm gas}/a$ with $c_{\rm P} \approx 50\,{\rm Hz \, (\upmu m/mbar)}$.
Note that the above holds for a spherical particle. For an anisotropic particle (e.g., a rod), the friction term is different along each of the axes, and depends upon the alignment of the particle \cite{martinetz2018gas}.

The mean free path of the gas molecules increases with decreasing pressure (e.g., $\bar{l}\sim 60\,{\rm \upmu m}$ at $1\,{\rm mbar}$). As a consequence, the particle no longer thermalizes with the gas since the impinging gas molecules no longer carry away enough thermal power to balance the optical absorption from the trapping laser. Due to the increased internal temperature $T_{\rm bulk}$ of the particle, the average energy of the gas molecules after a collision with the particle increases (section \ref{section:internal_temperature}).
The latter requires that the impinging molecules stick to the surface for a short time before they emerge again from the surface.
This type of collision is called \emph{diffusive reflection} \cite{millen2014nanoscale, epstein1924resistance}.
For diffusive reflection, the fluctuating forces originating from the impinging gas molecules are completely uncorrelated. This justifies the treatment of the impinging and the emerging gas molecules as two independent baths with individual bath temperatures, damping and heating rates.
The process by which a surface exchanges thermal energy with a gas is called \emph{accommodation}, which is characterized by the accommodation coefficient 
\begin{equation}\label{eq:accomdation_coeff}
  c_{\rm acc} 
  = 
  {
  T_{\rm em} - T_{\rm gas} 
  \over 
  T_{\rm bulk} - T_{\rm gas} 
  },
\end{equation}
where $T_{\rm em}$ is the temperature of the gas molecules emitted from the particle surface.
The accommodation coefficient quantifies the fraction of the thermal energy that the colliding gas molecule removes from the surface, such that $c_{\rm acc} =1$ means that the molecule fully thermalizes with the surface.
We do not distinguish between the  bulk and the surface temperature, since we assume that they are identical for nanoparticles.
Since the mean free path in a dilute gas is long, one can safely assume that an emitted molecule will not interact again with the particle before thermalizing with the environment.
Consequently, we can consider the particles that impinge on the particle surface and those that leave it as being in equilibrium with two independent baths \cite{millen2014nanoscale}.

The damping due to the impinging molecules is 
\begin{equation}\label{eq:gamma_im}
	\gamma_{\rm im} 
	= 
	\sqrt{
		\frac{
			8 m_{\rm gas}
		}{
			\pi k_{\rm B} T_{\rm gas}
		}
	} 
	\frac{P_{\rm gas} }{\rho_{\rm p} a }
\end{equation}
and the contribution from the emerging molecules is 
\begin{equation}
	\gamma_{\rm em} 
	= 
	{\pi \over 8}
	\sqrt{T_{\rm em} \over T_{\rm gas}}
	\gamma_{\rm im}.
\end{equation}
The total gas damping is then 
\begin{equation}\label{eq:damping_linear_hot}
	\frac{\gamma_{\rm gas}}{2\pi} 
	= 
	\frac{\gamma_{\rm im}}{2\pi} 
	+ 
	\frac{\gamma_{\rm em} }{2\pi}
\end{equation}
and the total force noise due to the gas is 
\begin{equation}\label{eq:force_noise_gas}
	S_{\rm ff}^{\rm gas} 
	= 
	\frac{m k_{\rm B}}{\pi} 
	\left(
		\gamma_{\rm im}T_{\rm gas}
		+
		\gamma_{\rm em}T_{\rm em}
	\right).
\end{equation}
When the temperature of the emerging molecules equals the gas temperature (i.e., $T_{\rm em} =T_{\rm gas}$), equation~\eqref{eq:damping_linear_hot} reduces to equation~\eqref{eq:gas_damping_lin}.

\subsubsection{Radiation damping and shot noise}

Once the pressure is low enough (i.e., $\leq 10^{-7}\,{\rm mbar}$), gas damping is negligible and the particulate nature of the light field becomes the primary source of dissipation, in the absence of technical noise sources such as laser phase and amplitude fluctuations \cite{jain2016direct}.
Photons arrive at discrete times, where the number of photons arriving per time interval $\Delta t$ is given by $\sqrt{\Delta t P_{\rm opt}\left/(\hbar \omega_{\rm opt}) \right.} $. The recoil from the fluctuating number of photons impinging on the nanoparticle can be modeled as an effective bath with the characteristics \cite{tebbenjohanns2019optimal, novotny2017radiation}
\begin{equation}\label{eq:radiation_damping}
	\frac{\gamma_\text{rad}}{2\pi} 
	= 
	c_{\rm dp}
	\frac{
		P_{\rm scatt}
	}{
		2\pi m c^2
	}
\quad{\rm and}\quad
	S_{\rm ff}^{\rm rad} 
	= c_{\rm dp} 
	\frac{
		\hbar \omega_{\rm L} P_{\rm scatt}
	}{
		2\pi c^2
	},
\end{equation}
where $c_{\rm dp}$ depends on the direction of motion of the particle with respect to the polarization of the laser (in particular, $c_{\rm dp}=1/5$ for motion along the direction of polarization, $c_{\rm dp}=2/5$ for motion perpendicular to the polarization, and $c_{\rm dp}=(2/5+A^2)$ along the direction of propagation, where $A$ is the radiation pressure contribution defined in Ref.~\cite{tebbenjohanns2019optimal}). The scattered power is $P_{\rm scatt} = \sigma_{\rm scatt} I_{\rm opt}$, where $\sigma_{\rm scatt} = |\alpha|^2k_{\rm opt}^4/(6\pi\epsilon_0^2)$ and $I_{\rm opt}$ is the laser intensity. The effective temperature of this bath can be calculated via equation~\eqref{eq:temp_def} and is $\sim \hbar \omega_{\rm L} / k_{\rm B}$.

\subsubsection{Displacement noise}

In the reference frame of the trap (i.e., where the trap appears stationary), acceleration of the trap center $a_\text{dis}$ appears as a force  $F_\text{dis} = m a_\text{dis}$ due to the particle's inertia $m$.
Therefore, fluctuations of the trap center with displacement spectral density $S_\text{dd}$ lead to the effective force spectral density 
\begin{equation}
	S_{\rm ff}^\text{dis} (\omega)
	= 
	m^2 \omega^4 S_\text{dd} (\omega),
\end{equation}
where we have used the fact that the acceleration spectral density $S_{\rm aa}(\omega)=\omega^4 S_{\rm dd}(\omega)$. 
There is no damping associated with displacement noise, since the source of the displacement noise is usually some active component like a pump.
Unlike the thermal force spectral density, the displacement noise spectral density is not flat but increases for lower frequencies. However, for the typical narrow resonances at $\Omega_0$ in optical levitation experiments, we can approximate it as white noise, $S_{\rm ff}^\text{dis} \approx m^2 \Omega_0^4 S_\text{dd}(\Omega_0)$.
The thermomechanical noise limited  acceleration sensitivity $S_{\rm aa}=k_{\rm B} T\gamma_0 / (\pi m)$ depends inversely on the particle mass. 
Hence, heavy particles make particularly good accelerometers \cite{monteiro2017optical}.
However, since the maximal trap stiffness is limited, heavier particles tend to have lower frequencies at which displacement noise dominates over thermomechanical noise. At the optimal particle size both noise sources contribute equally.

Displacement noise is a technical noise source and as such not a fundamental limitation. It is primarily noticeable in setups where the external forces change considerably over the range of movement of the trap center. That is in situations where the force has a strong spatial dependence, for example when the trapped particle interacts with a standing wave of an optical cavity or with the evanescent field from a tapered fiber \cite{gonzalez2019theory}. If the trap position changes randomly with respect to the cavity or the fiber, the particle experiences fluctuating optical forces from the cavity or evanescent field, respectively.

\subsubsection{Laser intensity noise}

Laser intensity fluctuations are another technical noise source \cite{gehm1998dynamics, gonzalez2019theory}.
Low frequency drifts of the intensity randomly change the particle's oscillation frequency, since the particle frequency scales with $\Omega_0\propto \sqrt{P_{\rm opt}}$. This can be a serious limitation in force-sensing experiments, where an external force is transduced into an oscillation amplitude. If the frequency changes, the particle's response to an external force changes even if this force is constant, resulting in a noisy measurement.

Intensity fluctuations at twice the motional frequency $2\Omega_0$ parametrically excite the particle and thereby raise the particle's energy.
If the phase of the particle oscillation is not actively stabilized to the parametric modulation, it self-locks (entrains) to the external modulation \cite{gieseler2014nonlinear}. Hence, an intensity modulation at $2\Omega_0$ without active stabilization always leads to heating as will be discussed in section~\ref{sec:5.7:feedback}.
Parametric excitation is a nonlinear process and its strength depends on the motional energy of the particle and therefore can be suppressed by reducing the particle's energy with feedback cooling.
In free-space experiments, laser-intensity noise can be reduced sufficiently such that it is  smaller than the fundamental photon shot noise \cite{jain2016direct}.
In cavity experiments, however, laser-phase noise is transduced into intensity noise and  has been identified as a major obstacle towards reaching the ground state with sideband cooling \cite{meyer2019resolved, aspelmeyer2012quantum}.

\subsection{Calibration}\label{sec:5.6:calibration}

Precise calibration is as important for quantitative measurements in the underdamped regime as it was in the overdamped regime (section~\ref{sec:3:calibration}).
As before, the thermal energy $k_{\rm B} T$ serves as a reference to calibrate the detector signal (section~\ref{sec:5.6.1:calibration_detector}). However, in contrast to the underdamped case, where the displacement signal from the particle is very broadband and centered around zero, in the underdamped regime the signal is sharply peaked and centered around a finite frequency. 
The weak coupling to the thermal bath in high vacuum also allows to use the energy scale of a single motional quantum $\hbar \Omega_0$ as a calibration reference. Remarkably, despite the huge difference in energy scale of 7-8 orders of magnitude one can achieve almost perfect agreement \cite{tebbenjohanns2019motional} (section~\ref{sec:5.6.1:sideband_thermometry}).
In high vacuum charges on the particle are preserved. Section~\ref{sec:5.6.2:calibration_charge} explores how to control and measure them. The precise knowledge of the charge then allows allows to precisely determine the particle's mass (section \ref{sec:5.6.3:calibration_mass}). 
Since a particle in vaccum is well isolated, its internal temperature raises well above the temperature of the environment in an optical tweezers. In section~\ref{section:coupling_com_internal}, we will see how to measure the internal temperature by analyzing the particles center of mass motion.  

\subsubsection{Detector calibration}\label{sec:5.6.1:calibration_detector}

For the most part optical trapping is concerned with the external degrees of freedom, and in particular with the center-of-mass motion. Quantitative measurements require a precise calibration of the detector signal that relates the measured signal (typically in volts) to the displacements of the particle (in meters).
Under the assumption that the detector signal $v(t)$ is proportional to the particle displacement $q(t)$, the calibration factor is defined as the proportionality factor $c = v/q$  with units of $[{\rm V\,m^{-1}}]$ (or $[{\rm bit\,m^{-1}}]$ for a digitized detector signal).
Using the fact that the expectation value of the potential energy of a harmonic oscillator $E_\text{pot}= 
{1\over2}m\Omega_0^2\langle q(t)^2\rangle = {1\over2}k_{\rm B} T$, the calibration factor is simply given by 
\begin{equation}\label{eq:calibration_position}
	c
	=
	\sqrt{m\Omega_0^2 \langle v(t)^2\rangle \left/k_{\rm B} T\right.},
\end{equation}
which does not depend on $\gamma_0$ and thus holds both for the overdamped and underdamped regimes.

In the underdamped regime, the PSD is highly peaked around the resonance frequencies. This allows to distinguish noise and cross-talk efficiently in the spectral domain. Therefore, it is often preferred to fit the PSD to the \emph{single-sided} harmonic oscillator PSD
\begin{equation}\label{eq:PSD_fit_function}
	S_{\rm vv}(f) 
	=
	2 \frac{\langle v(t)^2\rangle}{\pi}\frac{f_0^2 g}{(f^2-f_0^2)^2+f^2 g^2}
\end{equation}
to extract the variance of the particle displacement $\langle v(t)^2\rangle$ from the area under the PSD.
In the following, we adopt here the notation from Hebestreit \cite{hebestreit2017thermal}, where we convert from angular to ordinary frequencies, i.e., $\omega = 2\pi f$, $\gamma = 2\pi g$, and $S(f) = 2\pi S(\omega > 0)$.

The experimental \emph{single-sided} PSD is obtained from the discretely-sampled time trace $v(t)$ as
\begin{equation}\label{eq:PSD_experimental}
	S_{\rm vv}(f) 
	= 
	2 \frac{N}{f_{\rm s}}
	\left|
		\frac{1}{N}
		\mathcal{F}\left(v(t)\right)
	\right|^2 
	\quad\mbox{for}\quad f>0,
\end{equation}
where $N$ is the number of samples, $f_{\rm s}$ is the sampling frequency, and $\mathcal{F}\left(v(t)\right)$ is the discrete Fourier transform of $v(t)$.

Fitting the experimental PSD to expression~\eqref{eq:PSD_fit_function} allows to extract the variance $\langle v^2(t) \rangle$, the damping constant $g=\gamma_{\rm gas}/(2\pi)$, and the resonance frequency $f_0=\Omega_0/2\pi$. Then, the damping constant allows to determine the particle radius from the pressure and, therefore, the mass for known mass density (section~\ref{sec:5.6.3:calibration_mass}). 
Once the mass is known, we can calculate the calibration factor according to equation~\eqref{eq:calibration_position}.
 The biggest uncertainty in determining $c$ stems from the error in the particle mass. However, in many experiments we are only interested in the particle energy and we can define the mass-independent energy calibration factor 
\begin{equation}\label{eq:calibration_energy}
 	C
	= 
	{
		c^{2} 
	\over 
		m
	} 
	= 
	{
		\Omega_0^2 
		\langle v^2 \rangle 
	\over 
		k_{\rm B} T
	}.
\end{equation}

The above calibration method relies on two key assumptions. First, we consider the optical potential to be quadratic such that the particle behaves like a harmonic oscillator. Second, we assume that the detector signal is linear in the particle displacement, that is that the calibration factor does not depend on the particle
position. The latter depends significantly on the details of the experiments and will not be futher discussed here. However, it is possible to account for a nonlinear detector response. For calibration of a position dependent calibration factor see \cite{rondin2017direct} and for detector response calibration and discussion see \cite{hebestreit2017thermal}.

\begin{figure}[hbt]
	\begin{center}	
	\includegraphics[width=8cm]{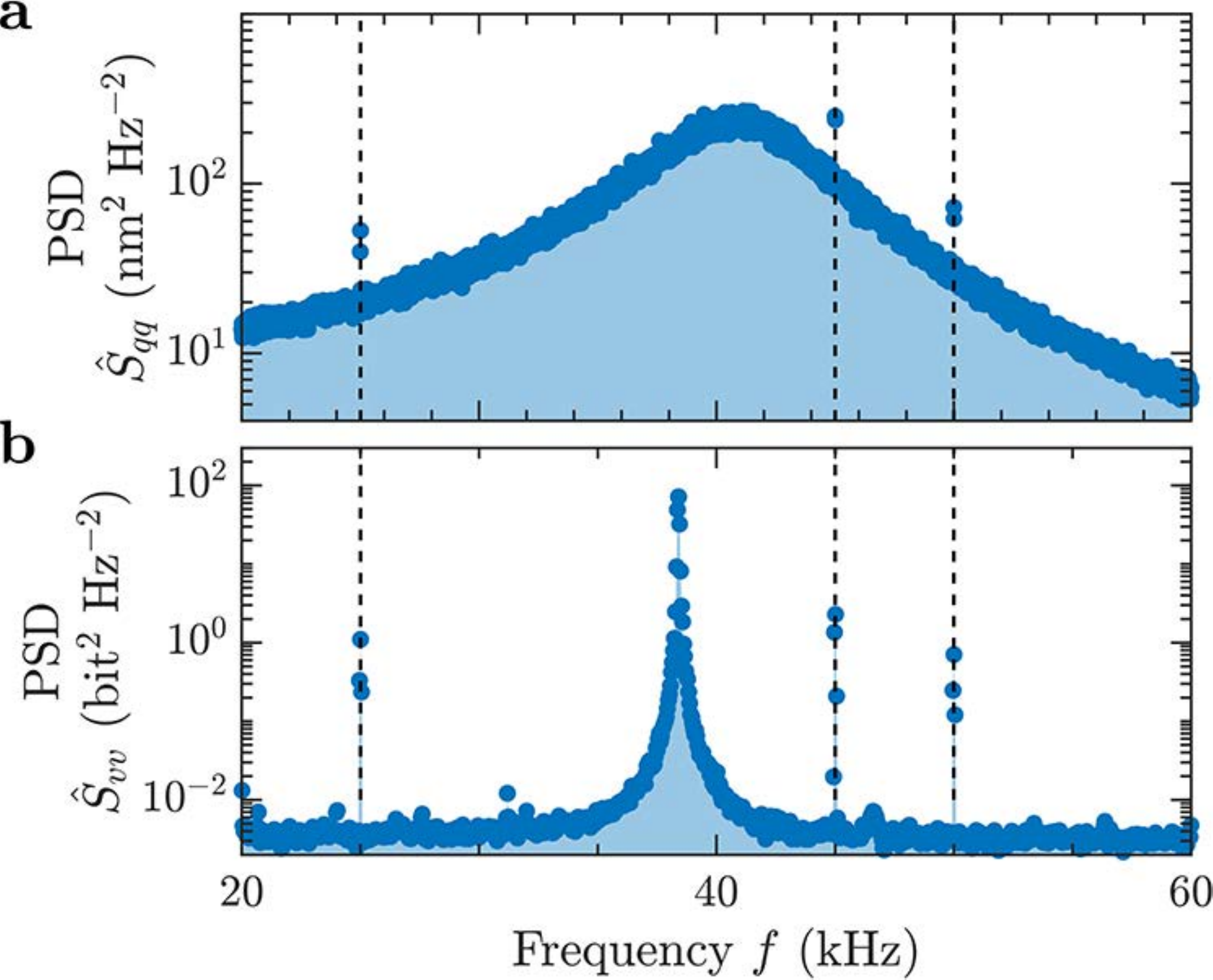}
	\caption{
  	{\bf Calibrated and uncalibrated $z$-detector signals.}
	(a) Calibrated $z$-detector signal at $10\pm1 \,{\rm mbar}$ with electric driving response. The area under that peak corresponds to the thermal temperature $k_{\rm B} T$.
	(b) Uncalibrated feedback-cooled oscillator signal with electric drive for recalibration of the signal.
  	}
	\label{fig:42}
	\end{center}
\end{figure}

The finite potential depth of the optical trap results in anharmonicities in the particle motion for large enough displacements. Due to the symmetry of the optical potential, the lowest nonlinear term is the \emph{cubic} or \emph{Duffing nonlinearity}\footnote{There are some subtelties in the detection along the propagation direction of the trapping laser, where this symmetry is broken by the scattering force, leading to small but finite quadratic nonlinear terms 
\cite{gieseler2014dynamics}.}.
The Duffing nonlinearity softens the optical potential at large particle displacement \cite{gieseler2013thermal}. 
This distorts and broadens the lineshape of the PSD, and causes an overestimation of the energy when following the calibration procedure described above.
To minimize the impact on the shape of the PSD, the calibration is usually carried out at moderate pressures ($\sim 10\,{\rm mbar}$), where the linewidth ($\gamma_{\rm gas}/(2\pi) \sim 10\,{\rm kHz}$) is narrow enough to reject noise and cross-talk, but still wide enough that the nonlinear broadening ($\gamma_{\rm NL}/(2\pi) \sim 3\,{\rm kHz}$) is negligible \cite{hebestreit2017thermal} (Fig.~\ref{fig:42}a).

The energy underestimation is accounted for by considering the \emph{kinetic} energy instead of the \emph{potential} energy \cite{hebestreit2018calibration}.
The kinetic energy is given by $E_\text{kin} = {1\over2}m\langle \dot{q}^2\rangle = {1\over2}k_{\rm B} T$ even for a nonlinear potential.
The variance $\langle \dot{q}^2\rangle$ can be obtained by integrating over the velocity PSD, which is related to the position-PSD by 
\begin{equation}
	S_{\dot{q}\dot{q}}(f) 
	= 
	4\pi^2f^2 S_{qq}(f).
\end{equation}
Hence, we do not have to measure $\dot{q}$ directly, which is experimentally challenging since it would require sufficient oversampling. 
Experimentally, we compute the variance by numerical integration over the voltage PSD 
\begin{equation}
	\langle \dot{v}^2\rangle 
	= 
	\int_{f_0-b/2}^{f_0+b/2} {\rm d}f\, 4\pi^2 f^2 S_{\rm vv}(f) 
	= 
	\frac{c^2}{m}k_{\rm B} T = Ck_{\rm B} T.
\end{equation}
The integration bandwidth $b$ is chosen such that we integrate most of the signal and don't integrate too much of the noise floor.
In the measured displacement PSD, we usually have a white noise floor. If we multiply this PSD by $f^2$ to get the velocity PSD, the noise floor therefore acquires a $f^2$ dependence as well. Integrating this noise will eventually cause a significant error in the calibration and energy estimate.

Drifts in the setup can cause the calibration to change over time. 
In particular, the calibration procedure detailed above is carried out at moderate pressures ($\sim {\rm mbar}$), while most experiments are carried out at much lower pressure many hours or even days after performing the calibration.
Hence, it is desirable to check the calibration even at much lower pressure and to correct for potential drift (see Fig.~\ref{fig:42}b).

Hebestreit suggests a method to track changes in calibration at arbitrary pressure \cite{hebestreit2017thermal}.
The main idea is to measure the response of the particle to a known force, for example the electrical force from a capacitor around the optical trap.
This method assumes that the electrical fields and the charge on the particle do not change through out the experiment. This can be achieved as will be discussed in section~\ref{sec:5.6.2:calibration_charge}.
At the initial pressure where the calibration is carried out, a sinusoidal signal is applied to the capacitor.
This leads to additional peaks in the response of the particle (Fig.~\ref{fig:42}). The power in the response peak is 
\begin{equation}
	\langle v^2_e\rangle 
	= 
	\frac{C}{2m^2} \frac{\mathcal{Q}_{\rm p}^2 E_0^2}{(\Omega_0^2-\omega_{\rm d}^2)^2+\gamma^2\omega_{\rm d}^2},
\end{equation}
where $\mathcal{Q}_{\rm p}$ is the total charge on the particle and $E_0$ is the electric field amplitude in the direction of oscillation at the location of the particle.
At a later time (e.g., after pumping down to a lower pressure), we repeat this measurement, yielding $\langle v^2_e\rangle'$.
At lower pressure, feedback is generally active. Hence, the response function of the particle changes. The new values of $\Omega_0'$ and $\gamma'$ can be extracted from a fit to the thermal noise peak.
Assuming that $\mathcal{Q}_{\rm p}E_0/m$ remain constant, we find the new calibration factor as
\begin{equation}\label{eq:recalibration}
	C' 
	= 
	C 
	\frac{
		\langle v^2_e\rangle'
	}{
		\langle v^2_e\rangle
	}
	\frac{
		(\Omega_0'^2-\omega_{\rm d}^2)^2+\gamma'^2\omega_{\rm d}^2
	}{
		(\Omega_0^2-\omega_{\rm d}^2)^2+\gamma^2\omega_{\rm d}^2
	}.
\end{equation}
We can also arrive at equation~\eqref{eq:recalibration} by considering the height of the peak in the PSD instead of the area under the peak. It's value is $\langle v^2_e\rangle \tau$, where $2\tau$ is the measurement time (equation~\eqref{eq:spectra_with_drive}).

\subsubsection{Sideband thermometry}\label{sec:5.6.1:sideband_thermometry}

\begin{figure}[hbt]
	\begin{center}	
	\includegraphics[width=12cm]{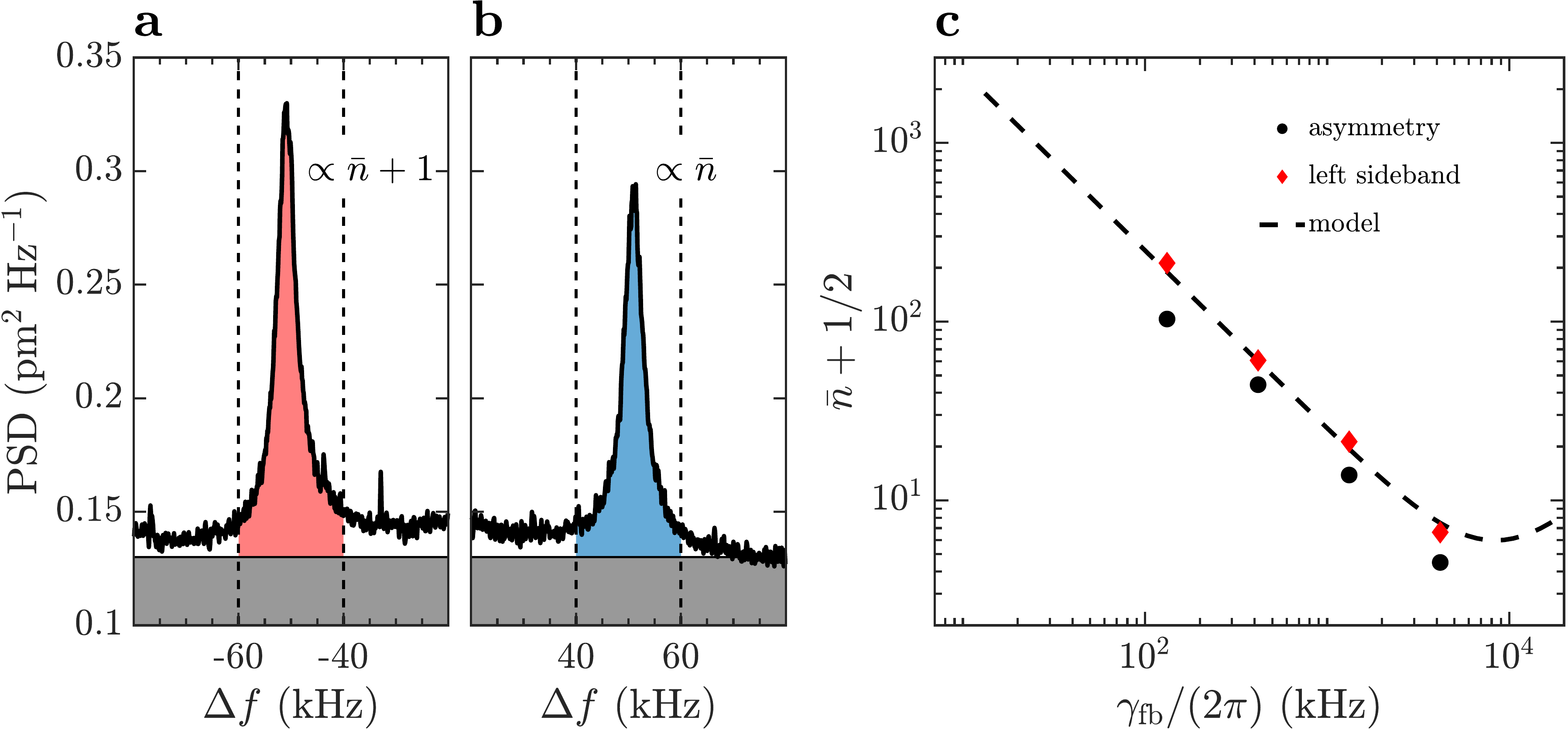}
  	\caption{
	{\bf Sideband thermometry.}
	(a-b) Motional sideband asymmetry measured with heterodyne measurement. The frequency difference $\Delta f$ is measured relative to the (absolute) local oscillator shift of $1\,{\rm MHz}$. The gray solid areas show the measurement noise floor, limited by technical laser noise. The vertical dashed lines indicate the integration bandwidth (see equation~\ref{eq:asymmetry}).
	(c) Mean occupation number as a function of feedback gain. The red diamonds are obtained by integrating the red sideband of the heterodyne spectrum. The black circles show the mean occupation number extracted according to equation~\eqref{eq:occupation_number_sidebands}. The black dashed line corresponds to a parameter-free model (see also section~\ref{sec:5.7.1:Force_feedback}).
	Figures adapted from Ref.~\cite{tebbenjohanns2019motional}.
  	}
	\label{fig:43:sidebands}
\end{center}
\end{figure}

In the previous section we used the thermal fluctuations of the particle to calibrate the detector signal. 
Similarly, we can use the quantum fluctuations for calibration. However, at ambient temperature, quantum fluctuations of a nanoparticle in an optical tweezers are 8-9 orders of magnitude weaker than the thermal fluctuations. As a consequence, they can only be observed in high vacuum, where the coupling to the thermal bath is sufficiently weak  \cite{tebbenjohanns2019motional}.

When light is reflected off the particle, the particle motion is imprinted on the scattered light as a phase modulation, which can be understood as a Doppler shift due to the particle motion.
To lowest order, the phase modulation leads to red and blue sidebands at $\omega_L\pm \Omega_0$, where $\omega_L$ is the laser frequency (Figs.~\ref{fig:43:sidebands}(a-b)).
Classically, the power in each sideband is proportional to the variance of the motion $\langle x^2 \rangle$, which in turn is proportional to the temperature of the thermal bath. This is the basis of the calibration procedure described in the previous section.

In the quantum mechanical formulation, the interaction of light with the trapped particle can be described as a Raman process.
The red sideband (Fig.~\ref{fig:43:sidebands}(a)) corresponds to Stokes scattering, which raises the population of the mechanical oscillator by a single quantum of mechanical energy, while simultaneously creating a low energy (red) photon. The blue sideband (Fig.~\ref{fig:43:sidebands}(b)) corresponds to anti-Stokes scattering, which lowers the oscillator's population by one mechanical quantum, while creating a high energy (blue) photon. 
The rate at which blue sideband photons are created is proportional to $\bar{n}$, just as in the classical case.
Importantly, anti-Stokes scattering is impossible by an oscillator in its quantum ground state.
In contrast, the rate at which red sidebands are produced is proportional to $\bar{n} + 1$.

The mean occupation number is given by the Bose-Einstein distribution 
\begin{equation}
  \bar{n} = \frac{1}{e^{\frac{\hbar \Omega_0}{k_{\rm B} T}}-1}
\end{equation}
For thermal energies large compared to the energy quantum ($k_{\rm B} T\gg \hbar\Omega_0$), the mean energy of the harmonic oscillator is $\bar{n} \hbar \Omega_0 = k_{\rm B} T  \propto \langle x^2\rangle$, in accordance with classical statistical mechanics and independent of $\hbar$.
The ratio of powers in the anti-Stokes and Stokes sidebands is given by $\bar{n}\left/(\bar{n}+1)\right. = \exp\left[-\hbar \Omega_0 \left/ (k_{\rm B} T)\right.\right]$ (Figs.~\ref{fig:43:sidebands}(c)) and vanishes for high temperatures. 
For thermal energies comparable to a single mechanical quantum excitation (also called phonon), the ratio can serve as a temperature measurement calibrated relative to the quantum of energy of the system \cite{clerk2010introduction}. 

This so-called sideband thermometry requires to measure the Stokes and anti-Stokes sidebands \cite{purdy2015optomechanical, safavi2012observation, weinstein2014observation,  tebbenjohanns2019motional}, which can be achieved with a heterodyne measurement. In a heterodyne measurement the scattered light interferes with a laser beam (local oscillator) with a frequency $\omega_{\rm LO}$  which is frequency-detuned from the light that is incident on the particle.
This produces a strong beat signal at the difference frequency $\Delta_\text{LO} = \omega_\text{LO}-\omega_L$, with sidebands at $\Delta_\text{LO}\pm \Omega_0$. For $\omega_\text{LO}<\omega_L$, the left sideband corresponds to the red sideband (Stokes) and the right to the blue (anti-Stokes). For $\omega_\text{LO}>\omega_L$, the roles are inverted.
The difference frequency is chosen such that $|\Delta_\text{LO}| \gg \Omega_0$, for example $|\Delta_\text{LO}|\sim 1\text{MHz}$ for typical trapping frequencies $\sim 100\text{kHz}$.
The measured sideband asymmetry 
\begin{equation}\label{eq:asymmetry}
  R_{\mp} = \frac{\int S_\text{vv}^{r, het}(f) {\rm d}f}{\int S_\text{vv}^{l, het}(f) {\rm d}f}
\end{equation}
can be used to determine the mean occupation number. 
Here, $ S_\text{vv}^{l, het}$ ($ S_\text{vv}^{r, het}$) is the power spectral density of the left (right) sideband and the subscript $R_{-}$ ($R_+$) is for $\omega_\text{LO}<\omega_L$ ($\omega_\text{LO}>\omega_L$) (see Fig.~\ref{fig:43:sidebands}).

In principle, $R_-$ or $R_+$ alone is sufficient to determine the mean occupation number.
However, classical artifacts due to a frequency dependent transfer function of the measurement device give rise to an asymmetry that is not related to the quantum dynamics of the particle.
Measuring both $R_-$ and $R_+$ allows to eliminate this artifact and to determine the actual mean occupation number as follows
\begin{equation}\label{eq:occupation_number_sidebands}
  \bar{n} = \frac{\sqrt{R_-/R_+}}{1-\sqrt{R_-/R_+}}.
\end{equation}

Fig.~\ref{fig:43:sidebands} shows the measured sidebands of a nanoparticle cooled to an effective temperature of $\bar{n}=4$ \cite{tebbenjohanns2019motional}.
Clearly the area under the red sideband is larger than under the blue sideband. This is the first room temperature measurement of the sideband asymmetry of a massive object and clear signature of its quantum ground state.
Remarkably, the inferred occupation number is in very satisfactory agreement with the traditional calibration method detailed in section~\ref{sec:5.6.1:calibration_detector}despite the huge difference in energy scales of the two methods.

\subsubsection{Charge calibration}\label{sec:5.6.2:calibration_charge}

\begin{figure}[b]
	\begin{center}
	\includegraphics[width=12cm]{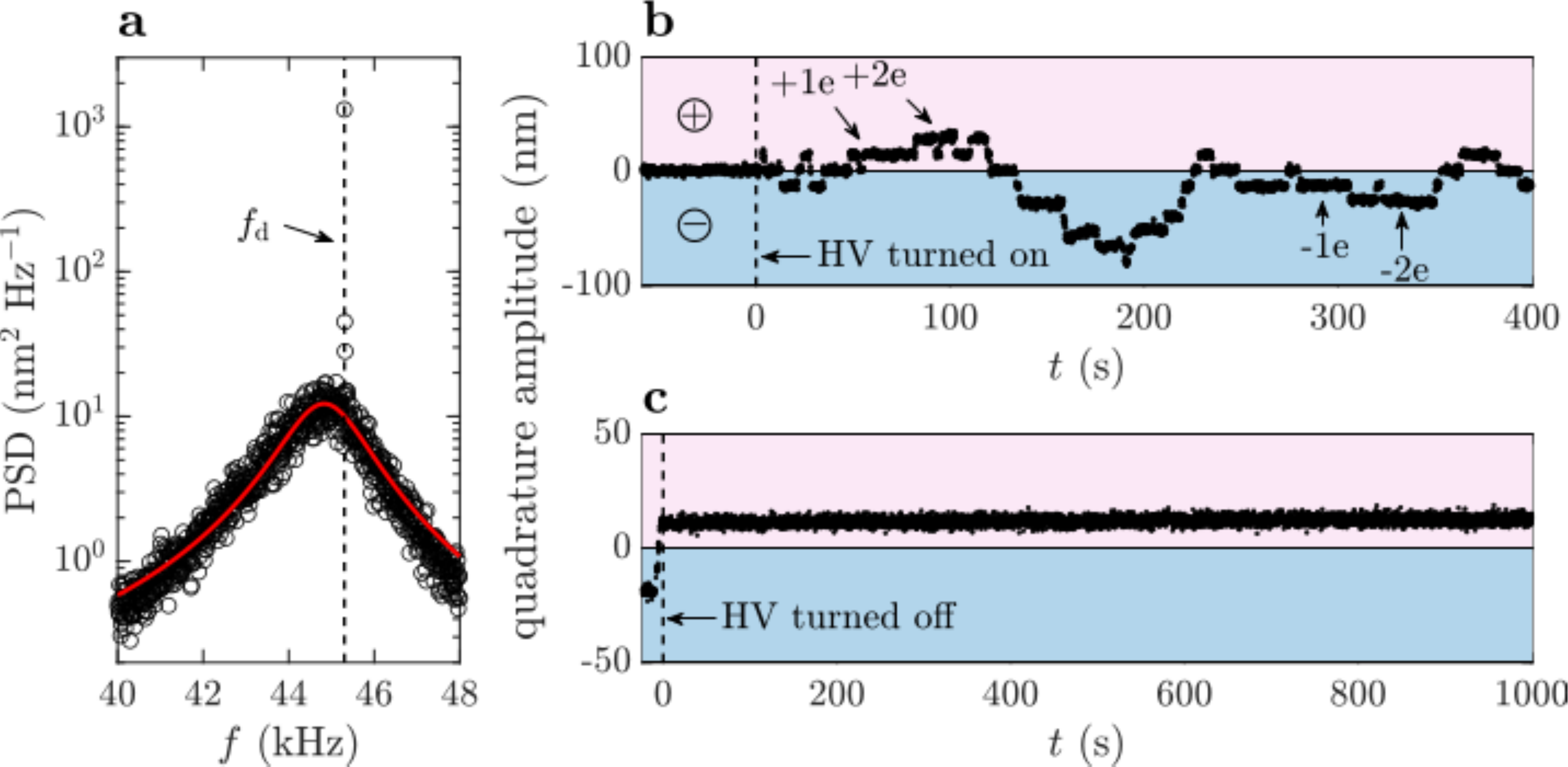}
 	\caption{
	{\bf Charge calibration.}
	(a) Power spectral density of the motion along the optical axis of a charge-carrying particle at a pressure of $1.9\,{\rm mbar}$ in the presence of a drive tone $f_{\rm d}=\omega_{\rm d}/2\pi$ applied to the capacitor. The solid line is a Lorentzian function fit to the data.
	(b) Quadrature component of particle oscillation in response to a driving voltage, demodulated in a bandwidth of $7\,{\rm Hz}$. The high-voltage discharge is turned on at $t = 0$. The oscillation amplitude changes in discrete steps while the high voltage is on.
	(c) Preparation of charge state. The high voltage is turned off at $t = 0$, while the particle carries a net charge of $1e$. The charge stays constant over the remainder of the measurement.
	Figure adapted from Ref.~\cite{frimmer2017controlling}.
	}
	\label{fig:44}
	\end{center}
\end{figure}

Generally, particles carry several tens of elementary charges. Under vacuum, the particle is completely isolated from the environment and therefore its charge state is preserved indefinitely.
Knowing the charge exactly provides a reliable handle for exerting a known force on the particle by applying an known electric field.

In optical levitation experiments, particles are loaded at high ($\sim 1\,{\rm atm}$) pressure to provide enough damping.
After the particle has been trapped, the vacuum chamber is evacuated to reduce air damping.
As the pressure decreases, residual water desorbs slowly from the particle followed by a sudden change in size, which is attributed to the removal of the final water layers (section~\ref{sec:5.4.1:nebuliser}).
During this final stage, the charge of the particle changes by tens of elementary charges \cite{ricci2019levitodynamics}, but once the particle has undergone this sudden change, its charge state is preserved indefinitely even when the pressure is increased again because the water is gone.

A relatively simple method to control the charges on the particle is to apply an electrical discharge near the particle.
The discharge provides free charges that can be adsorbed by the particle \cite{frimmer2017controlling}.
It can be applied through a bare wire of about 5 cm length at about 5 cm away from the optical trap. The grounded vacuum chamber serves as the counterelectrode.
A DC voltage  ($\sim 7\,{\rm kV}$) is sufficient to ionize gas molecules inside the vacuum chamber and to produce a corona discharge at moderate pressures ($\sim 1\,{\rm mbar}$). Note that an ion pressure gauge can produce also free charges, which lead to unwanted changes in the charge state of the particle, if it is not properly shielded \cite{frimmer2017controlling}.

The ionization process is triggered by random events such as absorption of a ultraviolet photon or cosmic rays that spontaneously ionize neutral air molecules.
The free electron and positive ion are accelerated toward opposite directions by the high voltage, which also prevents them to recombine.
The electron acquires enough kinetic energy to ionize other molecules, because of its small mass. The secondary electrons ionize more molecules, which then results in an electron avalanche.

Emission of (mainly) UV photons after electron-ion recombination causes the characteristic violet glow of the corona, as shown in Fig.~\ref{fig:36:Picture_Setup}.
The corona discharge is positive or negative, depending on the polarity of the high DC voltage, which therefore allows to select the ratio of positive-to-negative ionized molecules.
Eventually, free charges are adsorbed on the particle surface, thereby changing its net charge monotonically, except for few unfavorable events.
These charges can be both electrons and ionized air molecules.
Accordingly, by simply turning off the high-voltage source when the particle carries the desired number of charges, the levitated particle can be brought to any desired net-charge state from zero to a few elementary charges \cite{frimmer2017controlling, ricci2019levitodynamics}.

The charge state of the particle is monitored \emph{in situ} by driving the particle
motion with an electric field at frequency $\omega_{\rm d}$.
The electric field is created by applying a voltage $U(t) = U_0 \cos(\omega_{\rm d} t)$  across a capacitor that is centered around the particle.
A simple capacitor can be formed by grounding the metal housing of the microscope objective and applying a voltage to the metal holder of the collection lens \cite{frimmer2017controlling}.
This method does not require any additional mechanical components in the setup and therefore does not obstruct access to the particle .
Alternatively, dedicated electrodes that fit right at the sides of the optical trap can be beneficial to produce a more homogeneous field at the location of the particle \cite{ricci2019levitodynamics}, since field inhomogeneities can lead to a finite net charge density.
When a voltage $U_0 \sim 10\,{\rm V}$ is applied to the capacitor at a frequency $\omega_{\rm d}$ close to the particle's resonance frequency, the particle's response to the driving field shows as a distinct peak in the single-sided PSD \cite{ricci2019levitodynamics} on top of the thermal peak (see Figs.~\ref{fig:42} and \ref{fig:44}a):
\begin{equation}\label{eq:spectra_with_drive}
	S_{qq}(\omega) 
	= 
	2 m^{-2}
	|\chi(\omega)|^2 
	\left[
		S_{\rm ff}
		+ 
		\frac{
			F_0^2 \tau
		}{
			4 \pi
		}
		\text{sinc}^2
		\left[
			(\omega-\Omega_0)\tau
		\right]
	\right],
\end{equation}
where the thermal force spectral density is $S_{\rm ff}  = m k_{\rm B} T \gamma_0 / \pi$, $\chi(\omega) = \left[\Omega_0^2-\omega^2+i\omega\gamma_0\right]^{-1}$ is the mechanical response function, $F_0 = \mathcal{Q}_{\rm p} E_0$, $E_0\sim 500\,{\rm V\,m^{-1}}$ being the amplitude of the electric field modulation at the location of the particle, $\mathcal{Q}_{\rm p}$ is the charge on the particle, and $2 \tau$ is the duration of the time trace that is used to calculate the position spectral density $ S_{qq}(\omega) = 2\tau|\tilde{q}|^2$ (equation~\eqref{eq:PSD_experimental}).

The detector signal is demodulated in quadrature with the drive.
The demodulated signal can assume both positive and negative values, since the response to the driving field flips phase by $\pi$ when the polarity of the charge changes.
Positive (negative) signal amplitudes can then be associated with positive (negative) net charge, when the transfer functions of the electronics is accounted for.
Note that for a small detuning of the drive frequency $\omega_{\rm d}$ from the particle's resonance, there is a small phase offset from the response function of the particle. 

While the corona discharge is active, the demodulated signal features discrete steps, which are the signature of single elementary-charge transfers to and from the nanoparticle (see Fig.~\ref{fig:44}).
Thus, any desired net-charge state from zero to a few elementary charges can be achievd by simply selecting the polarity of the high-voltage source, counting the number of steps, and turning off the high-voltage source when the particle carries the desired number of charges. 
Importantly, the particle's charge state is preserved when the pressure is reduced and stays indefinitely  while the high-voltage source is turned off, even over a period of several days \cite{frimmer2017controlling}.

\subsubsection{Mass Calibration}\label{sec:5.6.3:calibration_mass}

\begin{figure}[b]
	\begin{center}
	\includegraphics[width=12cm]{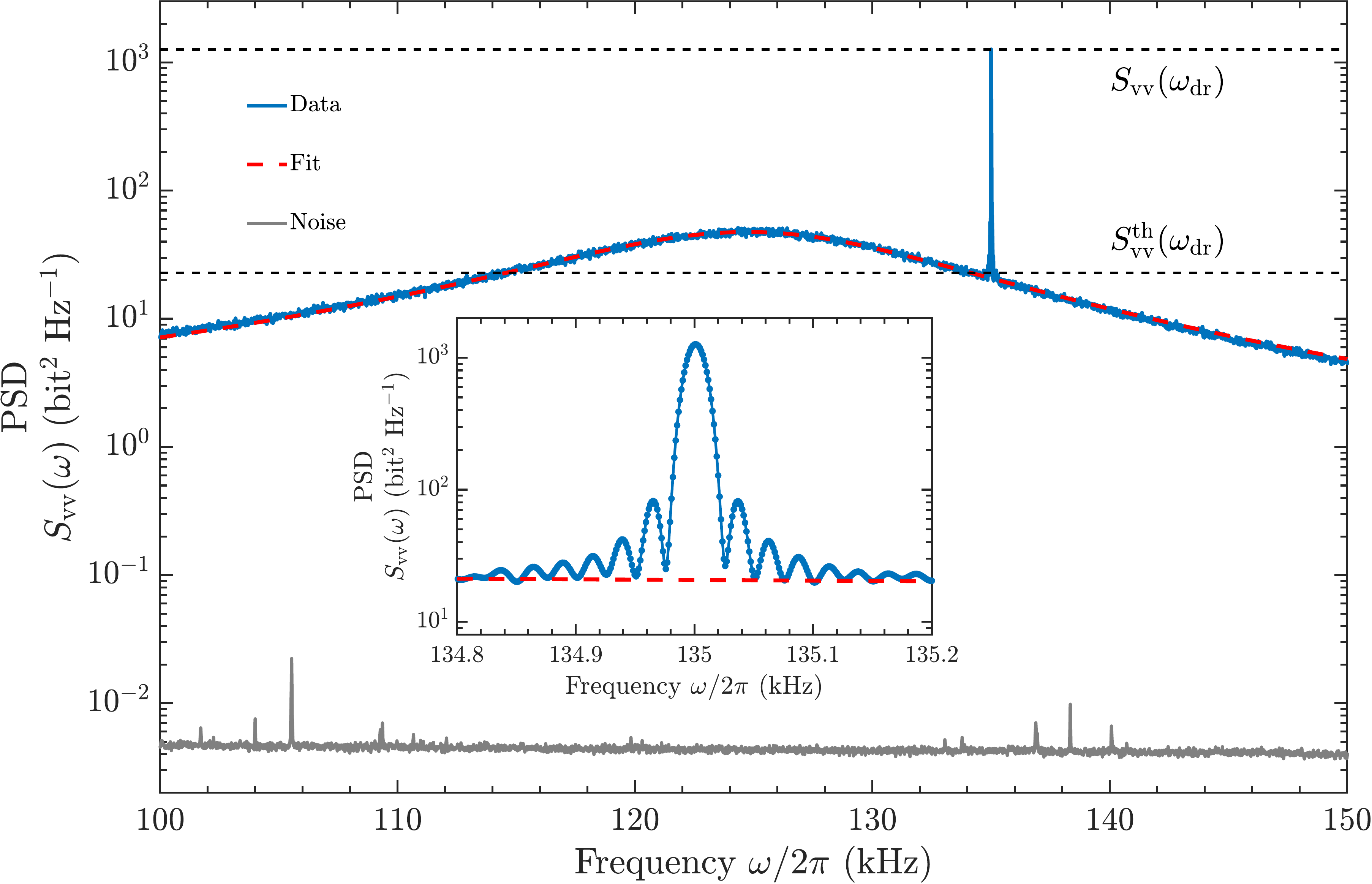}
	\caption{
	{\bf Mass calibration.}
	Power spectral density $S_{\rm vv}(\omega)$ of a thermally and harmonically driven resonator at $P = 50\,{\rm mbar}$. The broad peak centered at $\Omega_0/2\pi= 125\,{\rm kHz}$ corresponds to the thermally driven state. We fit it with a Lorentzian function (red) to extract $S_{\rm vv}^{\rm th}(\omega_{\rm d})$ together with $g/(2\pi) = 31.8\,{\rm kHz}$ and the corresponding uncertainties. The narrowband peak at $\omega_{\rm d}/(2\pi) = 135\,{\rm kHz}$, also shown in detail in the inset, depicts the electrical excitation from which we retrieve $S_{\rm vv}^{\rm d}(\omega_{\rm d})$. Gray data points at the bottom of the plot is the measurement noise, which is $\sim 40\,{\rm dB}$ below the particle's signal.
	Figure adapted from Ref.~\cite{ricci2019levitodynamics}.
	}
  	\label{fig:45:mass_calibration}
	\end{center}
\end{figure}

The exact value of the particle's inertial mass often plays an important role for precision measurements. For example, the common method for displacement calibration uses the thermal force noise as a reference, where the strength of the force depends on the mass, as we have seen in section~\ref{sec:5.6.1:calibration_detector}. Likewise, nanomechanical mass measurements rely on knowing the mass of the bare resonator \cite{chaste2012nanomechanical}.

The manufacturer's size specification gives a good first estimate of the particle size with an error up to $\sim 25\%$, which stems from the typical size variation ($\sim 5\%$) and the uncertainty in the specified mass density ($\sim 10\% $) of commercially available mono-disperse particles.

The mass of individual particles can be inferred from kinetic gas theory, which relates the measured damping rate $\gamma_0$ to the particle size \cite{beresnev1990motion}. In particular, for a spherical particle and at pressures where the mean free path is much larger than the radius of the particle, there is a simple linear relationship \cite{gieseler2014dynamics},
\begin{equation}
  	a
	= 
	\frac{2.223}{\rho_{\rm p}}
	\sqrt{\frac{m_{\rm gas}}{ k_{\rm B} T}} 
	\frac{P_{\rm gas}}{\gamma_0},
\end{equation}
which relates the pressure to the particle radius $a$ (equation~\eqref{eq:gas_damping_lin}). 
At room temperature the mean free path is $\sim 60\,{\rm \upmu m } \cdot P_{\rm gas}^{-1} {\rm mbar^{-1}}$.
Hence, the linear approximation is generally valid for nanoparticles in vacuum. 
The calibration measurements are generally performed at $\sim 10\,{\rm mbar}$, where individual peaks are clearly resolved and nonlinear broadening of the peaks is still negligible.
To remove excess water from the particle, which leads to an overestimate of the actual particle size, the pressure should be 
cycled to below $0.1\,{\rm mbar}$ and back (section~\ref{sec:5.4.1:nebuliser}).
From the measured radius, one can then infer the mass $m = {4\over3}\pi a^3 \rho_{\rm p}$ with a typical uncertainty of $\sim 35\%$ \cite{ricci2019levitodynamics}. 

The mass measurement can be significantly improved either by an independent measurement of the voltage calibration (equation~\eqref{eq:calibration_position}) or by applying a known force.
For the former approach, the wavelength of the light provides a well-defined length scale that allows to relate the phase to a displacement in an interferometer. The interferometer is calibrated by changing the relative length of the two interferometer arms and measuring the fringe pattern.
For the latter approach, electrical fields can be estimated quite accurately ($\sim 1\%$ error) from a numerical simulation of the electrode geometry \cite{ricci2019levitodynamics}.
Hence, the electrical force on the particle can be estimated with the same accuracy, because the number of charges can be determined exactly (section~\ref{sec:5.6.2:calibration_charge}).
Since the thermal force depends on the mass but the electrical force does not, their ratio (equation~\eqref{eq:spectra_with_drive}),
\begin{equation}\label{eq:RS}
	R_S 
	= 
	\frac{
		F_0^2\tau/2
	}{
		\mathcal{F}_L
	}
	= 
	\frac{
		S^{\rm th}_{\rm vv}(\omega_{\rm d})-S^d_{\rm vv}(\omega_{\rm d})
	}{
		S^{\rm th}_{\rm vv}(\omega_{\rm d})
	},
\end{equation}
provides a direct method to determine the mass from the measured spectra (Fig.~\ref{fig:45:mass_calibration}).
In equation~\eqref{eq:RS}, $S^d_{\rm vv}(\omega_{\rm d})$ is the peak of the measured spectrum at the drive frequency $\omega_{\rm d}$ when the sinusodial electrical force $F_{\rm el} = F_0\cos(\omega_{\rm d} t)$ is applied and $S^{\rm th}_{\rm vv}(\omega_{\rm d})$ is the thermal background at $\omega_{\rm d}$ (Fig.~\ref{fig:45:mass_calibration}).
Note that the spectra do not need to be calibrated to apply this procedure.
The mass is then simply given by \cite{ricci2019levitodynamics} 
\begin{equation}
	m 
	= 
	\frac{
		\mathcal{Q}_{\rm p}^2 E_0^2\tau
	}{
		8k_{\rm B}T\gamma_0 R_S
	}.
\end{equation}

\subsubsection{Internal temperature}\label{section:internal_temperature}

In optical tweezers experiments, the main focus is on the external degrees of freedom of the particles in the trap (i.e., the translation of the particle and its rotation). 
However, any particle will absorb some of the trapping light, which raises its \emph{internal} bulk temperature.
The power absorbed by a particle with volume $V$ from a laser beam with intensity $I_0$ is 
\begin{equation}
	P_{\rm abs} 
	= 
	6\pi 
	\frac{I_0}{\lambda} 
	V 
	\text{\rm Im}
	\left[
		\frac{
			n^2(\lambda)-1
		}{ 
			n^2(\lambda)^2+2
		}
	\right].
\end{equation}
For example, for silica, the refractive index is $n=1.45+{\rm i} 6 \cdot 10^{-8}$ at $\lambda = 1064\,{\rm nm}$. 
Note that absorption can be higher in nanoparticles due to additional absorbers such as defects and impurities.

At high pressure, the particle efficiently thermalizes with the surrounding gas by convective cooling with cooling power
\begin{equation}
	P_{\rm con}(T_{\rm bulk}) 
	= 
	-c_{\rm acc}
	\sqrt{2\over3\pi}
	(\pi a^2) 
	P_{\rm gas} 
	v_{\rm rms} 
	\frac{\gamma_{\rm sh}+1}{\gamma_{\rm sh}-1}
	\left(
		{T_{\rm bulk} \over T_{\rm gas}}-1
	\right),
\end{equation}
where $v_{\rm rms} = \sqrt{3k_{\rm B} T_{\rm gas}/m_{\rm gas}}$ is the root mean square velocity of the gas molecules, and $\gamma_{\rm sh} = 7/5$ is the specific heat ratio for an ideal diatomic gas.
Clearly, as the pressure $P_{\rm gas}$ decreases, convective cooling by the surrounding medium vanishes and the only heat exchange with the environment is radiative.
The emitted thermal radiation is $P_{\rm bb}^{\rm em} = -P_{\rm bb}(T_{\rm bulk})$, where
\begin{equation}
	P_{\rm bb}(T) 
	= 
	\frac{72\zeta(5)}{\pi^2}
	\frac{V}{c^3\hbar^4}
	{\rm Im}
	\left[
		\frac{n^2_{\rm bb}-1}{n^2_{\rm bb}+2}
	\right]
	\left(
		k_{\rm B} T
	\right)^5
\end{equation}
is the the black body radiated power for a dipolar emitter, being $\zeta(5)\approx 1.04$ the Riemann zeta function \cite{chang2010cavity}.
Conversely, the particle also absorbs thermal radiation from the environment at temperature $T_{\rm env}$ with $P_{\rm bb}^{\rm abs} = P_{\rm bb}(T_{\rm env})$.

The particle bulk temperature can be found from the requirement that at equilibrium all contributions add up to zero, i.e., 
\begin{equation}
	P_{\rm abs} 
	+
	P_{\rm bb}^{\rm abs}
	+
	P_{\rm con}(T_{\rm bulk}
	+
	P_{\rm bb}^{\rm em}(T_{\rm bulk}) 
	= 
	0.
\end{equation}
The emitted or incident power due to black-body radiation depends on the dielectric properties of the particle in the entire black-body spectrum, and their calculation therefore involves the integration of the black-body emission rates over all emission frequencies.
As an approximation, one assumes a constant permittivity of the particle in the entire spectral range of black-body emission.
This can be estimated as the average value of the refractive index over the black-body spectrum. For example, for silica this gives $n_{\rm bb} = 1.5 + {\rm i} 0.1$.

\begin{figure}[hbt]
	\begin{center}
	\includegraphics[width=12cm]{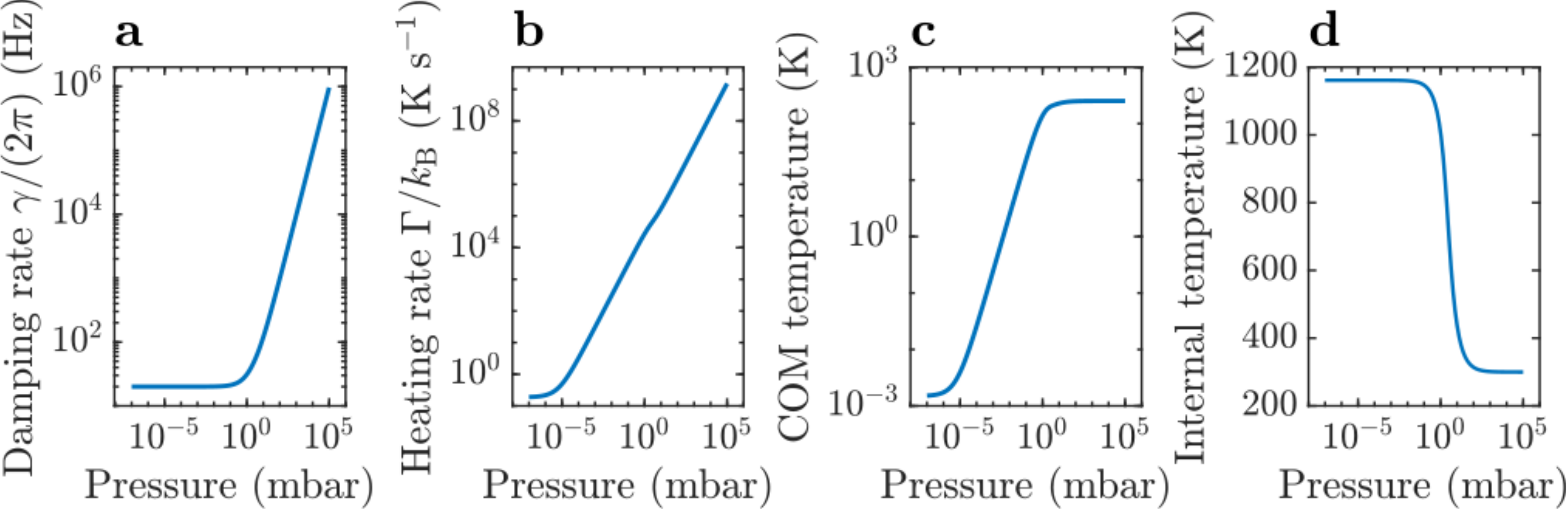}
	\caption{
	{\bf Heating and temperatures of an optically levitated particle.}
	(a) Damping and (b) heating rate as a function of pressure for feedback $\gamma_{\rm fb} / 2\pi= 20\,{\rm Hz}$.
	(c) Center of mass temperature as a function pressure. At low pressures recoil heating dominates and the rates and the center of mass temperature become pressure independent.
	(d) Internal temperature as a function of pressure. At a few millibars, there is a sharp transition between convective cooling and radiative cooling via black body radiation and the internal temperature transitions from the ambient temperature to a much higher value.
The internal temperature leads to a slightly higher center-of-mass heating rate visible as a small bump in (b).
	}
	\label{fig:41}
	\end{center}
\end{figure}

The internal temperature of a particle trapped in vacuum can raise significantly above the environmental temperature, to the point where the particle disintegrates if the absorption is too high \cite{rahman2016burning} (Fig. ~\ref{fig:41}). In fact, this problem has prevented optical levitation of nanodiamonds in high vacuum  \cite{frangeskou2018pure}. Even silica particles with low absorption can reach temperatures  $\sim 1000\,{\rm K}$ \cite{hebestreit2018measuring}.
As an example, figure~\ref{fig:41}d shows the internal temperature as a function of pressure.

\subsubsection{Coupling between internal temperature and external degrees of freedom}
\label{section:coupling_com_internal}

As we discussed earlier, the bulk temperature couples to the external degrees of freedom through the residual gas in the vacuum chamber (section \ref{sec:5.5:dissipation}).
This coupling can be explained by a simple two-bath model, where the gas molecules that interact with the particle are divided into two families: molecules that impinge on the particle being in thermal equilibrium with the vacuum chamber, and molecules that emerge from the hot surface of the particle at a higher temperature. 
From equation~\eqref{eq:force_noise_gas}, the thermal force noise increases with increasing bulk temperature. This allows to determine the \emph{internal} bulk temperature from the fluctuation of the center-of-mass of the particle by solving
\begin{equation}\label{eq:internal_temp_Tcm}
	T_{\rm cm} 
	= 
	\frac{
		T_{\rm gas}^{3/2}+\frac{\pi}{8}T^{3/2}_{\rm em}
	}{
		\sqrt{T_{\rm gas}}+\frac{\pi}{8}\sqrt{T_{\rm em}}
	}
\end{equation}
for $T_{\rm em}$, where the center-of-mass temperature is defined as $T_{\rm cm} = \Omega_0^2\langle v^2 \rangle\left/ (C k_{\rm B})\right.$, $C$ being the energy calibration factor and $\langle v^2 \rangle$ the variance of the voltage fluctuation on the detector (section~\ref{sec:5.6.1:calibration_detector}).
Then, from $T_{\rm em}$ one determines the temperature of the particle as
\begin{equation}
	T_{\rm bulk} 
	= 
	c_{\rm acc}^{-1}
	\left(
		T_{\rm em}-T_{\rm gas}
	\right)
	+
	T_{\rm gas}.
\end{equation}

\begin{figure}[t]
	\begin{center}	
  	\includegraphics[width=12cm]{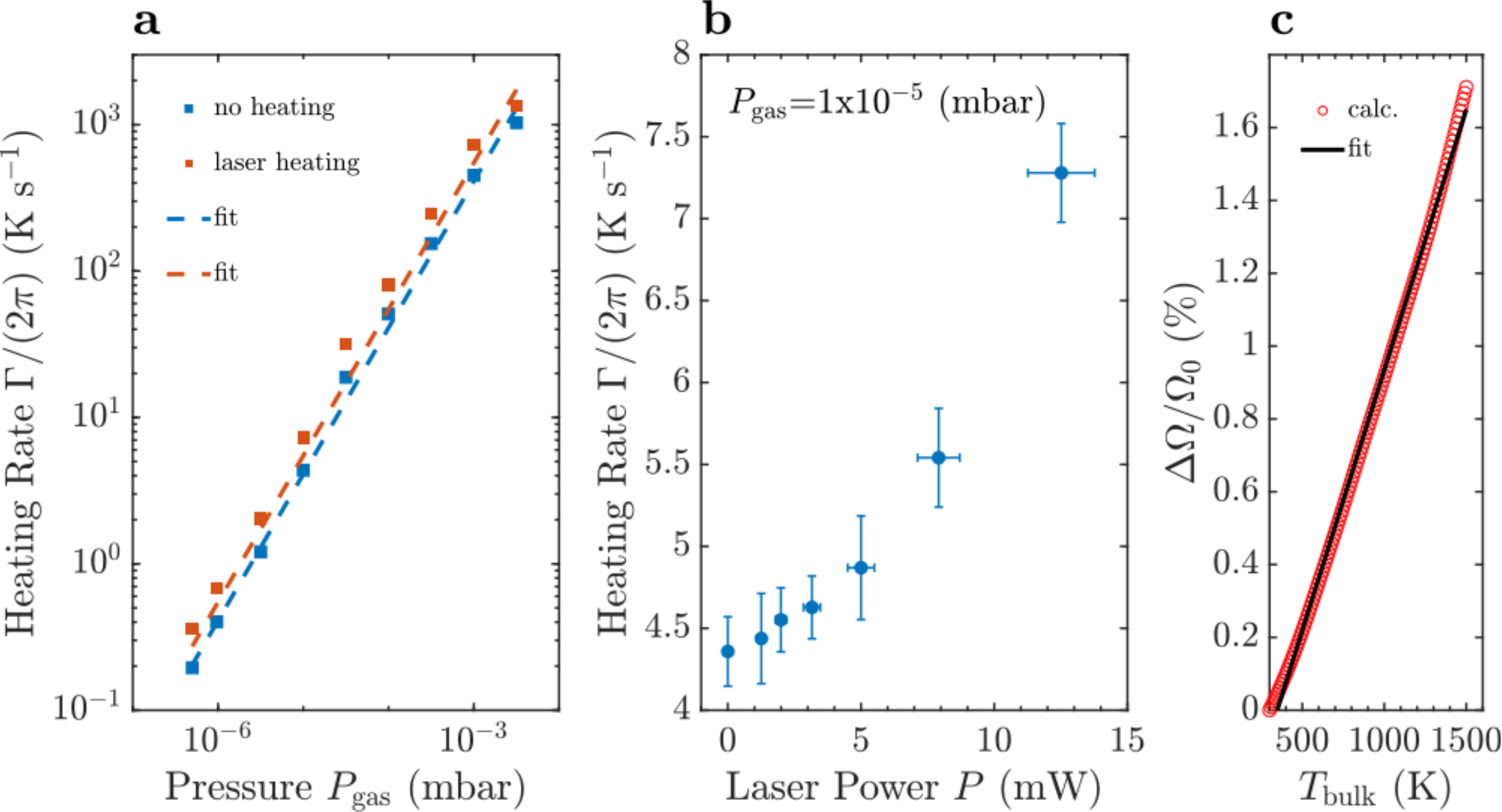}
  	\caption{
	{\bf Particle heating.}
	(a) Heating rates as a function of pressure extracted from relaxation measurements without additional heating of the internal temperature (blue) and with heating by a ${\rm CO_2}$ laser intensity of $0.47\,{\rm \upmu W\,\upmu m^{-2}}$ (orange). The dashed lines are linear fits to the data points. Error bars are smaller than the marker size.
	(b) Heating rates at different intensities of the ${\rm CO_2}$ laser measured at a pressure of $1\cdot 10^{-5}\,{\rm mbar}$. The error bars indicate the standard deviation of the measurements.
		(c) Calculated temperature dependence of oscillation frequency. When the particle is heated with the ${\rm CO_2}$ laser, its oscillation frequency increases due to changes in the particle's material
properties. This leads to a nearly linear relation between relative frequency change and increase of the internal particle temperature.
	Adapted from Ref.~\cite{hebestreit2018measuring}.
  	}
	\label{fig:46:Particle_Heating}
	\end{center}
\end{figure}

The method described above requires the particle's center-of-mass to reach its equilibrium steady state. This is impractical at low pressures where the particle is easily lost without feedback stabilization.
At low pressures, the thermal noise force from the interaction with the gas can be determined measuring the heating rate \cite{hebestreit2018measuring}.
Thereby, the center-of-mass motion is first reduced with feedback cooling (section~\ref{sec:5.7:feedback}). When the feedback is switched off, the particle energy increases at a rate
\begin{equation}\label{eq:heating_rate}
	\Gamma 
	= 
	\gamma_{\rm im} T_{\rm gas}
	+ 
	\gamma_{\rm em} T_{\rm em}
	= 
	\gamma_{\rm im}
	\frac{
		T_{\rm gas}^{3/2}
		+
		\frac{\pi}{8}
		T_{\rm em}^{3/2}
	}{
		\sqrt{T_{\rm gas}}
	}.
\end{equation}
This is a linear function of the gas pressure (see also equation~\eqref{eq:gamma_im} and Fig.~\ref{fig:46:Particle_Heating}a). Hence, the slope of the measured heating rates as a function of the gas pressure yields $T_{\rm em}$ from which we can then determine $T_{\rm bulk}$ \cite{hebestreit2017thermal}.

Note that knowledge of the accommodation factor $c_{\rm acc}$ is required to determine the particle temperature.
It can be determined by measuring $T_{\rm em}$ at two different particle temperatures.
The particle temperature can be raised by increasing the laser power to which the particle is exposed.
Ideally, the additional heating is generated with a weakly focused secondary laser at a wavelength that is strongly absorbed by the particle, e.g., a ${\rm CO_2}$ laser for silica particles, which has  6 to 7 orders more absorption than that typical trapping laser.
Thereby, the internal temperature of the particle increases without changing the particle response, as would be the case when increasing the power of the trapping laser. 
(The scattering force is negligible because the intensity is many orders of magnitude smaller.)
From the two measurements, we get the difference $\Delta T_{\rm em} = c_{\rm acc} \Delta T_{\rm bulk}$.
The temperature change $\Delta T_{\rm bulk}$ is measured independently from the observation that the particle oscillation frequency
\begin{equation}
	\Omega_0
	\propto 
	\sqrt{\kappa \over m}
	\propto
	\sqrt{\alpha' \over m}
	\propto
	\sqrt{
		{1 \over \rho_{\rm p}}
		{n^2-1 \over n^2+2}
	}
\end{equation}
depends on temperature because both the particle refractive index $n$ and its density $\rho_{\rm p}$ are temperature dependent. 
For silica, this leads to a linear increase of the oscillation frequency with the internal particle temperature.
The relative temperature change is given by
\begin{equation}
	\frac{\Delta \Omega_0}{\Omega_0} 
	= 
	c_{\rm T} \Delta T_{\rm bulk}
\end{equation}
with $c_{\rm T} = (1.43 \pm 0.01) \cdot 10^{-5}\,{\rm K^{-1}}$ (Fig.~\ref{fig:46:Particle_Heating}c).
Using this method, Hebestreit et al. \cite{hebestreit2018measuring, hebestreit2017thermal} measured the accommodation coefficient to be $c_{\rm acc} = 0.61 \pm 0.07$. This allowed them to determine the accommodation coefficient of water on silica for the first time combining this measurement with the measurement of the partial pressures (Fig.~\ref{fig:39:vacuum_species}).

\subsection{Feedback control}\label{sec:5.7:feedback}

One of the primary goals in optical levitation is to reduce the center-of-mass energy and cool the motional degrees of freedom of a levitated particle to its quantum mechanical ground state ($\sim 5\,{\rm \upmu K}$ for $\Omega_0 \approx 100\,{\rm kHz}$); the original proposals suggested the use of optical cavities and to cool the particle motion via the optomechanical interaction \cite{chang2010cavity,romero2010toward, barker2010cavity}. Initial experimental attempts were limited to relatively high pressures due to particle loss, where the interaction with the residual gas is too strong to achieve ground-state cooling \cite{kiesel2013cavity}. Recently, cavity cooling by coherent scattering was demonstrated as a robust route to three-dimensional cavity cooling primarily limited by fluctuations of the trap center with respect to the cavity, i.e., displacement noise \cite{windey2019cavity, delic2019cavity}. 
The first demonstration of quantum mechanical motion has been achieved with
feedback cooling
\cite{tebbenjohanns2019motional} with a residual occupation of $\bar{n}=4$, and recently an occupation below $\bar{n}=1$ has been claimed with cavity cooling \cite{delic2020cooling}.

With feedback cooling the oscillation amplitude is low enough that cross-coupling between the degrees of freedom can be neglected. This justifies the approximation where the particle center-of-mass motion is treated as three independent harmonic oscillators. Each degree of motion obeys the following equation of motion (c.f. equation~\eqref{eq:eq_of_motion}):
\begin{equation}\label{eq:eom_harmonic_oscillator}
	\ddot{q}(t)
	+
	\gamma_0 \dot{q}(t)
	+
	\Omega_0^2 q(t)
	=
	m^{-1}
	\left[
		\mathcal{F}_{\rm L}(t)
		+
		F_{\rm fb}(t)
	\right],
\end{equation}
where $F_{\rm fb}(t)$ is the feedback force (equation~\eqref{eq:eq_of_motion}). For most purposes, the stochastic contribution $\mathcal{F}_{\rm L}$ and the bare damping $\gamma_0$ result from interactions with the background gas. Additional contributions are discussed in section~\ref{sec:5.5:dissipation}.

The feedback damps the particle motion at a rate ($\gamma_{fb}/2\pi$). The additional damping reduces the fluctuations of the position of the particle.
As a consequence, feedback cooling reduces the response time to external perturbations and increases the confinement of the particle.
Since in a thermal equilibrium the variance of the position is proportional to the temperature of the environment, we define the effective temperature for the center-of-mass motion (equation~\eqref{eq:temp_def}) as 
\begin{equation}\label{eq:Teff_feedback}
	T_{\rm cm}  
	=
	{m\Omega_0^2 \over k_\text{B}}
	\left\langle q^2(t)\right\rangle 
	= 
	{\gamma_0 \over \gamma_0+\gamma_{\rm fb}} T,
\end{equation}
which assumes that the particle position and energy distributions reflect the  distributions of a harmonic oscillator in thermal equilibrium at temperature $T_{\rm cm}$. It is important to point out, however, that under the action of feedback the particle is \emph{not} in thermal equilibrium and the assumption of a simple thermal distribution might be violated, in particular for nonlinear feedback schemes 
(section~\ref{sec:5.7.2:parametric_nonlinear_fb}).

\subsubsection{Force feedback}\label{sec:5.7.1:Force_feedback}

From an inspection of equation~\eqref{eq:eom_harmonic_oscillator}, it is apparent that a feedback term proportional to the particle velocity results in damping.
Instead, a feedback term proportional to the particle position allows to 
optimize the transient times \cite{conangla2019optimal}.
Thus, the feedback force of a linear feedback controller reads
\begin{equation}
F_{\rm fb} = -k_{\rm d}\dot{q}-k_{\rm p} q,
\end{equation}
where $k_{\rm d}$ is the derivative feedback gain and $k_{\rm p}$ is the proportional feedback gain. 
Purely velocity-dependent feedback (i.e., $k_{\rm p}=0$) is also called \emph{cold damping}.

Experimentally such a force can be implemented with radiation pressure from a weakly focused beam. This technique was used in Ashkin and Dziedzic's original experiments to stabilize a particle against gravity \cite{ashkin1977feedback}, and later by Li et al. to feedback-cool microspheres in high vacuum \cite{li2011millikelvin}. The latter experiment required three laser beams in addition to the trapping laser to exert radiation pressure forces along the three spatial dimensions. 

More recent experiments use electrostatic forces instead \cite{iwasaki2019electric, conangla2019optimal, tebbenjohanns2019cold}. Electrodes near the optical trap apply a force $\mathbf{F}_\mathcal{Q}= \mathcal{Q}_{\rm p} \mathbf{E}$, where $\mathcal{Q}_{\rm p}$ is the total charge on the particle and $\mathbf{E}$ the electric field at the location of the particle. Often the electrode geometry is chosen such that they form a plate capacitor around the particle so that the field is homogeneous and given by $\mathbf{E}=\left(V_{\rm elec}/d_{\rm elec}\right)\mathbf{n}_{\rm elec}$, where $V_{\rm elec}$ is the applied voltage, $d_{\rm elec}$ the distance between the electrodes and $\mathbf{n}_{\rm elec}$ a unit vector pointing from one electrode to the other. The electrostatic force along direction $\mathbf{e}_i$ is then given by $F_\mathcal{Q}= \mathbf{F}_\mathcal{Q}\cdot \mathbf{e}_i$ and the electrode geometry can be chosen such that there is a force along all degrees of freedom.
Implementation of the velocity-dependent force requires to measure the instantaneous velocity of the particle. In practice, however, this is achieved with a bandpass filter that acts on the position time trace $q(t)$ and produces a time delay such that the signal is in phase with the velocity $\dot{q}(t)$ \cite{tebbenjohanns2019cold}. For example if the particle motion is approximately harmonic $q(t)\sim \sin(\Omega_0 t)$, a delay $\tau = \pi/(2\Omega_0)$ leads to a signal $\propto \cos(\Omega_0 t)$ that is proportional to the particle velocity $\dot{q}(t)$.

In the frequency domain, the equation of motion with linear feedback reads 
\begin{equation}\label{eq:eom_harmonic_oscillator_Fourier}
	\tilde{q}
	\left[
		-\omega^2
		+
		{\rm i}\omega\gamma_0
		+
		\Omega_0^2
	\right]
	=
	m^{-1}
	\left[
		\tilde{\mathcal{F}}_L
		+
		\tilde{f}_{\rm fb}
	\right],
\end{equation}
where the tilde indicates the Fourier transform and $\tilde{f}_{\rm fb}(\omega)=H(\omega)\left[\tilde{q} + \tilde{q}_n \right]$ with filter funcion $H(\omega)=-{\rm i} k_{\rm d}\omega-k_{\rm p}$. 
Explicitly accounting for the noise in the measurement of the position $q_n$ we arrive at
\begin{equation}
	\tilde{q}(\omega) 
	= 
	m^{-1}\chi_{\rm fb}(\omega)\tilde{\mathcal{F}}_L 
	- 
	\chi_{\rm fb}(\omega)
	\left[
		\Omega_{\rm fb}^2 
		+
		{\rm i} \omega\gamma_{\rm fb}
	\right]
	\tilde{q}_n
\end{equation}
where the susceptibility $\chi_{\rm fb}(\omega) = \left[\Omega_0^2+\Omega^2_{\rm fb}-\omega^2+i\omega(\gamma_0+\gamma_{\rm fb})\right]^{-1}$,
and we introduced the feedback damping $\gamma_{\rm fb}= k_{\rm d}/m$ and frequency shift $\Omega^2_{\rm fb}= k_{\rm p}/m$.
The in-loop detector signal is $\tilde{q}_{\rm IL} = \tilde{q}+\tilde{q}_n$.
Hence, the measured in-loop two-sided PSD is \cite{tebbenjohanns2019cold, conangla2019optimal}
 \begin{align}
 	S^{\rm IL}_{qq}(\omega) 
	&= 
	|\chi_{\rm fb}|^2 
	m^{-2}S_{\rm ff}
	+  
	|\chi_{\rm fb}|^2 
	\left[
		(\Omega_0^2-\omega^2)^2
		+
		\omega^2\gamma_0^2
	\right] 
	S_{\rm nn}
	\\
	&= 
	\frac{ 
		m^{-2}S_{\rm ff}
		+
		\left[(
			\Omega_0^2-\omega^2)^2
			+
			\omega^2\gamma_0^2
		\right]
		S_{\rm nn}
	}{
		(\Omega^2_0+\Omega_{\rm fb}^2-\omega^2)^2
		+
		(\gamma_0+\gamma_{\rm fb})^2\omega^2
	},
\end{align}
where $S_{qq}= \langle|\tilde{q}|^2\rangle$ and $S_{\rm nn}$ is the noise PSD of the in-loop detector (Fig.~\ref{fig:47}). 
The in-loop measurement noise is reintroduced into the feedback loop, which drives the particle. The correlations between the particle's position and the measurement noise lead to \emph{noise squashing}, where the apparent signal is squashed below the noise floor \cite{poggio2007feedback} (Fig.~\ref{fig:47}b).

In contrast, an out-of-loop detector provides an independent measurement of the actual particle motion $\tilde{q}_{\rm OoL} = \tilde{q}+\tilde{q}_\nu$:
\begin{align}
	S^\text{OoL}_{qq}(\omega) 
	&= 
	|\chi_{\rm fb}|^2 m^{-2}S_{\rm ff}
	+  
	|\chi_{\rm fb}|^2 
	\left[
		\Omega_{\rm fb}^4
		+
		\omega^2\gamma_{\rm fb}^2
	\right] 
	S_{\rm nn}
	+
	S_{\nu\nu}
	\\
	&= 
	\frac{ 
		m^{-2}S_{\rm ff}
		+
		\left[
			\Omega_{\rm fb}^4
			+
			\omega^2\gamma_{\rm fb}^2
		\right]
		S_{\rm nn}
	}{
		(\Omega^2_0+\Omega_{\rm fb}^2-\omega^2)^2
		+
		(\gamma_0+\gamma_{\rm fb})^2\omega^2
	}
	+
	S_{\nu\nu},
\end{align}
where $S_{\nu\nu}$ is the noise PSD of the out-of-loop detector (Fig.~\ref{fig:47}).

Integration over the PSD yields the position variance\footnote{We make use of
\begin{equation*}
	\frac{1}{2\pi}
	\int_{-\infty}^\infty 
	\left[
		(\omega^2-\omega_0^2)^2
		+
		\gamma^2\omega^2
	\right]^{-1}
	{\rm d}\omega 
	= 
	\left[2\omega_0^2\gamma\right]^{-1}
	\quad\text{and}\quad
  	\frac{1}{2\pi}
	\int_{-\infty}^\infty 
	\omega^2
	\left[
		(\omega^2-\omega_0^2)^2
		+
		\gamma^2\omega^2
	\right]^{-1}
	{\rm d}\omega 
	= 
	\left[2\gamma\right]^{-1}.
\end{equation*}
}
\begin{align}
	\langle q^2\rangle  
	&= 
	\int_{-\infty}^\infty  
	\left[S^\text{OoL}_{qq}(\omega) - S_{\nu\nu}\right] 
	{\rm d}\omega
	\nonumber \\
	&= 
	\frac{ 
		\pi m^{-2}S_{\rm ff} 
		+ 
		\Omega_{\rm fb}^4 S_{\rm nn}
	}{
		(\Omega^2_0+\Omega_{\rm fb}^2) 
		(\gamma_0+\gamma_{\rm fb})
	}
	+
	\frac{
		\pi \gamma_{\rm fb}^2S_{\rm nn}
	}{
		(\gamma_0+\gamma_{\rm fb})
	} 
	\nonumber \\
	&\approx
  	\frac{1}{\gamma_{\rm fb}}
	\frac{\pi S_{\rm ff}}{m^2\Omega^2_0}
	+ 
	\gamma_{\rm fb} 
	\pi S_{\rm nn}.
  	\label{eq:linear_feedback_position_variance}
\end{align}
In the approximation we set $k_{\rm p}=\Omega_{\rm fb}=0$ and $\gamma_{\rm fb}\gg\gamma_0$ and we used $S_{\rm ff} = k_{\rm B} T m \gamma_0 / \pi$.
The first term in equation~\eqref{eq:linear_feedback_position_variance} scales with the inverse of the feedback cooling rate. This term resembles the desired action of the feedback, which is to reduce the impact of the heating term given by the fluctuating force. The second term is proportional to the feedback damping rate, which is multiplied by the measurement noise. This term resembles the undesired but inevitable effect of the control-loop heating the particle by feeding back measurement noise. Consequently, there exists an optimum feedback cooling rate 
$\gamma_{\rm opt} = \sqrt{S_{\rm ff}\left/S_\text{nn}\right.}\left/(m\Omega_0)\right.$, 
where the mode temperature reaches its minimum value 
$ \langle q^2\rangle_{\rm min} = 2\pi \sqrt{S_{\rm ff} S_\text{nn}}\left/(m\Omega_0)\right.$.
With cold damping, optically levitated nanoparticles have been cooled to 4
occupational quanta, corresponding to an effective temperature of $11\,{\rm \upmu K}$ \cite{tebbenjohanns2019motional} (Fig.~\ref{fig:47}(b) and Fig.~\ref{fig:43:sidebands}).

Classically, the measurement is not perturbative and the position variance can be reduced to arbitrarily low values, because, at least in principle, the noise $S_{\rm nn}$ can be made arbitrarily small (equation~\eqref{eq:linear_feedback_position_variance}).
However, quantum mechanics imposes that every measurement is accompanied by a measurement back-action, leading to a finite contribution $S_{\rm nn}$ even in an ideal measurement. In the case of an optical measurement this back action is the random arrival of photons and the resulting random momentum kicks that lead to a stochastic back-action force \cite{jain2016direct}. 
Therefore, in a regime where the measurement back-action is the dominant contribution to the force noise, the impression back-action product is $S_{\rm ff}S_{\rm nn} = (\hbar/(4\pi))^2 / \eta_{\rm det}$, where $\eta_{\rm det}<1$ is the measurement efficiency \cite{clerk2010introduction, tebbenjohanns2019cold}.
The impression back-action product is a manifestation of the Heisenberg uncertainty principle and describes the tradeoff between measurement imprecision and measurement back-action.
At the optimal feedback gain, the effective phonon occupation number $n_{\rm cm} =  m\Omega_0 \langle q^2\rangle_{\rm min} / \hbar-1/2$ depends only on the detection efficiency as $n_{\rm min} = (\eta_{\rm det}^{-1/2}-1)/2$.
At the Heisenberg limit ($\eta_{\rm det}=1$), when the imprecision noise is minimized by optimally detecting all photons scattered by the levitated
particle, the particle's center-of-mass motion could, in principle, be brought to its quantum ground state $n_{\rm min} = 0$.
Hence, to reach the ground state with feedback cooling, the experiment should be operated in the photon-recoil-limited regime ($P_{\rm gas}\leq 10^{-8}\,{\rm mbar}$) and the collection of scattered photons needs to be optimized.
A promising approach is to collect the scattered photons with a high-finesse optical cavity, which collects a fraction of $(\eta_{\rm P}/(\eta_{\rm P}+1))$ of the overall scattered power due to the Purcell effect, where $\eta_{\rm P} = (6\mathcal{F}_c\lambda^2) / (\pi^3 w_0^2)$ is the Purcell factor for a cavity with finesse $\mathcal{F}_{\rm c}$ and beam waist $w_0$ \cite{purcell1995spontaneous}.

\subsubsection{Parametric feedback cooling}\label{sec:5.7.2:parametric_nonlinear_fb}

While cooling levitated particles with linear feedback is very powerful, its experimental implementation along all three motional degrees of freedom is challenging.
All-optical linear feedback requires three auxiliary laser beams to apply radiation pressure in all directions \cite{li2011millikelvin}. Electrical feedback relies on a finite net charge on the levitated object. 

The method of choice to control charge-neutral optically levitated particle is parametric feedback cooling \cite{gieseler2012subkelvin}, which allows to control all three spatial degrees of freedom with the same laser that is used for trapping, thereby significantly reducing the experimental complexity.

The basic concept of parametric feedback is to introduce a modulation in time of the intensity of the trapping laser. In particular, applying a modulation $\delta I(t) = I_0 (\eta_{\rm para}/\Omega_0) q\dot{q}$ results in the parametric feedback force
\begin{equation}\label{eq:force_parametric_feedback}
	F_{\rm fb}^{\rm param} = -k_{\rm para} q^2 \dot{q},
\end{equation}
where $k_{\rm para} = \Omega_0 m \eta_{\rm para}$ is the parametric feedback gain.
When the particle moves away from the trap center, the intensity modulation $\delta I(t)$ stiffens the trap. When the particle falls back to the trap center the feedback softens the trap. Thus, by synchronizing the laser intensity modulation with the particle motion energy is extracted from the center-of-mass motion of the particle \cite{gieseler2014dynamics}. This principle is analogous to the way a child on a swing gains motional energy by modulating its center-of-mass.

Constructing the feedback signal as in equation~\eqref{eq:force_parametric_feedback} provides exactly the right phase relation, such that the trap stiffens when the particle moves away from the trap center and softens when is falls back, thereby cooling the particle. 
However, latencies in the experimental implementation require an additional phase shifter to compensate for unwanted phase shifts. 
With an additional phase shift we can also amplify the particle motion.
Note that, in the absence of active feedback, the particle's oscillation self-locks (entrains its phase) to the modulation in such a way that the motion is amplified \cite{yariv1989quantum}. Cooling therefore requires active stabilization of the modulation phase.

Experimentally, the detector signal is first frequency-doubled and then phase-shifted, where the right phase is determined empirically by optimizing the cooling performance, i.e., minimizing the position variance $\langle q^2\rangle$.
The QPD or split-detection measurement yields a position signal for each spatial direction. The feedback signal is produced for each axis independently and the three signals are summed together, which allows to effectively cool all spatial degrees of freedom with a single laser beam (section~\ref{sec5.3:setup}).
For a particle oscillating at frequency $\Omega_0$ with $q(t) \propto \cos(\Omega_0 t)$, parametric feedback generates a signal $\propto \sin(2\Omega_0)$.
Thus, the feedback signal appears on the in-loop detector signal predominantly at $2\Omega_0$.
The contribution to the detector signal at $\Omega_0$ is much smaller, on the order of $\sim \eta_{\rm para} \langle q^2\rangle$ and, thus, much smaller than the contribution from the particle motion $\langle q^2\rangle$. Hence, noise squashing of the in-loop detector signal does not occur and the trapping laser can also be used for particle detection. 

Due to the nonlinear nature of the parametric feedback, there is no closed expression for the detector PSD. In particular, the position distribution is not a thermal distribution because large amplitude fluctuations are damped more strongly than small fluctuations \cite{gieseler2014dynamic, gieseler2015non}.
However, we can still assign an effective temperature to the center-of-mass motion $k_{\rm B} T_{\rm cm} = m\Omega_0^2\langle q^2\rangle$. For strong feedback ($\eta_{\rm para}\gg m\Omega_0\gamma_0\left/(k_{\rm B} T)\right.$), the position variance is \cite{gieseler2014dynamic, gieseler2015non}
\begin{equation}\label{eq:variance_parametric_feedback}
	\langle q^2 \rangle 
	= 
	\sqrt{
		\frac{
			4k_{\rm B} T \gamma_0
		}{
			\pi \eta_{\rm para}m \Omega_0^3}
		}.
\end{equation}
In contrast to linear-feedback cooling, the effective temperature or position variance scales with the square root of the feedback gain and not linearly. This can be understood intuitively.
The modulation depth and therefore the feedback force is proportional to the energy of the particle. Hence, as the feedback reduces the particle motion it also becomes less effective.
Nonetheless, the quantum mechanical theory suggests that ground state-cooling with parametric feedback cooling is possible \cite{rodenburg2016quantum}, even though direct feedback seem to be more promising \cite{zhong2017quantum}.

\subsubsection{Parametric PLL cooling}

\begin{figure}[bt]
	\begin{center}	
  	\includegraphics[width=12cm]{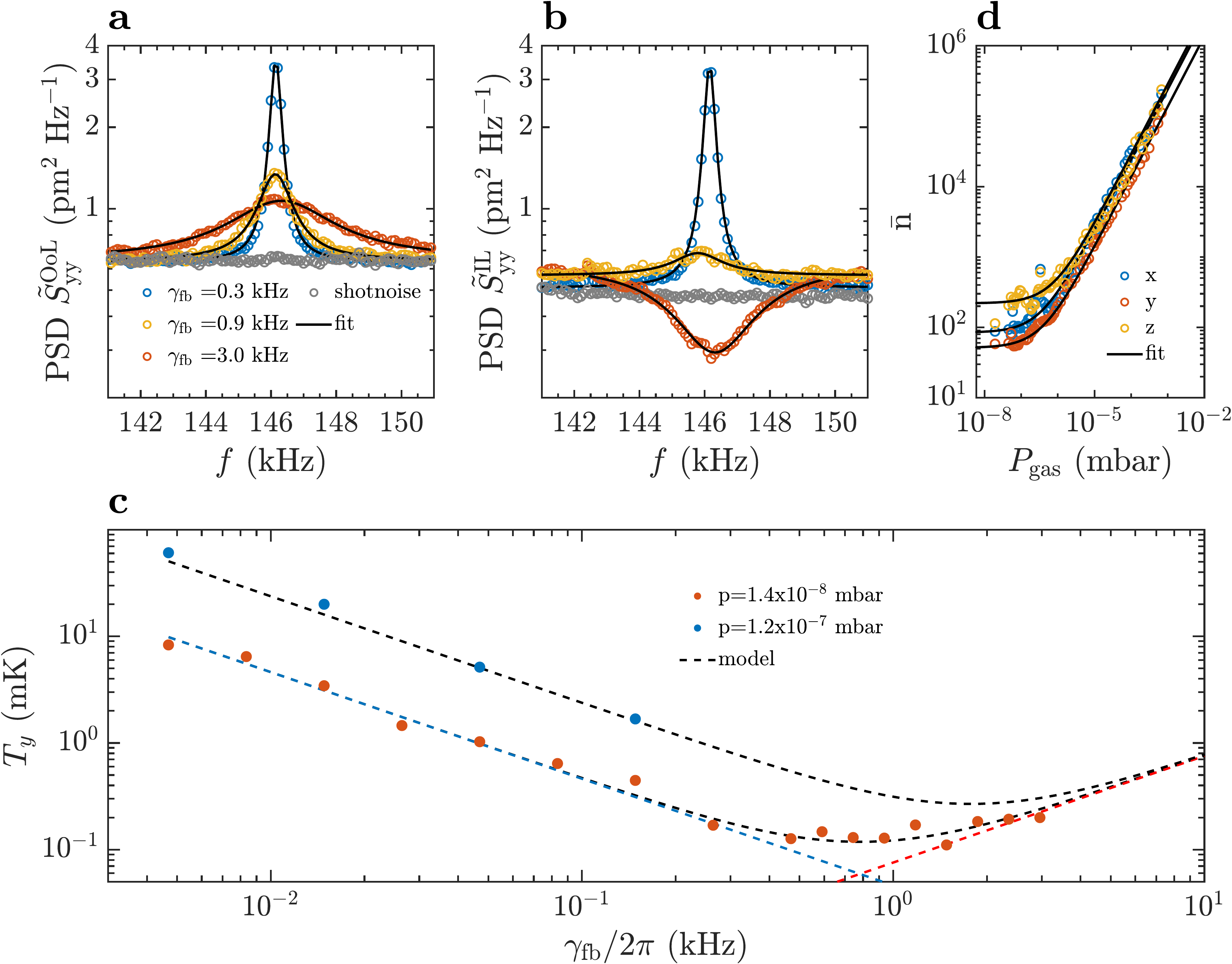}
  	\caption{
	{\bf Feedback cooling.}
	(a) Single-sided power spectral densities $\tilde{S}^{\rm Ool}_\text{yy}$ of the motion of the nanoparticle measured by the out-of-loop detector for different feedback damping rates $\gamma_{\rm fb}$. The solid lines are Lorentzian fits to the data. The gray data points denote the measured shotnoise level on the out-of-loop detector.
	(b) Power spectral densities $\tilde{S}^{\rm IL}_\text{yy}$ measured by the in-loop detector for the same settings as in (a). In contrast to (a), for a large feedback gain the measured signal drops below the noise floor because of correlations between the detector noise and the feedback signal (noise squashing) .
	(c) Mode temperature $T_y$ derived from the out-of-loop signal as in (a) function of feedback gain $\gamma_{\rm fb}$. The red (blue) circles denote the measured values at a pressure of $1.4 \cdot 10^{-8}\,{\rm mbar}$ ($1.2 \cdot 10^{-7}\,{\rm mbar}$.
	(d) Steady state under parametric PLL feedback cooling. Mean occupation number along the three principal axes $(x, y, z)$ as a function of gas pressure measured under constant feedback cooling. At low enough pressures, photon recoil becomes the main source of heating and therefore the occupation number remains constant.
  Adapted from Ref.~\cite{tebbenjohanns2019cold} (a-c) and from Ref.~\cite{jain2016direct} (d).
	}
	\label{fig:47}
	\end{center}
\end{figure}

Parametric feedback extracts energy from or pumps energy into the particle center-of-mass motion by modulating the trapping laser and therefore the trap stiffness at twice the particle oscillation frequency.
A phase locked loop (PLL) synthesizes an output signal whose phase is related to the phase of a sinusoidal input signal. Thereby, a PLL can track an input frequency and it can generate a frequency that is an integer multiple of the input frequency.
Hence, tracking the detector signal with a PLL and feeding the PLL output at the second harmonic into the optical modulator results in a feedback term
\begin{equation}\label{eq:force_PLL_feedback}
	F_{\rm fb}^{\rm PLL} 
	= 
	-k_{\rm PLL} 
	\cos(2\left[\Omega_0t + \phi_{\rm PLL}\right]) q,
\end{equation}
where $k_{\rm PLL} = m\Omega_0^2 \eta_{\rm PLL}$ is the PLL feedback gain and $\eta_{\rm PLL}$  is the modulation depth.
The PLL feedback loop keeps the phase $\phi_{\rm PLL}$ fixed with respect to the randomly changing phase of the particle oscillation.
Similarly to the case of the $q\dot{q}$ parametric feedback, a phase of $\phi_{\rm PLL} = -\pi/4$ reduces the oscillation amplitude.
However, in contrast to the previous case, where the modulation depth is proportional to the variance of the particle position, the PLL feedback has a \emph{constant} amplitude.
As a consequence, the position variance with PLL parametric feedback is 
\begin{equation}\label{eq:variance_PLL_feedback}
	\langle q^2 \rangle 
	= 
	\frac{k_{\rm B} T}{m\Omega_0^2}
	\left[
		1
		-
		\frac{
			\eta_{\rm PLL}
			\Omega_0 
			\sin(2\phi_{\rm PLL})
		}{
			2\gamma_0
		}
	\right]^{-1}
	\approx 
	\frac{
		\gamma_0
	}{
		\eta_{\rm PLL}\Omega_0/2
	} 
	\frac{
		k_{\rm B} T
	}{
		m\Omega_0^2
	}.
\end{equation}
The approximation holds for $\phi_{\rm PLL} = -\pi/4$ and in the limit $\eta_{\rm PLL}\Omega_0\left/2\gamma_0\right.\gg 1$.
Comparing equation~\eqref{eq:variance_PLL_feedback} with equation~\eqref{eq:Teff_feedback}, we can identify the PLL damping as $\gamma_{\rm PLL} = \eta_{\rm PLL}\Omega_0/2$.
Interestingly, even though parametric feedback with a PLL is a nonlinear feedback, we find the same linear scaling as for linear feedback and also the energy distribution follows the Boltzmann distribution associated with a harmonic oscillator at thermal equilibrium with a bath at temperature $T_{\rm cm} = (\gamma_0/\gamma_{\rm PLL})T$ \cite{gieseler2015non}. 
Like in the case of linear feedback, noise in the feedback will ultimately limit how far one can cool. However, unlike in the case of linear feedback (equation~\eqref{eq:linear_feedback_position_variance}), there is no analytical solution for the minimum temperature for parametric feedback.
Using PLL parametric feedback simultaneous cooling of all three motional degrees of freedom to millikelvin temperatures ($n$ below 100) has been demonstrated \cite{jain2016direct} (Fig.~\ref{fig:47}(d)).

\subsection{Outlook}

Levitated particles have already opened the door to a wide range of interesting physics in the classical regime and are now entering into the quantum regime.
The next challenges are to achieve ground state cooling for all three center-of-mass translational degrees of freedom, to control the librational modes and precession of the particle, and to preparte truely non-classical states.

Even though our discussion has been focused on optical tweezers in high vacuum, their combination with cavity optomechanics will play an important role towards creating more exotic quantum mechanical states beyond the ground state.
Once a strong cavity optomechanical interaction has been realized, a range of traditional quantum optics experiments can be realized with these massive particles such as quantum state transfer, quantum squeezing, entanglement, and teleportation \cite{chang2010cavity, romero2011large, aspelmeyer2014cavity}.

The envisioned fundamental tests of quantum mechanics require to control the internal temperature of levitated particles, since high internal temperature will wash out quantum interference effects \cite{wan2016free}.
The internal temperature can be reduced with optical refrigeration that exploit optical dopants inside the particle \cite{kern2017optical, rahman2017laser}. Dopants can also be used to directly measure the particle temperature \cite{delord2017diamonds}.
Experimental data of the internal dynamics of levitated particles could challenge our understanding of thermalization dynamics on the nanoscale \cite{lopez2018internal} and it will be important to measure what is going on \emph{inside} the particle.

There is still plenty of room in the classical regime, where levitated particles will allow to test fundamental questions in single particle thermodynamics in the underdamped regime and provide new insights into the nonlinear dynamics of nanomechanical resonators.
Most attention has been given to the center-of-mass degrees of freedom and simple Gaussian beams. 
There remains a lot to be explored about the remaining degrees of freedom and complex light fields, for example beams carrying orbital angular momentum \cite{arita2017dynamics}.
In addition, new materials, material combinations in the form of doped particles or composite particles and new shapes will enable unprecedented light matter interactions \cite{spesyvtseva2016trapping}.
Optical levitation experiments will hugely benefit from the tools that have been developed for traditional optical tweezers experiments, such as structured light fields \cite{rubinsztein2016roadmap} and holographic tweezers \cite{grier2006holographic, burnham2006holographic}.
Those tools will then also allow to study collective particle dynamics, for example optical binding and forces between levitated particles \cite{dholakia2010colloquium}. This will be required the development of new techniques and calibration methods that are able to resolve the fast particle dynamics in high vacuum at a high spatial resolution over a large field of view.

Careful calibration with a solid understanding of all the statistical and systematic error sources will be paramount for precision measurements with levitated particles. In particular, experiments aiming at detecting near field forces at interfaces  require to approach a surface with a levitated particle \cite{diehl2018optical}. The surface will scatter the trapping light and, thus, these experiments need detection and calibration techniques that are robust in extreme experimental conditions that might even change over the course of a measurement.

\section*{Funding}

JG is supported by the European Commission H2020-MSCA-IF-2014 (SEQOO 655369).
IPC acknowledges financial support from the funding UNAM-DGAPA-PAPIIT-IA103417.
JRGS. acknowledges financial support from DGAPA-UNAM PAPIIT Grant No. IA103320.
AVA acknowledges financial support from the funding UNAM-DGAPA-PAPIIT-IN111919.
KVS acknowledges support from DGAPA-UNAM (grant PAPIIT IN113419 and PASPA grant for sabbatical leave).
AVA and GV acknowledge the Swedish Council for Higher Education through the Linnaeus-Palme International Exchange Program (contract 3.3.1.34.10235-2018).
GV acknowledges funding from the European Research Council (ERC) Starting Grant ERC-677511 ComplexSwimmers.

We thank Erik Hebestreit, Vijay Jain, Martin Frimmer, Felix Tebbenjohanns, Francesco Ricci for providing us with data and feedback and the Quidant and Novotny groups for stimulating discussions.
IPC gratefully thanks \'Edgar Rold\'an for illuminating discussions about Carnot cycles and microscopic engines.


\end{document}